\documentclass[journal]{IEEEtran}
\usepackage{amsmath}
\usepackage{mathtools, cuted}
\usepackage{setspace}
\usepackage{flushend, cuted}
\usepackage{lipsum}
\usepackage{lingmacros}
\usepackage{tree-dvips}
\usepackage{varioref}
\usepackage{colortbl}
\usepackage{url}
\usepackage{hyperref}
\usepackage{balance}
\usepackage{bm}
\usepackage{alphalph}
\usepackage{enumerate}
\usepackage{pgffor}

\usepackage{graphicx}
\usepackage{epstopdf}
\usepackage{color,soul}
\usepackage{amsfonts}
\usepackage{subfigure}
\usepackage{multicol}
\usepackage{multirow}
\usepackage{hhline}
\usepackage{amssymb}
\usepackage{newlfont}
\usepackage{mathrsfs}
\usepackage{mathtools}
\usepackage{mathrsfs}
\usepackage{cite}
\usepackage{multirow}
\usepackage{nomencl}   
\usepackage{fancyhdr}
\usepackage{caption}
\usepackage{slashbox}
\usepackage{float}
\usepackage{placeins}
\usepackage{tabularx}
\usepackage[table]{xcolor}
\usepackage{physics}
\usepackage{makecell}
\usepackage[linesnumbered,ruled]{algorithm2e}
\SetKwRepeat{Do}{do}{while}%

\DeclareRobustCommand{\frac}[3][0pt]{%
	{\begingroup\hspace{#1}#2\hspace{#1}\endgroup\over\hspace{#1}#3\hspace{#1}}}

\let\norm\undefined 
\DeclarePairedDelimiter\norm{\lVert}{\rVert}

\ifCLASSINFOpdf
\else
\fi
\begin{document}

\renewcommand{\theenumi}{\AlphAlph{\value{enumi}}}

\title{Exploiting Mechanics-Based Priors for Lateral Displacement Estimation in Ultrasound Elastography}
%
%
\author{Md~Ashikuzzaman, \textit{Graduate Student Member}, \textit{IEEE},
	    Ali~K.~Z.~Tehrani, \textit{Graduate Student Member}, \textit{IEEE},
        and~Hassan~Rivaz, \textit{Senior Member}, \textit{IEEE}
\thanks{Md Ashikuzzaman, Ali K. Z. Tehrani, and Hassan Rivaz are with the Department
of Electrical and Computer Engineering, Concordia University, Montreal,
QC, H3G 1M8, Canada.
 Email: m\_ashiku@encs.concordia.ca,~A\_Kafaei@encs.concordia.ca,~and~hrivaz@ece.concordia.ca}
}

%

\maketitle
\setlength{\abovedisplayskip}{1pt}
\setlength{\belowdisplayskip}{1pt}
\begin{abstract}
Tracking the displacement between the pre- and post-deformed radio-frequency (RF) frames is a pivotal step of ultrasound elastography, which depicts tissue mechanical properties to identify pathologies. Due to ultrasound's poor ability to capture information pertaining to the lateral direction, the existing displacement estimation techniques fail to generate an accurate lateral displacement or strain map. The attempts made in the literature to mitigate this well-known issue suffer from one of the following limitations: 1) Sampling size is substantially increased, rendering the method computationally and memory expensive. 2) The lateral displacement estimation entirely depends on the axial one, ignoring data fidelity and creating large errors. This paper proposes exploiting the effective Poisson's ratio (EPR)-based mechanical correspondence between the axial and lateral strains along with the RF data fidelity and displacement continuity to improve the lateral displacement and strain estimation accuracies. We call our techniques MechSOUL (Mechanically-constrained Second-Order Ultrasound eLastography) and $L1$-MechSOUL ($L1$-norm-based MechSOUL), which optimize $L2$- and $L1$-norm-based penalty functions, respectively. Extensive validation experiments with simulated, phantom, and \textit{in vivo} datasets demonstrate that MechSOUL and $L1$-MechSOUL's lateral strain and EPR estimation abilities are substantially superior to those of the recently-published elastography techniques. We have published the MATLAB codes of MechSOUL and $L1$-MechSOUL at \url{http://code.sonography.ai}.                                       
\end{abstract}


\IEEEpeerreviewmaketitle
                  
\section{Introduction}
Since its discovery in the 1950s, ultrasound has gradually established itself as one of the most commonly used medical imaging modalities thanks to its non-invasiveness, low expense, and portability. Elastography~\cite{ophir_91,varghese2009quasi} is an emerging clinical application of ultrasound that reveals tissue abnormalities by portraying hidden mechanical properties. Among different ultrasound elastography techniques~\cite{shear_wave_2018,deffieux2011effects,doherty2013acoustic}, the free-hand palpation quasi-static~\cite{hall2003vivo} one has drawn the special attention of researchers over the last three decades, because it is low cost and requires no additional hardware. Consequently, it has been employed in successful assessments of breast~\cite{rglue_ius,rglue}, liver~\cite{guest,tang2015ultrasound}, thyroid~\cite{lyshchik2005thyroid}, prostate~\cite{correas2013ultrasound}, lymph node~\cite{alam2008accuracy}, uterine~\cite{mazza2006mechanical}, blood vessels~\cite{li2019two}, and heart~\cite{konofagou2002myocardial,varghese2003ultrasonic}. Tracking the displacement (also known as time-delay estimation) between two radio-frequency (RF) frames collected before and after tissue deformation is the main step of quasi-static elastography. The estimated displacement field is spatially differentiated to obtain the strain maps, which show a color contrast between the healthy and abnormal tissues.

Several approaches have been followed thus far to solve the critical problem of displacement estimation. A common approach is to split the RF data into a certain number of windows and determine their displacements based on the peak normalized cross-correlation (NCC)~\cite{luo2010fast,al2022murine} or zero-phase crossing~\cite{intro12}. Although the window-based algorithms are straightforward, they are sensitive to noise and make a compromise between the tracking accuracy and the spatial resolution depending on the window size. Recently, machine learning-based techniques~\cite{gao2019learning,wu2018direct,tehrani2021mpwc} have been employed to accomplish this task. This newly-introduced class includes both supervised~\cite{kibria2018gluenet} and unsupervised~\cite{kz2020semi,tehrani2022bi,unsupervised_2022} training-based algorithms. Although the preliminary validation results of machine learning-based methods are promising, they are still in the feasibility stage. This paper focuses on regularized optimization-based or energy-based~\cite{intro14,mglue,glue,rpca_glue,islam2018new} algorithms, another established class of displacement tracking techniques that involve formulating and optimizing an energy function for obtaining the displacement fields. These techniques are mathematically complex but produce accurate and spatially smooth displacement and strain maps. 

While many strides in improving axial displacement estimation have been made, accurate lateral displacement estimation remains an elusive problem. The existing techniques' substandard lateral strain imaging capability originates from the wider point-spread function~\cite{qiong_2017} in this dimension. The lack of an echo carrier~\cite{intro26} and the low sampling rate~\cite{jianwen_2009} are two other mainstream contributors to the loss of lateral estimation accuracy. However, lateral strain carries important diagnostic information. In addition, an accurate lateral displacement estimation is vital for precise reconstructions of Young's modulus as well as poro- and rotation-elastograms~\cite{selladurai2018strategies}. Therefore, several attempts have been made to improve the lateral tracking quality. In~\cite{liu2017systematic}, the number of RF lines is increased by interpolating the acquired data in the lateral direction. RF data has been enhanced at subpitch locations using a conventional linear array transducer in~\cite{selladurai2018strategies}. A multi-angle acquisition scheme has been incorporated in~\cite{techavipoo2004estimation} to improve lateral estimation using beam-steered RF data. The data augmentation- and beam-steering-based techniques either require artificial enhancement of RF data or substantially increase the hardware and software complexities. Multi-step virtual source technique~\cite{mirzaei2020virtual}, which requires channel data acquisition and synthetic aperture beamforming for better lateral estimation, has been proposed. Other notable algorithms~\cite{babaniyi2017recovering,skovoroda1998nonlinear} derive good quality lateral estimates from accurate axial and noisy lateral priors depending on some mechanical correspondence. These techniques disregard RF data while calculating the lateral strain; therefore, the lateral estimate follows the axial one, which might lead to incorrect results. In fact, we show in some of our results that if the Poisson’s ratio (PR) and the elastic modulus vary independently, the lateral and axial strains are no longer correlated. Our proposed technique will exploit the data fidelity term to address this issue.

In this paper, we develop two novel speckle tracking techniques optimizing regularized cost functions that incorporate effective Poisson’s ratio (EPR), which is defined as the negative of the sample-wise ratio of the lateral and axial strains, to leverage the mechanical relation between different strain components. The proposed techniques aim to exploit the newly-introduced mechanical, first- and second-order continuity, and the RF data fidelity constraints simultaneously (see Fig. 1 of the Supplemental Video) to produce highly accurate lateral strain maps without hampering the axial strain quality. Another purpose of the proposed algorithm is to iteratively improve the EPR estimate, which can be used as a contrast mechanism in addition to the strain images. We name our techniques \textbf{MechSOUL}: Mechanically-constrained Second-Order Ultrasound eLastography and $\bm{L1}$\textbf{-MechSOUL}: $L1$-norm-based MechSOUL. The difference between these two proposed algorithms is that MechSOUL penalizes the $L2$-norms of the mechanical inconsistency and the displacement derivatives, whereas $L1$-MechSOUL employs the $L1$-norms. Note that in the case of an inhomogeneous tissue containing an inclusion, EPR is spatially varying (typically between 0.2 and 0.5) and technically different from the PR, which is a material property and spatially constant. Therefore, MechSOUL and $L1$-MechSOUL consider distinct EPR values for each RF sample and iteratively update the strain maps and the EPR distribution. It is worth mentioning that EPR-driven physical constraint has been used in a deep-learning-based tracking technique~\cite{tehrani2022lateral}, unlike which the proposed algorithms incorporate EPR in regularized optimization-based frameworks to improve lateral strain and EPR simultaneously. The performance of the proposed techniques has been validated against \textit{in silico}, phantom, and \textit{in vivo} datasets. Similar to our previous techniques~\cite{RAPID_TMI,soul,soulmate}, MechSOUL and $L1$-MechSOUL codes have been published at \url{http://code.sonography.ai}.                                                               

\begin{figure}
	\centering
	\subfigure[Differences among elastography techniques]{{\includegraphics[width=0.45\textwidth]{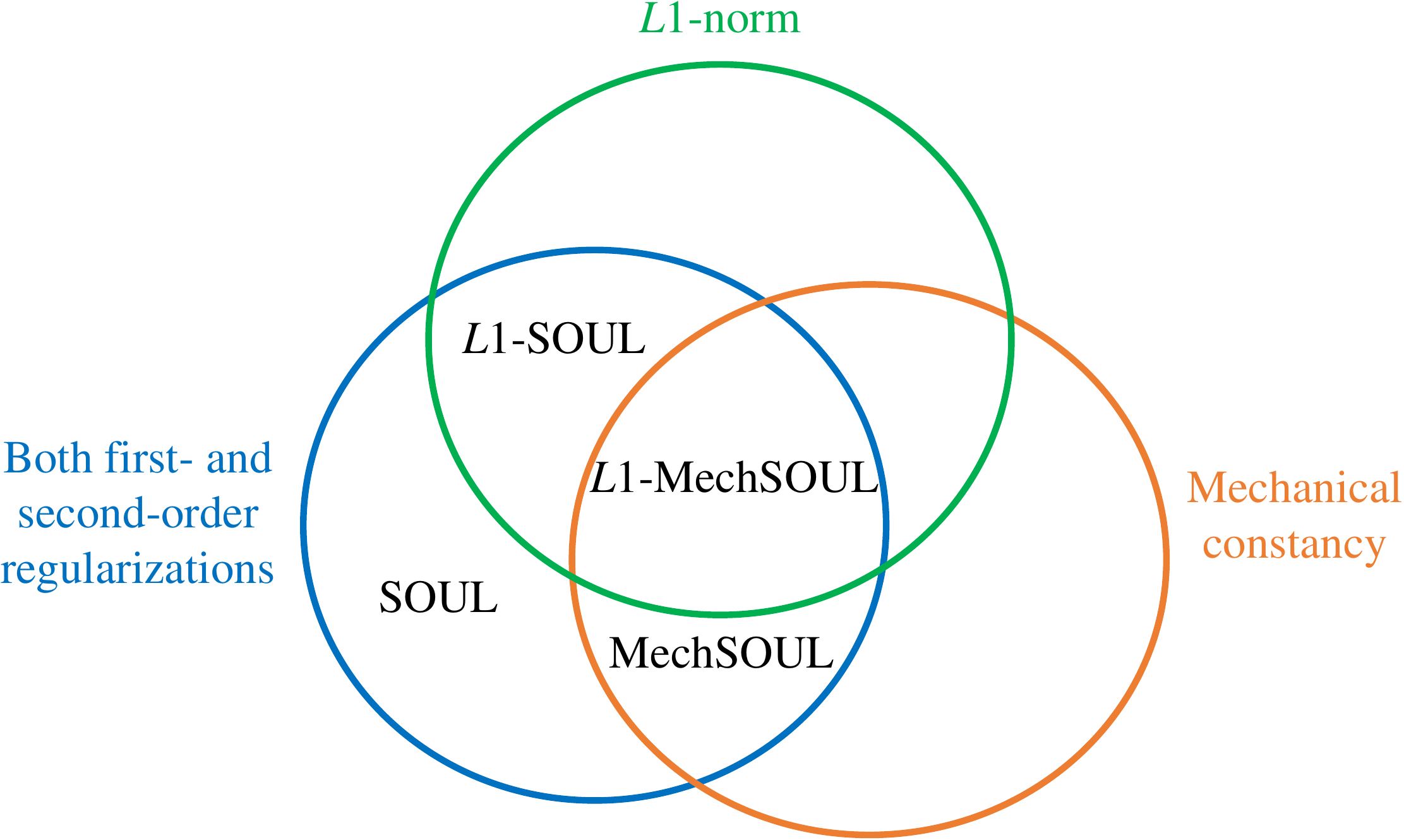}}}
	\subfigure[Lateral strain]{{\includegraphics[width=0.45\textwidth]{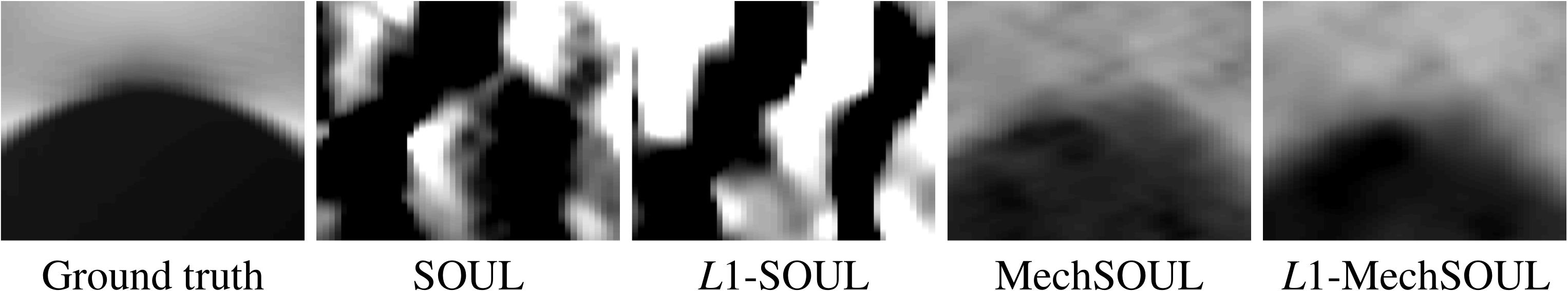}}}%
	\caption{Comparison among different strain imaging algorithms. (a) depicts the methodical differences among elastography techniques. (b) demonstrates the lateral strain imaging performance of four different tracking algorithms.}
	\label{method_illus}
\end{figure}

\section{Methods}
Our goal is to estimate the displacement field between two RF frames $I_{1}(i,j)$ and $I_{2}(i,j)$, $1 \leq i \leq m$, $1 \leq j \leq n$, collected before and after tissue deformation and spatially differentiate its components to obtain the axial and lateral strain fields. Dynamic Programming (DP)~\cite{DP} provides $a \in \mathbb{R}^{m \times n}$ and $l \in \mathbb{R}^{m \times n}$, the initial guesses for the axial and lateral displacement fields. The vital step of estimating $\Delta a \in \mathbb{R}^{m \times n}$ and $\Delta l \in \mathbb{R}^{m \times n}$, the refinement displacement fields, is performed by a continuous optimization technique. This section first describes SOUL~\cite{soul} and $L1$-SOUL~\cite{soulmate}, two such recently-published techniques, and then MechSOUL and $L1$-MechSOUL, the proposed algorithms.

\subsection{Second-Order Ultrasound eLastography (SOUL)}
SOUL optimizes $C_{l2}$, a non-linear cost function comprised of $L2$-norm data constancy as well as $L2$-norm first- and second-order continuity terms.    

\begin{equation}
\begin{aligned}
&C_{l2} (\Delta a_{1,1},...,\Delta a_{m,n},\Delta l_{1,1},...,\Delta l_{m,n}) =\\ 
&\norm{D_{I}(i,j,a_{i,j},l_{i,j},\Delta a_{i,j},\Delta l_{i,j})}_{2}^{2} + \gamma \norm{\partial_{y}a_{f}}_{2}^{2}+\\
&\alpha_{1} \norm{\partial_{y}a -\mathbf{\epsilon_{a}}}_{2}^{2} + \alpha_{2} \norm{\partial_{x}a -\mathbf{\epsilon_{a}}}_{2}^{2} + \beta_{1} \norm{\partial_{y}l -\mathbf{\epsilon_{l}}}_{2}^{2}+\\ 
&\beta_{2} \norm{\partial_{x}l -\mathbf{\epsilon_{l}}}_{2}^{2}+w\alpha_{1} \norm{\partial_{y}^{2}a}_{2}^{2} + w\alpha_{2} \norm{\partial_{x}^{2}a}_{2}^{2} +
w\beta_{1} \norm{\partial_{y}^{2}l}_{2}^{2} +\\ 
&w\beta_{2} \norm{\partial_{x}^{2}l}_{2}^{2} 
\end{aligned}
\label{eq:c_soul}
\end{equation}

\noindent
where $D_{I}$ denotes the data constancy term:

\begin{equation}
\begin{aligned}
&D_{I}(i,j,a_{i,j},l_{i,j},\Delta a_{i,j},\Delta l_{i,j})=\\
&[I_{1}(i,j)-I_{2}(i+a_{i,j}+\Delta a_{i,j},j+l_{i,j}+\Delta l_{i,j})]^{2}
\end{aligned}
\label{eq:data}
\end{equation}

The non-linearity present in the data function is removed by approximating $I_{2}$ by its first-order Taylor series expansion:

\begin{equation}
\begin{aligned}
&I_{2}(i+a_{i,j}+\Delta a_{i,j},j+l_{i,j}+\Delta l_{i,j}) \approx \\
&I_{2}(i+a_{i,j},j+l_{i,j})+\Delta a_{i,j}I_{2,a}^{'}+\Delta l_{i,j}I_{2,l}^{'} 
\end{aligned}
\label{eq:i2_taylor}
\end{equation}

$\gamma$, $\alpha_{1}$, $\alpha_{2}$, $\beta_{1}$, $\beta_{2}$, and $w$ are tunable parameters. $\mathbf{\epsilon_{a}}$ and $\mathbf{\epsilon_{l}}$ contain the axial and lateral bias parameters that prevent displacement underestimation~\cite{guest,soul,soulmate}. $\partial_{y}a_{f}$ stands for the axial derivatives of the RF lines' first samples. Considering that the imaginary sample prior to an RF line's first sample is zero, $(\partial_{y}a_{f})_{1,j}$ is defined as:

\begin{equation}
\begin{aligned}
(\partial_{y}a_{f})_{1,j} = a_{1,j}+\Delta a_{1,j}
\end{aligned}
\label{eq:dela_af}
\end{equation}

$(\partial_{y}a)_{i,j}$, $(\partial_{x}a)_{i,j}$, $(\partial_{y}l)_{i,j}$, and $(\partial_{x}l)_{i,j}$ denote the first-order axial and lateral displacement derivatives, whereas $(\partial_{y}^{2}a)_{i,j}$, $(\partial_{x}^{2}a)_{i,j}$, $(\partial_{y}^{2}l)_{i,j}$, and $(\partial_{x}^{2}l)_{i,j}$ refer to the second-order displacement derivatives.

\subsection{SOUL using $L1$-norm Regularization ($L1$-SOUL)}
Unlike SOUL, $L1$-SOUL minimizes a penalty function $C_{l1}$ consisting of $L2$-norm data and $L1$-norm continuity terms:
  
\begin{equation}
\begin{aligned}
&C_{l1} (\Delta a_{1,1},...,\Delta a_{m,n},\Delta l_{1,1},...,\Delta l_{m,n}) =\\ 
&\norm{D_{I}(i,j,a_{i,j},l_{i,j},\Delta a_{i,j},\Delta l_{i,j})}_{2}^{2} + \gamma_{s} \norm{\partial_{y}a_{f}}_{1}+\\
&w_{f}\alpha_{1s} \norm{\partial_{y}a -\mathbf{\epsilon_{a}}}_{1} + w_{f}\alpha_{2s} \norm{\partial_{x}a -\mathbf{\epsilon_{a}}}_{1} + w_{f}\beta_{1s} \norm{\partial_{y}l -\mathbf{\epsilon_{l}}}_{1}+\\ 
&w_{f}\beta_{2s} \norm{\partial_{x}l -\mathbf{\epsilon_{l}}}_{1}+w_{s}\alpha_{1s} \norm{\partial_{y}^{2}a}_{1} + w_{s}\alpha_{2s} \norm{\partial_{x}^{2}a}_{1} +\\
&w_{s}\beta_{1s} \norm{\partial_{y}^{2}l}_{1} + w_{s}\beta_{2s} \norm{\partial_{x}^{2}l}_{1} 
\end{aligned}
\label{eq:c_soulmate}
\end{equation}

\noindent
where $\gamma_{s}$, $\alpha_{1s}$, $\alpha_{2s}$, $\beta_{1s}$, $\beta_{2s}$, $w_{f}$, and $w_{s}$ are tunable parameters. To facilitate analytic optimization, $L1$-SOUL replaces the $L1$-norm with the total variation distance (TVD) approximating the absolute value function with its smooth version. Therefore, $L1$-norm is defined as: 

\begin{equation}
\begin{aligned} 
\norm{\cdot}_{1} = \sum\limits_{j=1}^{n} \sum\limits_{i=1}^{m} \sqrt{(\cdot)_{i,j}^2 + \eta^{2}}
\end{aligned}
\label{eq:l1norm_def}
\end{equation}

\noindent
where $\eta$ is a sharpness controlling parameter. As detailed in \cite{soulmate}, $L1$-SOUL iteratively optimizes Eq.~\ref{eq:c_soulmate} to obtain a sharp displacement map. 

\subsection{Mechanically-constrained SOUL (MechSOUL)}
SOUL and $L1$-SOUL are not suitable for generating high-quality lateral strain maps. MechSOUL resolves this limitation by adding a mechanically-inspired constraint to SOUL's cost function. This newly-added constraint takes the EPR into account to impose the mechanical relation between the axial and lateral components ($s_{yy}=\partial_{y}a$ and $s_{xx}=\partial_{x}l$) of the strain tensor. Note that optimizing a regularized cost function penalizing $s_{xx} + \nu s_{yy}$ is different from estimating $s_{yy}$ first and then multiplying it by $-\nu$ to find $s_{xx}$, where $\nu$ is the EPR. Because in our work, $s_{xx} + \nu s_{yy}$ is just a soft constraint in a cost function that contains data fidelity and spatial continuity terms as well. Therefore, the estimated lateral strain has the freedom to deviate from the axial strain's multiple depending on the RF data under investigation. A comprehensive analysis of this feature is presented in the Discussion Section. In addition, since EPR is expected to be spatially varying in real tissue, MechSOUL (and $L1$-MechSOUL) establishes an iterative scheme to employ a distinct EPR for each RF sample. The MechSOUL cost function is given by:           

\begin{equation}
\begin{aligned}
&C_{l2m} (\Delta a_{1,1},...,\Delta a_{m,n},\Delta l_{1,1},...,\Delta l_{m,n}) =\\ 
&C_{l2} (\Delta a_{1,1},...,\Delta a_{m,n},\Delta l_{1,1},...,\Delta l_{m,n}) +\\
&\sum\limits_{j=1}^n \sum\limits_{i=1}^m \alpha_{3}[(\partial_{x}l)_{i,j} + \nu_{i,j}(\partial_{y}a)_{i,j}]^{2}
\end{aligned}
\label{eq:c_mechsoul}
\end{equation}

\noindent
where $\alpha_{3}$ is the mechanical constancy weight, whereas $\nu_{i,j}$ stands for the EPR for sample $(i,j)$. We minimize $C_{l2m}$ by setting $\frac{\partial C_{l2m,i,j}}{\partial \Delta a_{i,j}}=0$ and $\frac{\partial C_{l2m,i,j}}{\partial \Delta l_{i,j}}=0$ and obtain:

\begin{equation}
(H+D_{l2}+D_{2l2}+M_{l2})\Delta d_{l2} = H_{1}\mu - (D_{l2}+D_{2l2}+M_{l2})d + b_{s2}
\label{eq:mechsoul_axb}
\end{equation}

\noindent
where $d \in \mathbb{R}^{2mn \times 1}$ and $\Delta d_{l2} \in \mathbb{R}^{2mn \times 1}$, respectively, stack the initial and the fine-tuning displacements. $D_{l2}$ and $D_{2l2}$, respectively, are sparse matrices of size $2mn \times 2mn$ containing functions of first- and second-order regularization parameters. $H \in \mathbb{R}^{2mn \times 2mn}$ and $H_{1} \in \mathbb{R}^{2mn \times 2mn}$, respectively, are symmetric tridiagonal and diagonal matrices comprising of data derivatives and their functions. $\mu \in \mathbb{R}^{2mn \times 1}$ contains the data residuals. $M_{l2} \in \mathbb{R}^{2mn \times 2mn}$ contains the functions of the EPR and the mechanical constancy weight. $b_{s2} \in \mathbb{R}^{2mn \times 1}$ denotes the adaptive regularization vector.

\subsection{Mechanically-constrained $L1$-SOUL ($L1$-MechSOUL)}

$L1$-MechSOUL is developed as the $L1$ version of MechSOUL. $L1$-MechSOUL modifies the cost function of $L1$-SOUL by adding the $L1$-norm of the aforementioned mechanical constraint. As described in Eq.~\ref{eq:l1norm_def}, the $L1$-norm is defined in terms of a differentiable approximation of the absolute value function. The $L1$-MechSOUL penalty function is given by:
   
\begin{equation}
\begin{aligned}
&C_{l1m} (\Delta a_{1,1},...,\Delta a_{m,n},\Delta l_{1,1},...,\Delta l_{m,n}) =\\ 
&C_{l1} (\Delta a_{1,1},...,\Delta a_{m,n},\Delta l_{1,1},...,\Delta l_{m,n}) +\\
&\sum\limits_{j=1}^n \sum\limits_{i=1}^m \alpha_{3s}\sqrt{[(\partial_{x}l)_{i,j} + \nu_{i,j}(\partial_{y}a)_{i,j}]^{2} + \eta_{m}^{2}}
\end{aligned}
\label{eq:c_l1mechsoul}
\end{equation}

\noindent
where $\alpha_{3s}$ and $\eta_{m}$ are mechanical and sharpness parameters, respectively. Optimizing $C_{l1m}$ in the same fashion as \cite{soulmate} leads to: 

\begin{equation}
(H+D_{l1}+D_{2l1}+M_{l1})\Delta d_{l1} = H_{1}\mu - (D_{l1}+D_{2l1}+M_{l1})d + b_{s1}
\label{eq:l1mechsoul_axb}
\end{equation}

\noindent
where $\Delta d_{l1} \in \mathbb{R}^{2mn \times 1}$ stacks the refinement displacements. $D_{l1}$ and $D_{2l1}$, respectively, are sparse matrices of size $2mn \times 2mn$ containing functions of first- and second-order continuity weights. $M_{l1} \in \mathbb{R}^{2mn \times 2mn}$ consists of the functions of the EPR and the mechanical parameter. $b_{s1} \in \mathbb{R}^{2mn \times 1}$ denotes the adaptive regularization vector.

Both MechSOUL and $L1$-MechSOUL initialize the EPRs with the organ- or material-specific nominal value of the PR (e.g., 0.3 for liver). The subsequent iterations update each sample's EPR using $\nu_{i,j}=-(s_{xx,i,j})/(s_{yy,i,j})$, where $s_{xx}$ and $s_{yy}$ are lateral and axial strains calculated in the previous iteration.

The estimated fine-tuning displacement fields are added to the DP initial guesses to obtain the final displacements, which are spatially differentiated using a least-square technique to estimate the axial and lateral strain fields. Fig.~\ref{method_illus} illustrates methodical differences among SOUL, $L1$-SOUL, MechSOUL, and $L1$-MechSOUL.

\begin{figure*}[h]
	\centering
	\subfigure[Ground truth]{{\includegraphics[width=0.13\textwidth]{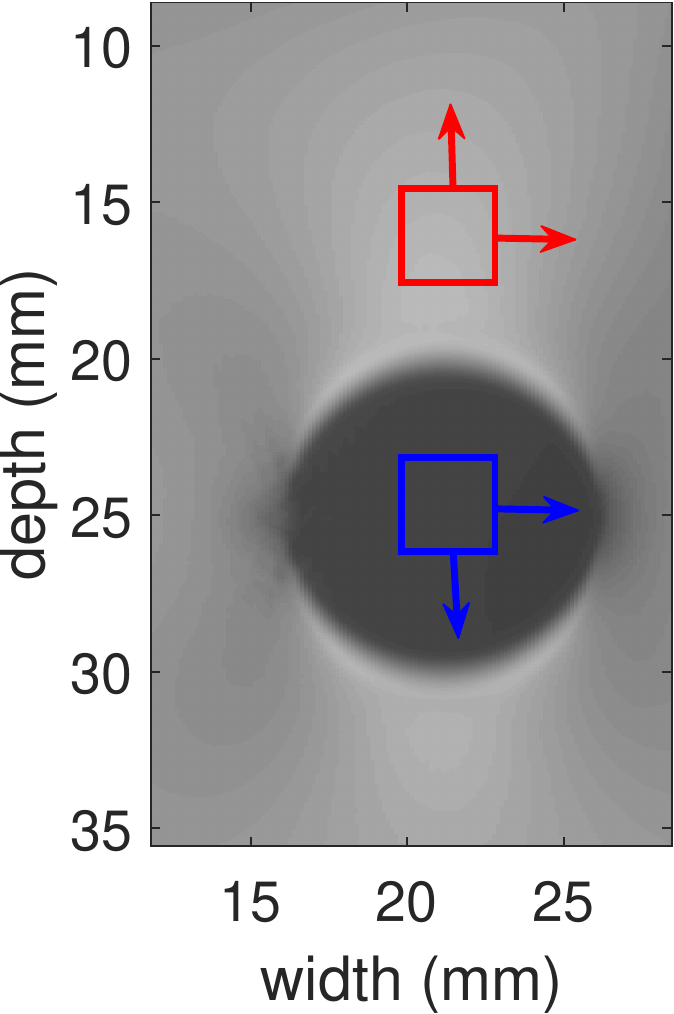}}}%
	\subfigure[NCC]{{\includegraphics[width=0.13\textwidth]{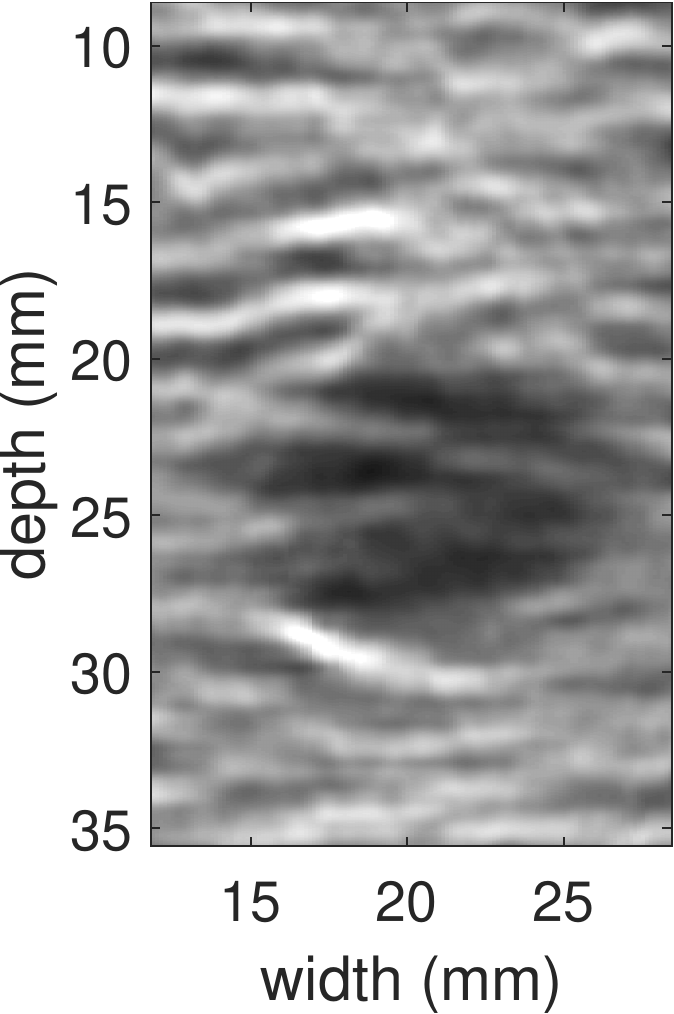}}}%
	\subfigure[NCC + PDE]{{\includegraphics[width=0.13\textwidth]{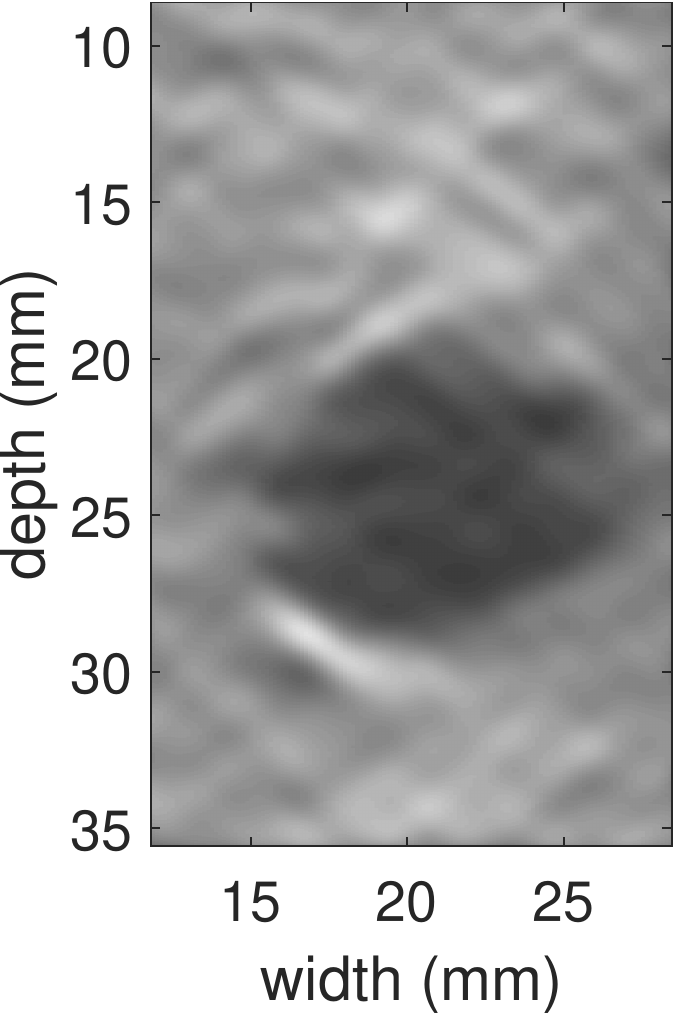}}}%
	\subfigure[SOUL]{{\includegraphics[width=0.13\textwidth]{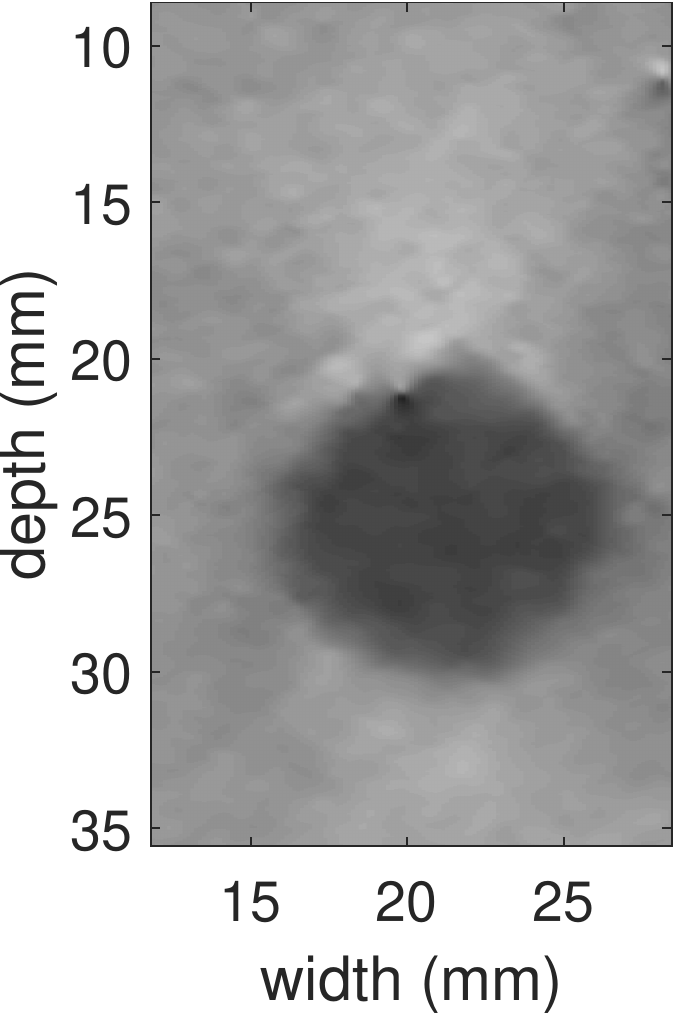}}}%
	\subfigure[$L1$-SOUL]{{\includegraphics[width=0.13\textwidth]{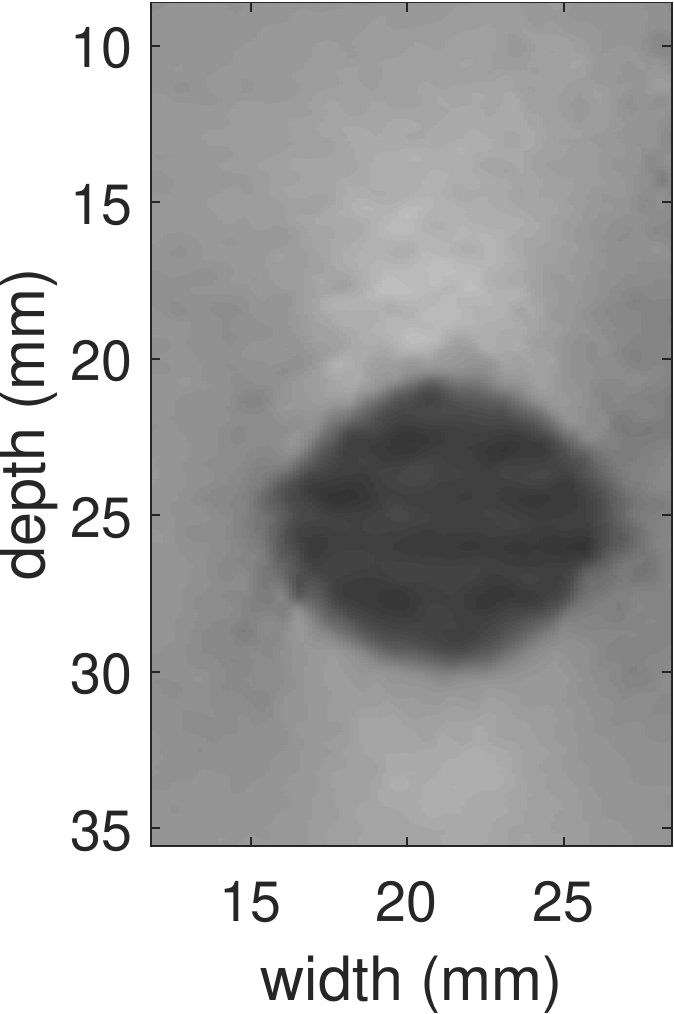}}}%
	\subfigure[MechSOUL]{{\includegraphics[width=0.13\textwidth]{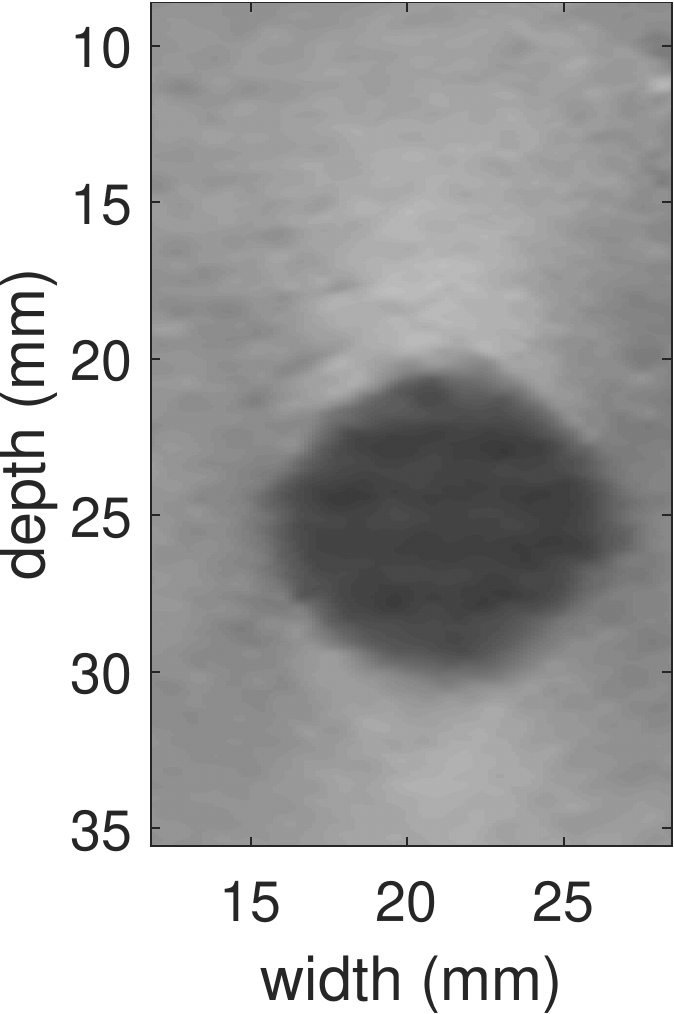}}}%
	\subfigure[$L1$-MechSOUL]{{\includegraphics[width=0.13\textwidth]{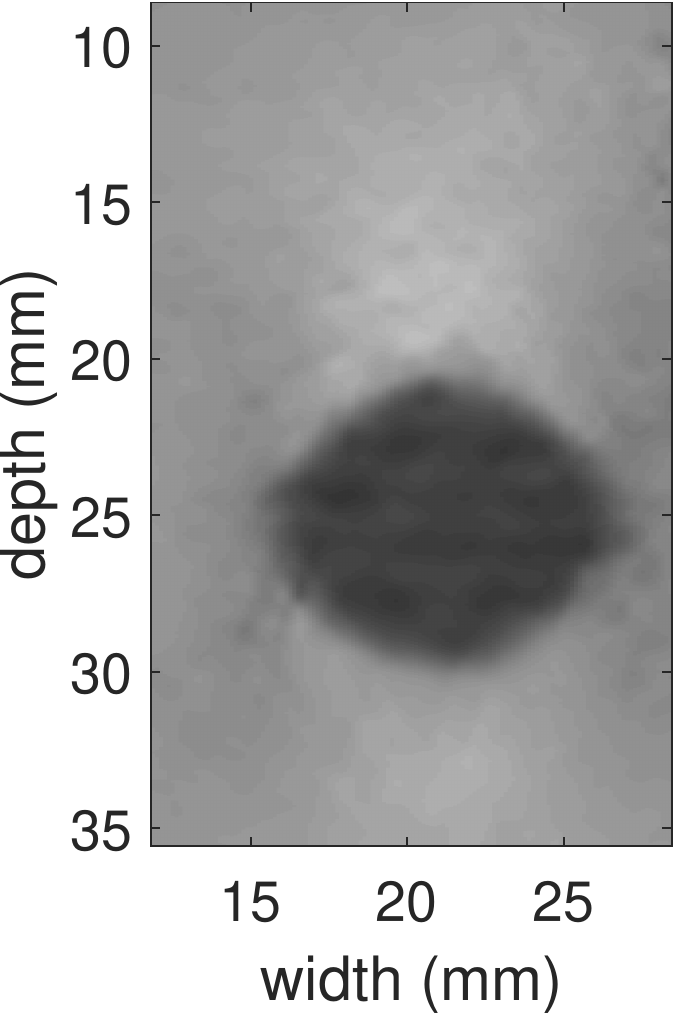}}}
	\subfigure[Ground truth]{{\includegraphics[width=0.13\textwidth]{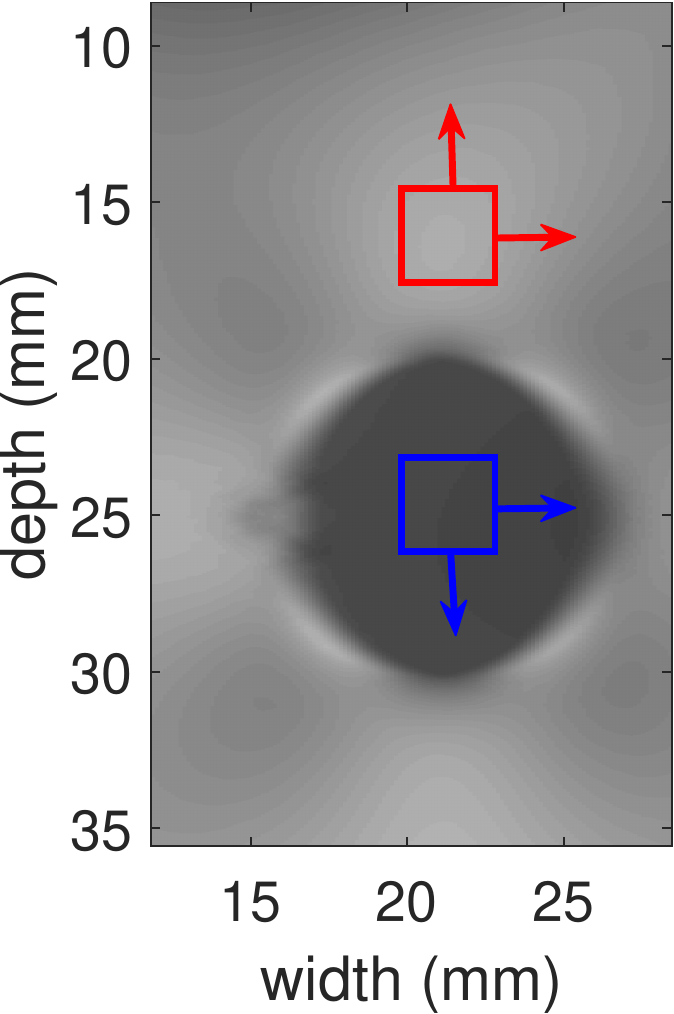}}}%
	\subfigure[NCC]{{\includegraphics[width=0.13\textwidth]{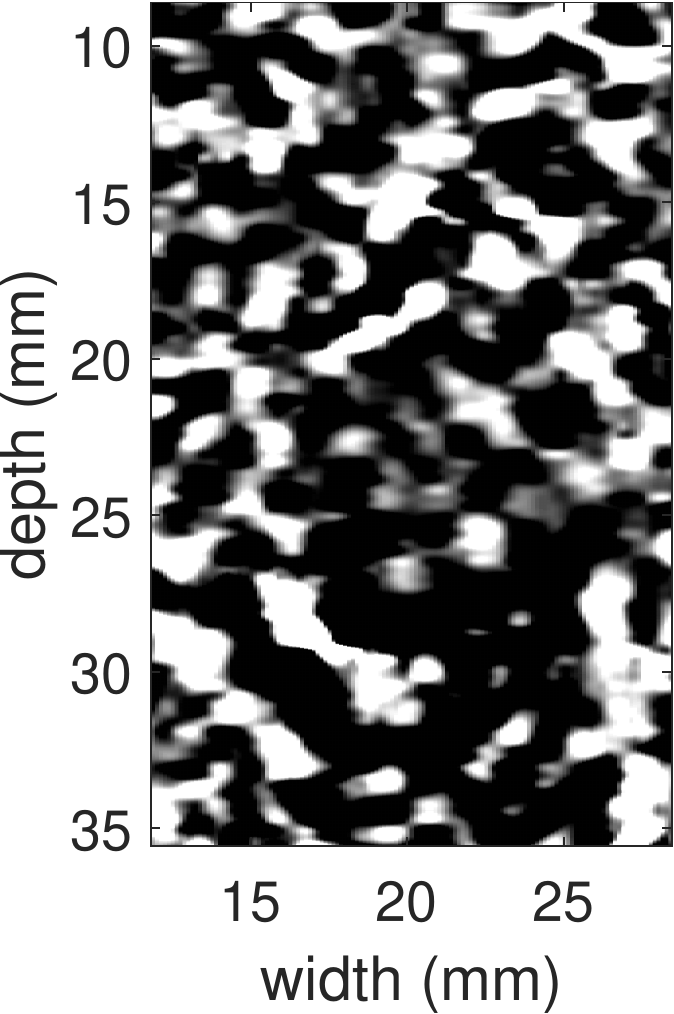}}}%
	\subfigure[NCC + PDE]{{\includegraphics[width=0.13\textwidth]{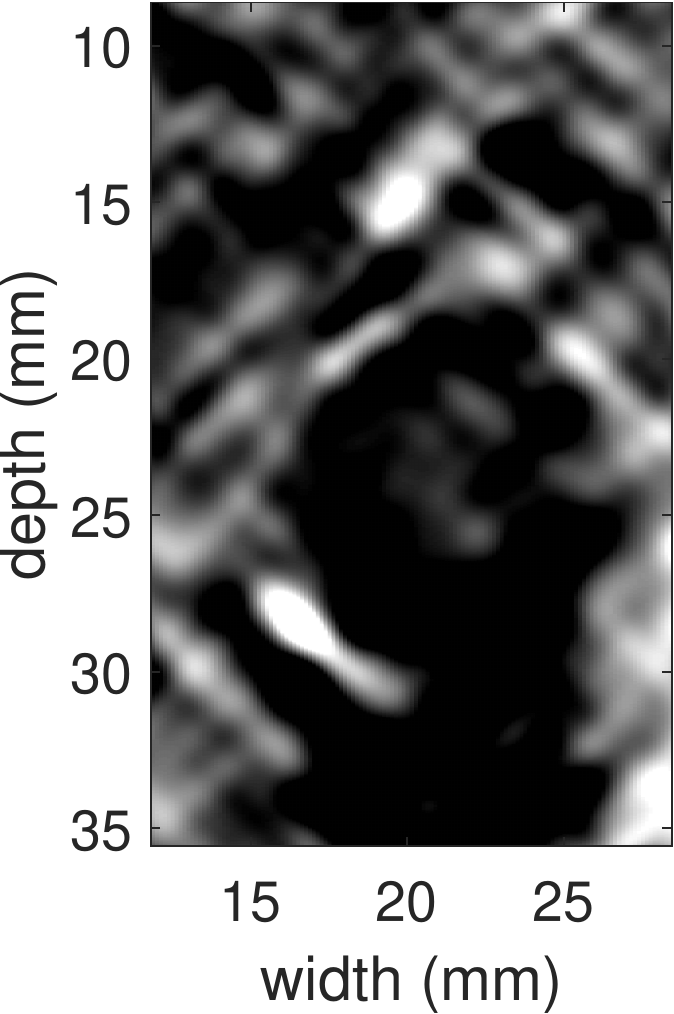}}}%
	\subfigure[SOUL]{{\includegraphics[width=0.13\textwidth]{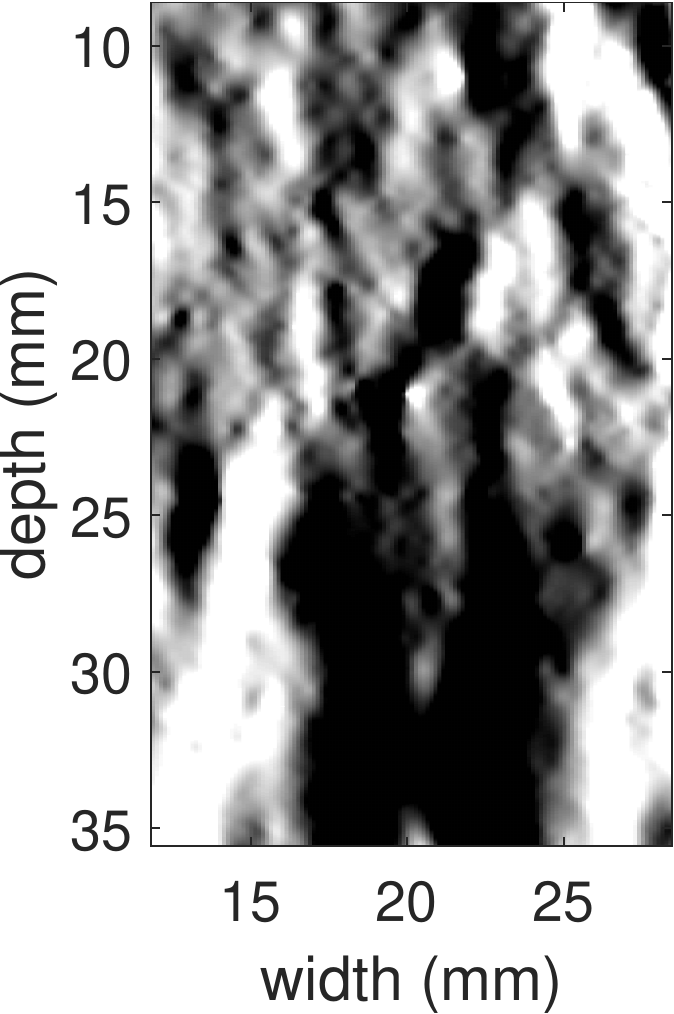}}}%
	\subfigure[$L1$-SOUL]{{\includegraphics[width=0.13\textwidth]{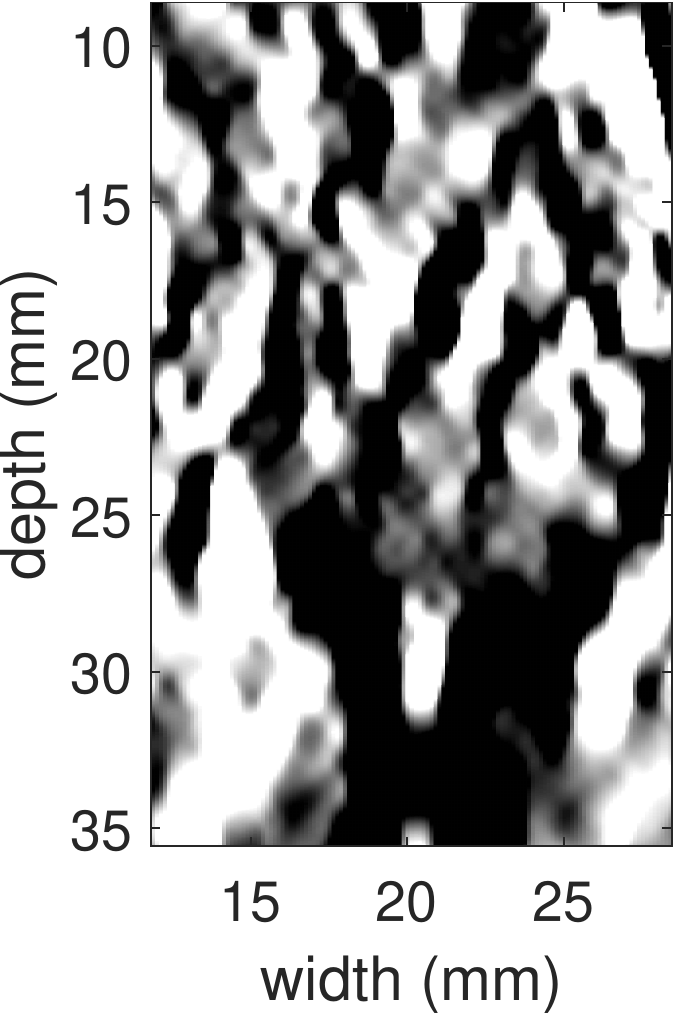}}}%
	\subfigure[MechSOUL]{{\includegraphics[width=0.13\textwidth]{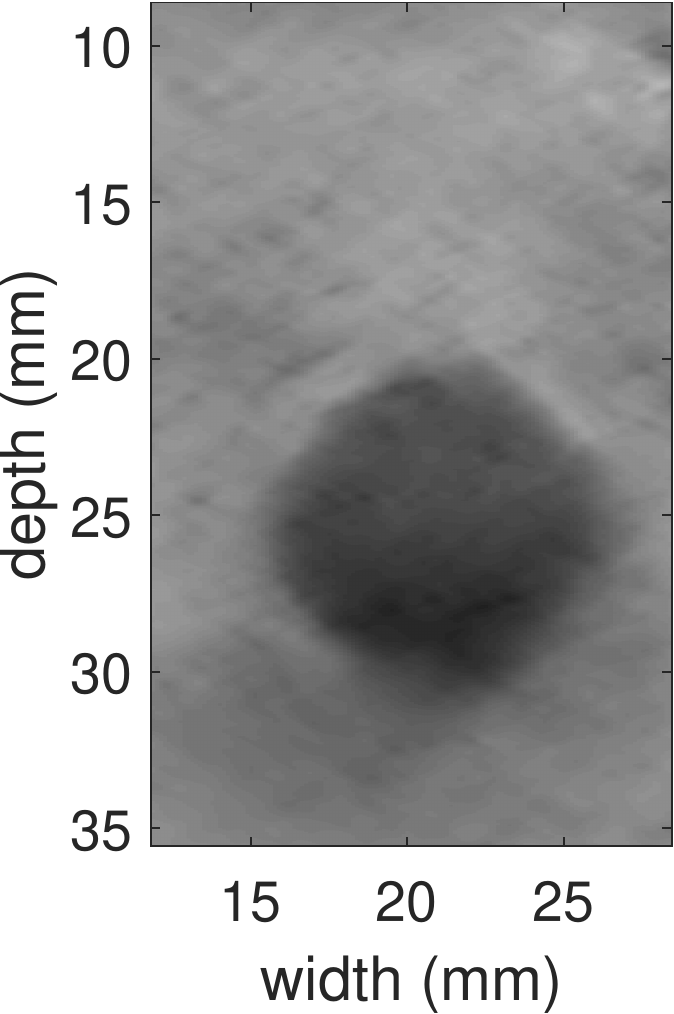}}}%
	\subfigure[$L1$-MechSOUL]{{\includegraphics[width=0.13\textwidth]{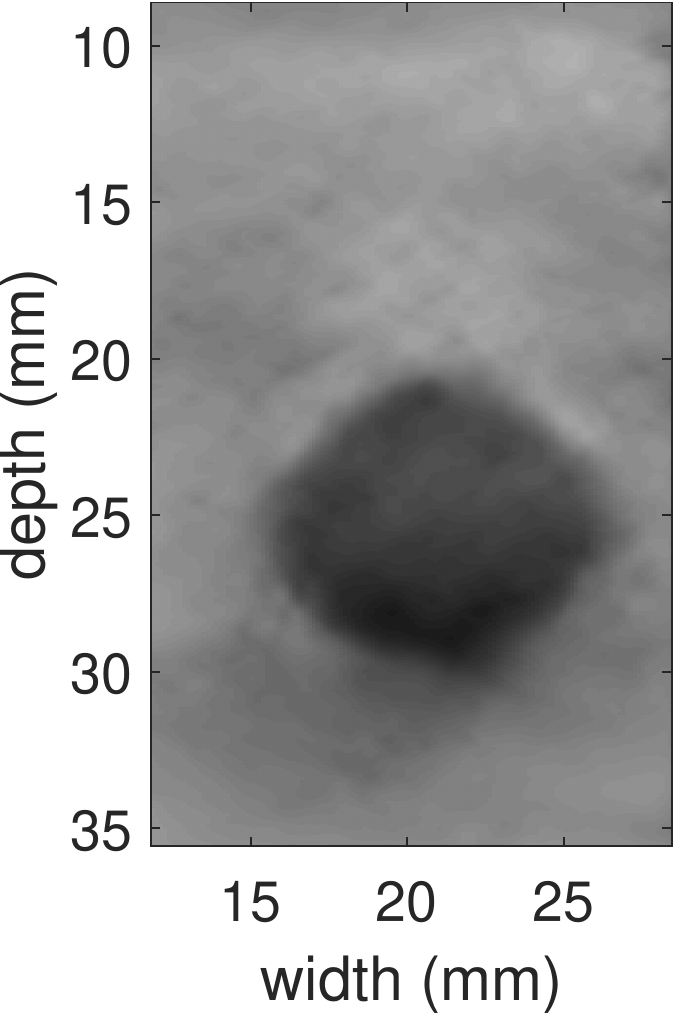}}}
	\subfigure[Ground truth]{{\includegraphics[width=0.13\textwidth]{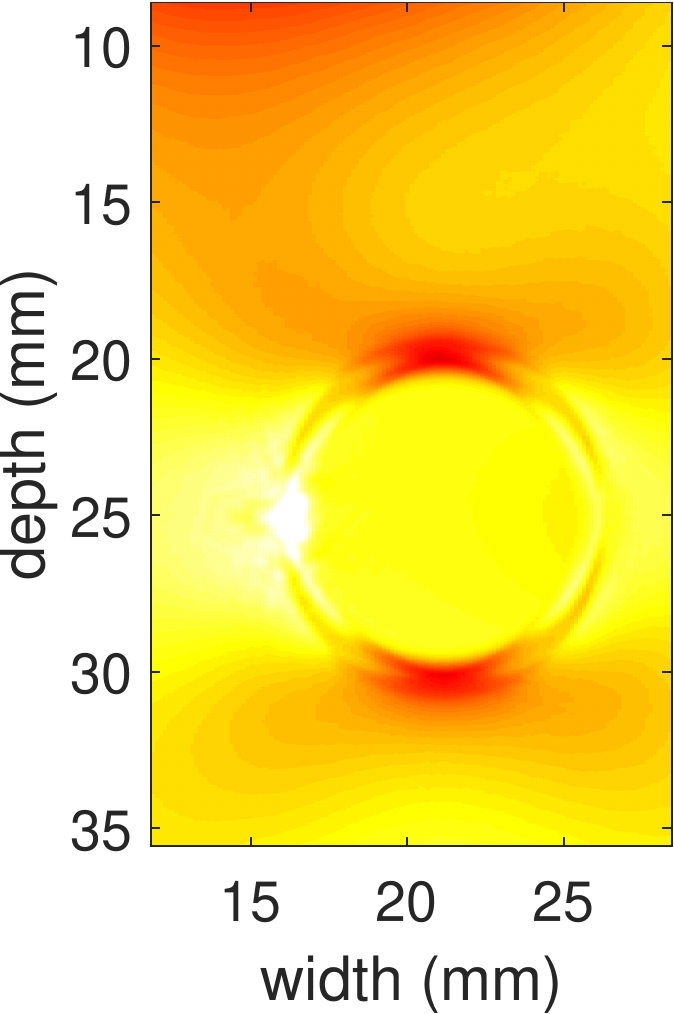}}}%
	\subfigure[NCC]{{\includegraphics[width=0.13\textwidth]{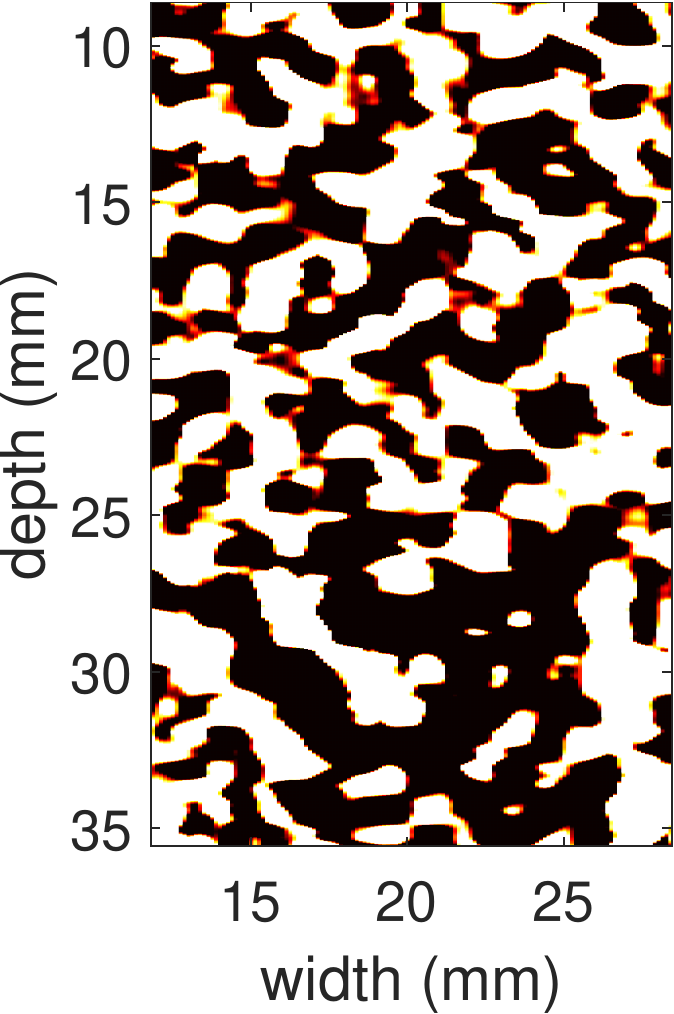}}}%
	\subfigure[NCC + PDE]{{\includegraphics[width=0.13\textwidth]{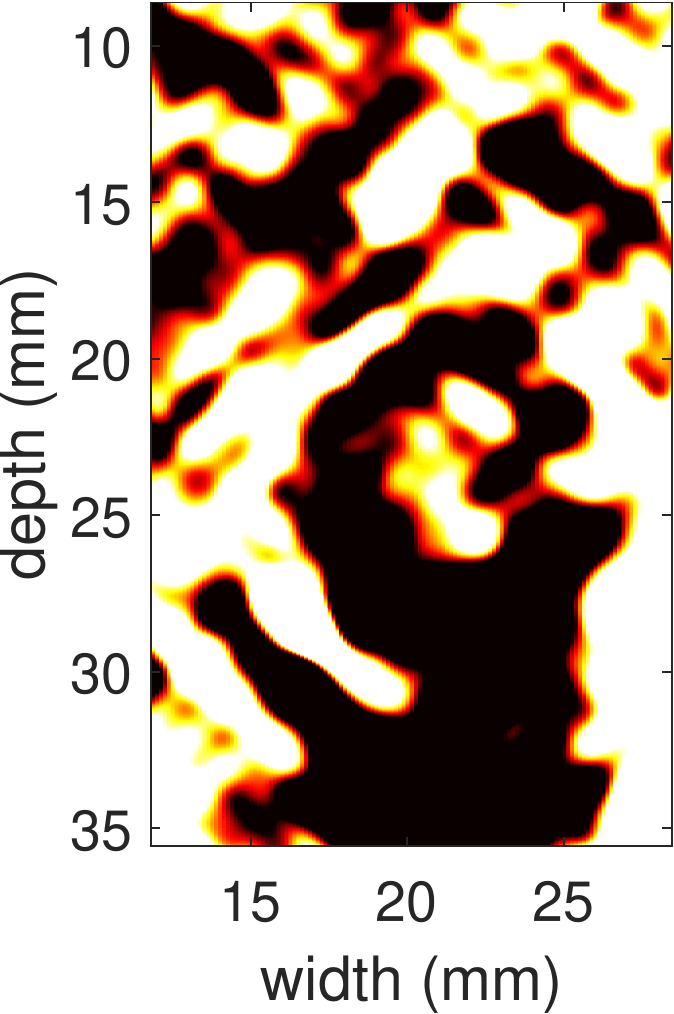}}}%
	\subfigure[SOUL]{{\includegraphics[width=0.13\textwidth]{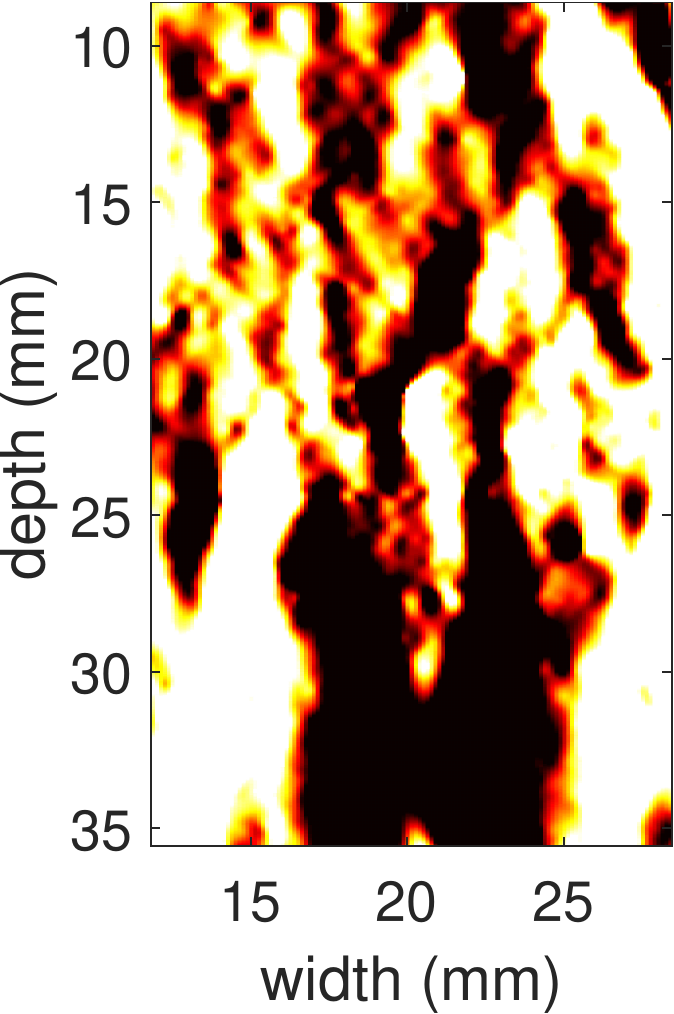}}}%
	\subfigure[$L1$-SOUL]{{\includegraphics[width=0.13\textwidth]{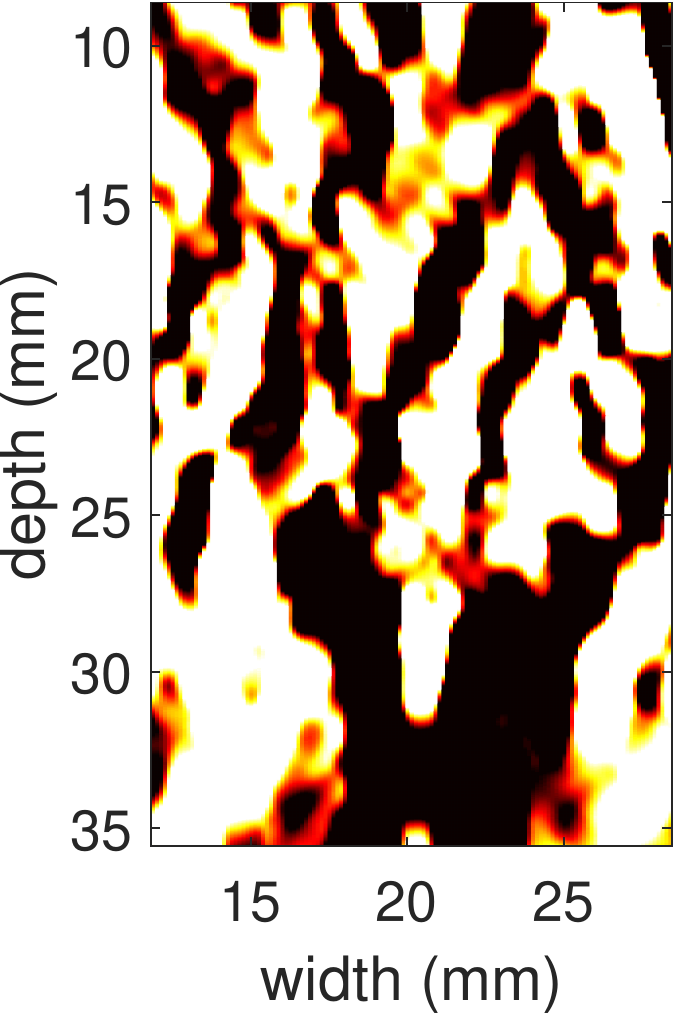}}}%
	\subfigure[MechSOUL]{{\includegraphics[width=0.13\textwidth]{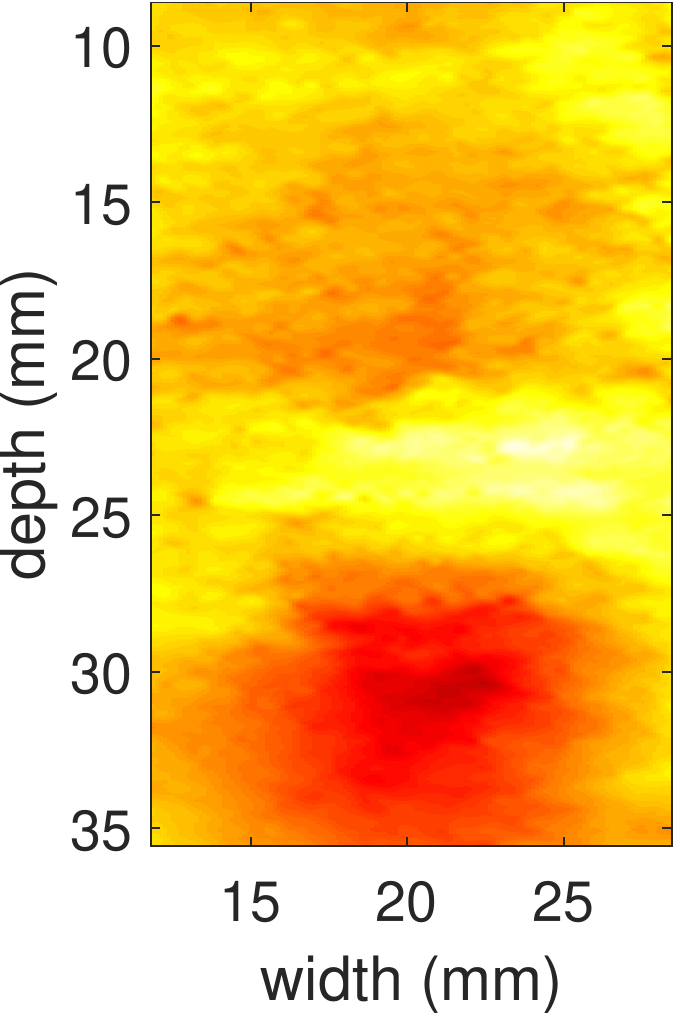}}}%
	\subfigure[$L1$-MechSOUL]{{\includegraphics[width=0.13\textwidth]{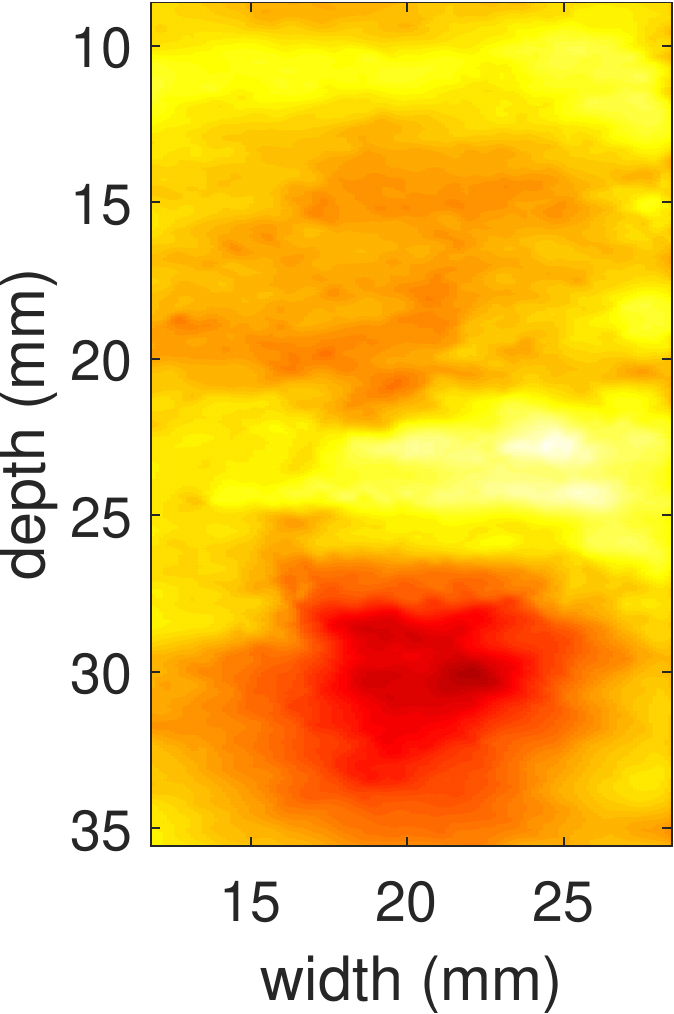}}}
	\subfigure[Axial strain]{{\includegraphics[width=0.20\textwidth]{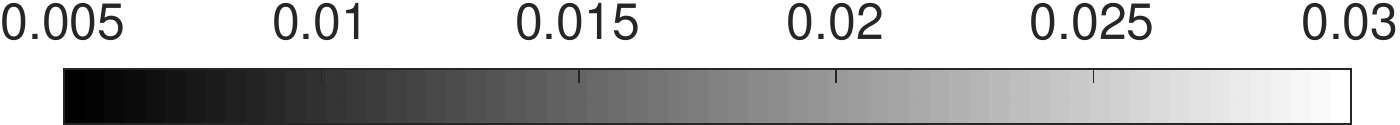}}}%
	\quad
	\subfigure[Colorbar for (i) and (j)]{{\includegraphics[width=0.20\textwidth]{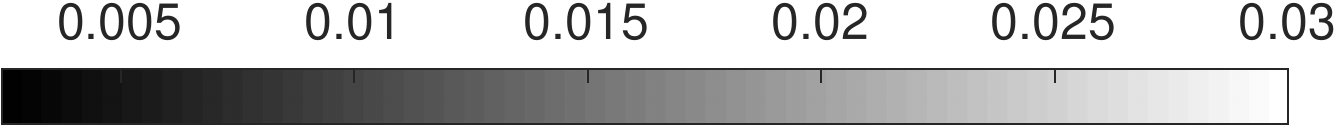}}}%
	\quad
	\subfigure[Colorbar for (k), (l), (m), (n)]{{\includegraphics[width=0.20\textwidth]{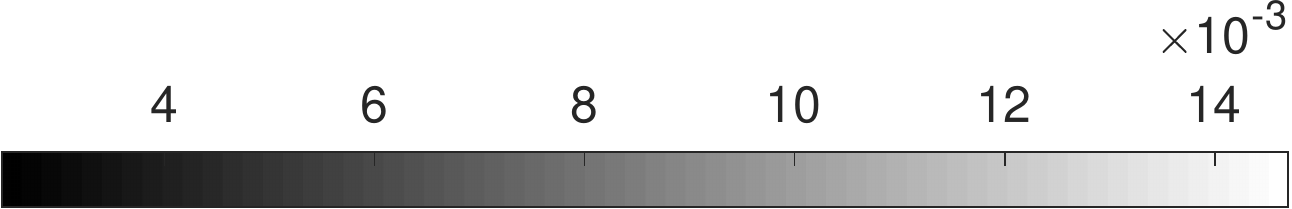}}}%
	\quad
	\subfigure[EPR]{{\includegraphics[width=0.20\textwidth]{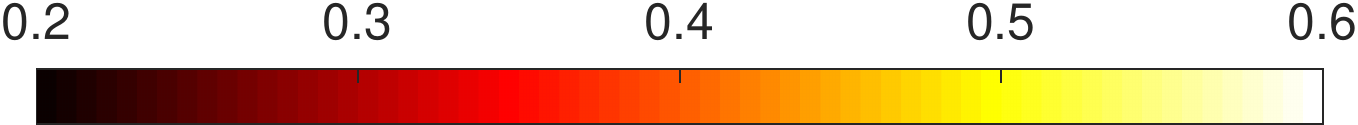}}}
	\caption{Results for the hard-inclusion simulated phantom. Rows 1 and 2 show the axial and lateral strains, respectively, whereas, row 3 shows the EPR maps. Columns 1 to 7 correspond to ground truth, NCC, NCC + PDE, SOUL, $L1$-SOUL, MechSOUL, and $L1$-MechSOUL, respectively.}
	\label{hard_simu}
\end{figure*}

\begin{figure*}
	\centering
	\subfigure[Ground truth]{{\includegraphics[width=0.13\textwidth]{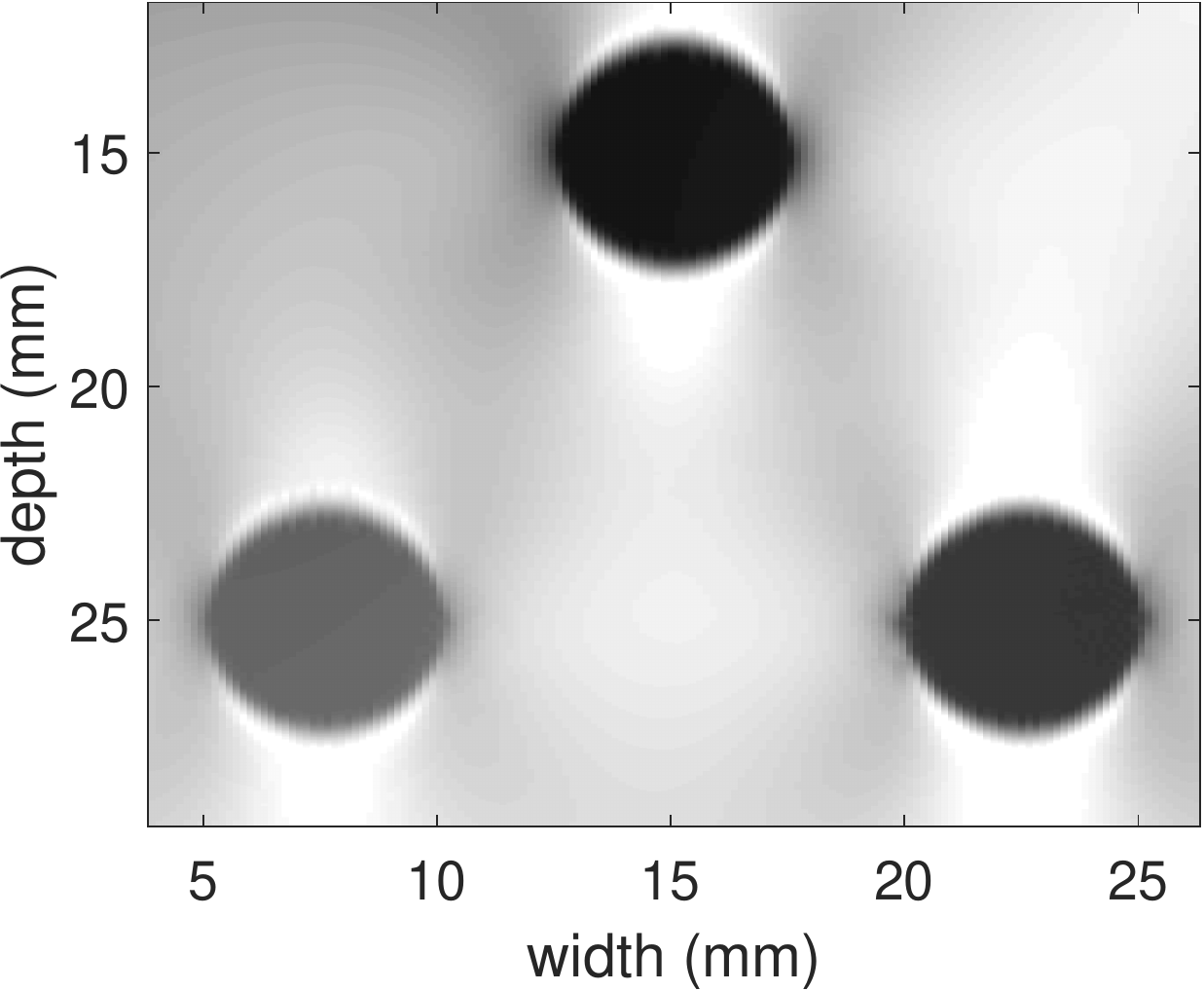}}}%
	\subfigure[NCC]{{\includegraphics[width=0.13\textwidth]{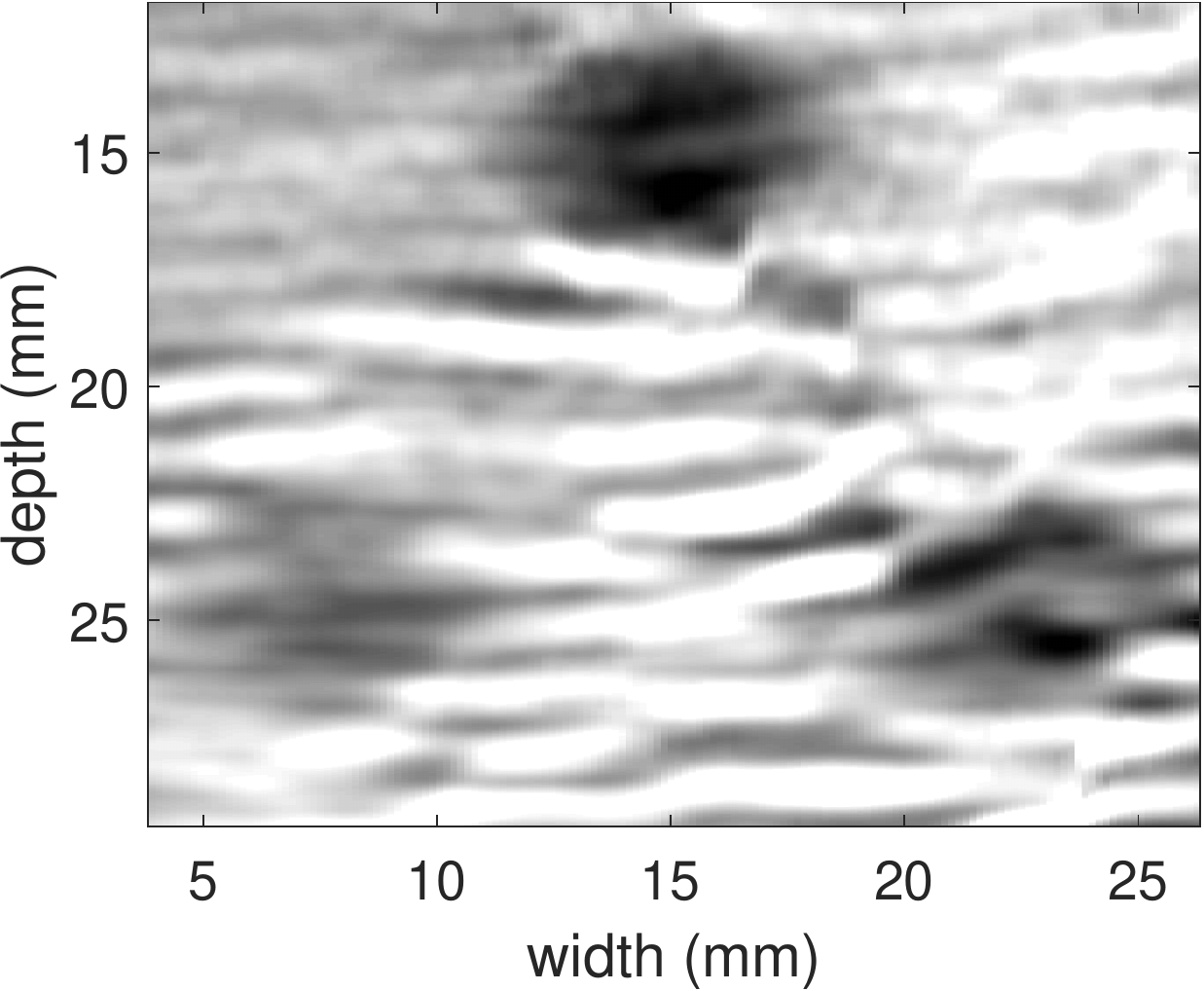}}}%
	\subfigure[NCC + PDE]{{\includegraphics[width=0.13\textwidth]{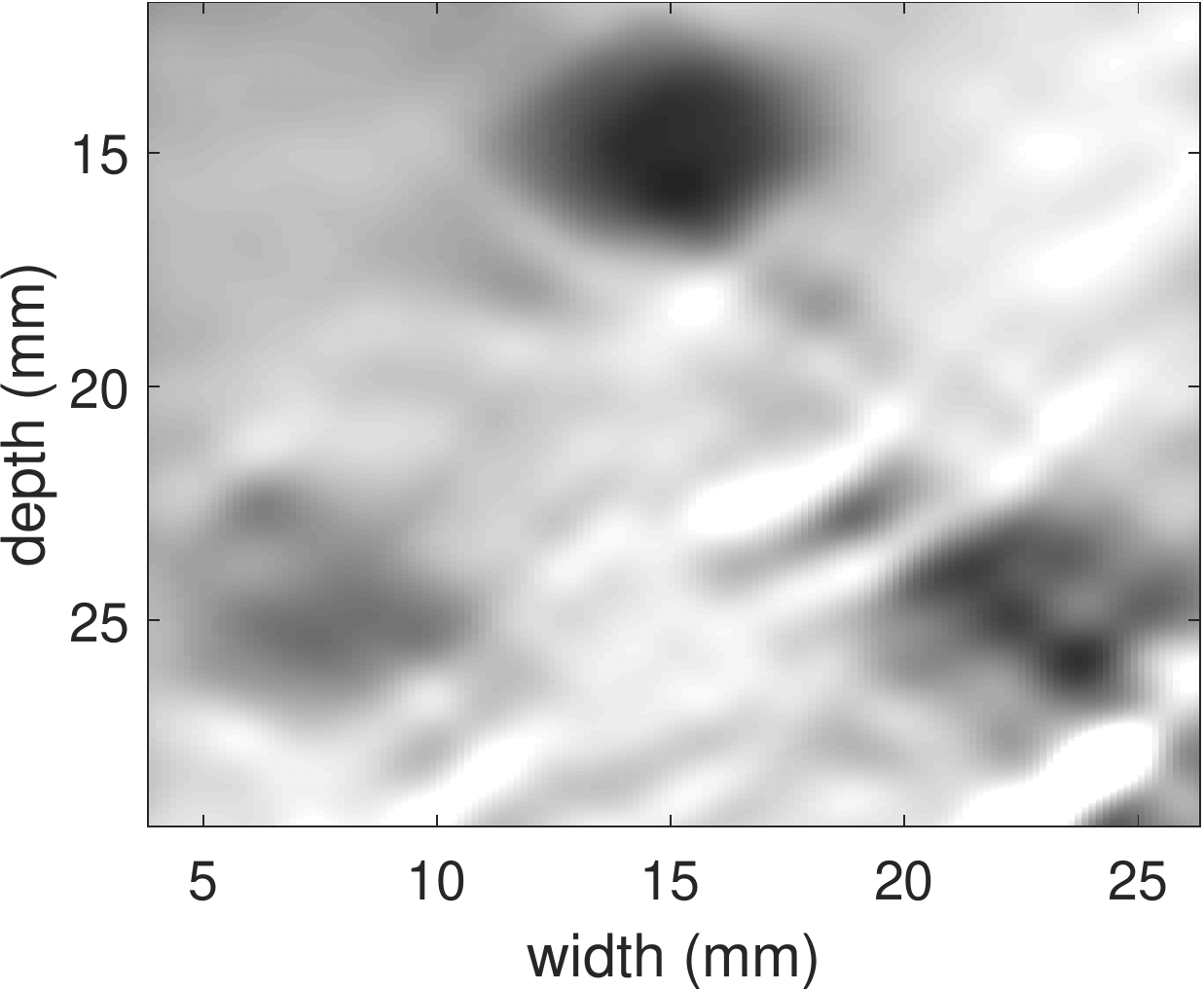}}}%
	\subfigure[SOUL]{{\includegraphics[width=0.13\textwidth]{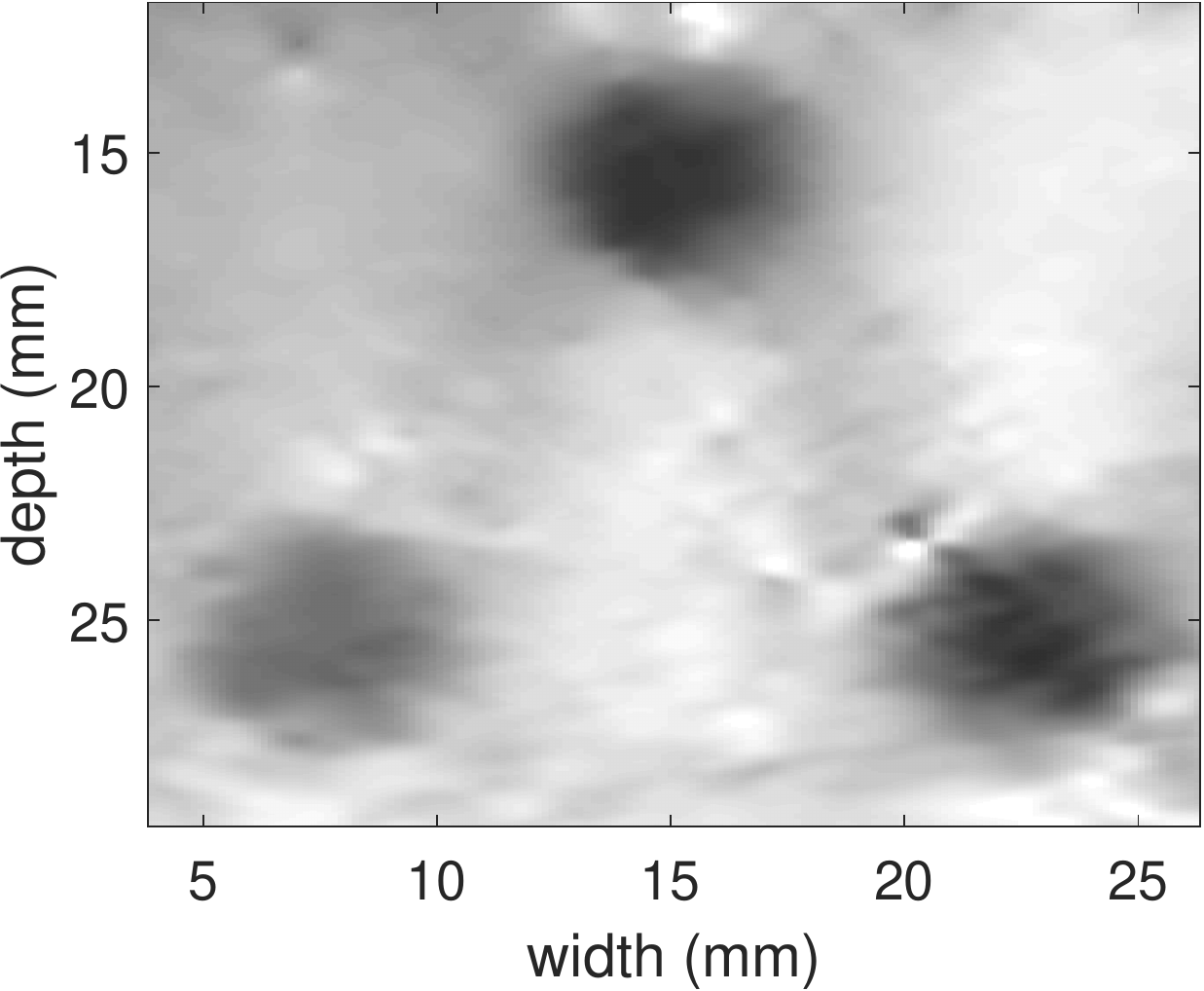}}}%
	\subfigure[$L1$-SOUL]{{\includegraphics[width=0.13\textwidth]{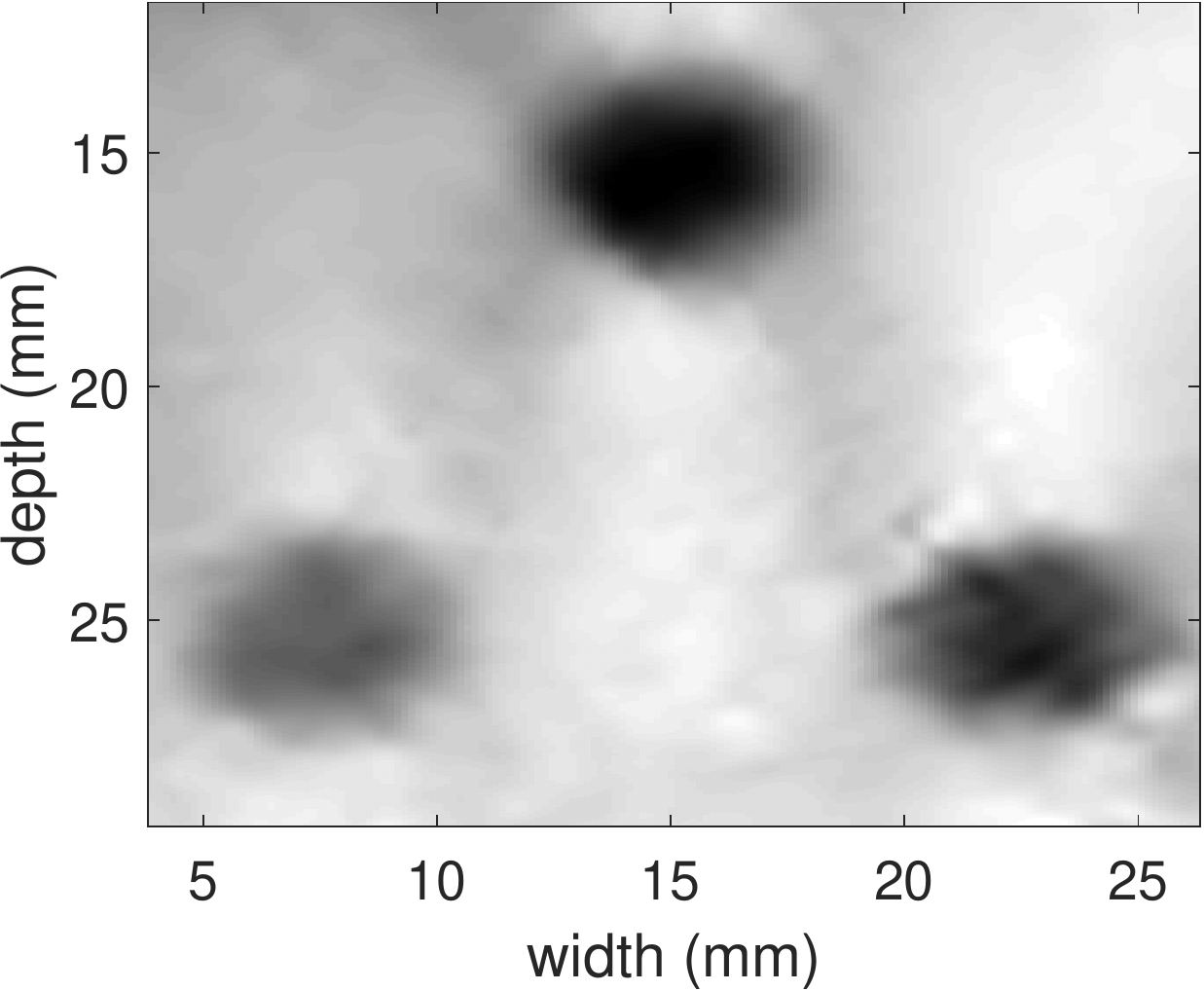}}}%
	\subfigure[MechSOUL]{{\includegraphics[width=0.13\textwidth]{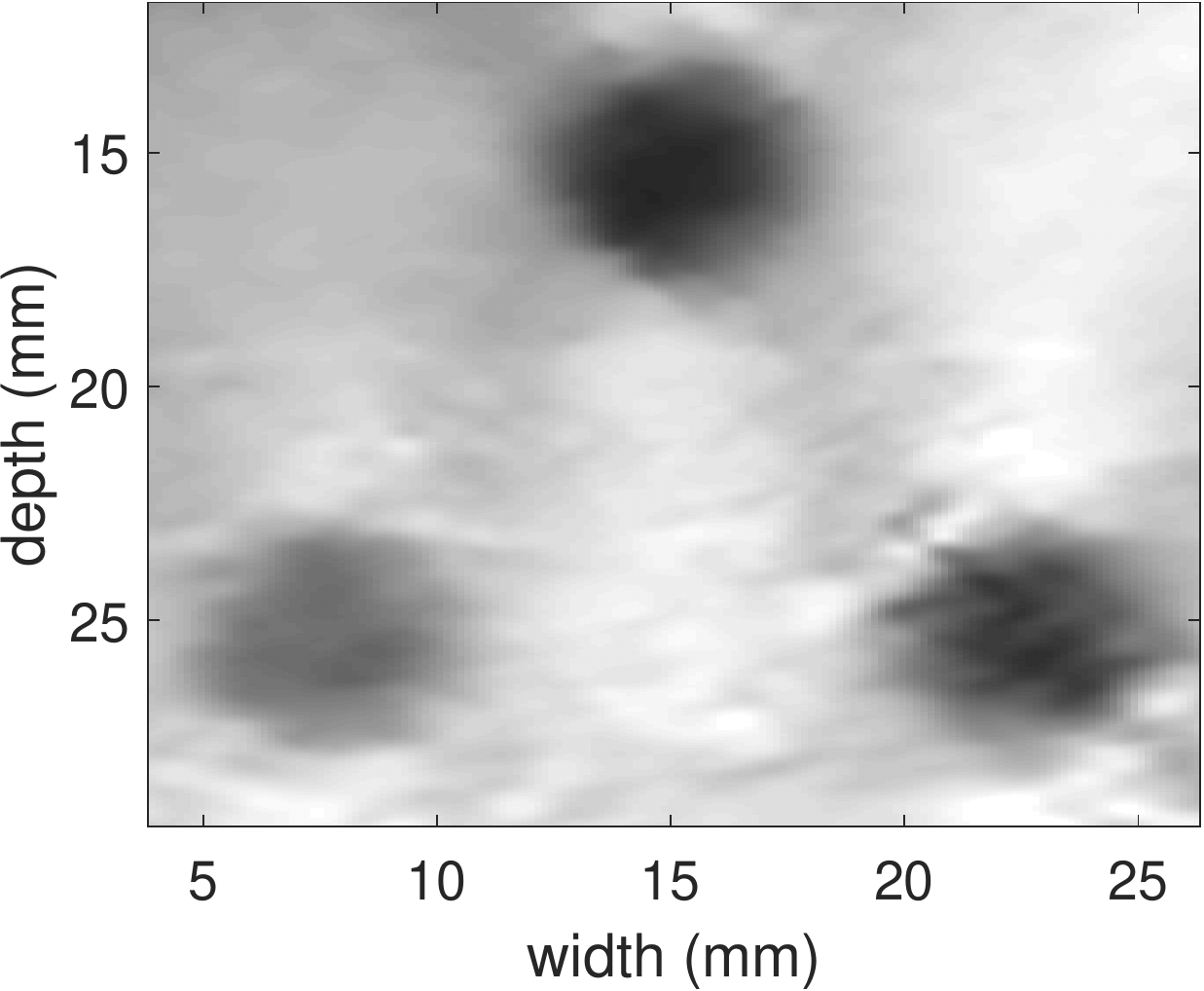}}}%
	\subfigure[$L1$-MechSOUL]{{\includegraphics[width=0.13\textwidth]{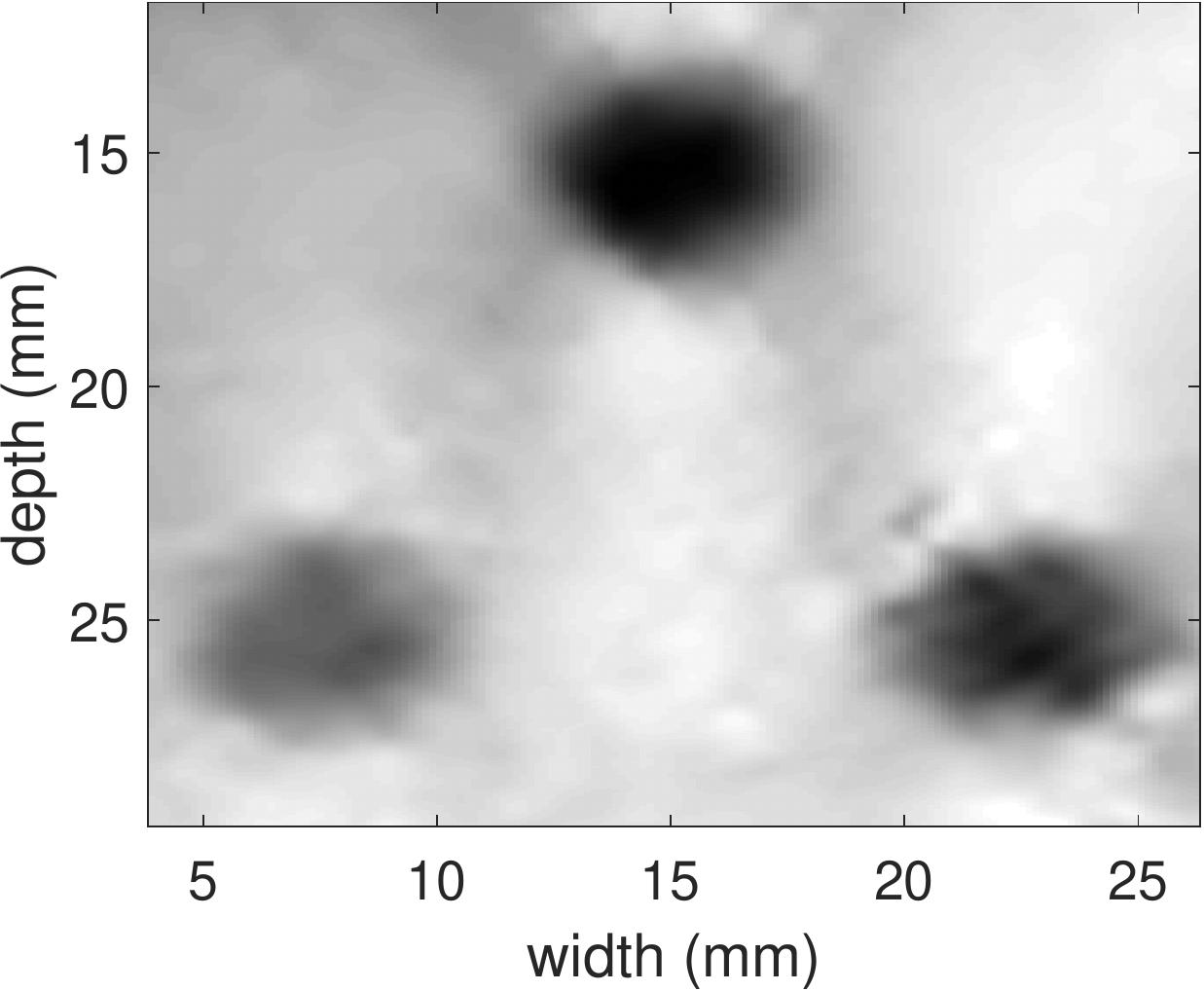}}}
	\subfigure[Ground truth]{{\includegraphics[width=0.13\textwidth]{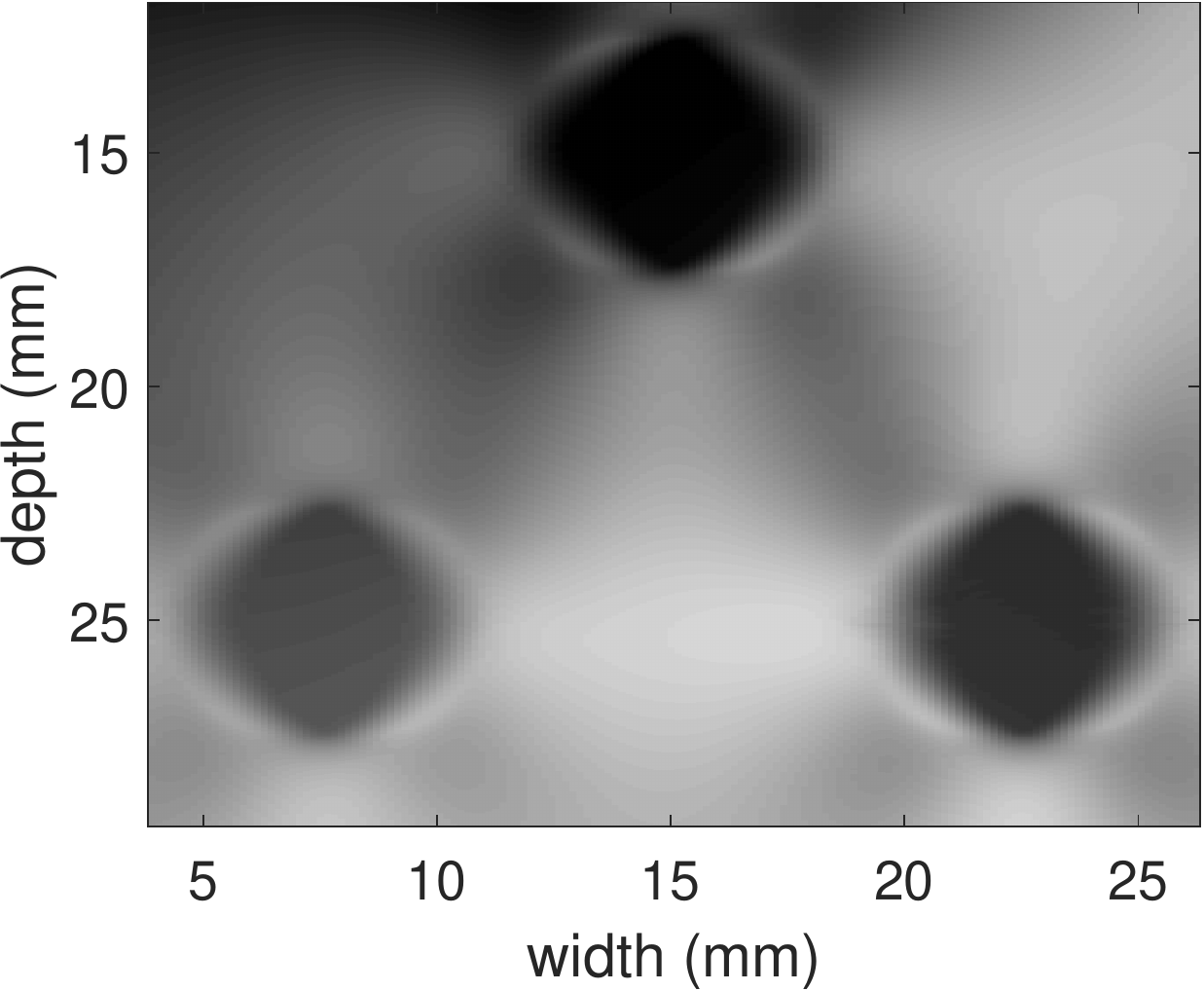}}}%
	\subfigure[NCC]{{\includegraphics[width=0.13\textwidth]{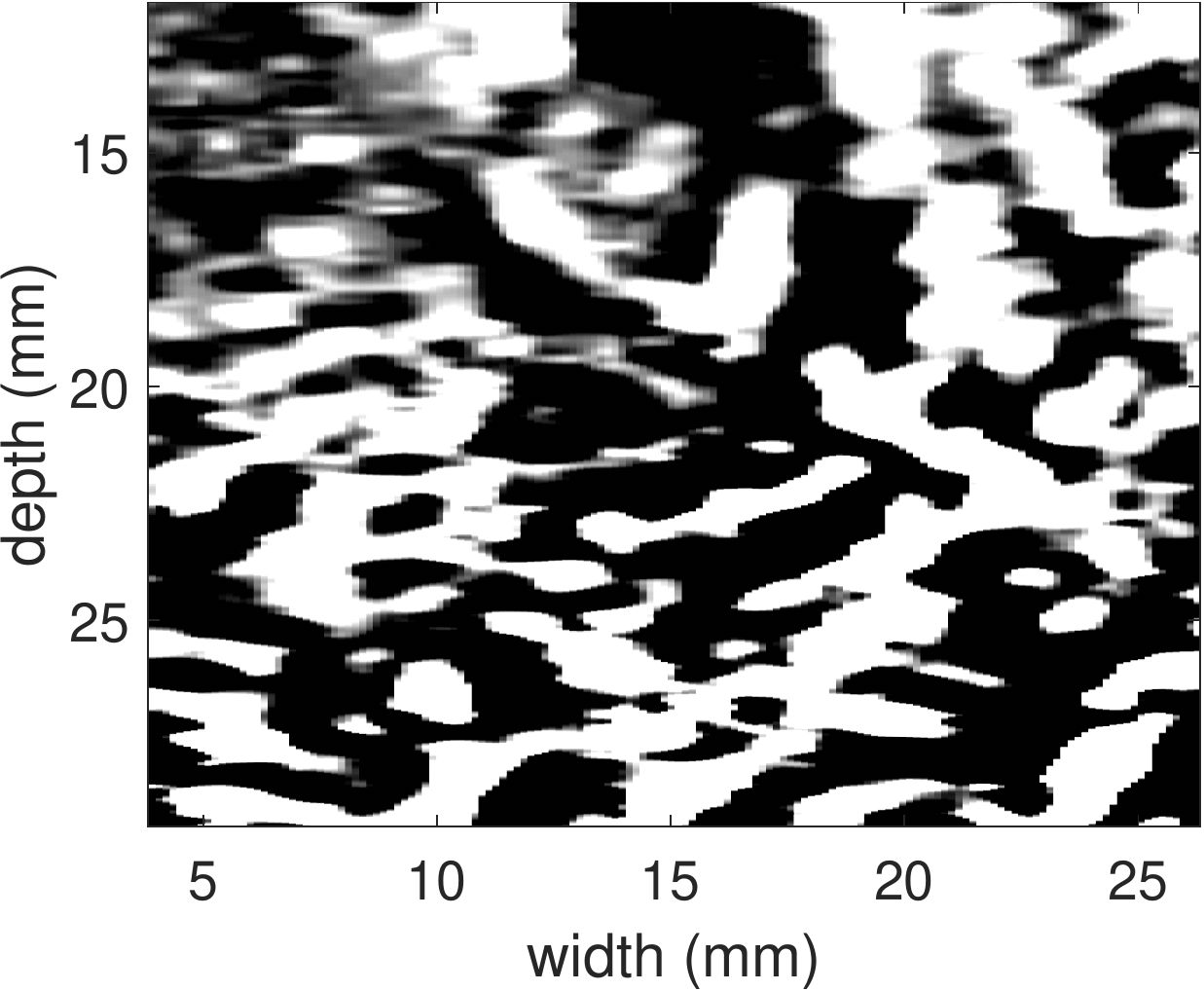}}}%
	\subfigure[NCC + PDE]{{\includegraphics[width=0.13\textwidth]{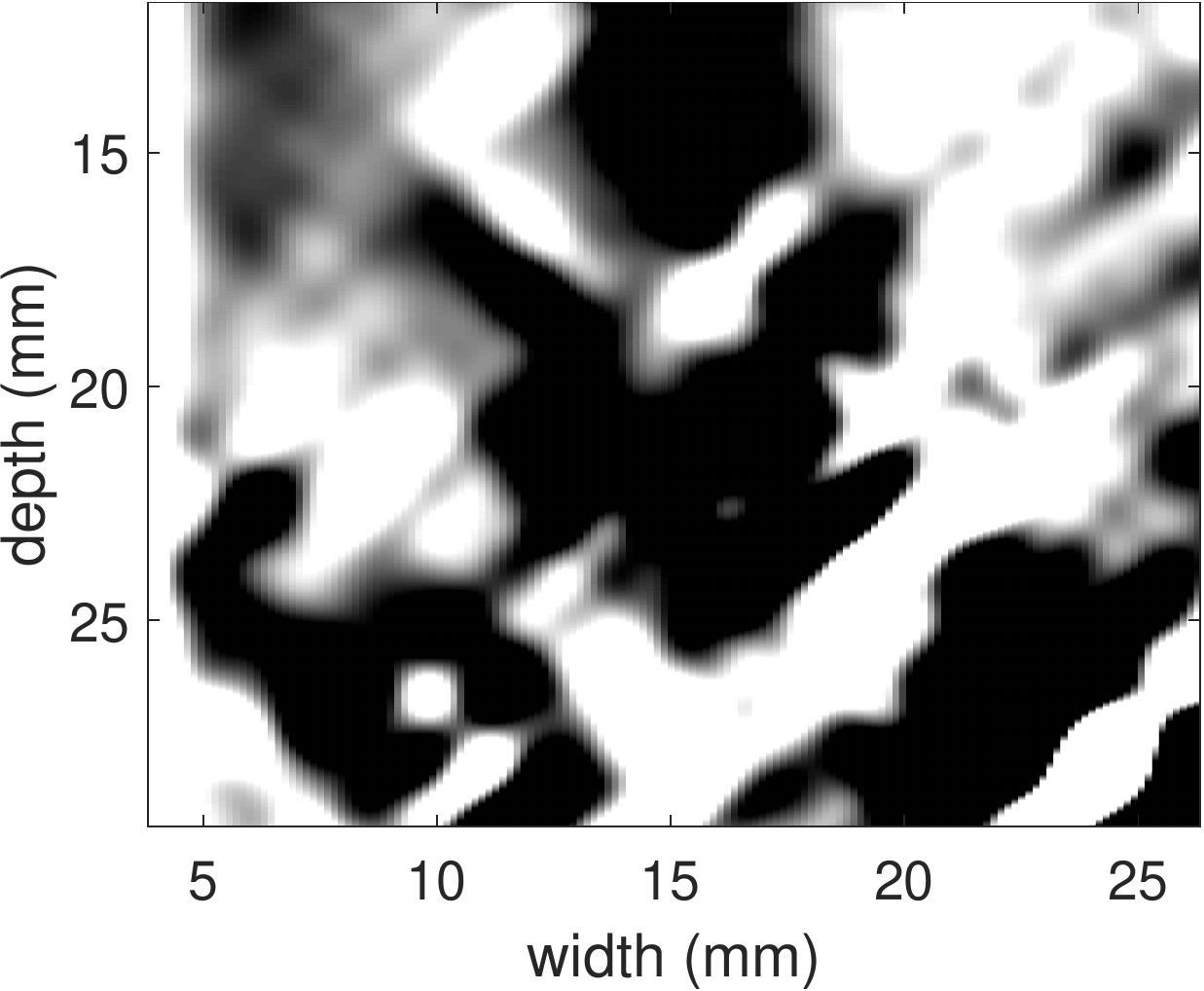}}}%
	\subfigure[SOUL]{{\includegraphics[width=0.13\textwidth]{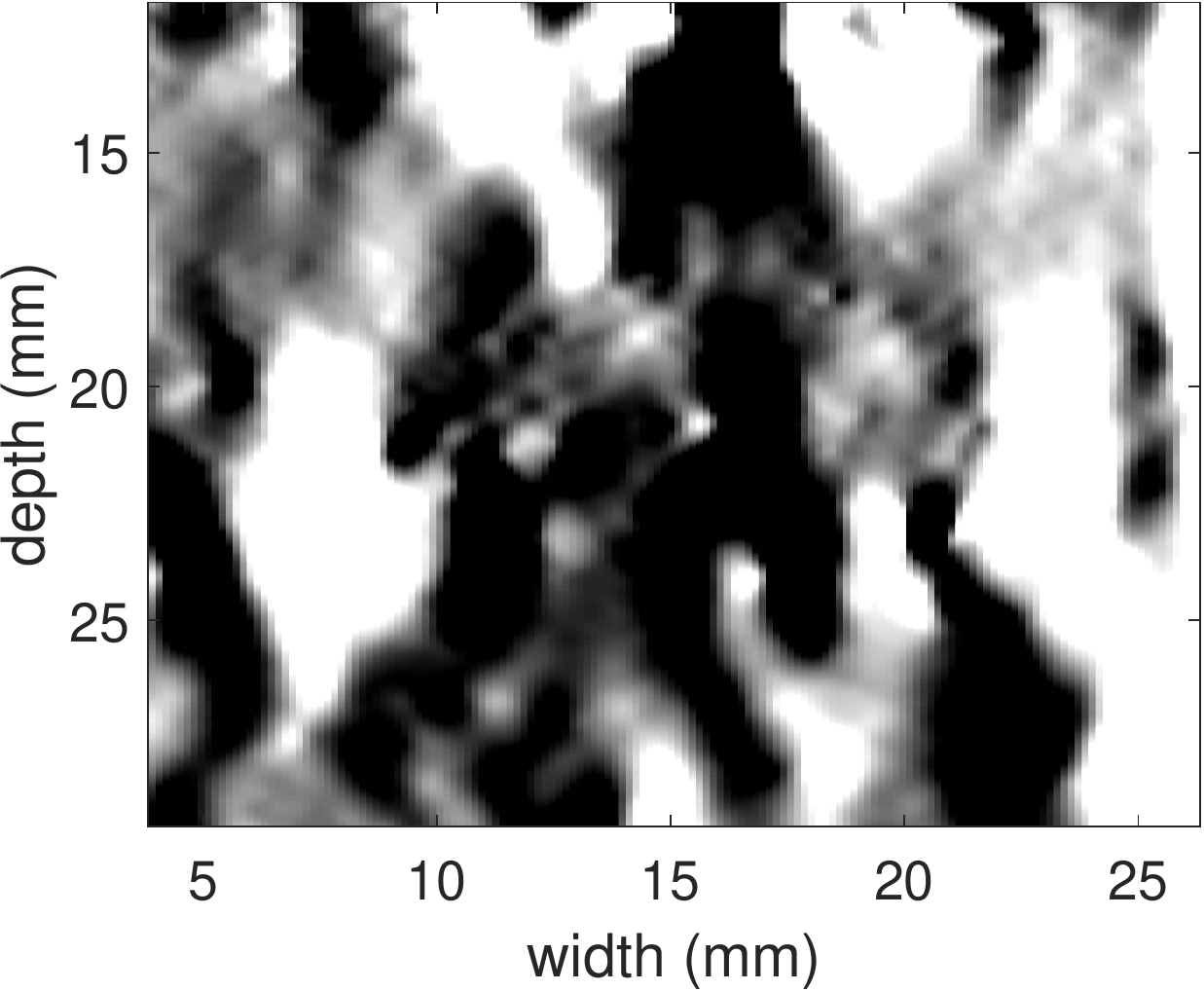}}}%
	\subfigure[$L1$-SOUL]{{\includegraphics[width=0.13\textwidth]{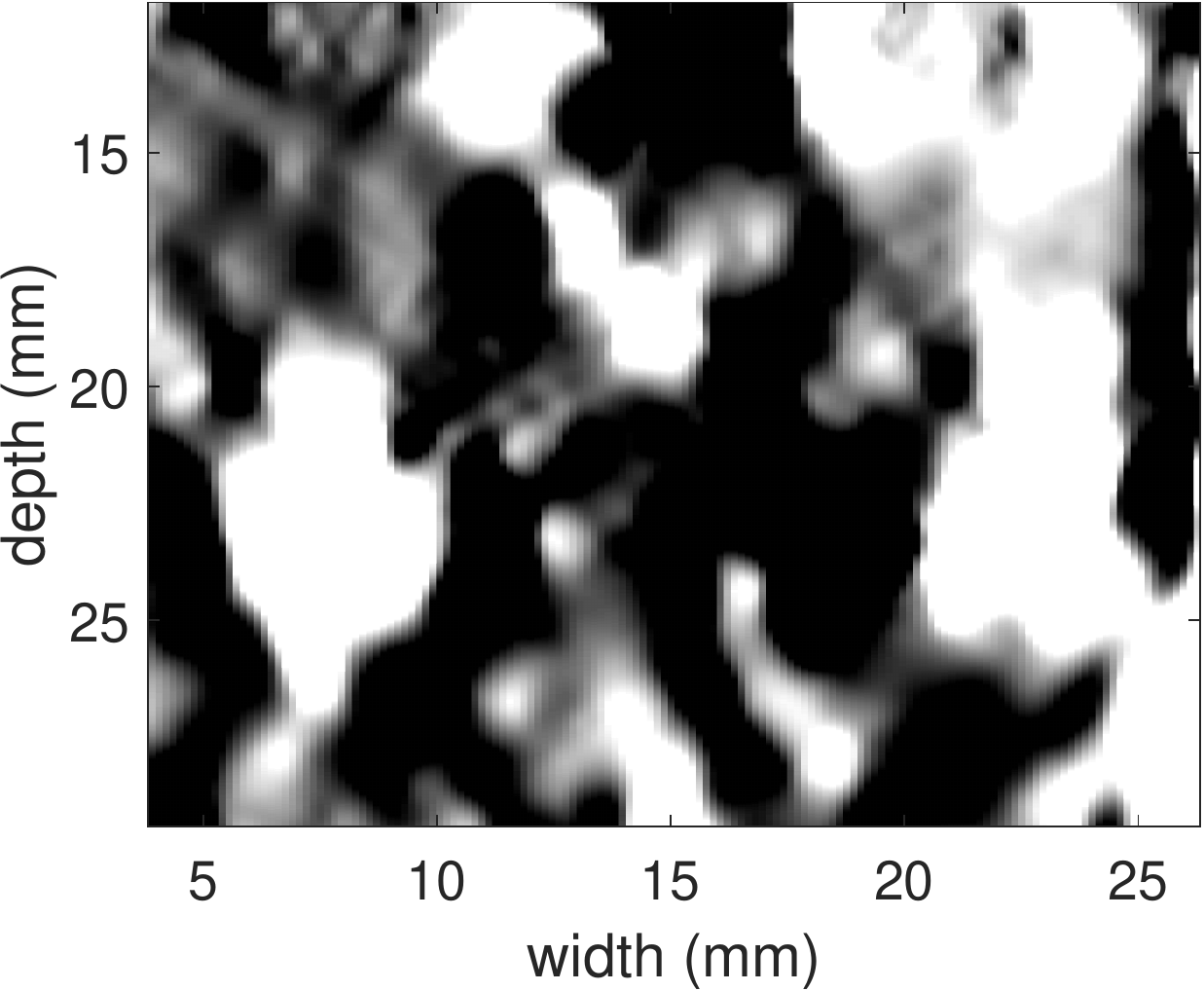}}}%
	\subfigure[MechSOUL]{{\includegraphics[width=0.13\textwidth]{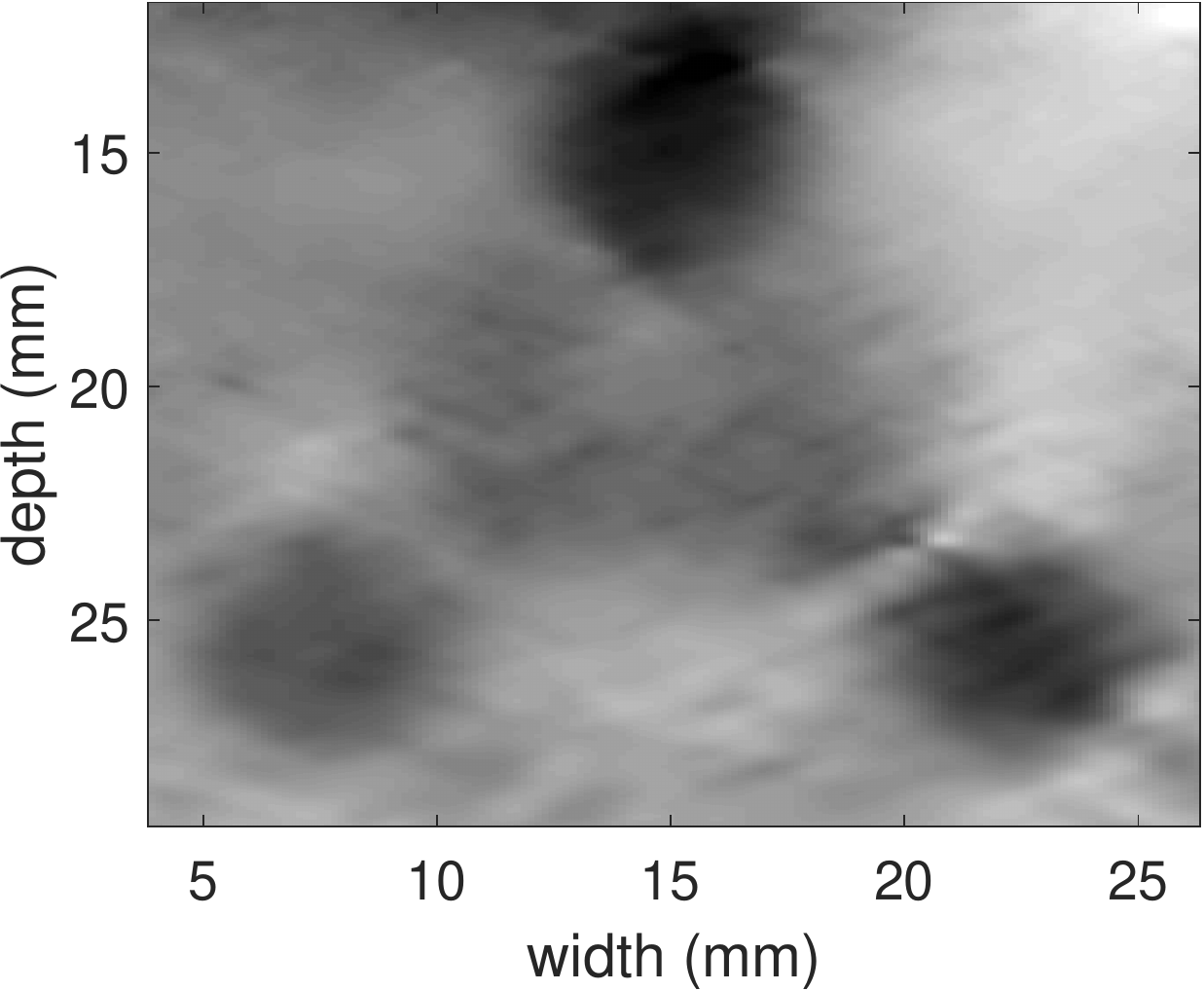}}}%
	\subfigure[$L1$-MechSOUL]{{\includegraphics[width=0.13\textwidth]{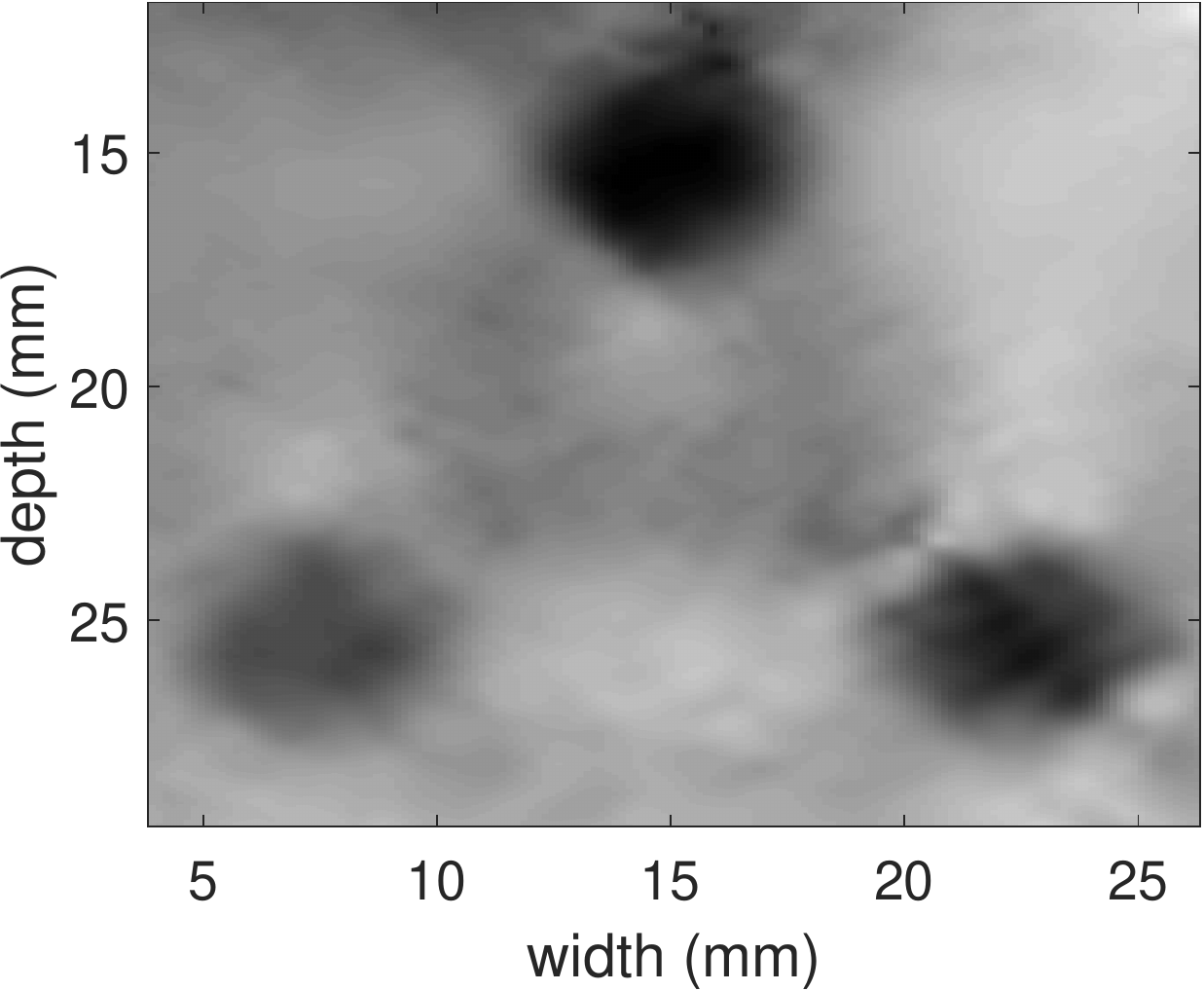}}}
	\subfigure[Ground truth]{{\includegraphics[width=0.13\textwidth]{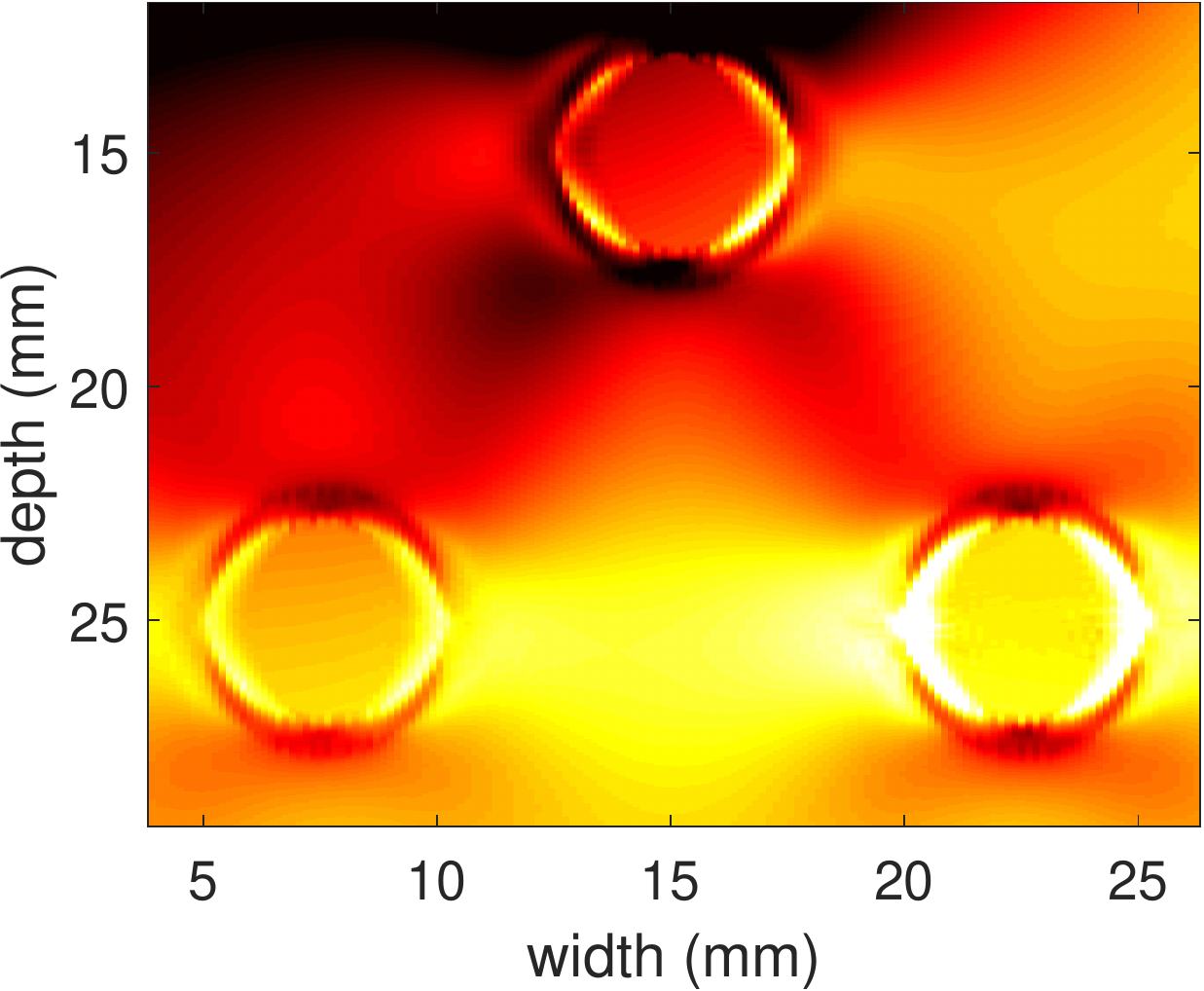}}}%
	\subfigure[NCC]{{\includegraphics[width=0.13\textwidth]{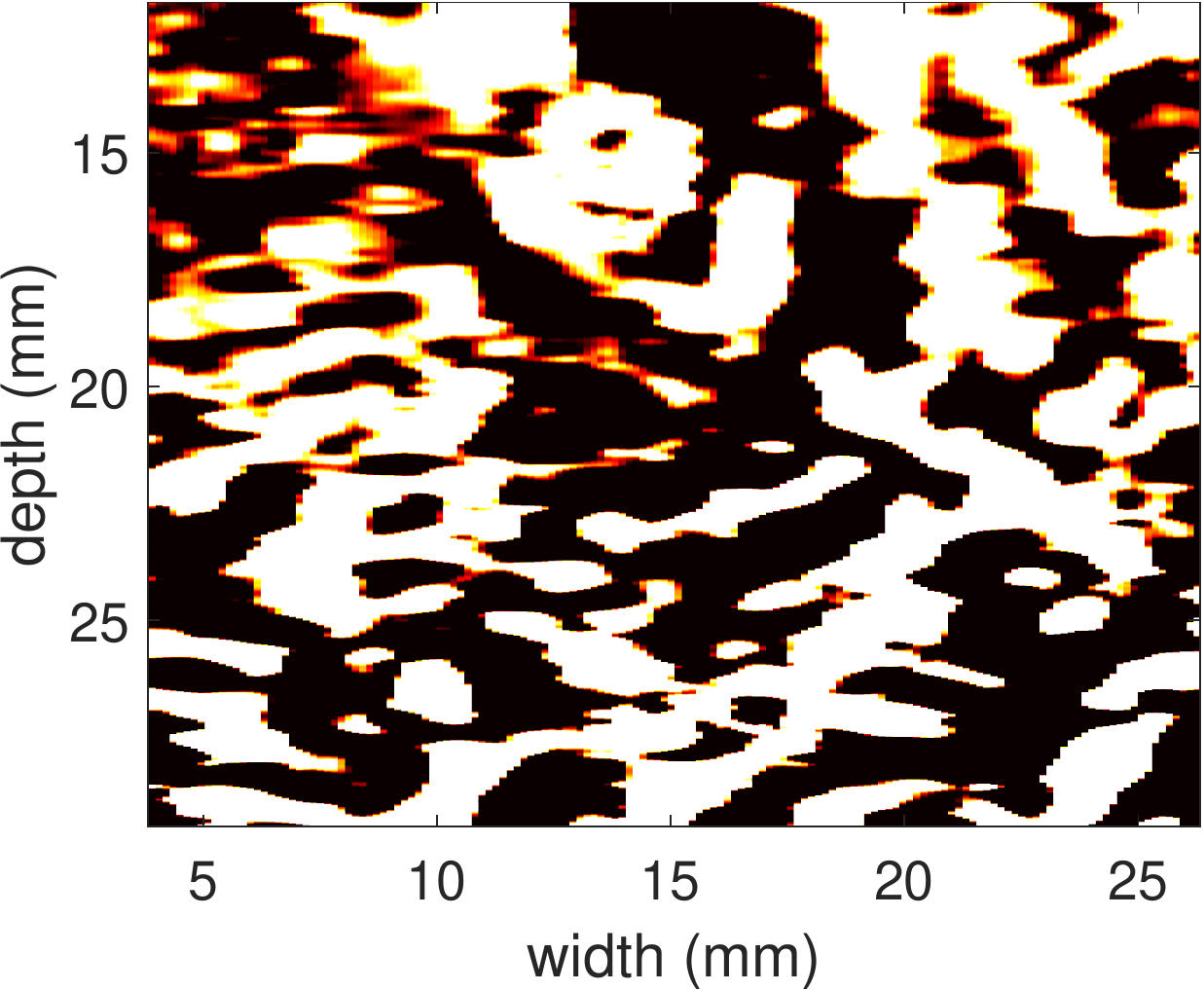}}}%
	\subfigure[NCC + PDE]{{\includegraphics[width=0.13\textwidth]{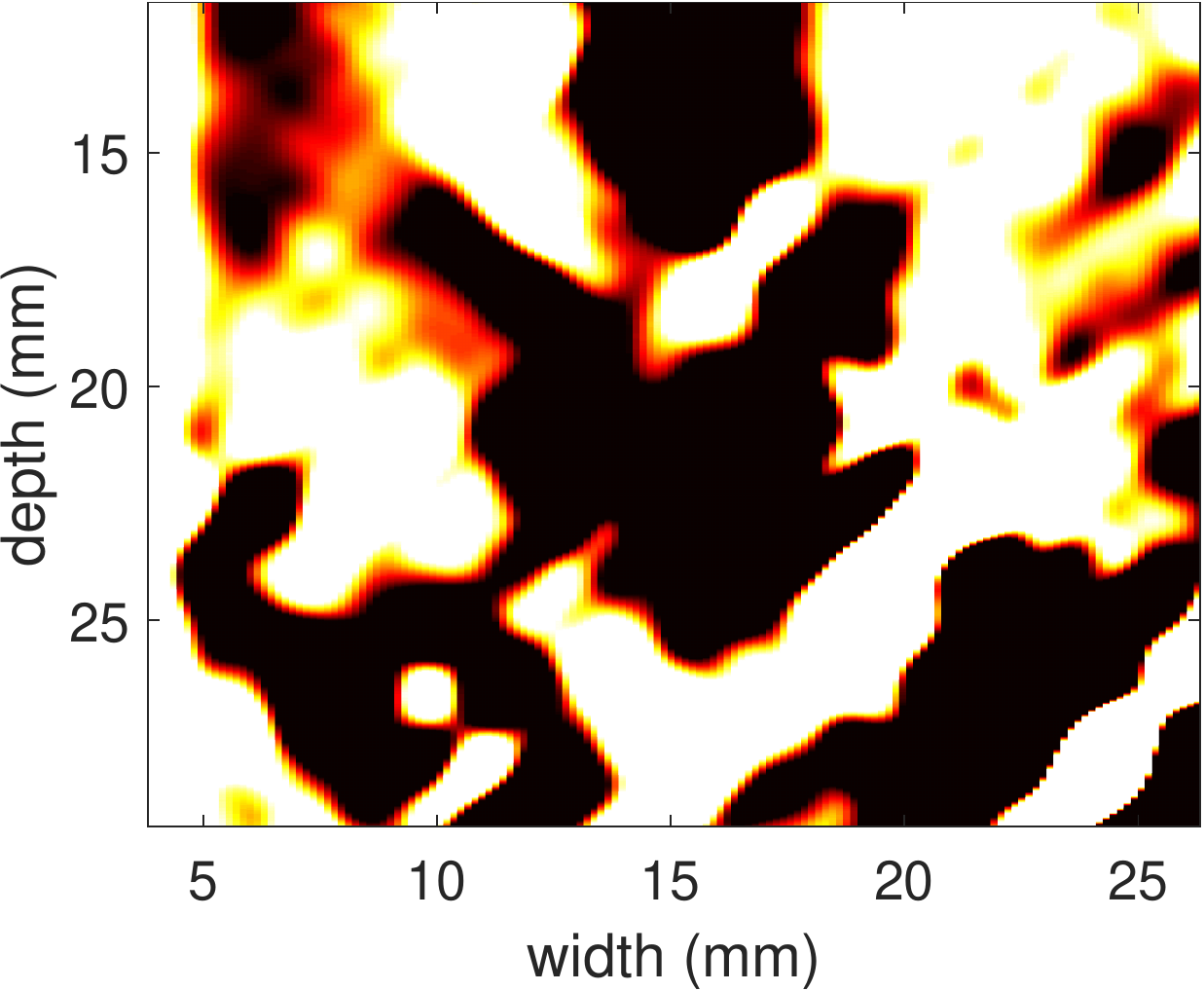}}}%
	\subfigure[SOUL]{{\includegraphics[width=0.13\textwidth]{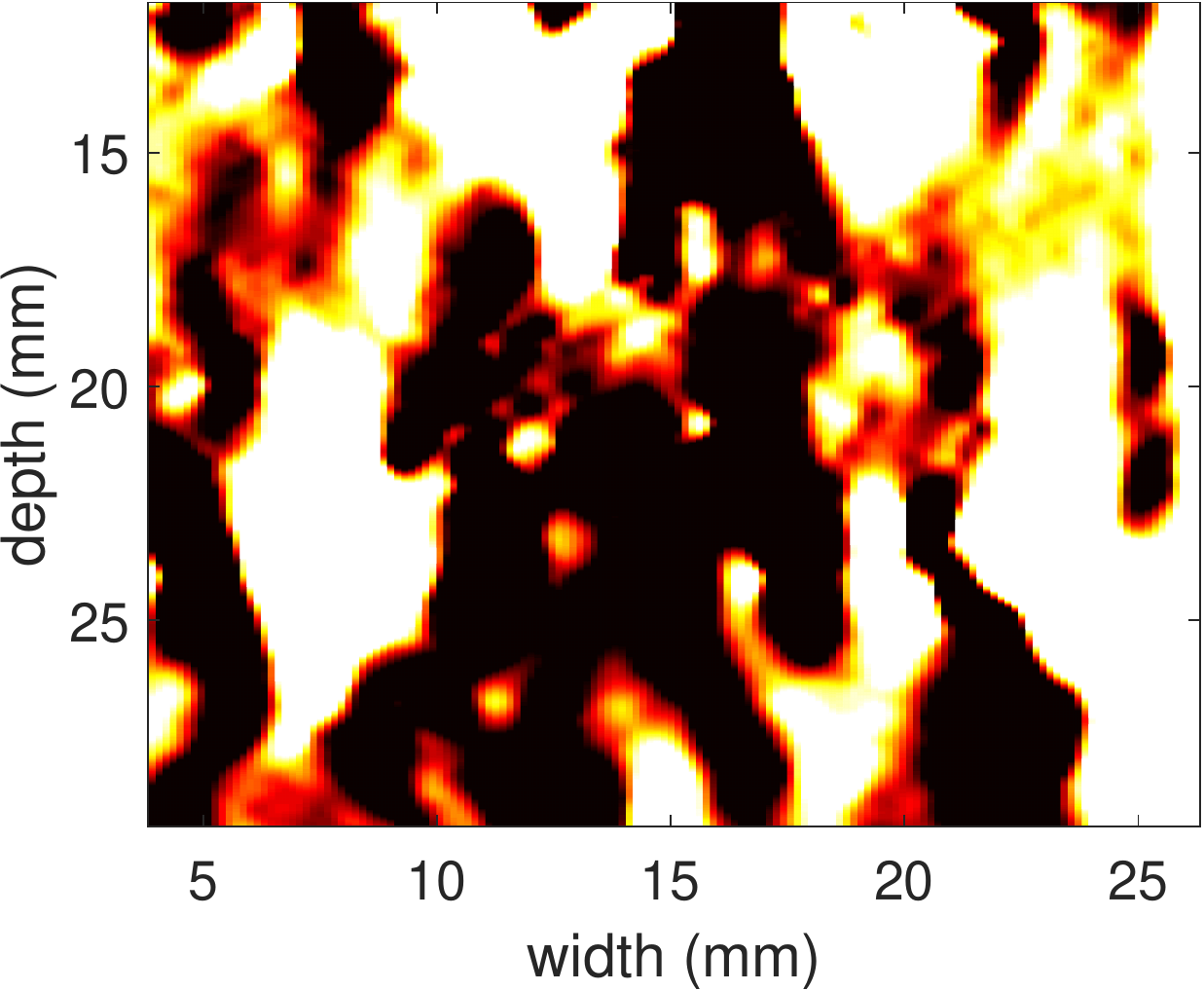}}}%
	\subfigure[$L1$-SOUL]{{\includegraphics[width=0.13\textwidth]{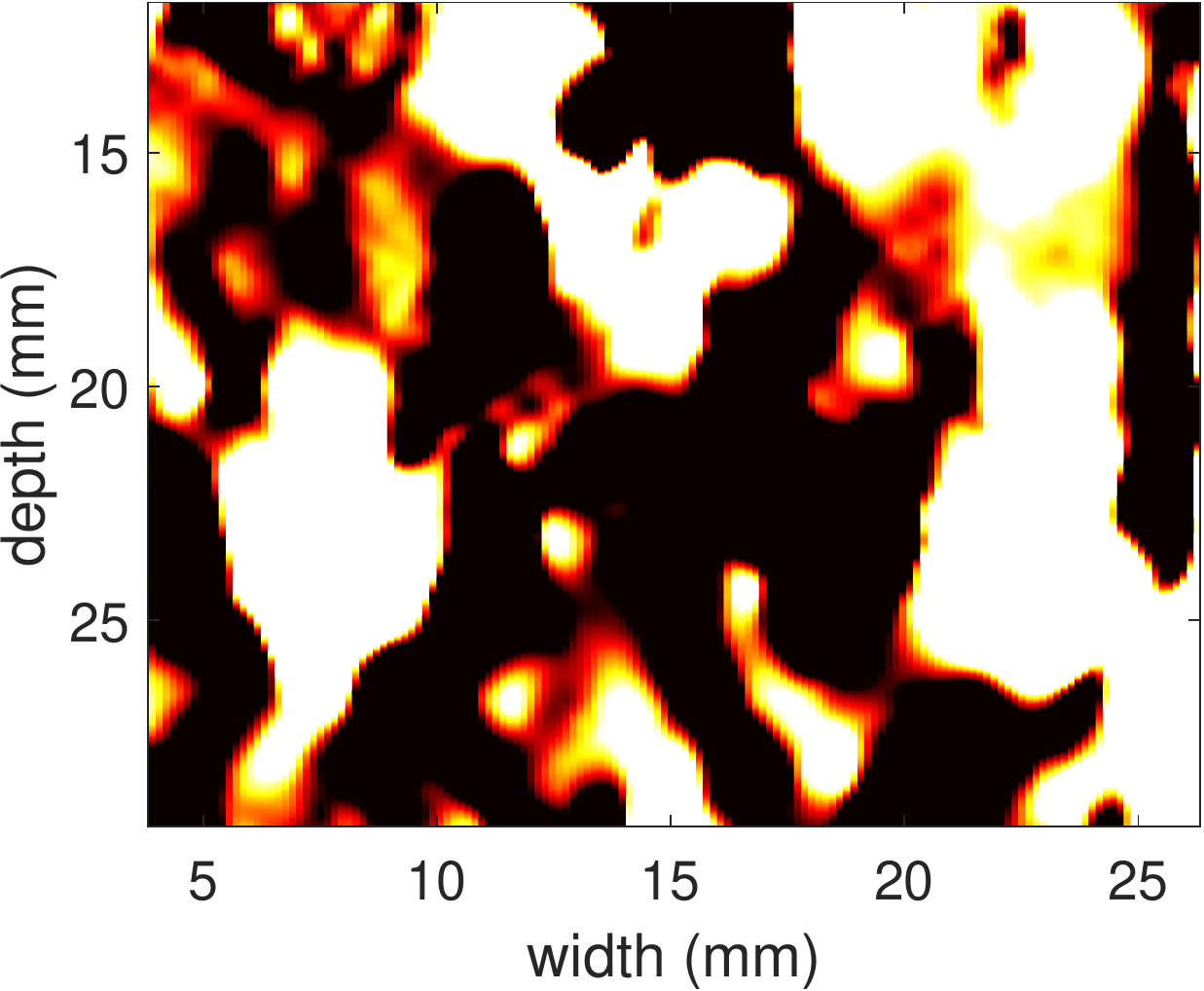}}}%
	\subfigure[MechSOUL]{{\includegraphics[width=0.13\textwidth]{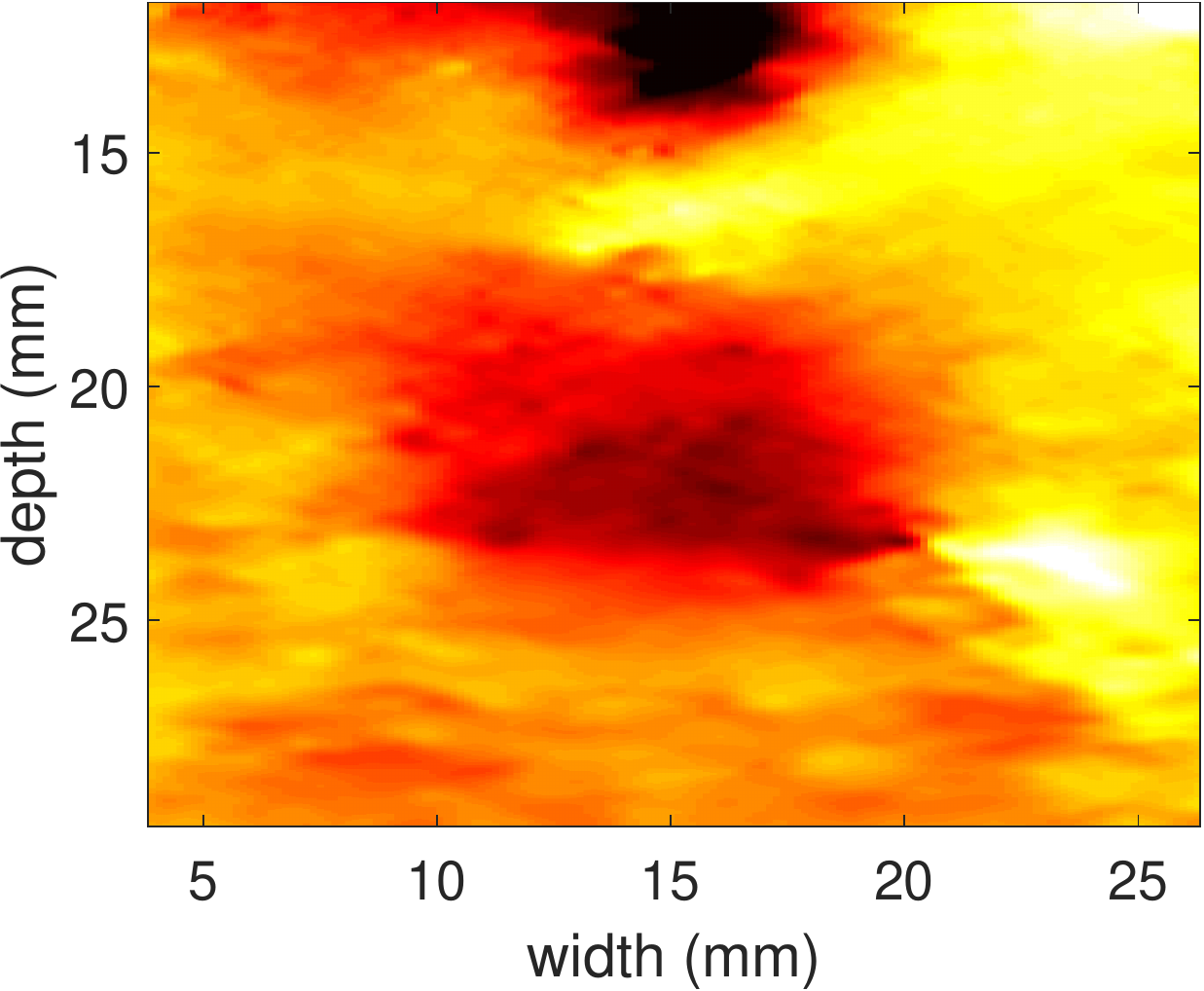}}}%
	\subfigure[$L1$-MechSOUL]{{\includegraphics[width=0.13\textwidth]{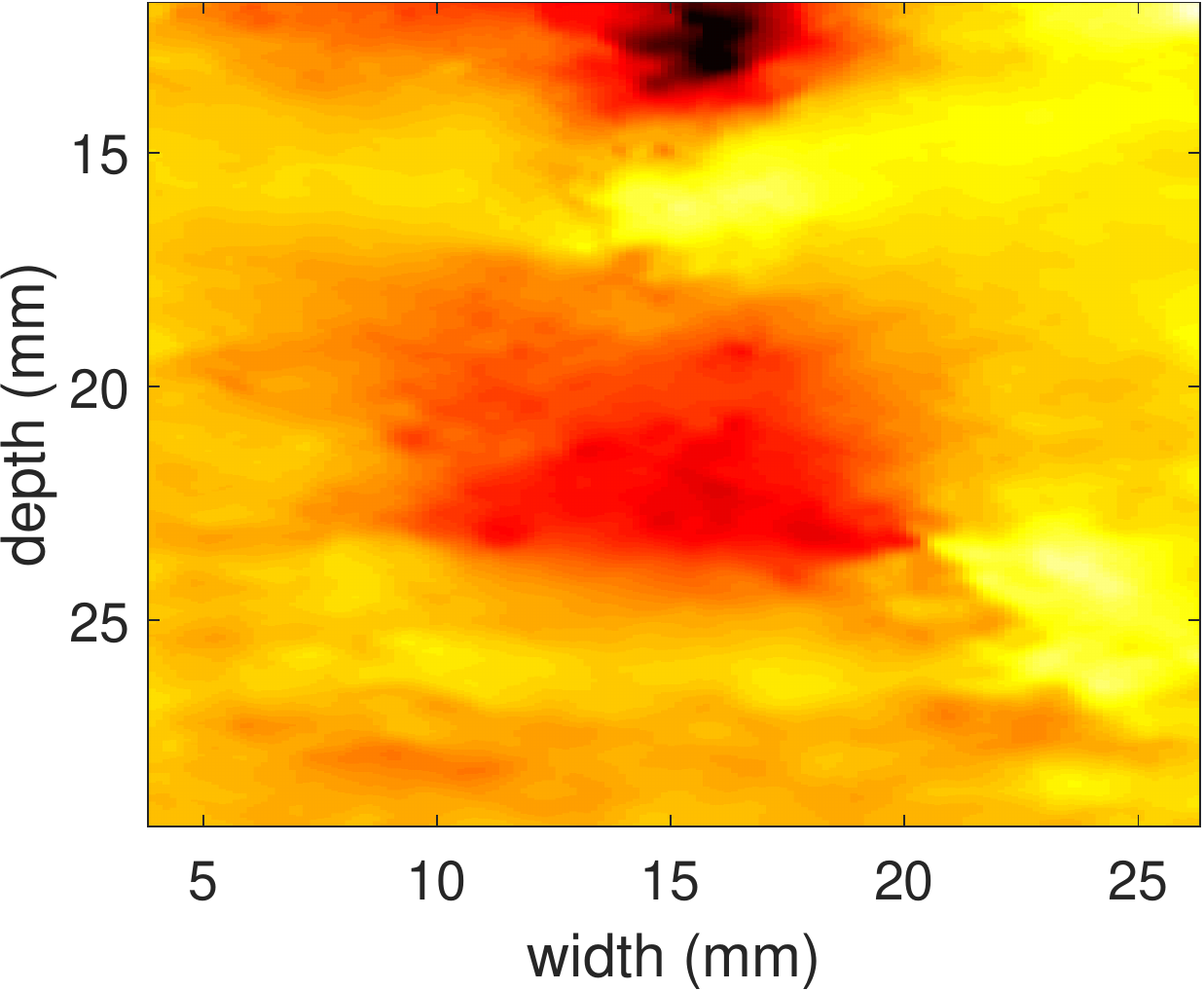}}}
	\subfigure[Axial strain]{{\includegraphics[width=0.27\textwidth]{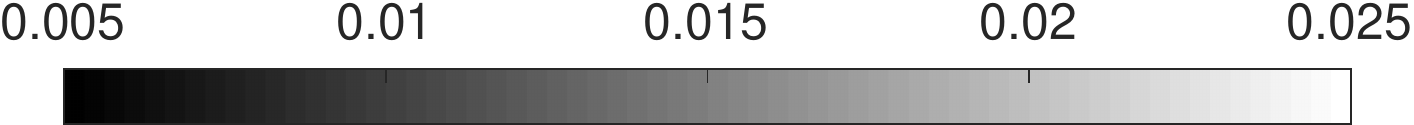}}}%
	\quad
	\subfigure[Lateral strain]{{\includegraphics[width=0.27\textwidth]{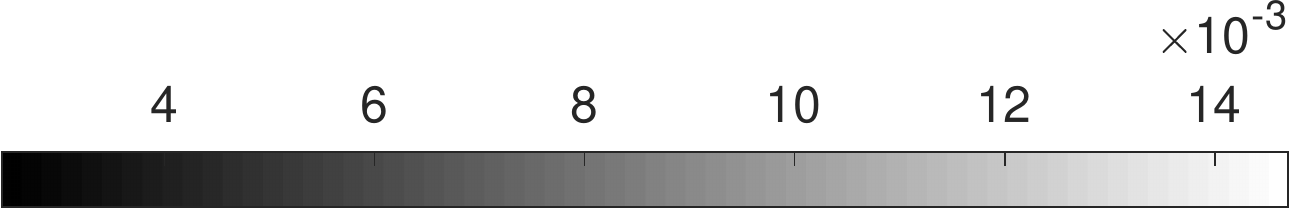}}}%
	\quad
	\subfigure[EPR]{{\includegraphics[width=0.27\textwidth]{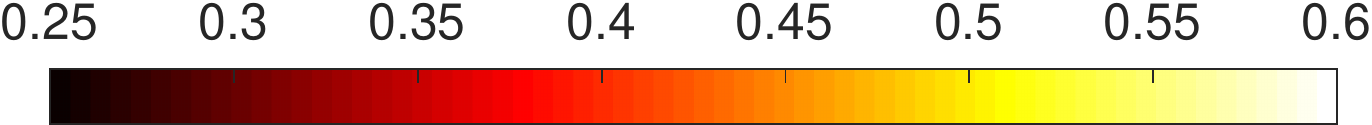}}}%
	\caption{Results for the multi-inclusion simulated phantom with an additional lateral boundary condition. Rows 1 and 2 depict the axial and lateral strains, respectively, whereas, row 3 presents the EPR maps. Columns 1 to 7, respectively, correspond to ground truth, NCC, NCC + PDE, SOUL, $L1$-SOUL, MechSOUL, and $L1$-MechSOUL.}
	\label{boundary_simu}
\end{figure*}

\begin{figure*}
	\centering
	\subfigure[Ground truth]{{\includegraphics[width=0.13\textwidth]{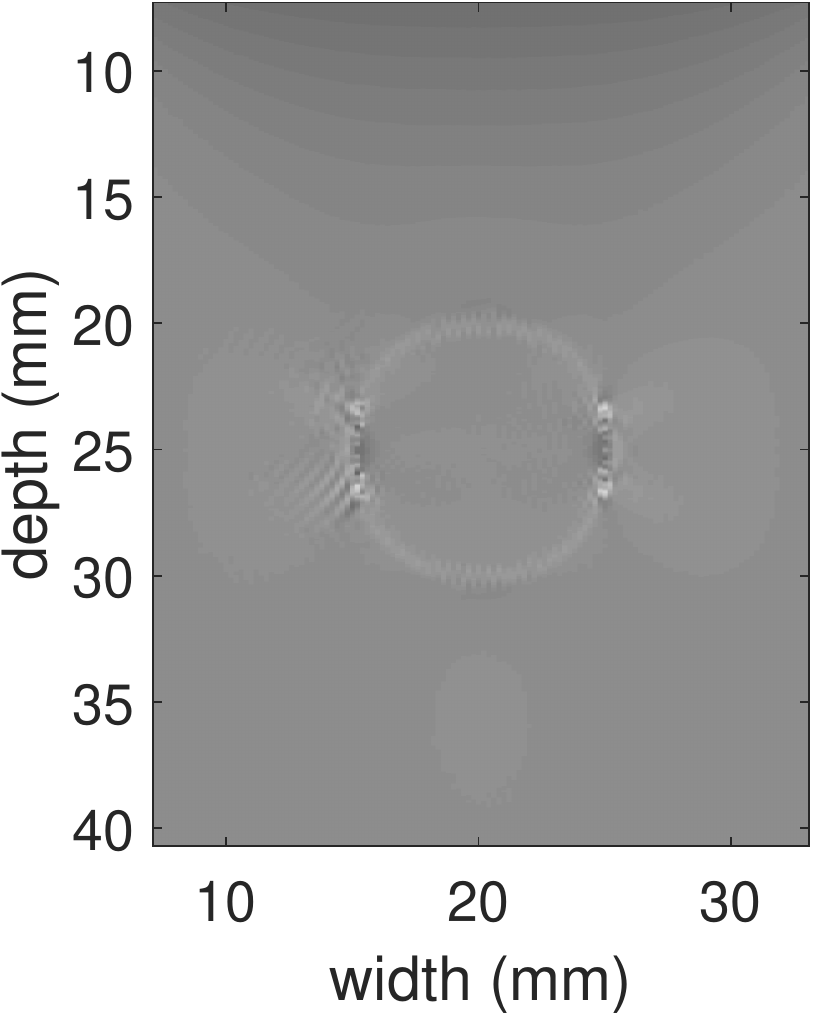}}}%
	\subfigure[NCC]{{\includegraphics[width=0.13\textwidth]{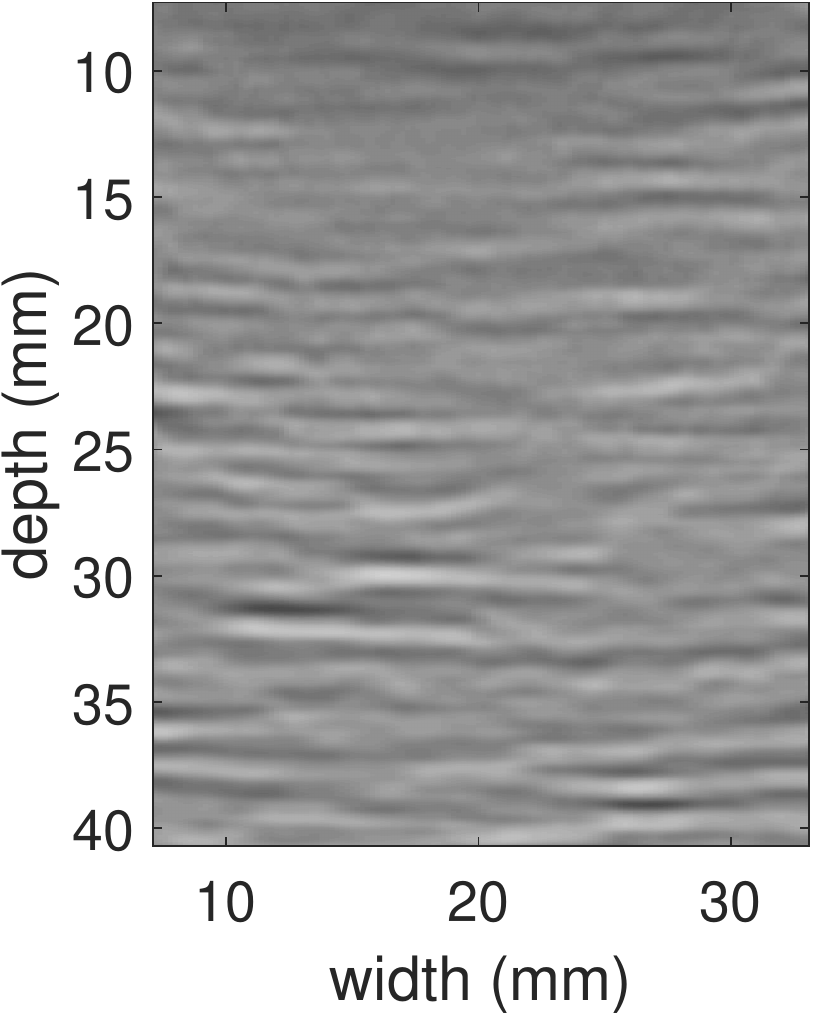}}}%
	\subfigure[NCC + PDE]{{\includegraphics[width=0.13\textwidth]{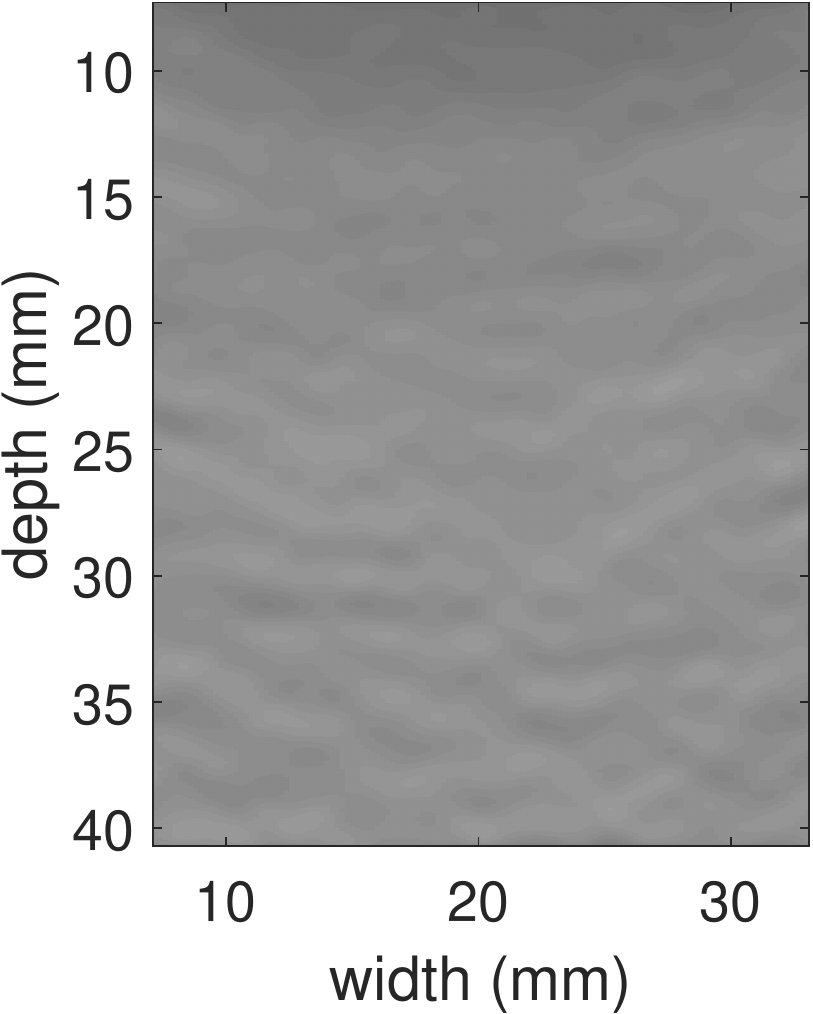}}}%
	\subfigure[SOUL]{{\includegraphics[width=0.13\textwidth]{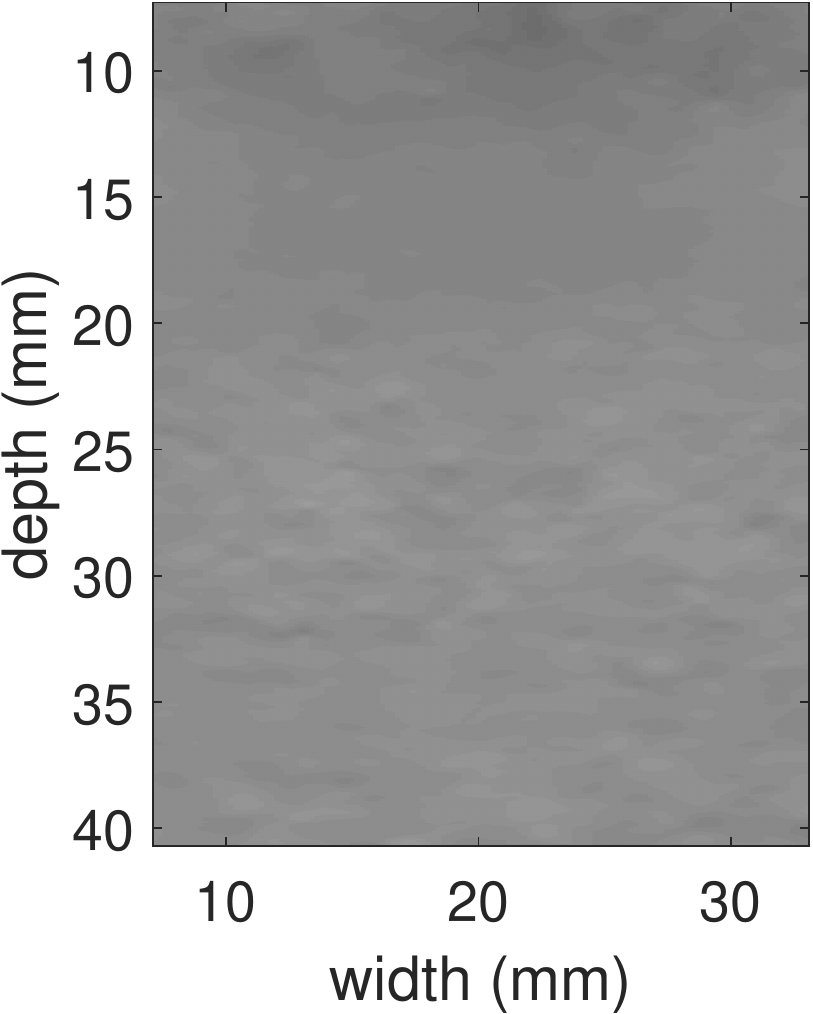}}}%
	\subfigure[$L1$-SOUL]{{\includegraphics[width=0.13\textwidth]{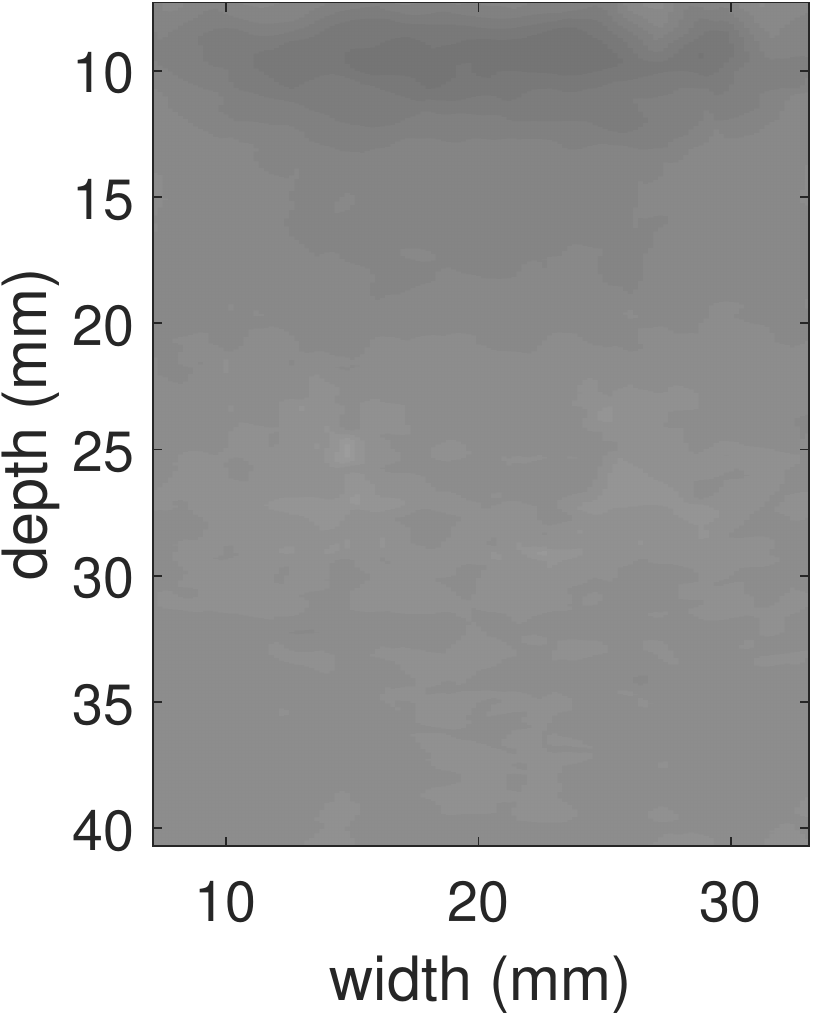}}}%
	\subfigure[MechSOUL]{{\includegraphics[width=0.13\textwidth]{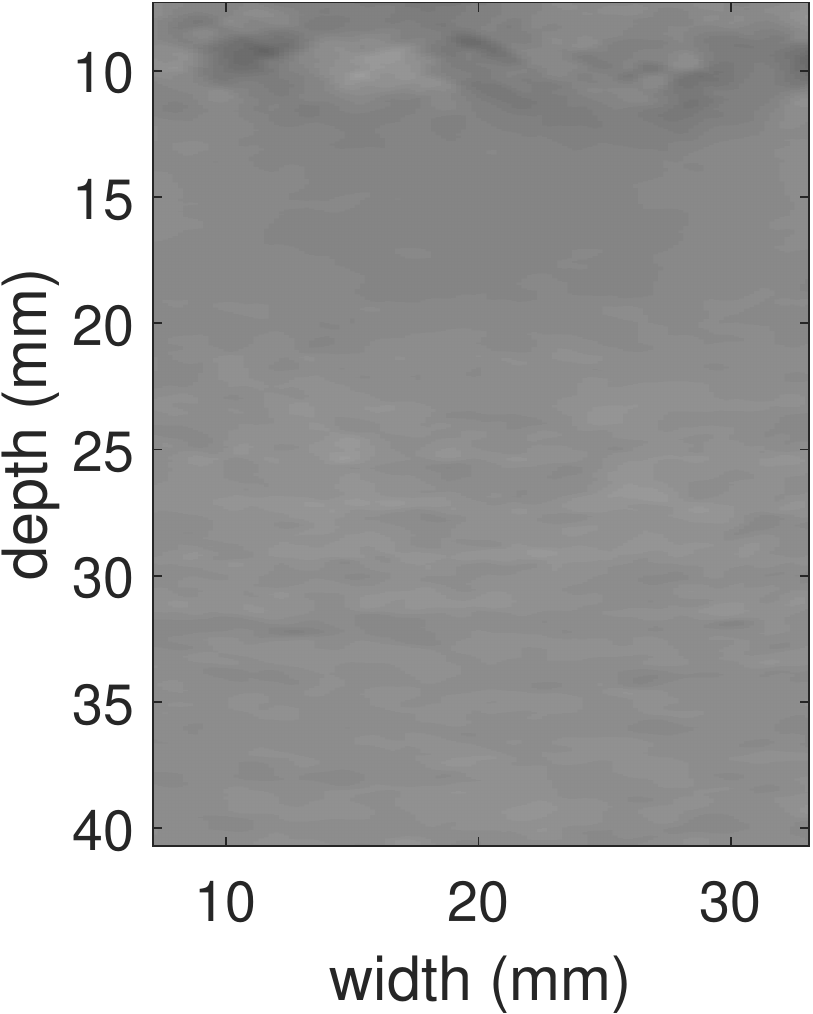}}}%
	\subfigure[$L1$-MechSOUL]{{\includegraphics[width=0.13\textwidth]{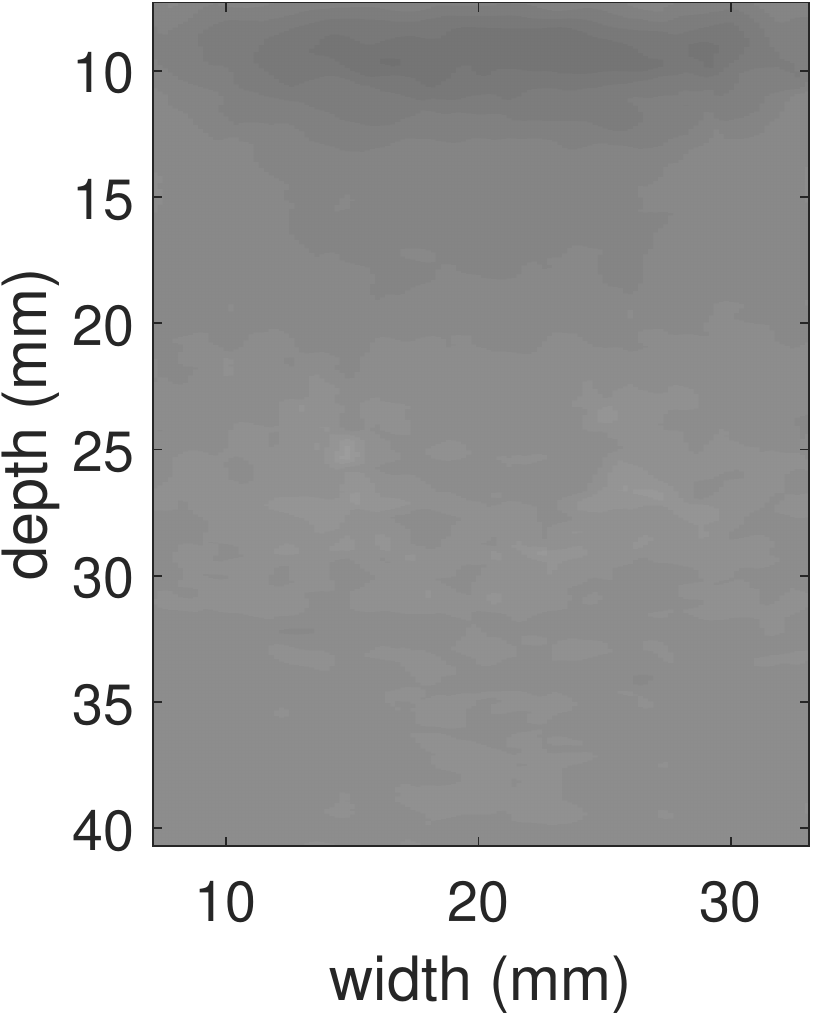}}}
	\subfigure[Ground truth]{{\includegraphics[width=0.13\textwidth]{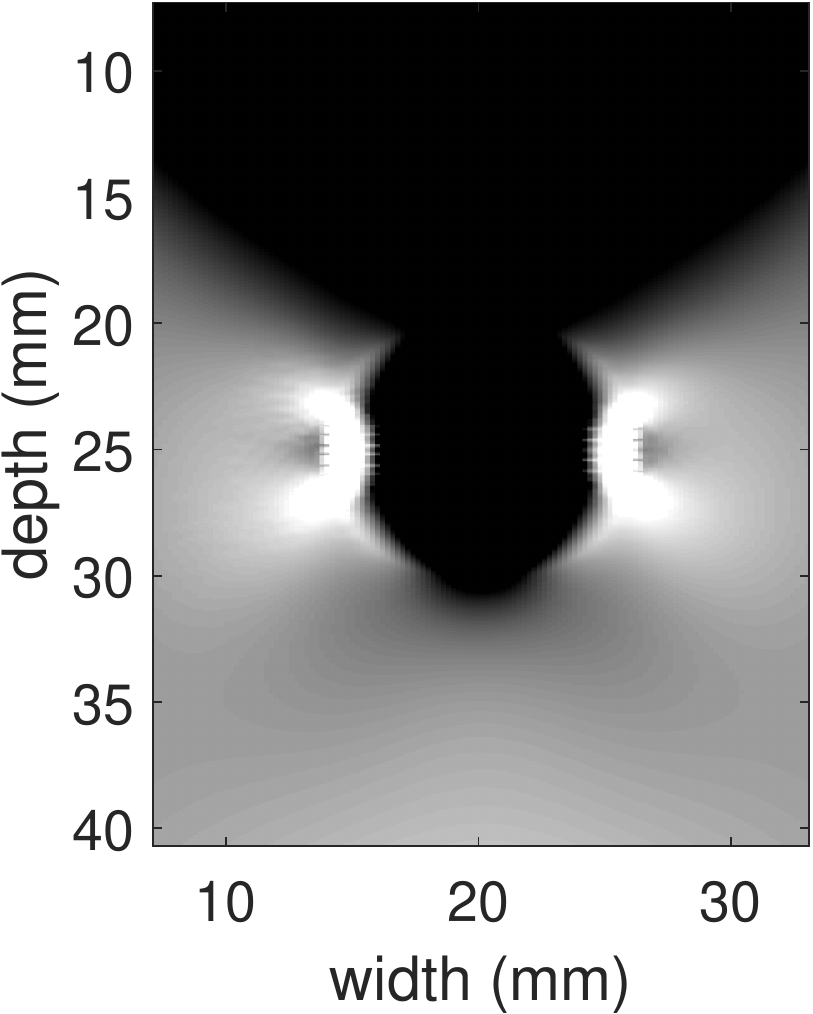}}}%
	\subfigure[NCC]{{\includegraphics[width=0.13\textwidth]{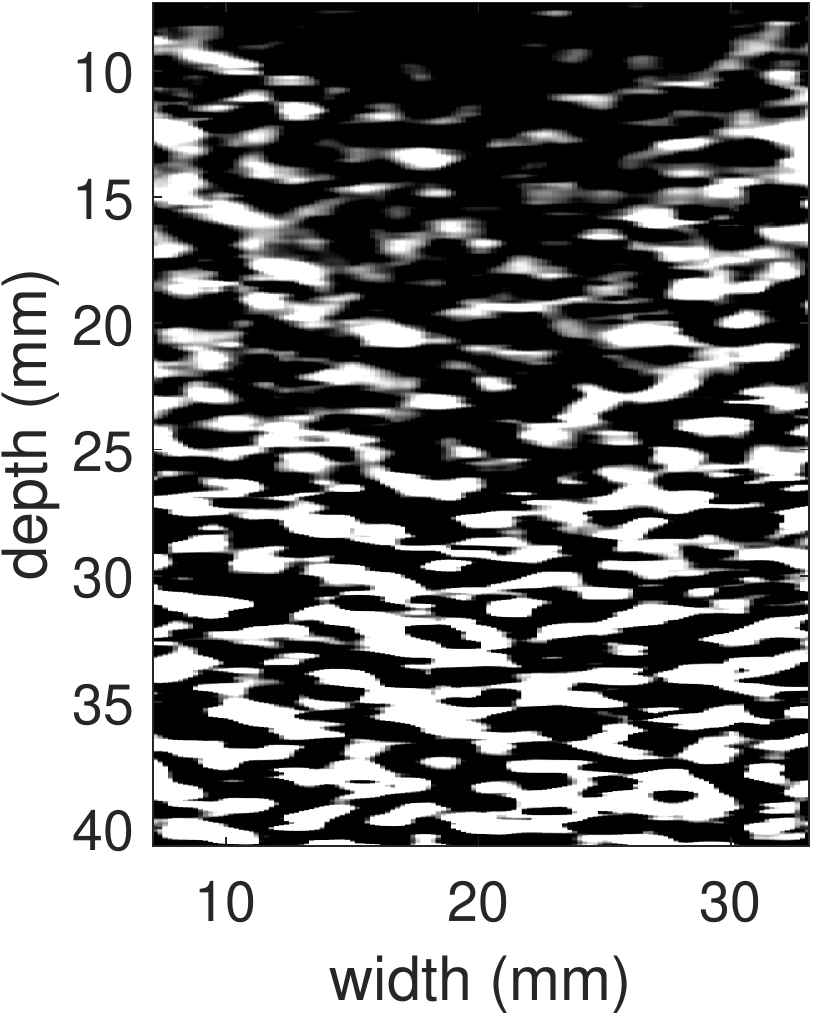}}}%
	\subfigure[NCC + PDE]{{\includegraphics[width=0.13\textwidth]{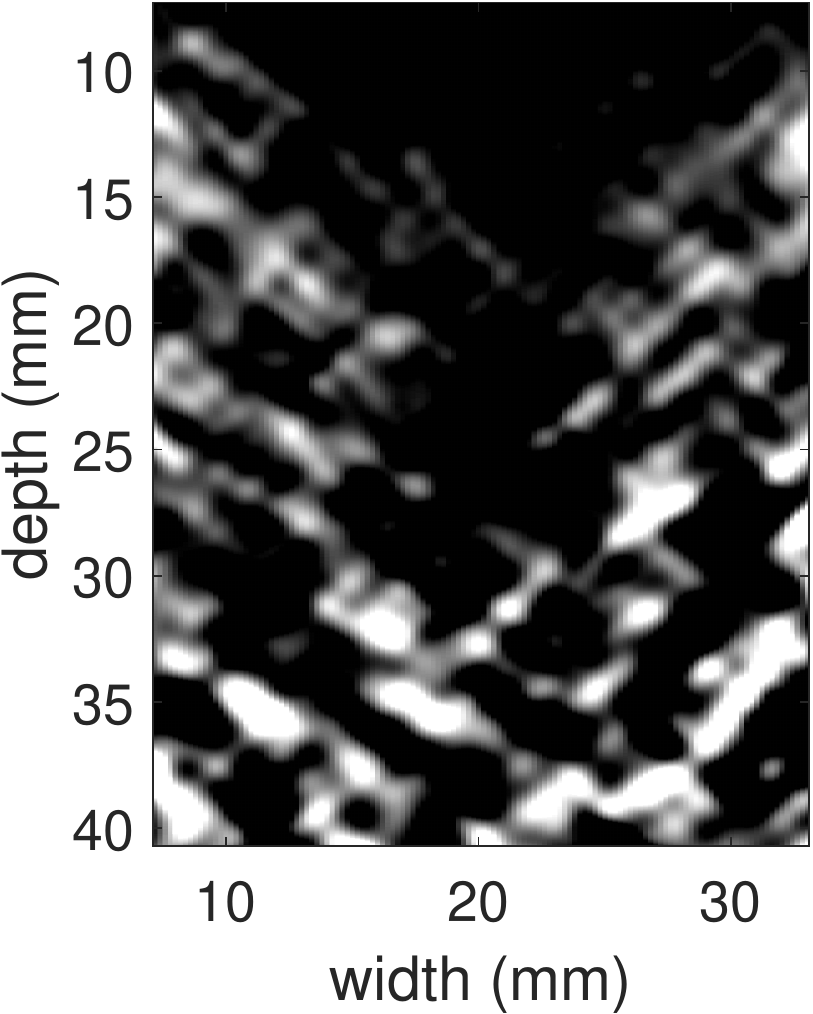}}}%
	\subfigure[SOUL]{{\includegraphics[width=0.13\textwidth]{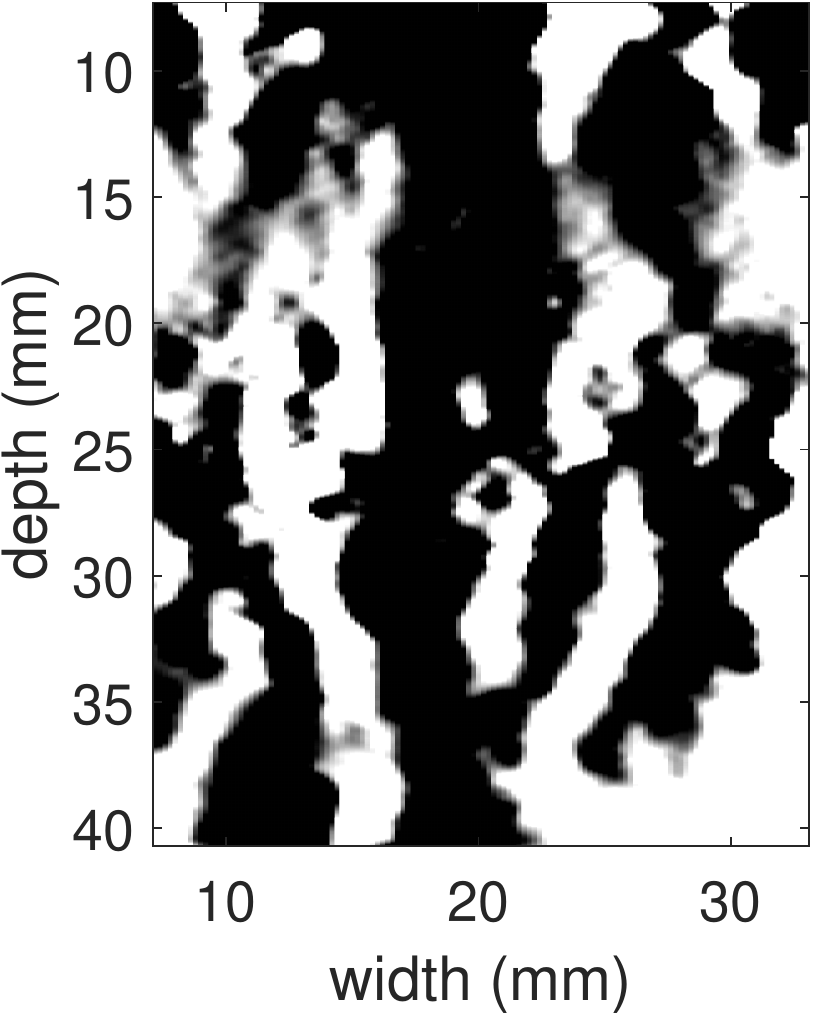}}}%
	\subfigure[$L1$-SOUL]{{\includegraphics[width=0.13\textwidth]{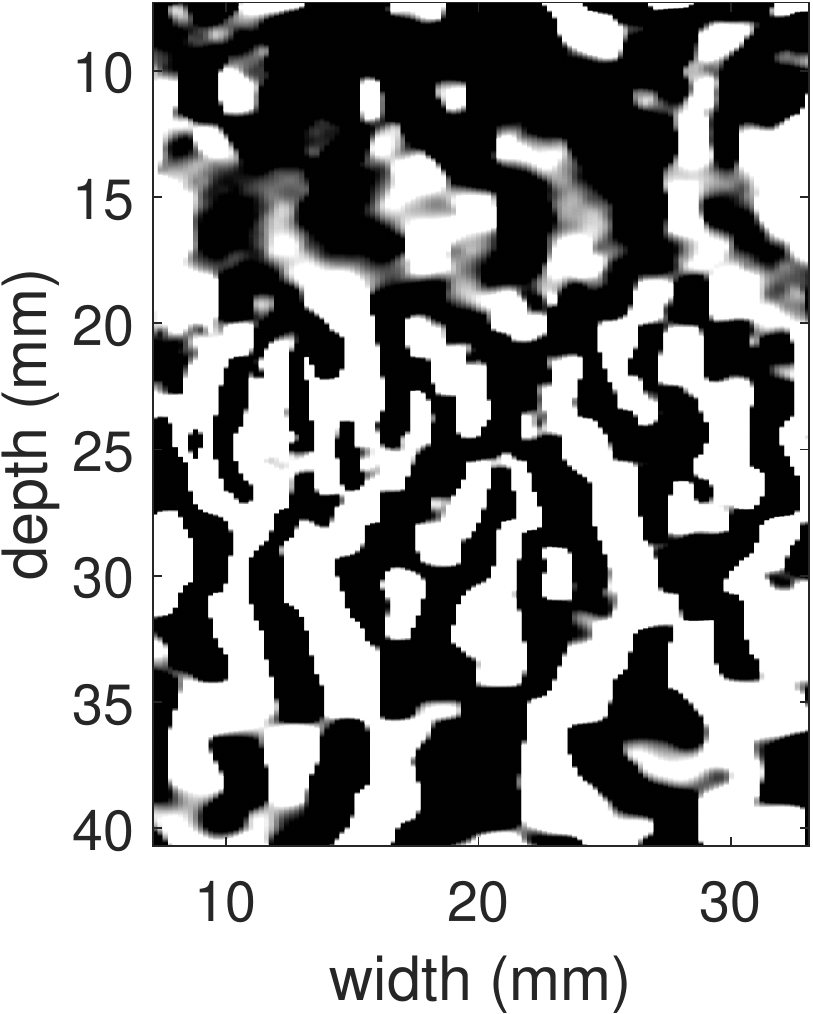}}}%
	\subfigure[MechSOUL]{{\includegraphics[width=0.13\textwidth]{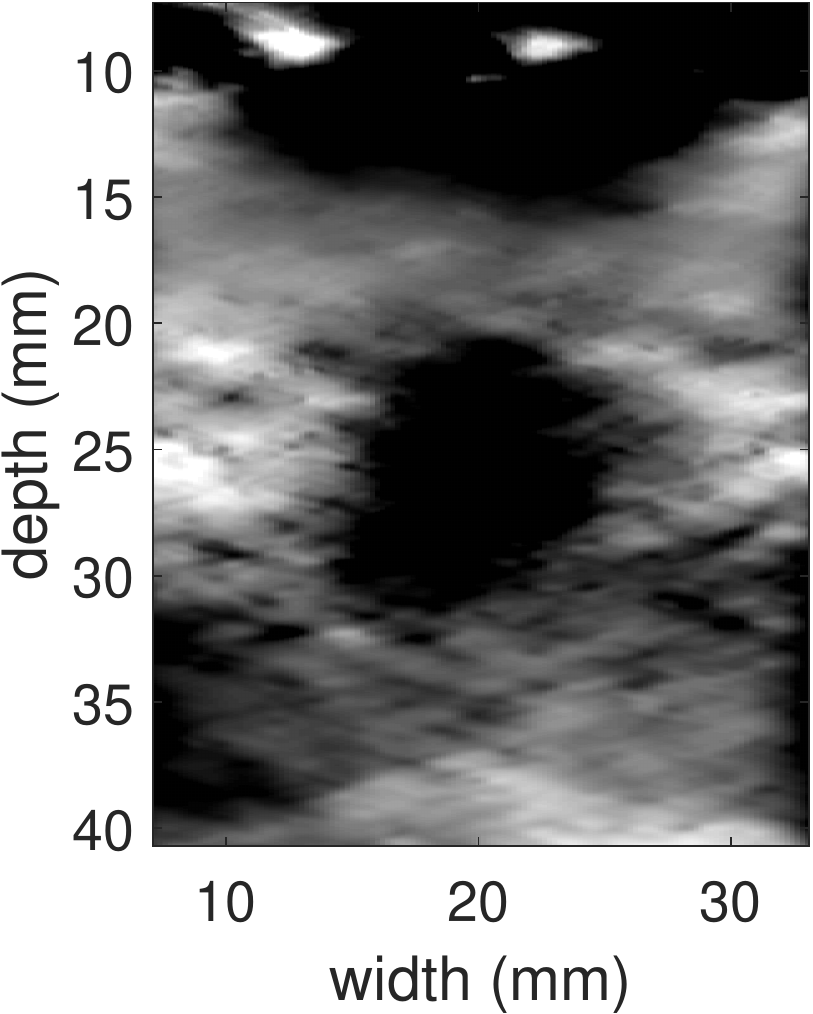}}}%
	\subfigure[$L1$-MechSOUL]{{\includegraphics[width=0.13\textwidth]{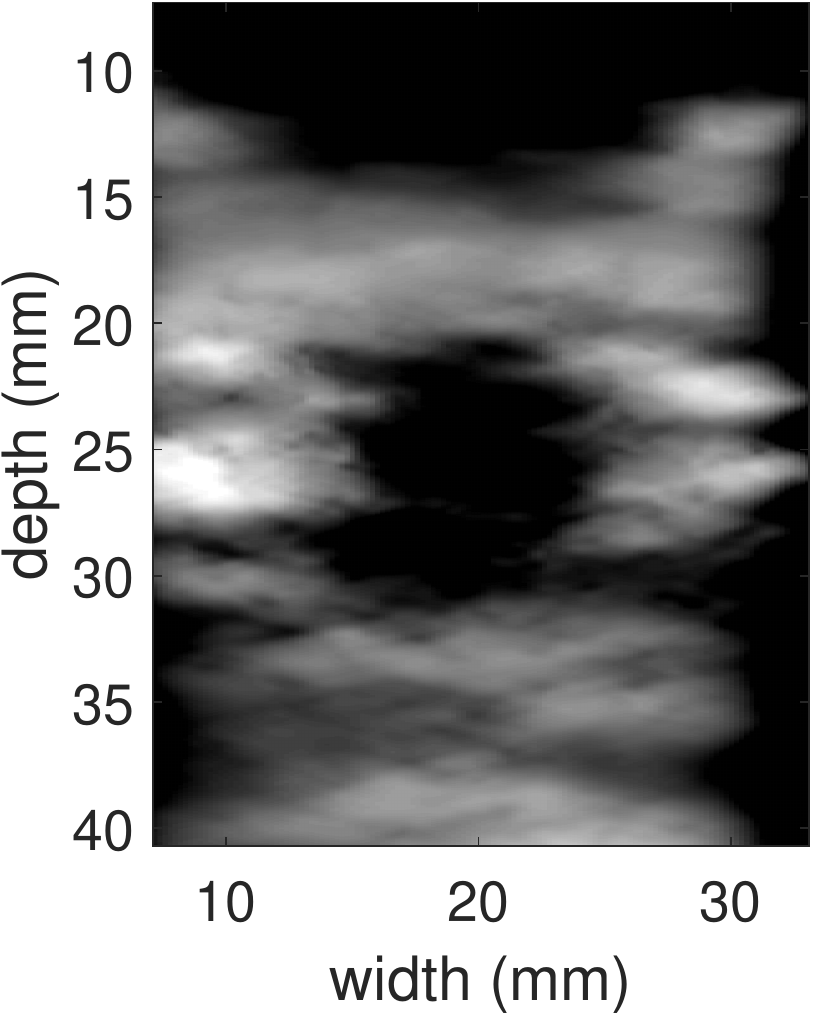}}}
	\subfigure[Ground truth]{{\includegraphics[width=0.13\textwidth]{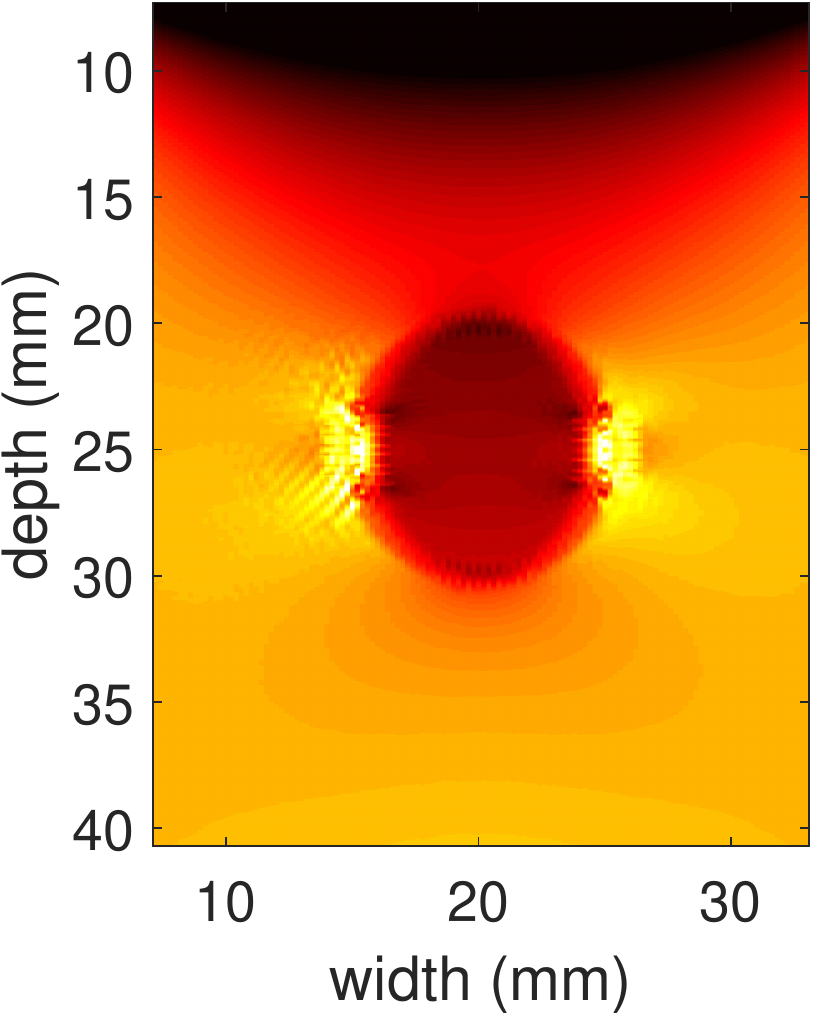}}}%
	\subfigure[NCC]{{\includegraphics[width=0.13\textwidth]{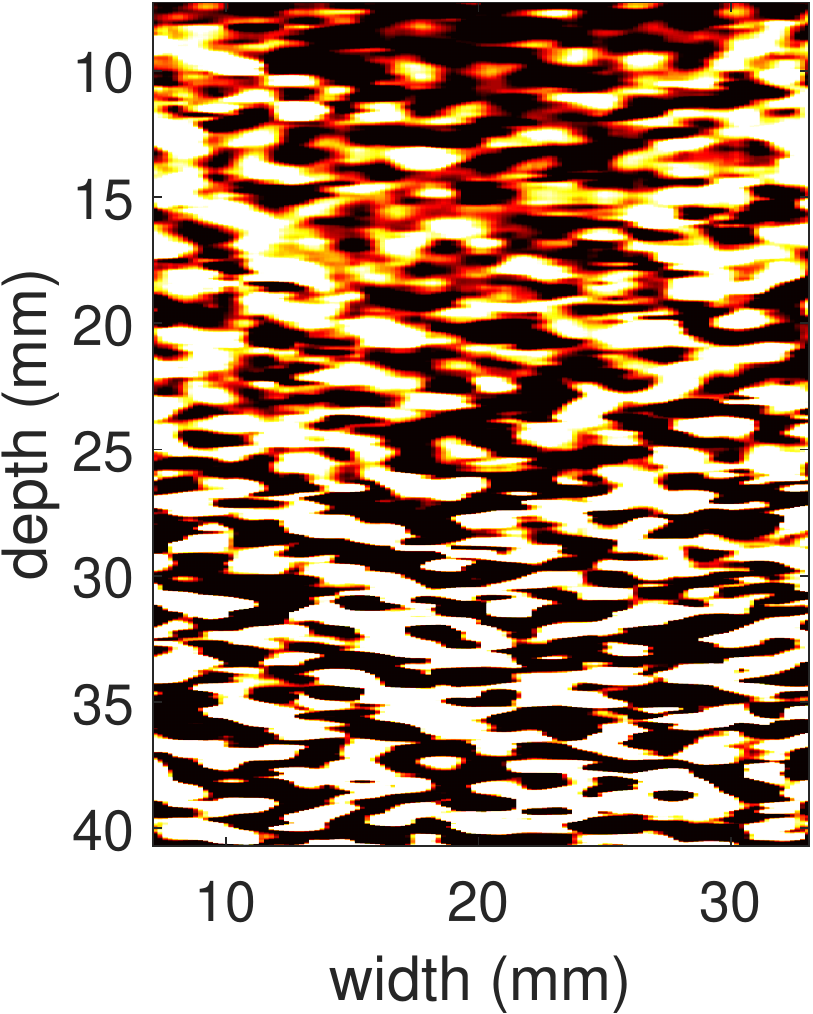}}}%
	\subfigure[NCC + PDE]{{\includegraphics[width=0.13\textwidth]{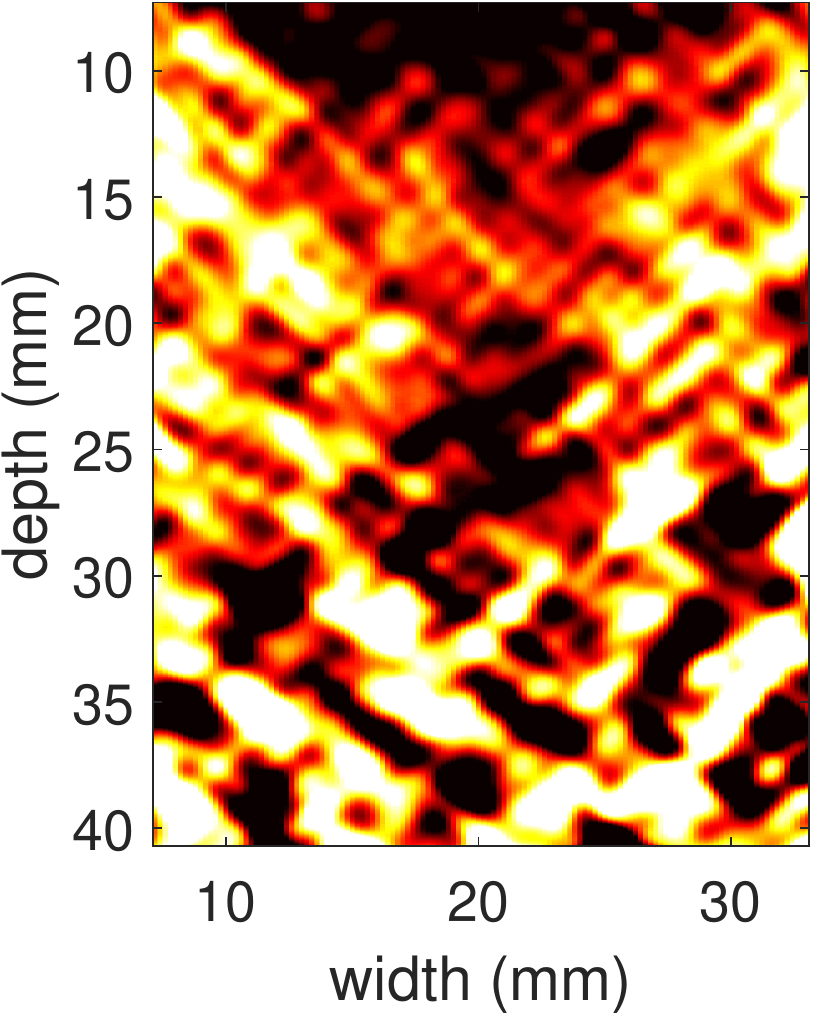}}}%
	\subfigure[SOUL]{{\includegraphics[width=0.13\textwidth]{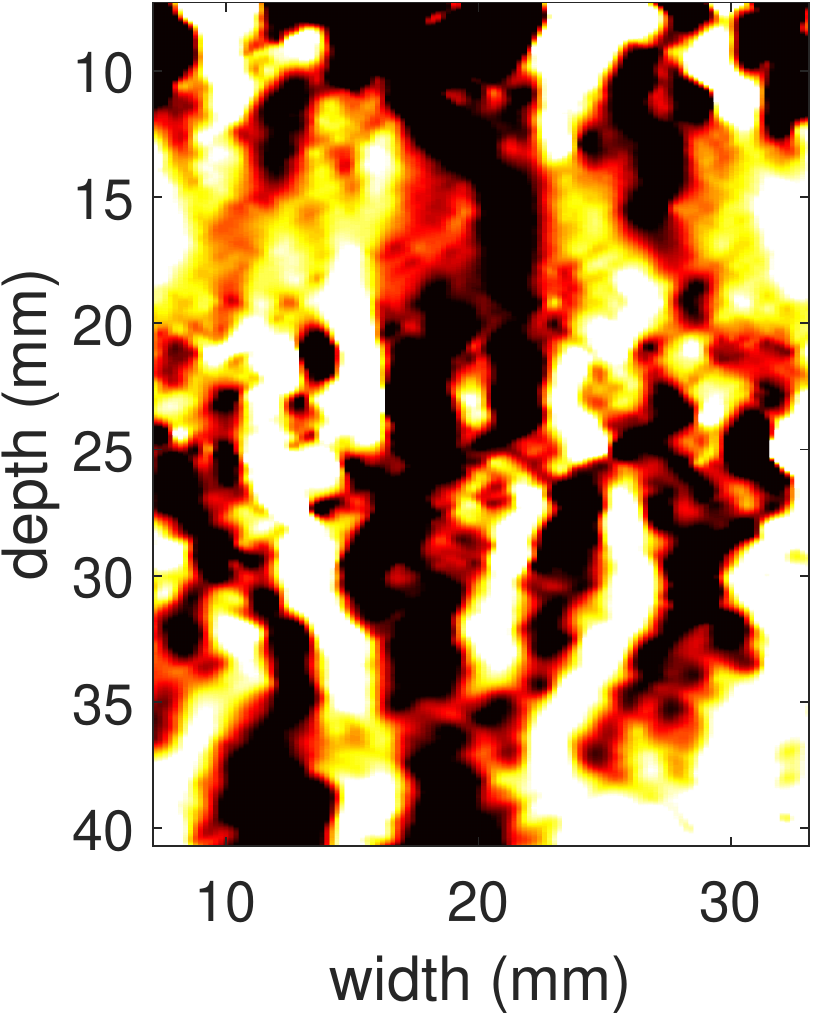}}}%
	\subfigure[$L1$-SOUL]{{\includegraphics[width=0.13\textwidth]{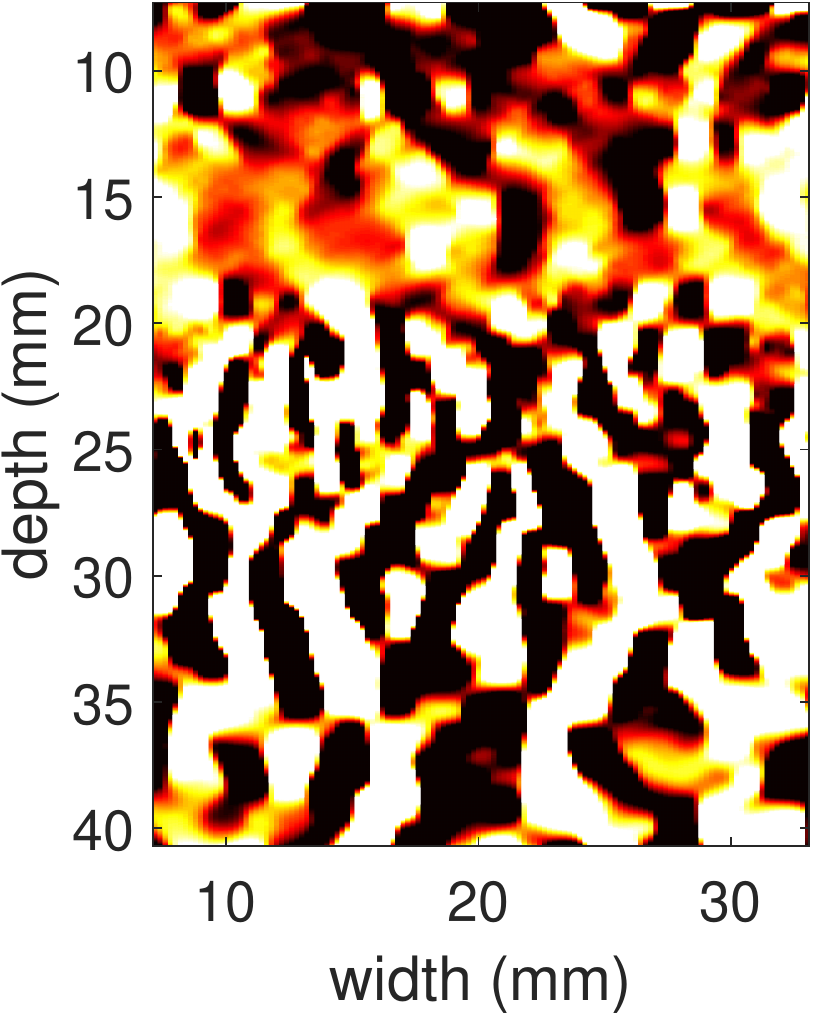}}}%
	\subfigure[MechSOUL]{{\includegraphics[width=0.13\textwidth]{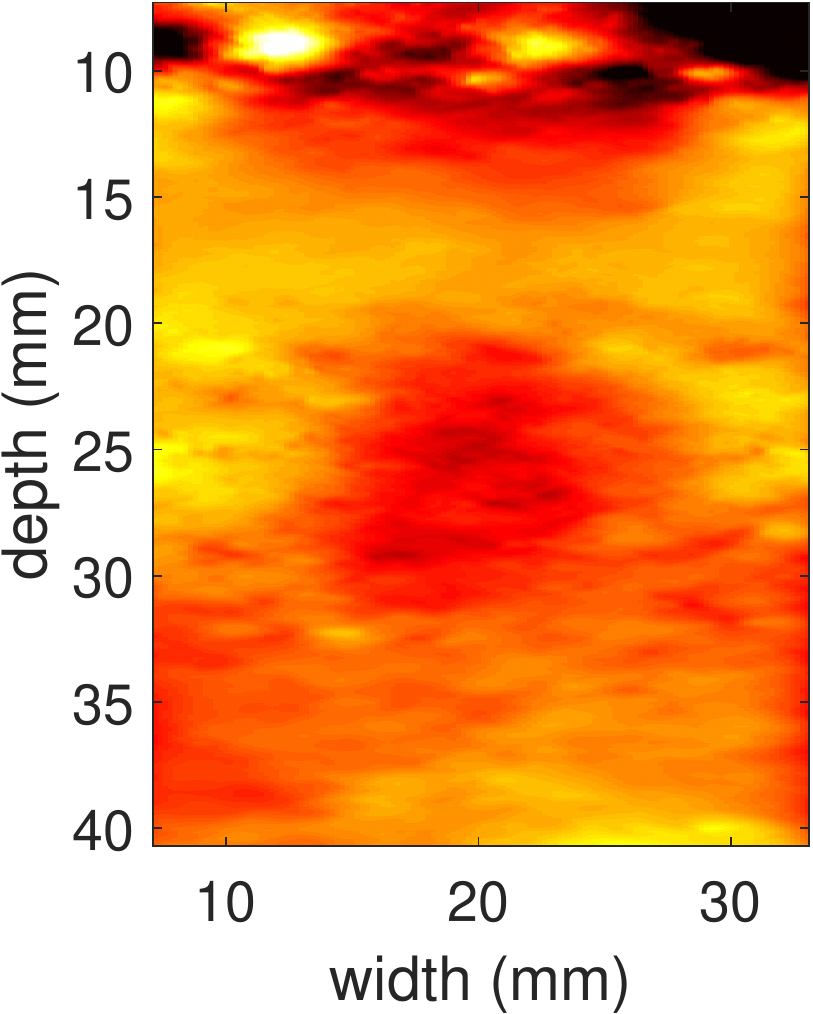}}}%
	\subfigure[$L1$-MechSOUL]{{\includegraphics[width=0.13\textwidth]{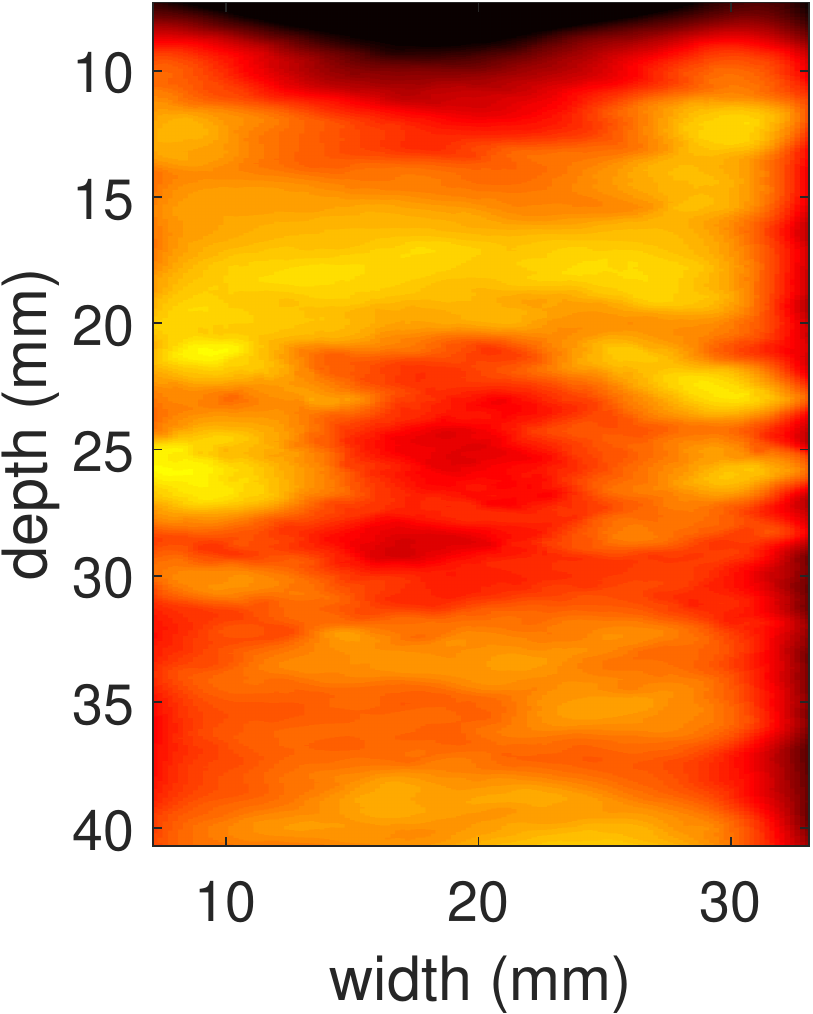}}}	
	\subfigure[Axial strain]{{\includegraphics[width=0.20\textwidth]{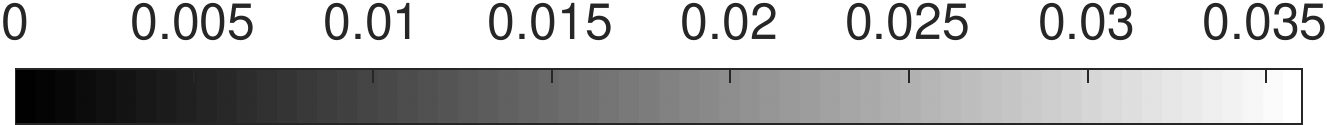}}}%
	\quad
	\subfigure[Color bar for (i) and (j)]{{\includegraphics[width=0.20\textwidth]{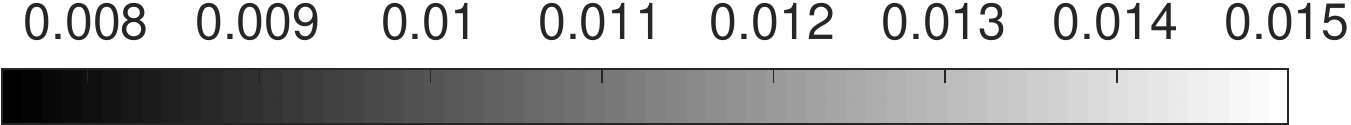}}}%
	\quad
	\subfigure[Color bar for (k), (l), (m), (n)]{{\includegraphics[width=0.20\textwidth]{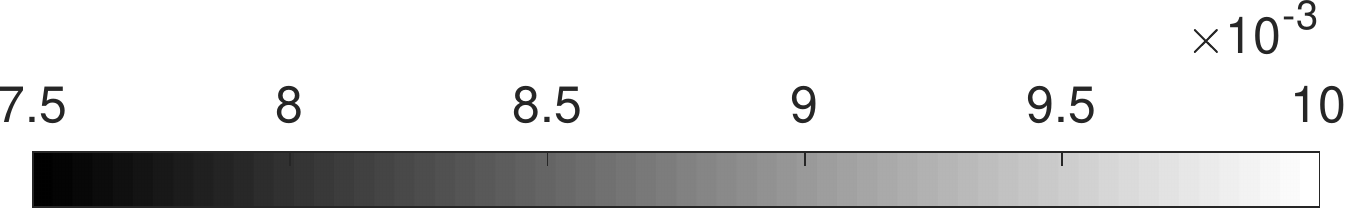}}}%
	\quad
	\subfigure[EPR]{{\includegraphics[width=0.20\textwidth]{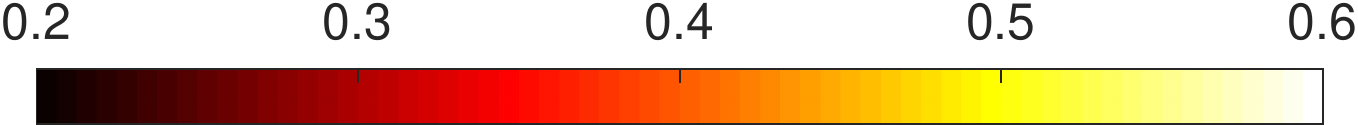}}}
	\caption{Results for the simulated phantom with different target and background PRs. Rows 1-3 show the axial and lateral strains and the EPR maps, respectively. Columns 1-7 correspond to FEM, NCC, NCC+PDE, SOUL, $L1$-SOUL, MechSOUL, and $L1$-MechSOUL, respectively.}
	\label{different_poisson_simu}
\end{figure*}

\begin{table}[tb]  
	\centering
	\caption{RMSE for the hard-inclusion simulated phantom. The best values are highlighted in bold.} 
	\label{table_rmse}
	\begin{tabular}{c c c c c c c c c} 
		\hline
		$ $  $ $&  Axial & Lateral & EPR\\
		\hline
		NCC & $3.2 \times 10^{-3}$ &  $2.59 \times 10^{-2}$ & 1.37\\
		NCC + PDE & $1.7 \times 10^{-3}$ &  $9.4 \times 10^{-3}$ & 0.52\\
		SOUL & $8.61 \times 10^{-4}$ &  $7.4 \times 10^{-3}$ & 0.39\\
		$L1$-SOUL & $7.36 \times 10^{-4}$ &  $1.29 \times 10^{-2}$ & 0.68\\
		MechSOUL & $7.41 \times 10^{-4}$ &  $\mathbf{1.1 \times 10^{-3}}$ & \textbf{0.05}\\
		$L1$-MechSOUL & $\mathbf{7.35 \times 10^{-4}}$  & $\mathbf{1.1 \times 10^{-3}}$ & 0.06\\
		\hline
	\end{tabular}
\end{table}

\begin{table}[h]  
	\centering
	\caption{PSNR (dB) for the hard-inclusion simulated phantom.}
	\label{table_psnr}
	\begin{tabular}{c c c c c c c c c} 
		\hline
		$ $  $ $&  Axial & Lateral & EPR\\
		\hline
		NCC & 49.94 &  31.73 & -2.74\\\
		NCC + PDE & 55.18 &  40.51 & 5.73\\
		SOUL & 61.30 &  42.62 & 8.08\\
		$L1$-SOUL & 62.66 &  37.82 & 3.33\\
		MechSOUL & 62.60 & 59.29 & \textbf{25.23}\\
		$L1$-MechSOUL & \textbf{62.67}  & \textbf{59.39} & 25.07\\
		\hline
	\end{tabular}
\end{table}

\begin{table}[tb]  
	\centering
	\caption{RMSE for the multi-inclusion simulated phantom with an additional lateral boundary condition.} 
	\label{table_rmse_boundary}
	\begin{tabular}{c c c c c c c c c} 
		\hline
		$ $  $ $&  Axial & Lateral & EPR\\
		\hline
		NCC & $3.6 \times 10^{-3}$ &  $4.8 \times 10^{-2}$ & 2.41\\
		NCC + PDE & $2.4 \times 10^{-3}$ &  $1.7 \times 10^{-2}$ & 0.93\\
		SOUL & $2.2 \times 10^{-3}$ &  $1.2 \times 10^{-2}$ & 0.61\\
		$L1$-SOUL & $\mathbf{1.9 \times 10^{-3}}$ &  $1.4 \times 10^{-2}$ & 0.77\\
		MechSOUL & $2 \times 10^{-3}$ &  $1.8 \times 10^{-3}$ & \textbf{0.09}\\
		$L1$-MechSOUL & $\mathbf{1.9 \times 10^{-3}}$  & $\mathbf{1.7 \times 10^{-3}}$ & \textbf{0.09}\\
		\hline
	\end{tabular}
\end{table}

\begin{table}[h]
	\centering
	\caption{PSNR (dB) for the multi-inclusion simulated phantom with an additional lateral boundary condition.}
	\label{table_psnr_boundary}
	\begin{tabular}{c c c c c c c c c} 
		\hline
		$ $  $ $&  Axial & Lateral & EPR\\
		\hline
		NCC & 48.95 &  26.41 & -7.65\\
		NCC + PDE & 52.47 & 35.44 & 0.64\\
		SOUL & 53.08 &  38.59 & 4.23\\
		$L1$-SOUL & \textbf{54.59} &  36.86 & 2.27\\
		MechSOUL & 53.81 & 54.78 & \textbf{20.88}\\
		$L1$-MechSOUL & \textbf{54.59}  & \textbf{55.15} & 20.87\\
		\hline
	\end{tabular}
\end{table}

\begin{table}[tb]  
	\centering
	\caption{RMSE for the different PR simulated phantom.}
	\label{table_rmse_diffp}
	\begin{tabular}{c c c c c c c c c} 
		\hline
		$ $  $ $&  Axial & Lateral & EPR\\
		\hline
		NCC & $2.1 \times 10^{-3}$ &  $1.3 \times 10^{-2}$ & 0.65\\
		NCC + PDE & $6.93 \times 10^{-4}$ &  $3.6 \times 10^{-3}$ & 0.18\\
		SOUL & $5.83 \times 10^{-4}$ &  $5.7 \times 10^{-3}$ & 0.30\\
		$L1$-SOUL & $5.72 \times 10^{-4}$ &  $1.16 \times 10^{-2}$ & 0.58\\
		MechSOUL & $8.42 \times 10^{-4}$ &  $1.7 \times 10^{-3}$ & 0.09\\
		$L1$-MechSOUL & $\mathbf{5.17 \times 10^{-4}}$  & $\mathbf{1.5 \times 10^{-3}}$ & \textbf{0.08}\\
		\hline
	\end{tabular}
\end{table}

\begin{table}[h]
	\centering
	\caption{PSNR (dB) for the different PR simulated phantom.}
	\label{table_psnr_diffp}
	\begin{tabular}{c c c c c c c c c} 
		\hline
		$ $  $ $&  Axial & Lateral & EPR\\
		\hline
		NCC & 53.50 &  37.74 & 3.79\\
		NCC + PDE & 63.19 &  48.83 & 15.15\\
		SOUL & 64.68 &  44.88 & 10.52\\
		$L1$-SOUL & 64.85 &  38.71 & 4.79\\
		MechSOUL & 61.49 & 55.59 & 21.32\\
		$L1$-MechSOUL & \textbf{65.73}  & \textbf{56.53} & \textbf{22.09}\\
		\hline
	\end{tabular}
\end{table}

\begin{table*}[tb]  
	\centering
	\caption{SNR and CNR values for the hard-inclusion simulated phantom dataset. The best values are highlighted in bold.}
	\label{table_hard_simu}
	\begin{tabular}{c c c c c c c c c c} 
		\hline
		\multicolumn{1}{c}{} &
		\multicolumn{3}{c}{SNR} &
		\multicolumn{1}{c}{} &
		\multicolumn{3}{c}{CNR} \\
		\cline{2-4} 
		\cline{6-8}
		$ $  $ $&    Axial & Lateral & EPR $ $  $ $&$ $  $ $ &$ $  $ $ Axial & Lateral & EPR\\
		\hline
		NCC &  7.61 $\pm$ 2.28 & 0.27 $\pm$ 0.31 & 0.28 $\pm$ 0.31 && 4.04 $\pm$ 1.14 & 0.34 $\pm$ 0.26 & 0.39 $\pm$ 0.29\\
		NCC + PDE &  18.04 $\pm$ 4.75 & 1.15 $\pm$ 1.11 & 1.20 $\pm$ 1.17 && 10.07 $\pm$ 2.14 & 1.07 $\pm$ 0.88 & 0.91 $\pm$ 0.92\\
		SOUL &  45.32 $\pm$ 9.68 & 2.15 $\pm$ 1.56 & 2.16 $\pm$ 1.57 && 22.28 $\pm$ 3.61 & 1.27 $\pm$ 0.76 & 0.57 $\pm$ 0.69\\
		$L1$-SOUL & \textbf{61.42} $\pm$ 22.90 &  1.14 $\pm$ 1.04 & 1.14 $\pm$ 1.03 && 26.38 $\pm$ 3.99 & 0.73 $\pm$ 0.66 & 0.76 $\pm$ 0.60\\
		MechSOUL & 51.40 $\pm$ 12.78 & \textbf{39.84} $\pm$ 12.41 & \textbf{44.88} $\pm$ 13.62 && 26.20 $\pm$ 4.67 & \textbf{13.01} $\pm$ 4.33 & \textbf{3.48} $\pm$ 2.48\\
		$L1$-MechSOUL & 60.59 $\pm$ 21.20 &  37.72 $\pm$ 13.73 & 43.16 $\pm$ 19.91 && \textbf{27.13} $\pm$ 4.10 & 12.67 $\pm$ 4.09 & 3.27 $\pm$ 2.40\\
		\hline
	\end{tabular}
\end{table*}

\begin{figure*}
	\begin{center}
		\subfigure[B-mode]{{\includegraphics[width=0.1428\textwidth]{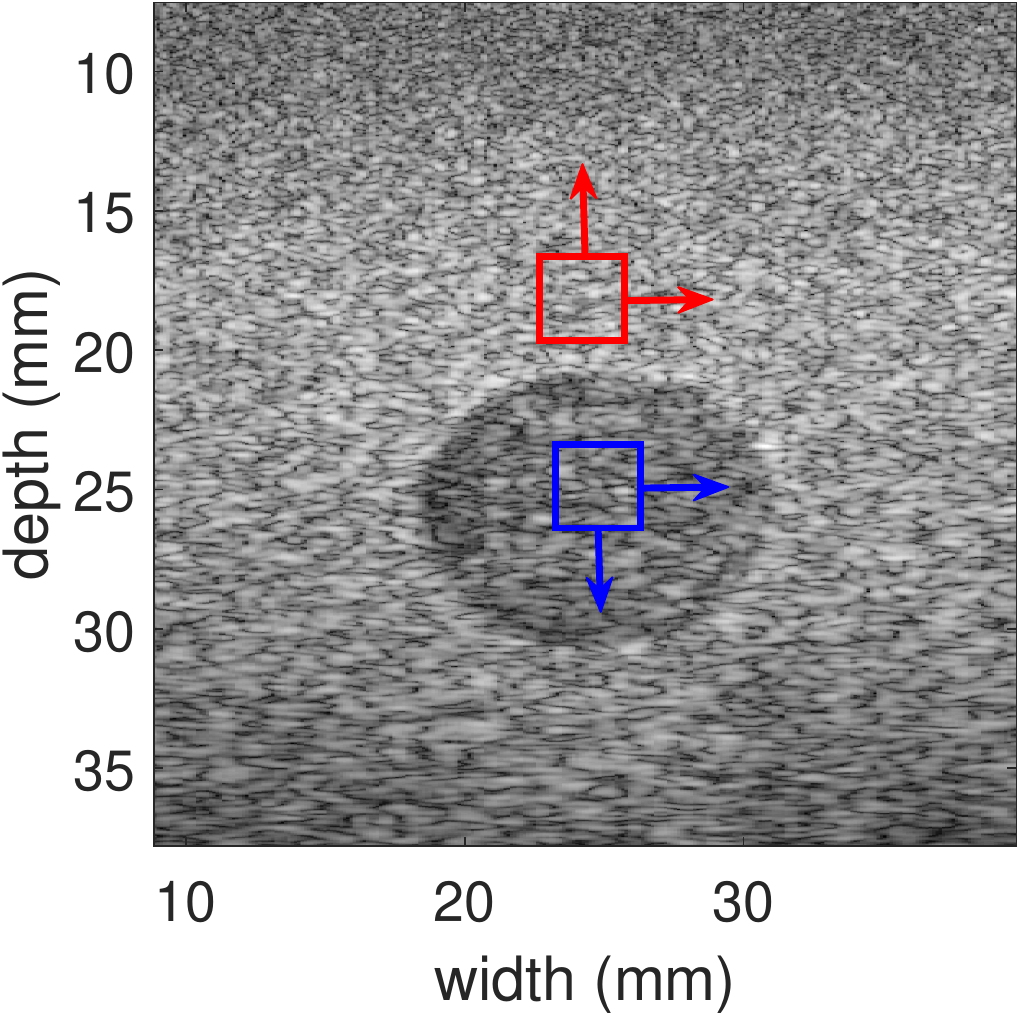}}}%
		\subfigure[NCC]{{\includegraphics[width=0.1428\textwidth]{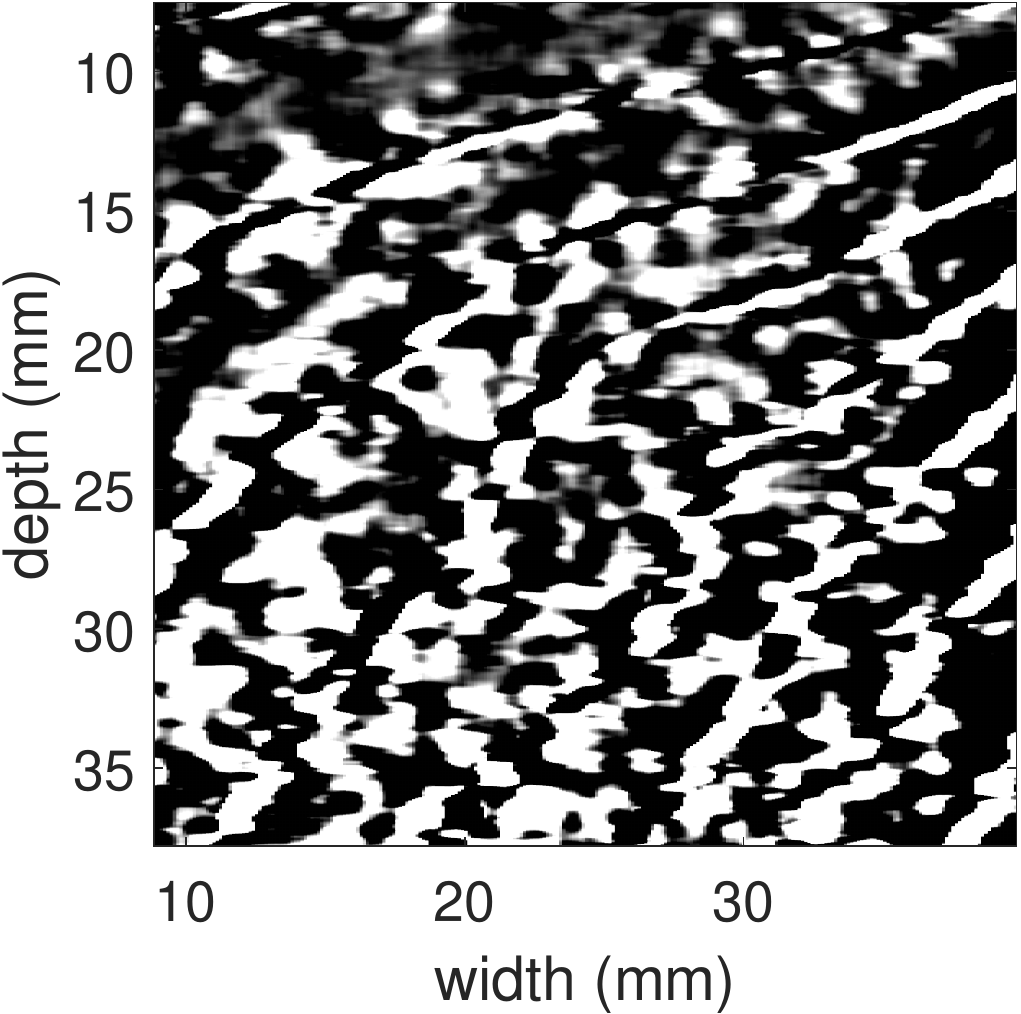}}}%
		\subfigure[NCC + PDE]{{\includegraphics[width=0.1428\textwidth]{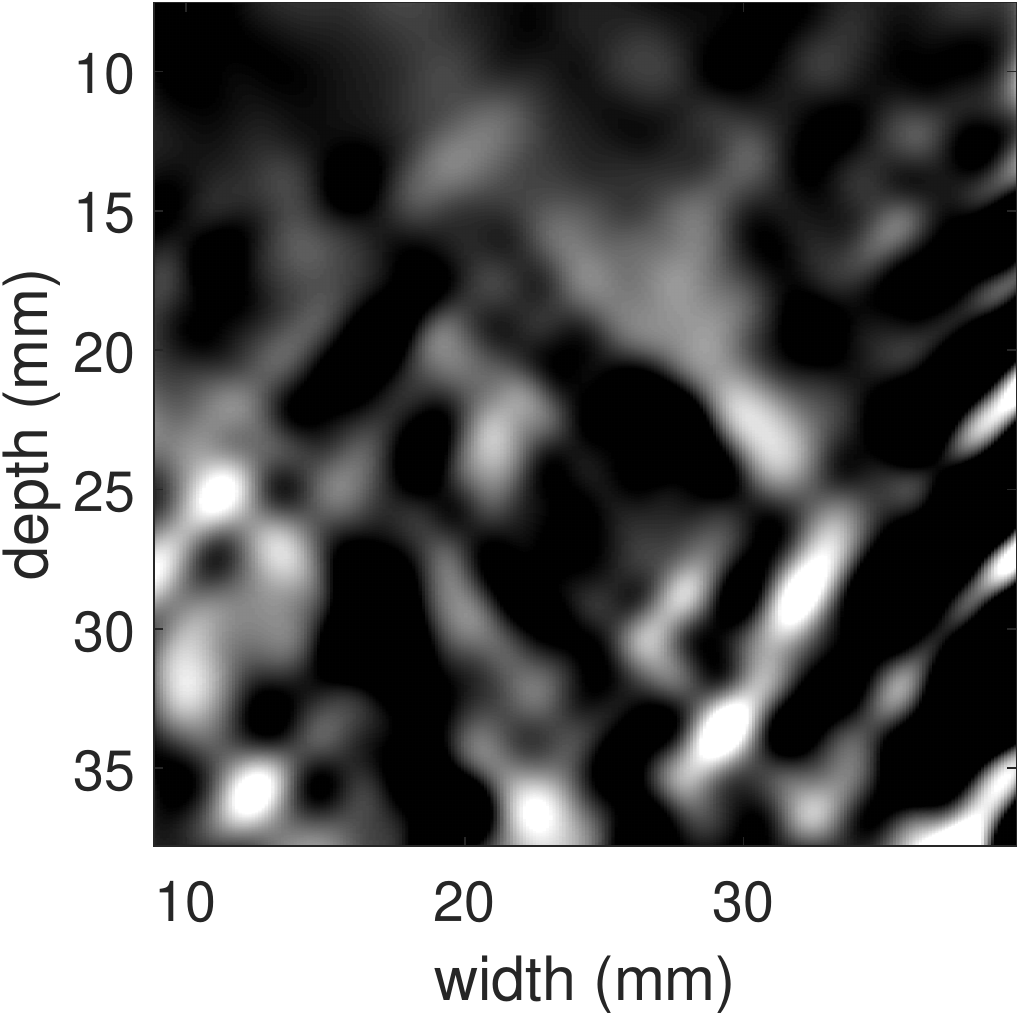}}}%
		\subfigure[SOUL]{{\includegraphics[width=0.1428\textwidth]{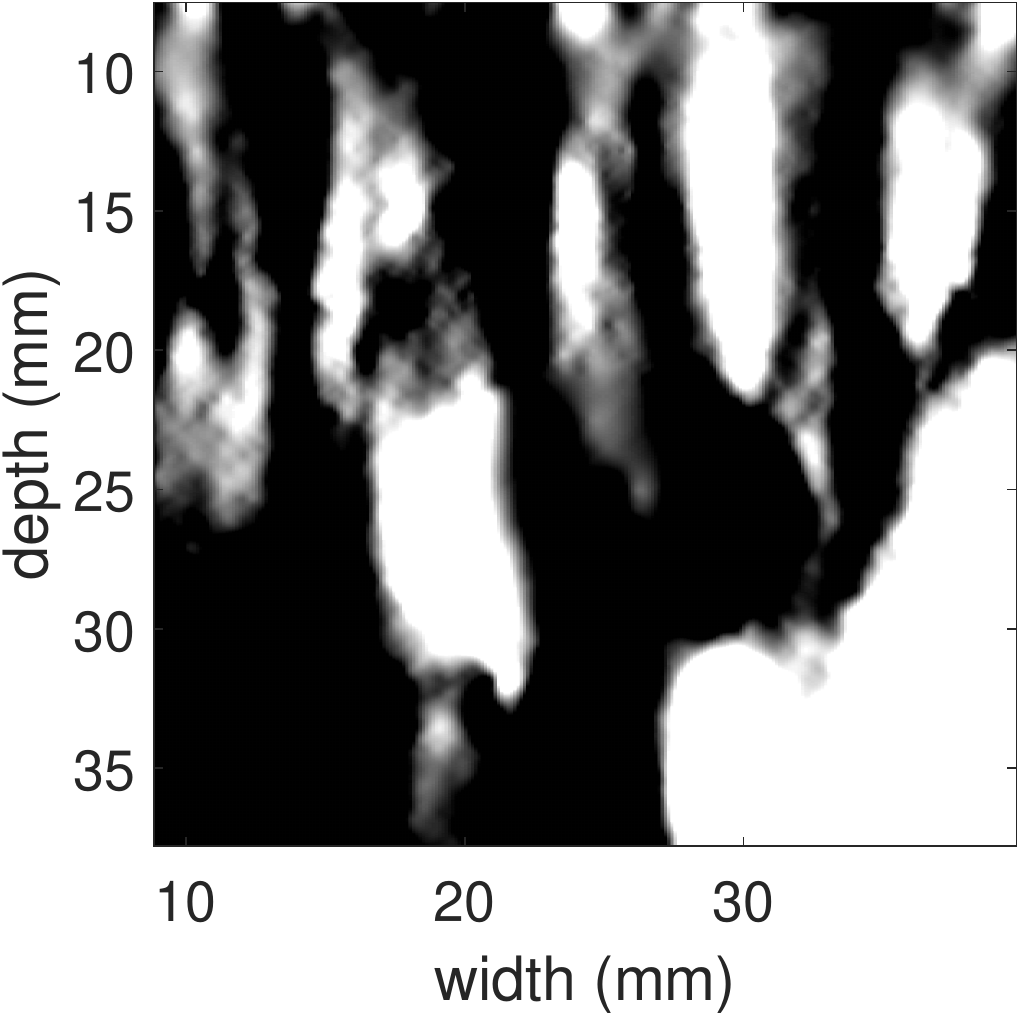}}}%
		\subfigure[$L1$-SOUL]{{\includegraphics[width=0.1428\textwidth]{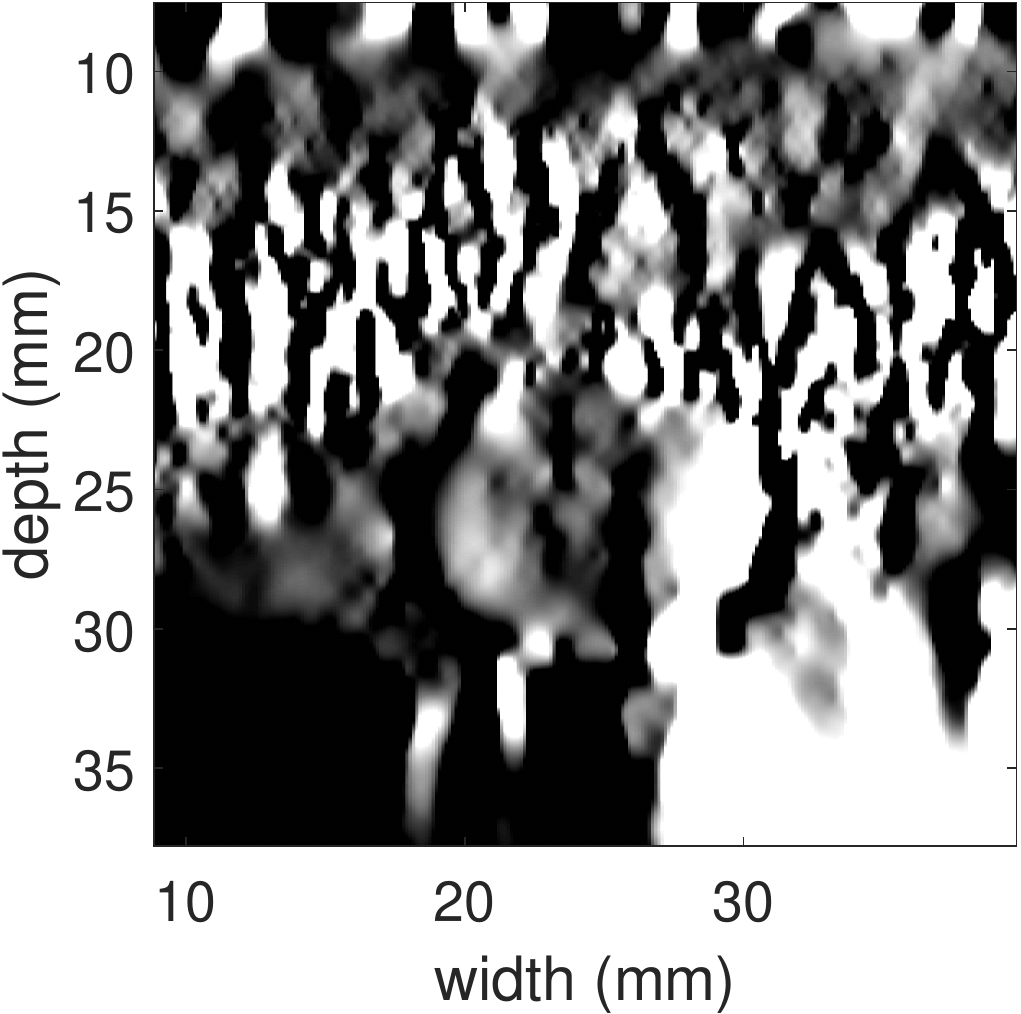}}}%
		\subfigure[MechSOUL]{{\includegraphics[width=0.1428\textwidth]{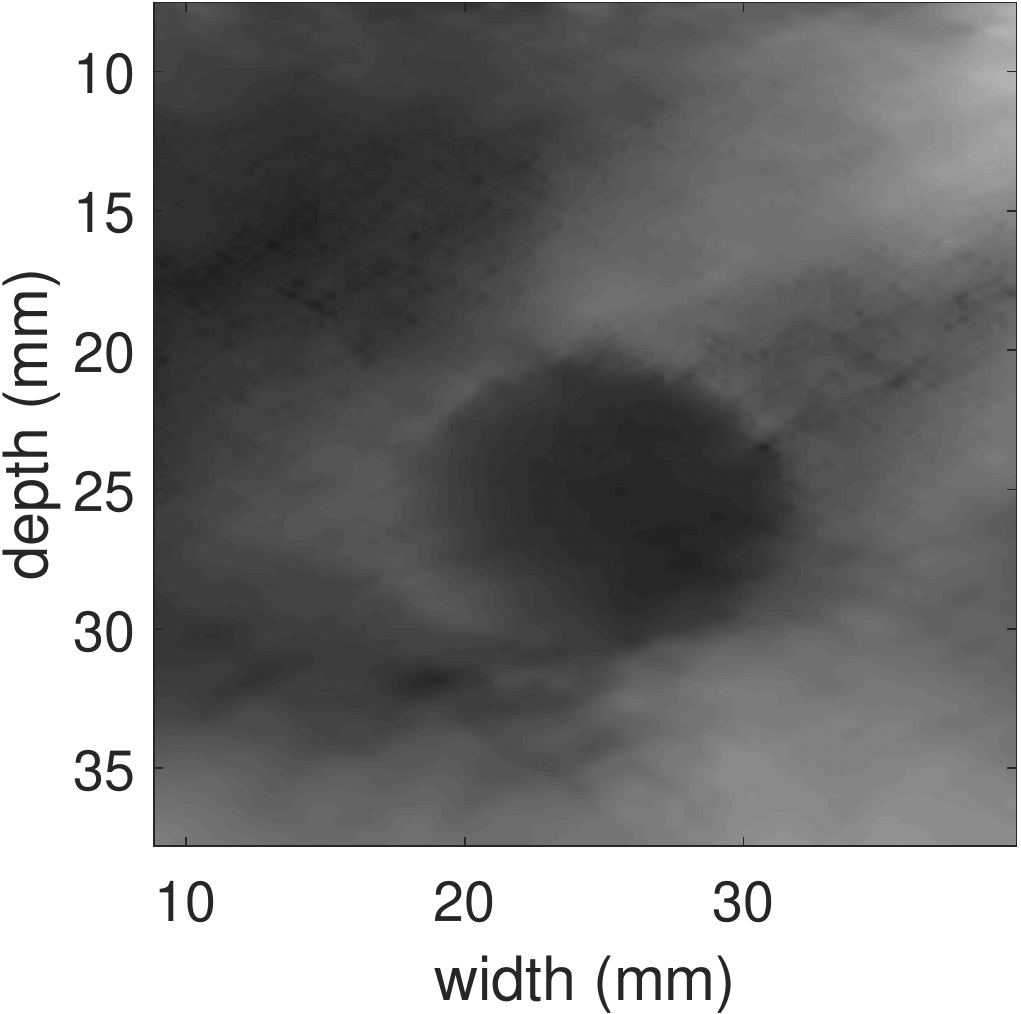}}}%
		\subfigure[$L1$-MechSOUL]{{\includegraphics[width=0.1428\textwidth]{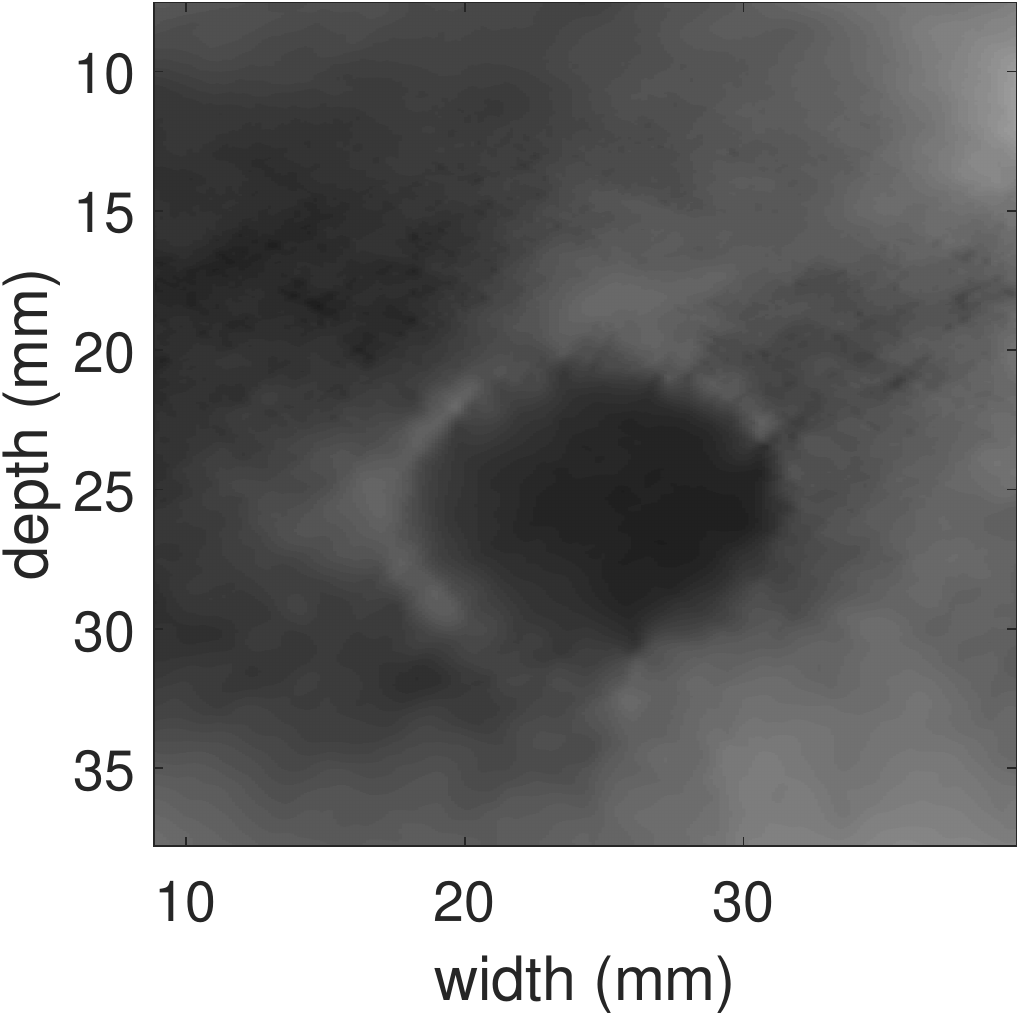}}}
		\begin{flushright}
			\subfigure[NCC]{{\includegraphics[width=0.1428\textwidth]{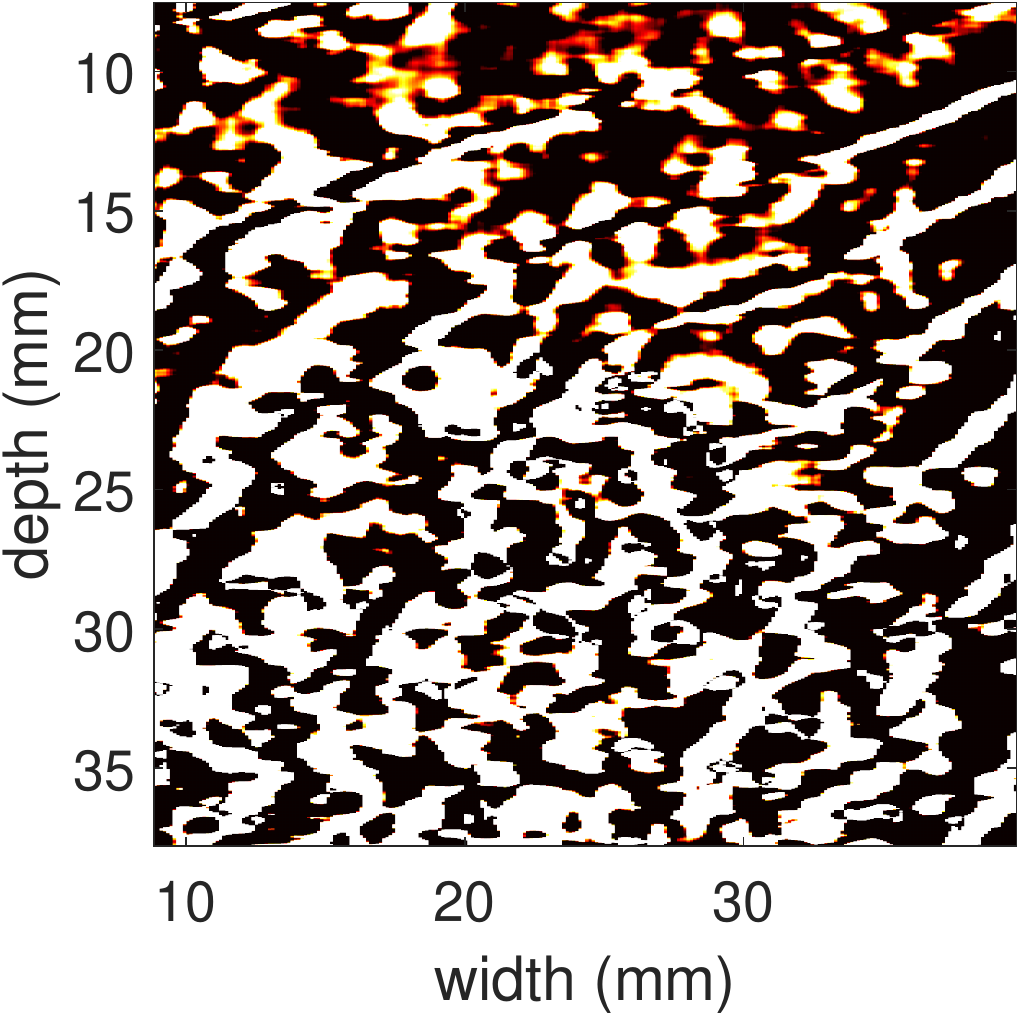}}}%
			\subfigure[NCC + PDE]{{\includegraphics[width=0.1428\textwidth]{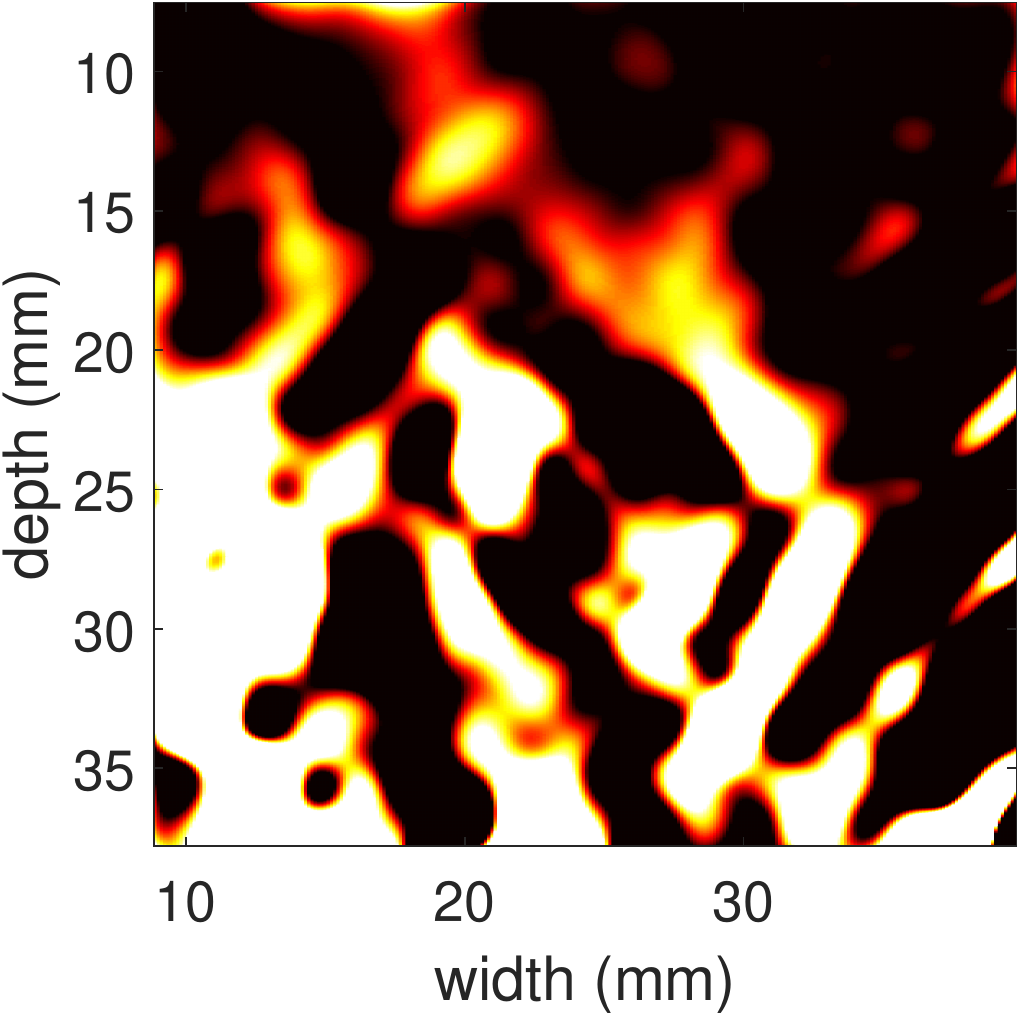}}}%
			\subfigure[SOUL]{{\includegraphics[width=0.1428\textwidth]{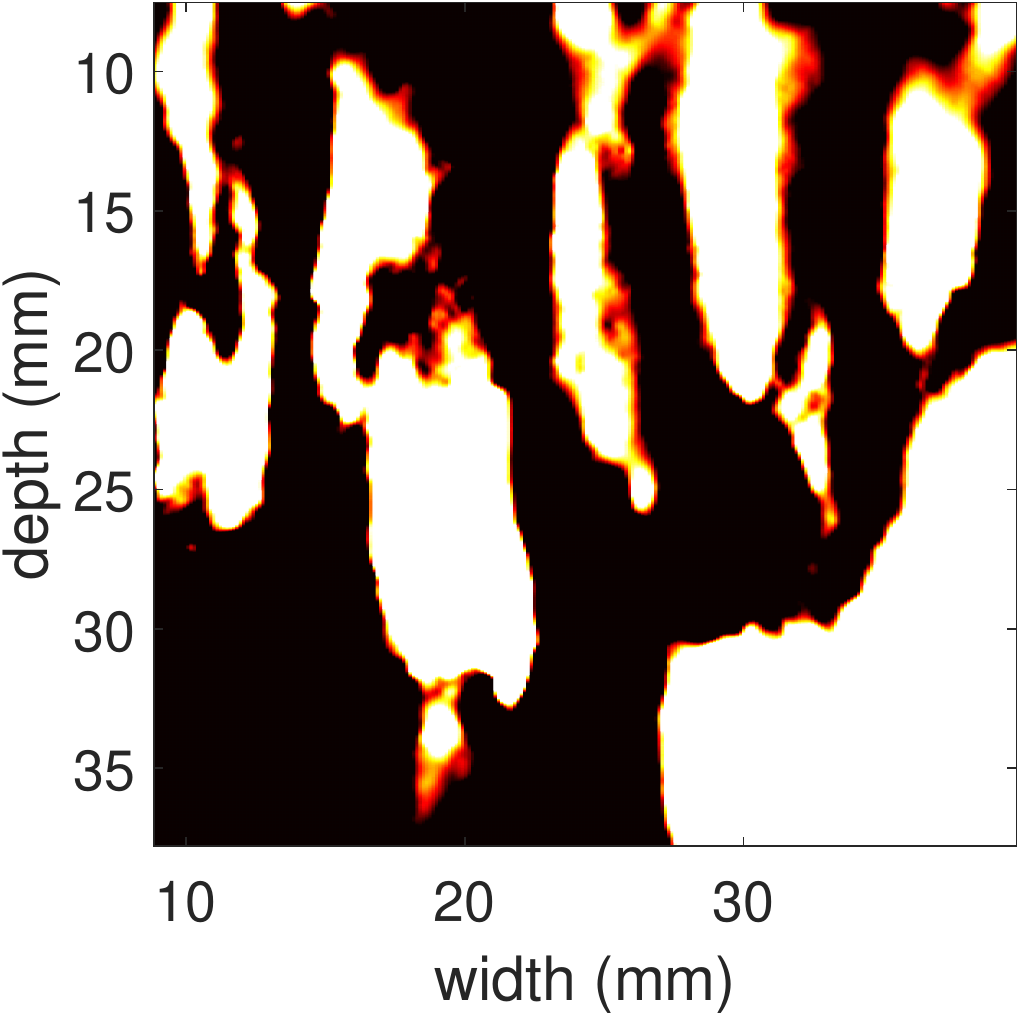}}}%
			\subfigure[$L1$-SOUL]{{\includegraphics[width=0.1428\textwidth]{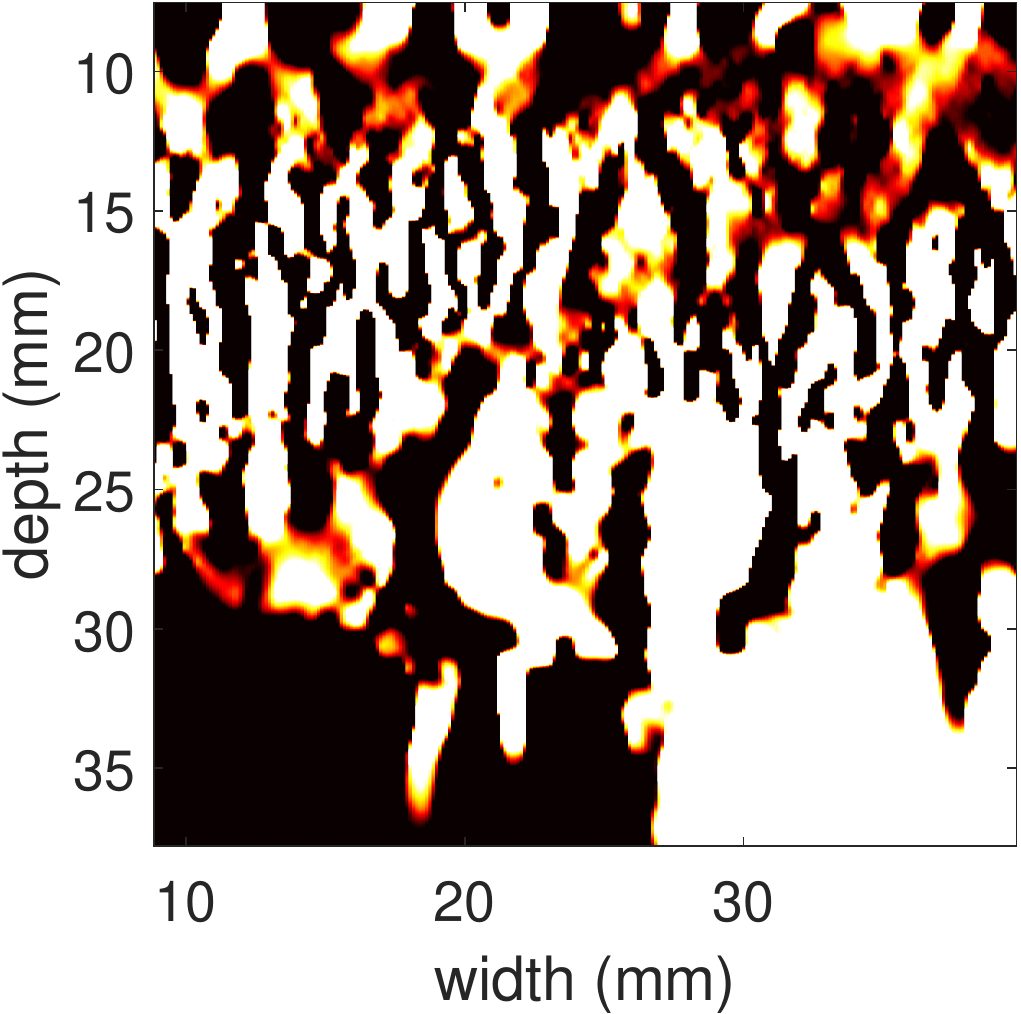}}}%
			\subfigure[MechSOUL]{{\includegraphics[width=0.1428\textwidth]{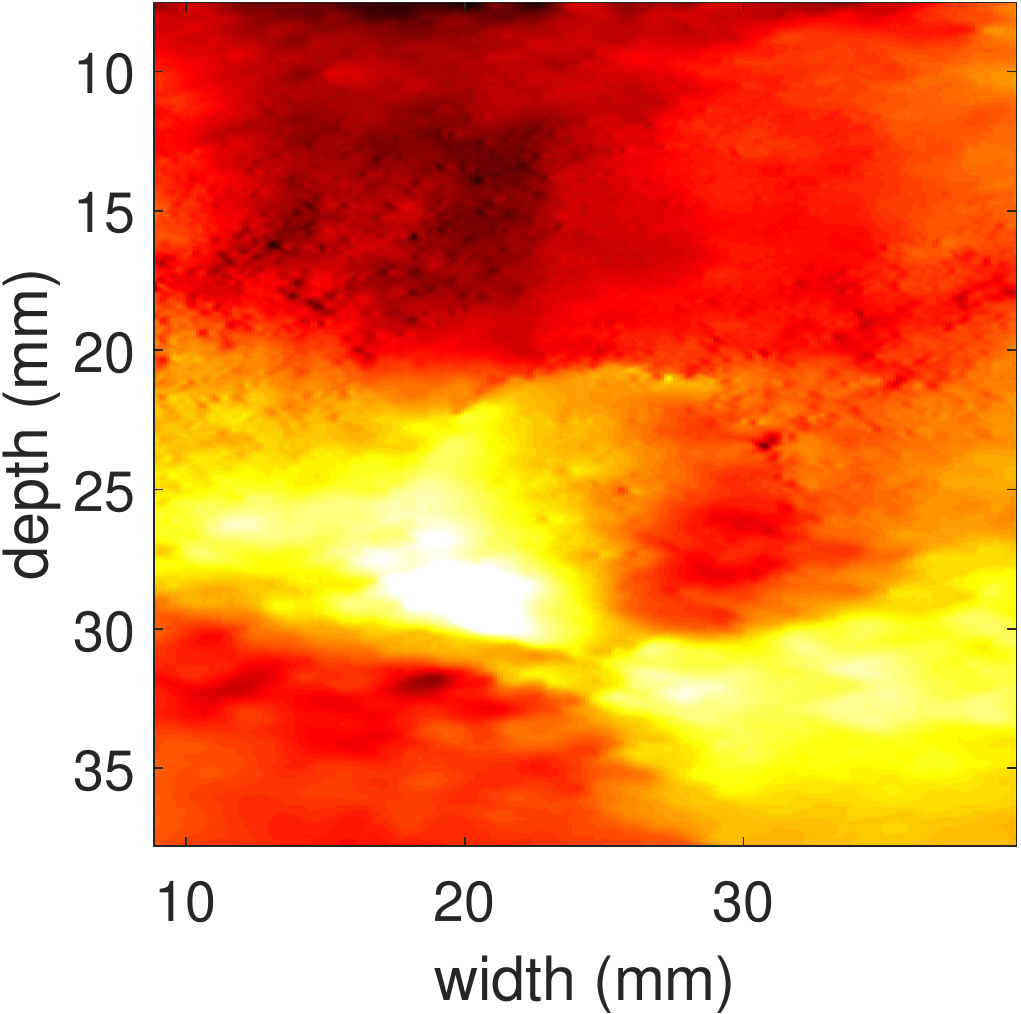}}}%
			\subfigure[$L1$-MechSOUL]{{\includegraphics[width=0.1428\textwidth]{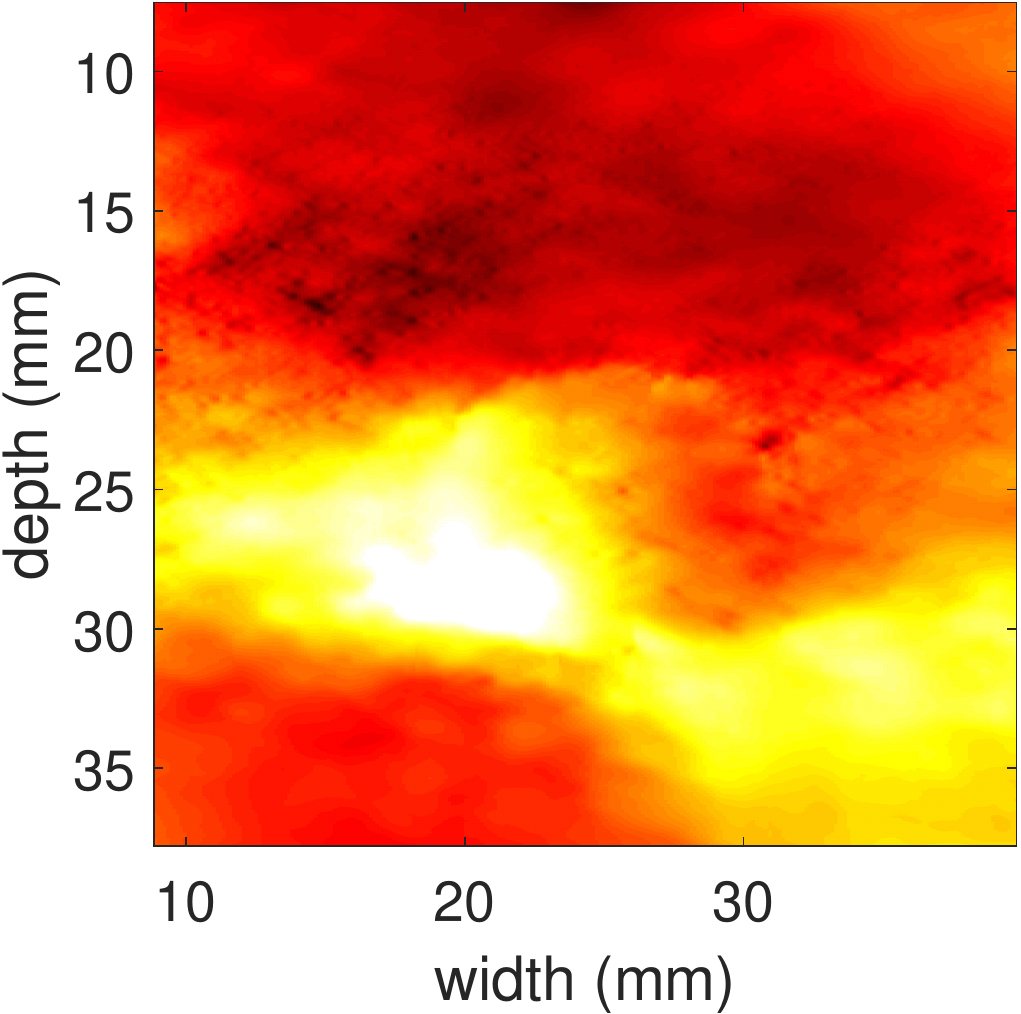}}}	
		\end{flushright}
		\subfigure[Lateral strain]{{\includegraphics[width=0.3\textwidth]{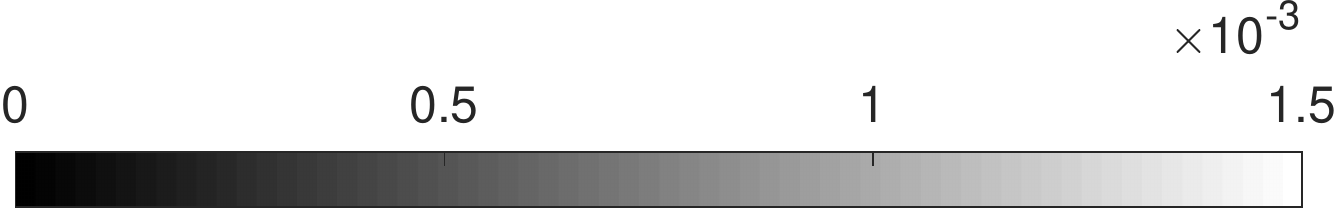}}}%
		\quad\subfigure[EPR]{{\includegraphics[width=0.3\textwidth]{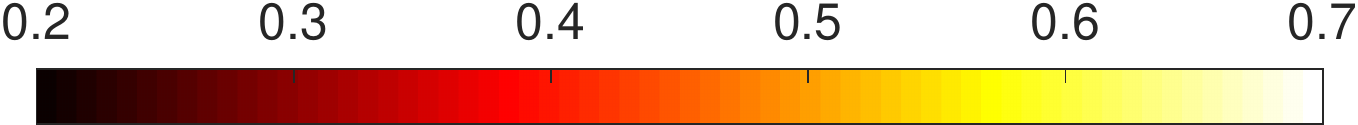}}}
	\end{center}
	\caption{Results for the experimental breast phantom. Rows 1 and 2 show the lateral strain images and the estimated EPR maps, respectively, whereas columns 1 to 7 correspond to B-mode, NCC, NCC + PDE, SOUL, $L1$-SOUL, MechSOUL, and $L1$-MechSOUL, respectively.}
	\label{phan}
\end{figure*}

\begin{table*}[tb]  
	\centering
	\caption{SNR and CNR values for the experimental phantom dataset. Physically impossible values are highlighted in red.}
	\label{table_phan}
	\begin{tabular}{c c c c c c c c c c c c} 
		\hline
		\multicolumn{1}{c}{} &
		\multicolumn{3}{c}{SNR} &
		\multicolumn{1}{c}{} &
		\multicolumn{3}{c}{CNR} \\
		\cline{2-4} 
		\cline{6-8}
		$ $  $ $&    Axial & Lateral & EPR $ $  $ $&$ $  $ $ &$ $  $ $ Axial & Lateral & EPR\\
		\hline
		NCC &  4.99 $\pm$ 3.51 & 0.13 $\pm$ 0.31 & 0.14 $\pm$ 0.28 && 2.02 $\pm$ 1.32 & 0.23 $\pm$ 0.17 & 0.03 $\pm$ 0.01\\
		NCC + PDE &  14.54 $\pm$ 7.11 & 1.77 $\pm$ 1.37 & 1.91 $\pm$ 1.55 && 9.39 $\pm$ 5.80 & 1.49 $\pm$ 1.10 & 0.86 $\pm$ 0.80\\
		SOUL &  14.94 $\pm$ 5.78 & \textcolor{red}{-0.44} $\pm$ 2.87 & \textcolor{red}{-0.30} $\pm$ 2.51 && 11.33 $\pm$ 4.74 & 2.39 $\pm$ 2.34 & 2.25 $\pm$ 1.59\\
		$L1$-SOUL & 15.48 $\pm$ 5.19 &  \textcolor{red}{-0.23} $\pm$ 2.03 & \textcolor{red}{-0.24} $\pm$ 2.00 && 13.25 $\pm$ 4.63 & 1.07 $\pm$ 1.35 & 0.71 $\pm$ 0.81\\
		MechSOUL & 15.96 $\pm$ 5.33 & 14.66 $\pm$ 7.65 & 22.00 $\pm$ 9.01 && 11.70 $\pm$ 4.69 & 7.16 $\pm$ 6.85 & 5.98 $\pm$ 2.90\\
		$L1$-MechSOUL & \textbf{17.69} $\pm$ 5.80 &  \textbf{16.72} $\pm$ 7.99 & \textbf{31.09} $\pm$ 14.92 && \textbf{14.14} $\pm$ 5.02 & \textbf{8.93} $\pm$ 6.20 & 6.28 $\pm$ 2.82\\
		\hline
	\end{tabular}
\end{table*}

\begin{figure*}[h]
	\begin{center}
		
		\includegraphics[width=0.9\textwidth]{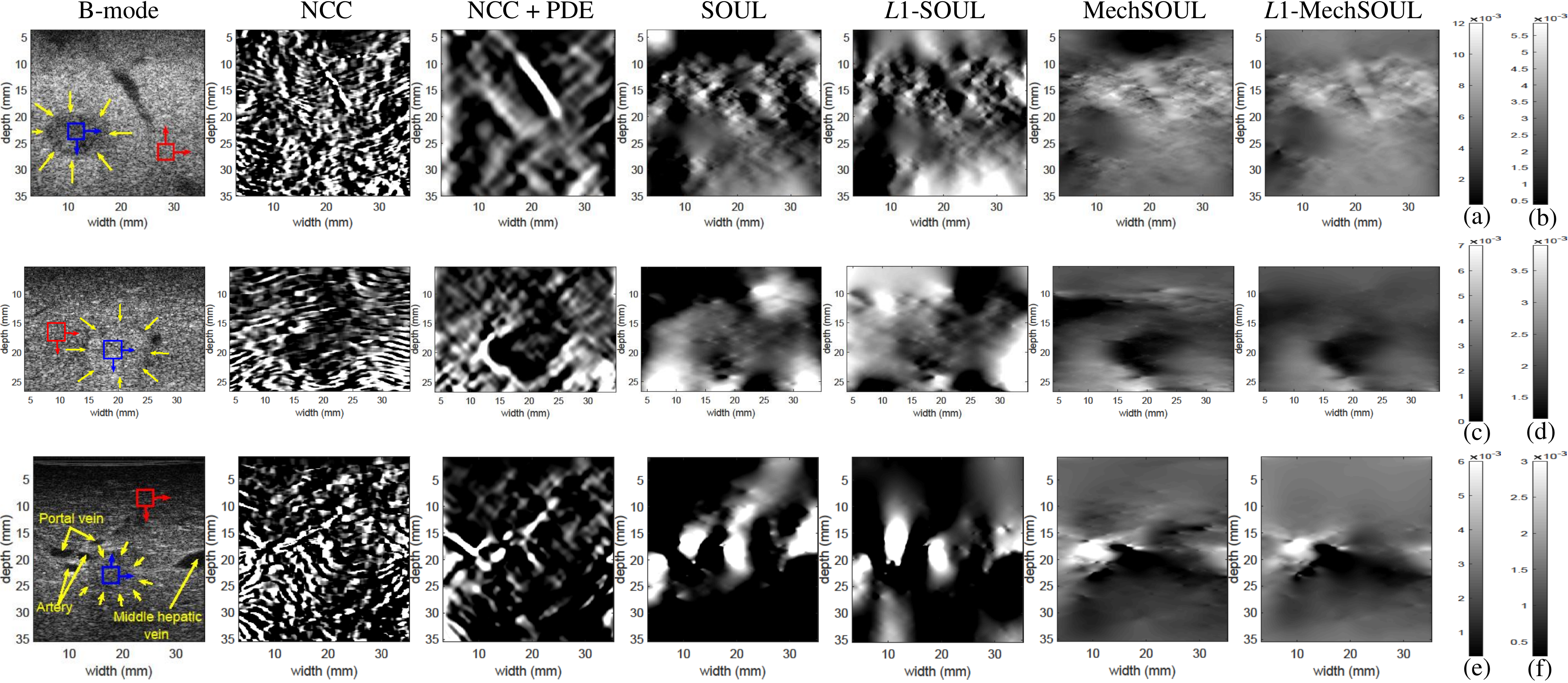}%
	\end{center}
	\caption{Lateral strain results for the liver datasets. Rows 1, 2, and 3 show the lateral strain estimates for patients 1, 2, and 3, respectively. Columns 1 to 7 in each row correspond to B-mode, NCC, NCC + PDE, SOUL, $L1$-SOUL, MechSOUL, and $L1$-MechSOUL, respectively. (a) Color bar (patient 1) for NCC, NCC + PDE, SOUL, and $L1$-SOUL (b) Color bar (patient 1) for MechSOUL and $L1$-MechSOUL (c) Color bar (patient 2) for NCC, NCC + PDE, SOUL, and $L1$-SOUL (d) Color bar (patient 2) for MechSOUL and $L1$-MechSOUL (e) Color bar (patient 3) for NCC and NCC + PDE (f) Color bar (patient 3) for SOUL, $L1$-SOUL, MechSOUL, and $L1$-MechSOUL.}
	\label{liver}
\end{figure*}

\begin{figure*}[h]
	\begin{center}
		\subfigure[NCC]{{\includegraphics[width=0.14\textwidth]{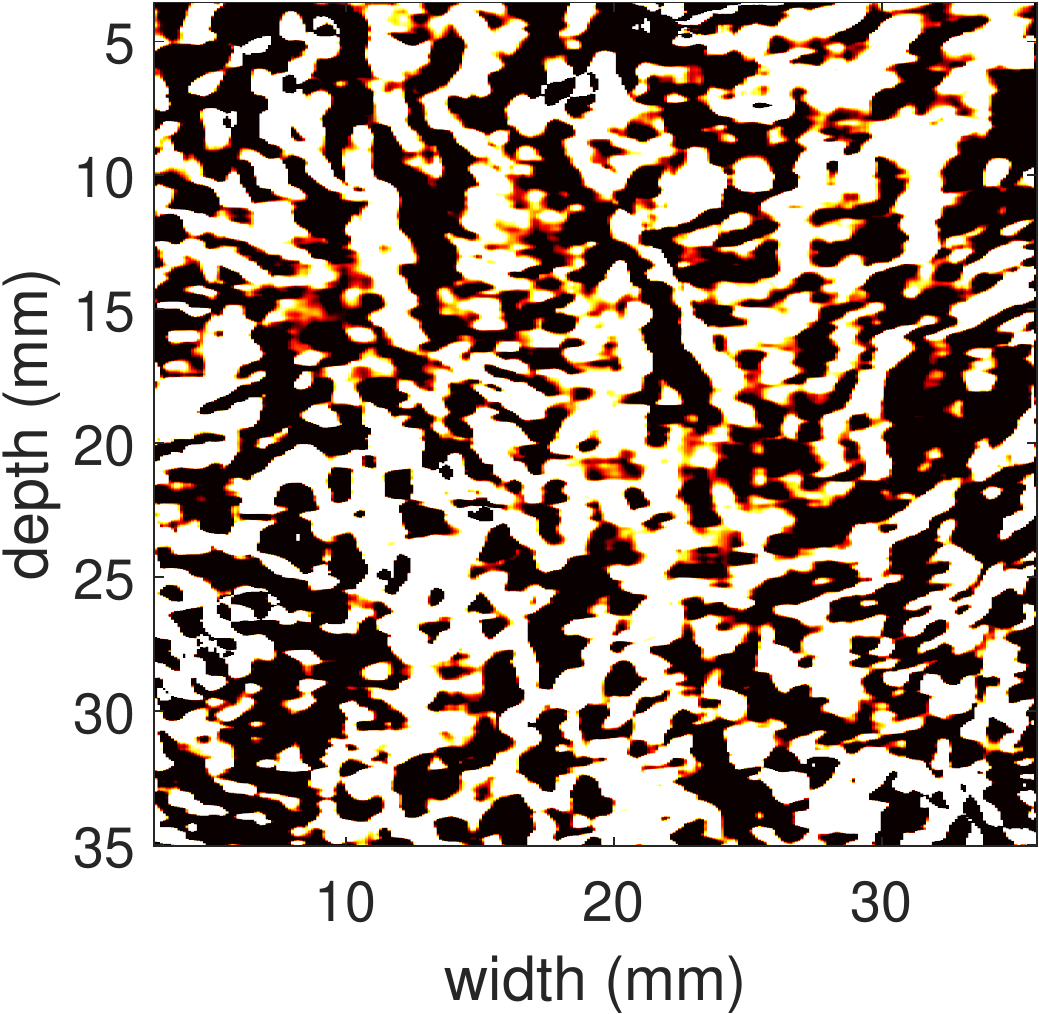}}}%
		\subfigure[NCC + PDE]{{\includegraphics[width=0.14\textwidth]{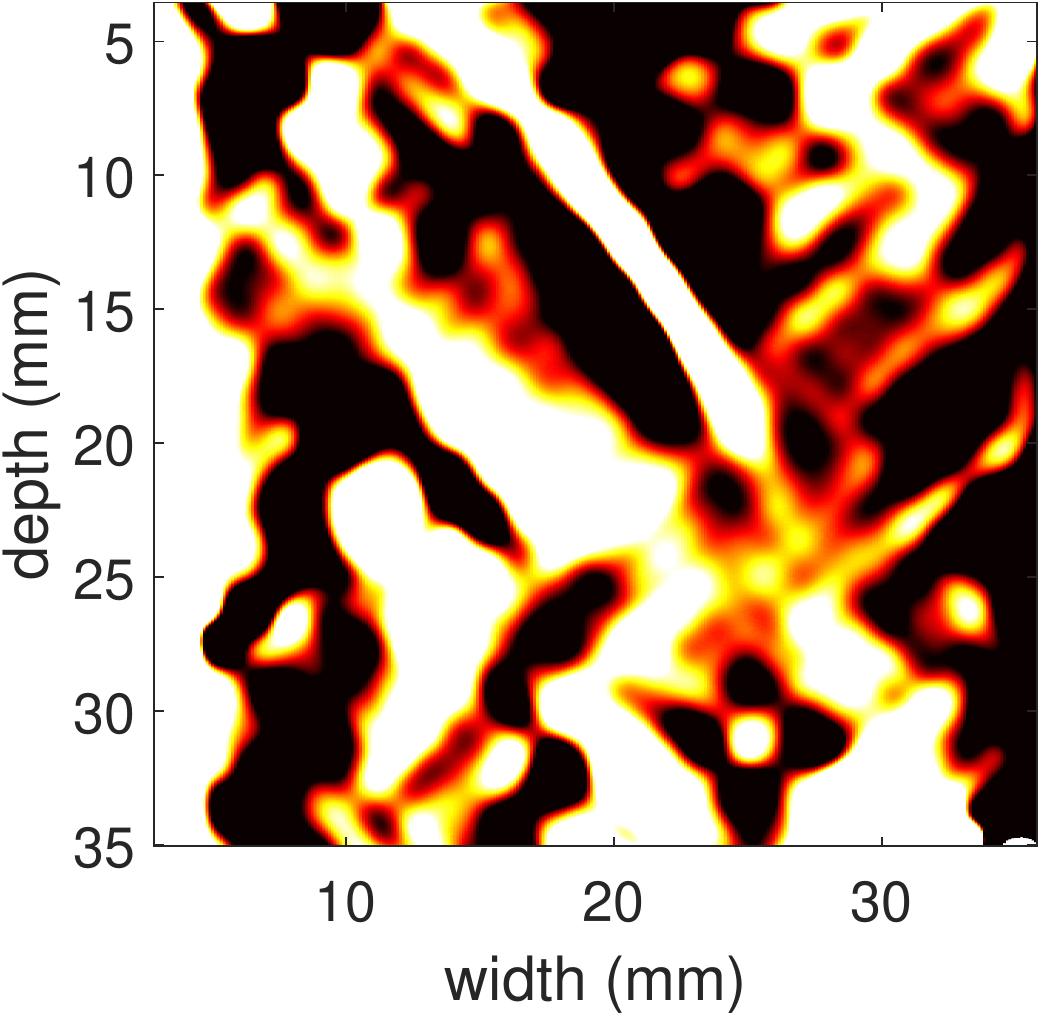}}}%
		\subfigure[SOUL]{{\includegraphics[width=0.14\textwidth]{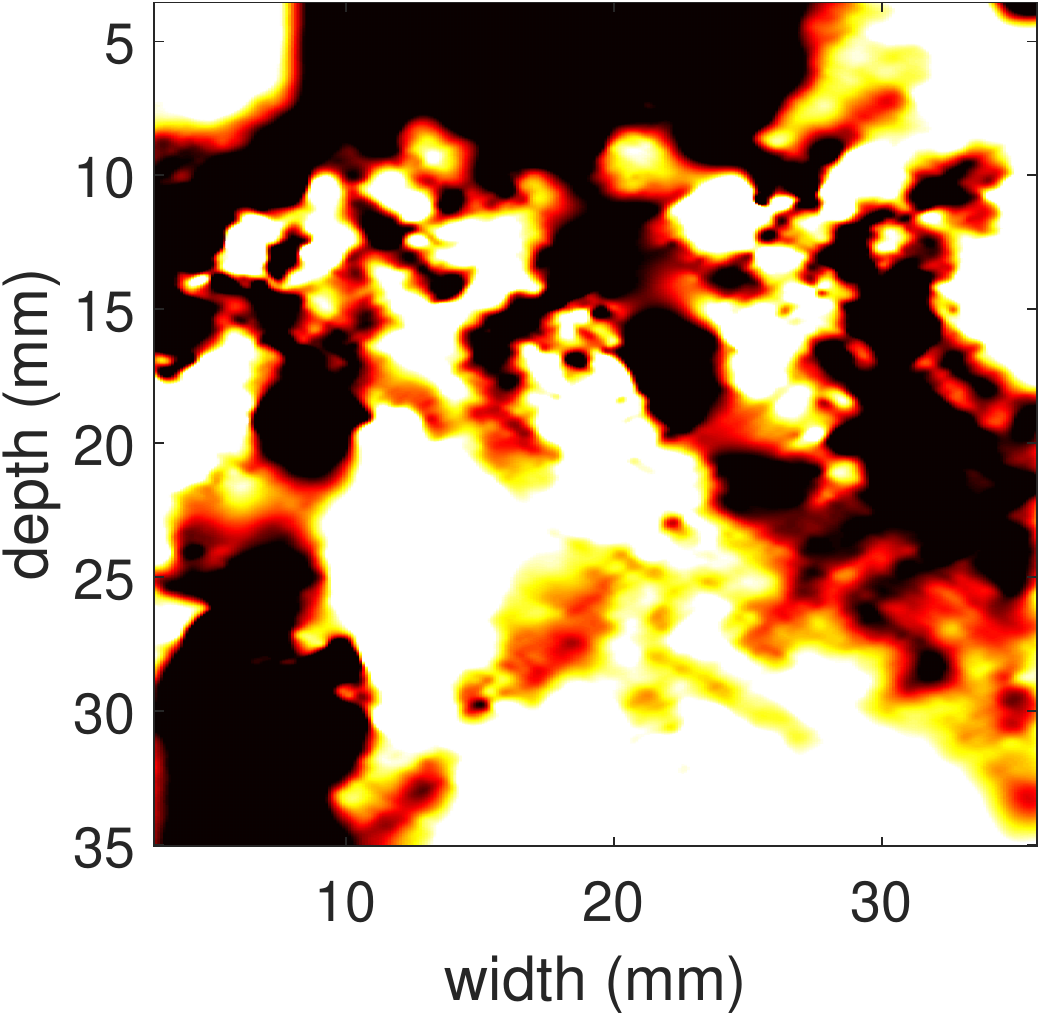}}}%
		\subfigure[$L1$-SOUL]{{\includegraphics[width=0.14\textwidth]{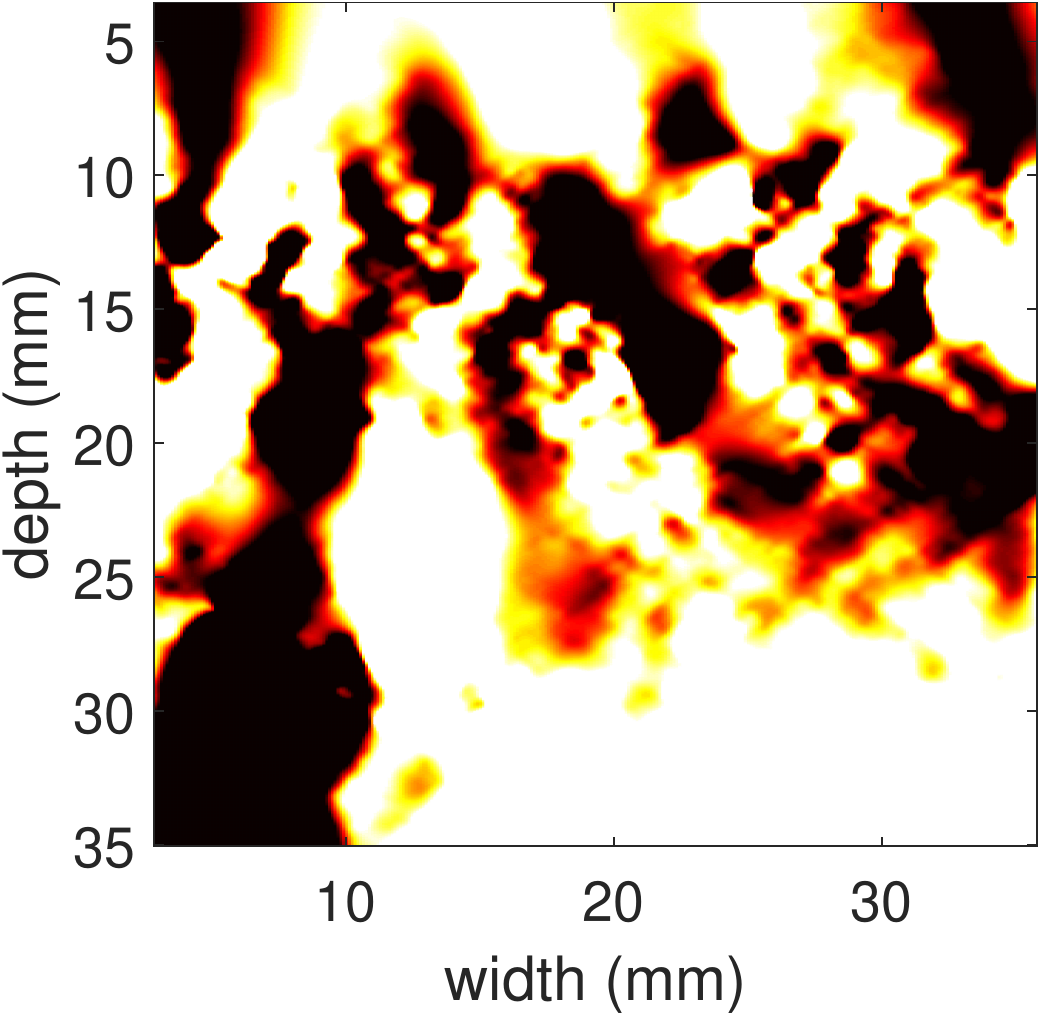}}}%
		\subfigure[MechSOUL]{{\includegraphics[width=0.14\textwidth]{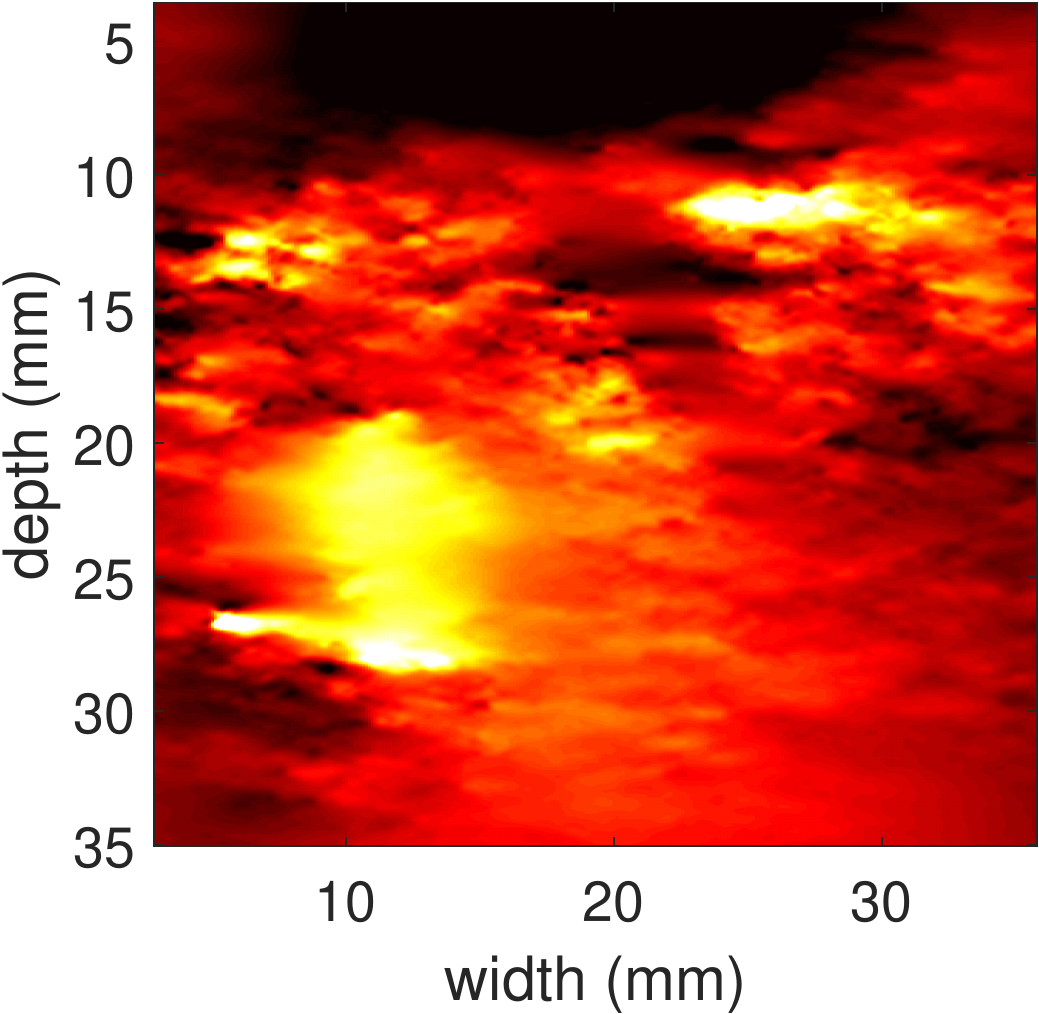}}}%
		\subfigure[$L1$-MechSOUL]{{\includegraphics[width=0.14\textwidth]{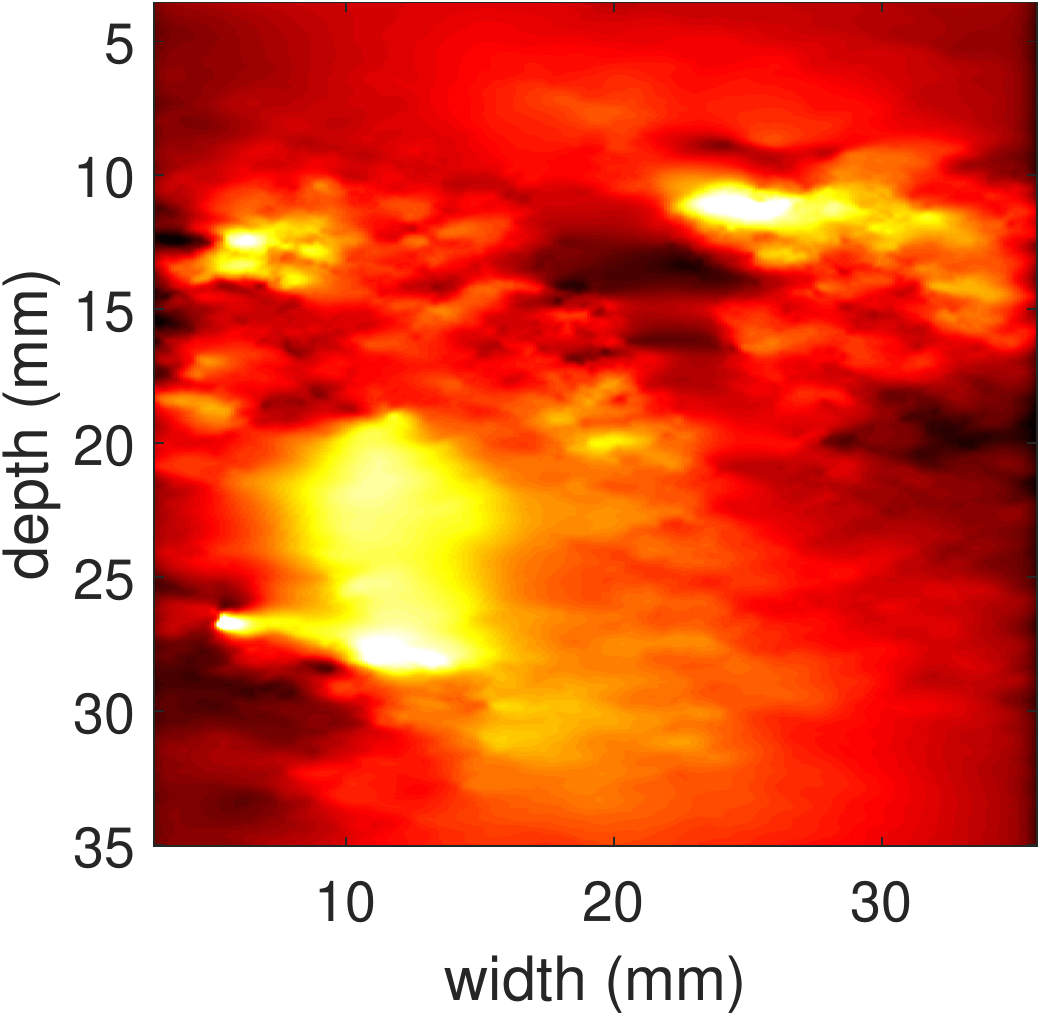}}}
		\subfigure[NCC]{{\includegraphics[width=0.14\textwidth]{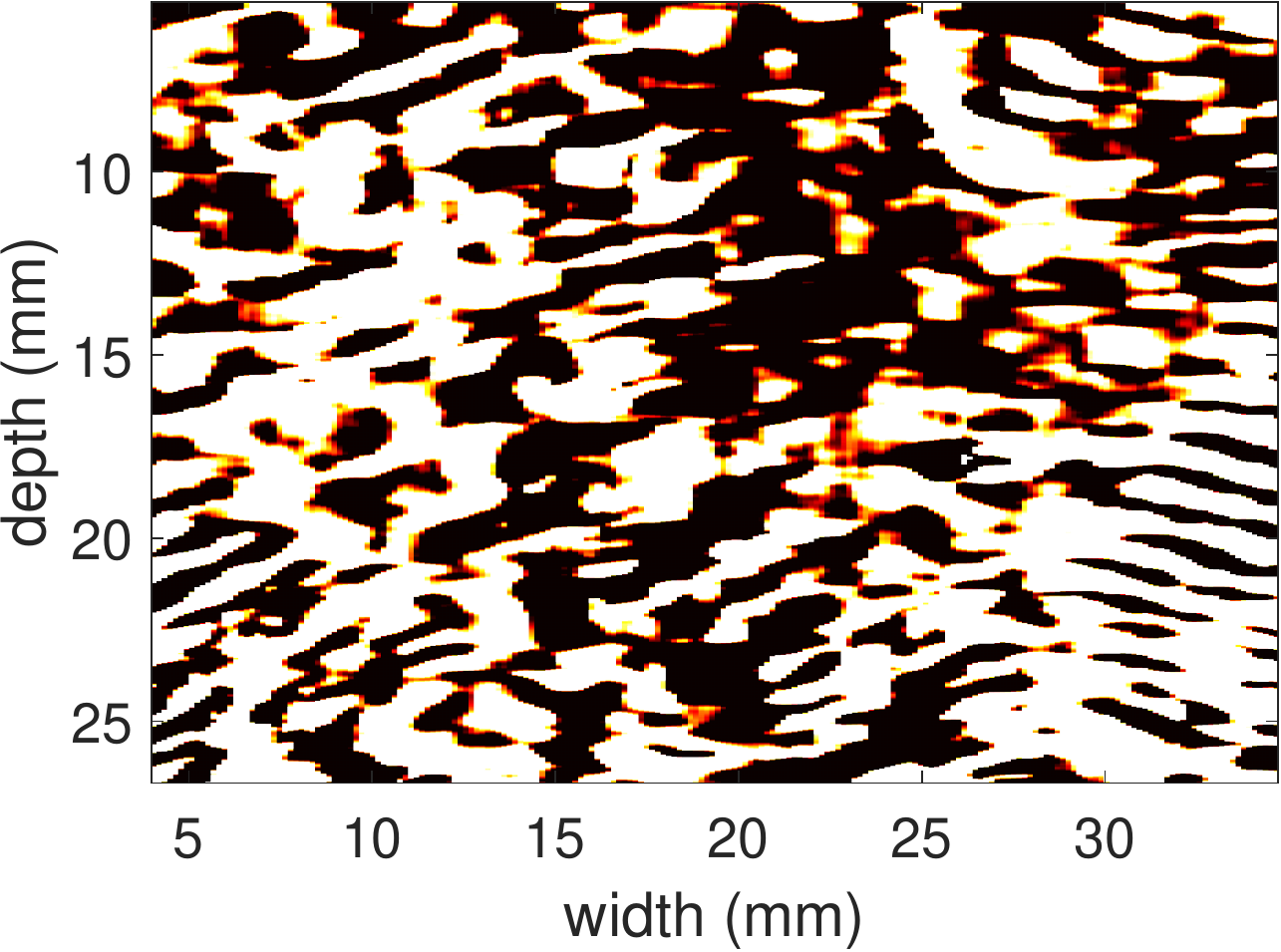}}}%
		\subfigure[NCC + PDE]{{\includegraphics[width=0.14\textwidth]{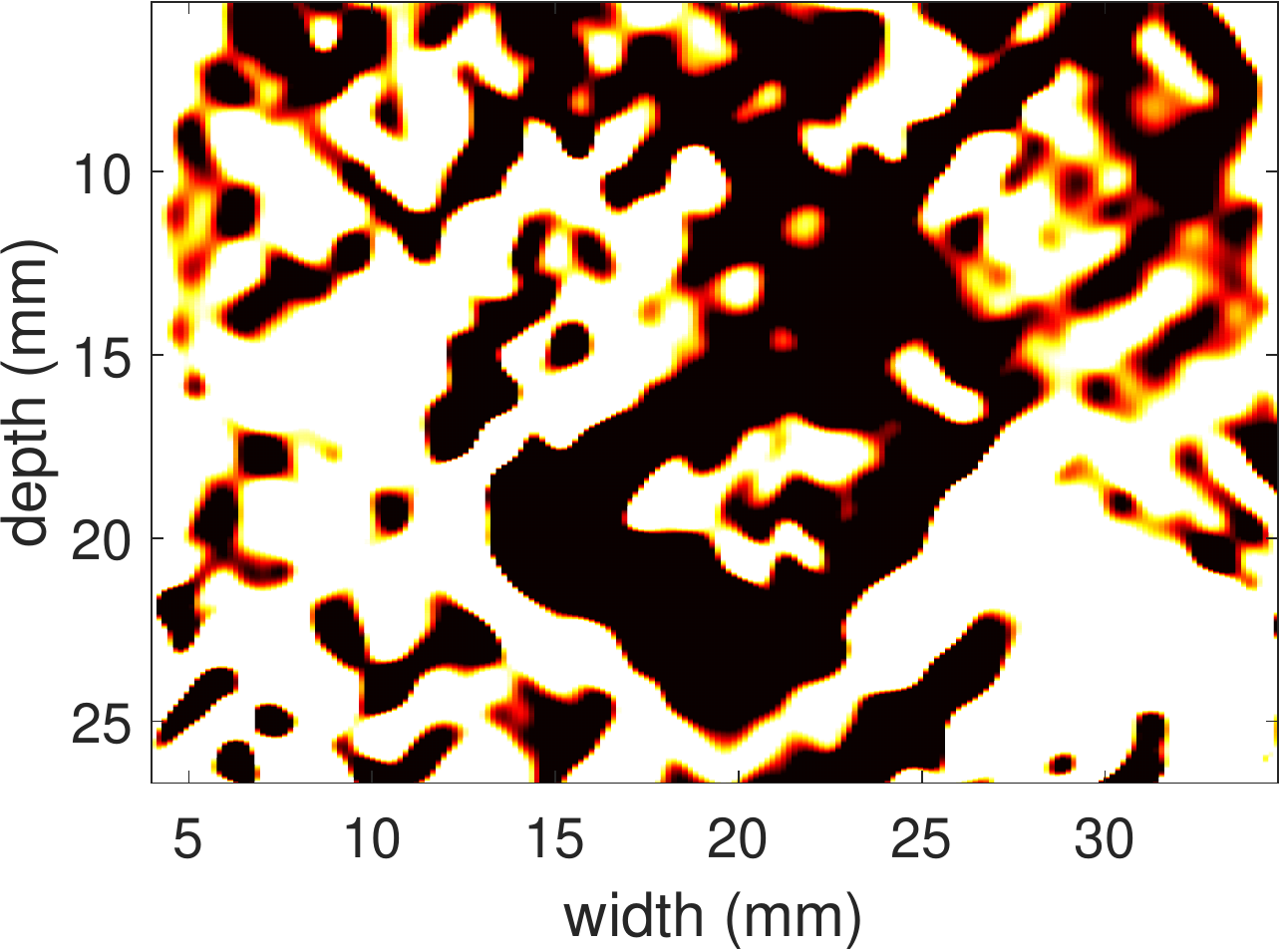}}}%
		\subfigure[SOUL]{{\includegraphics[width=0.14\textwidth]{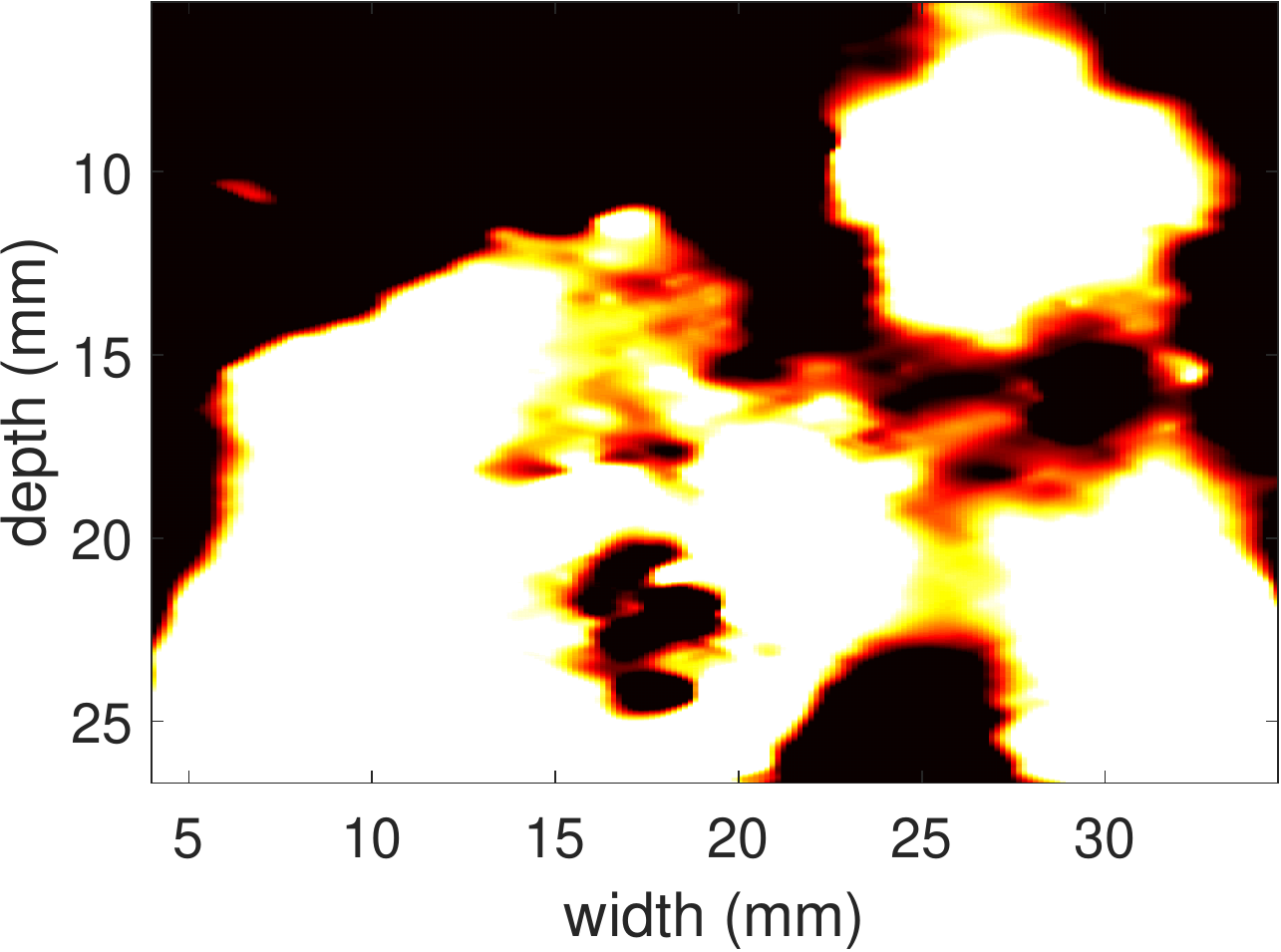}}}%
		\subfigure[$L1$-SOUL]{{\includegraphics[width=0.14\textwidth]{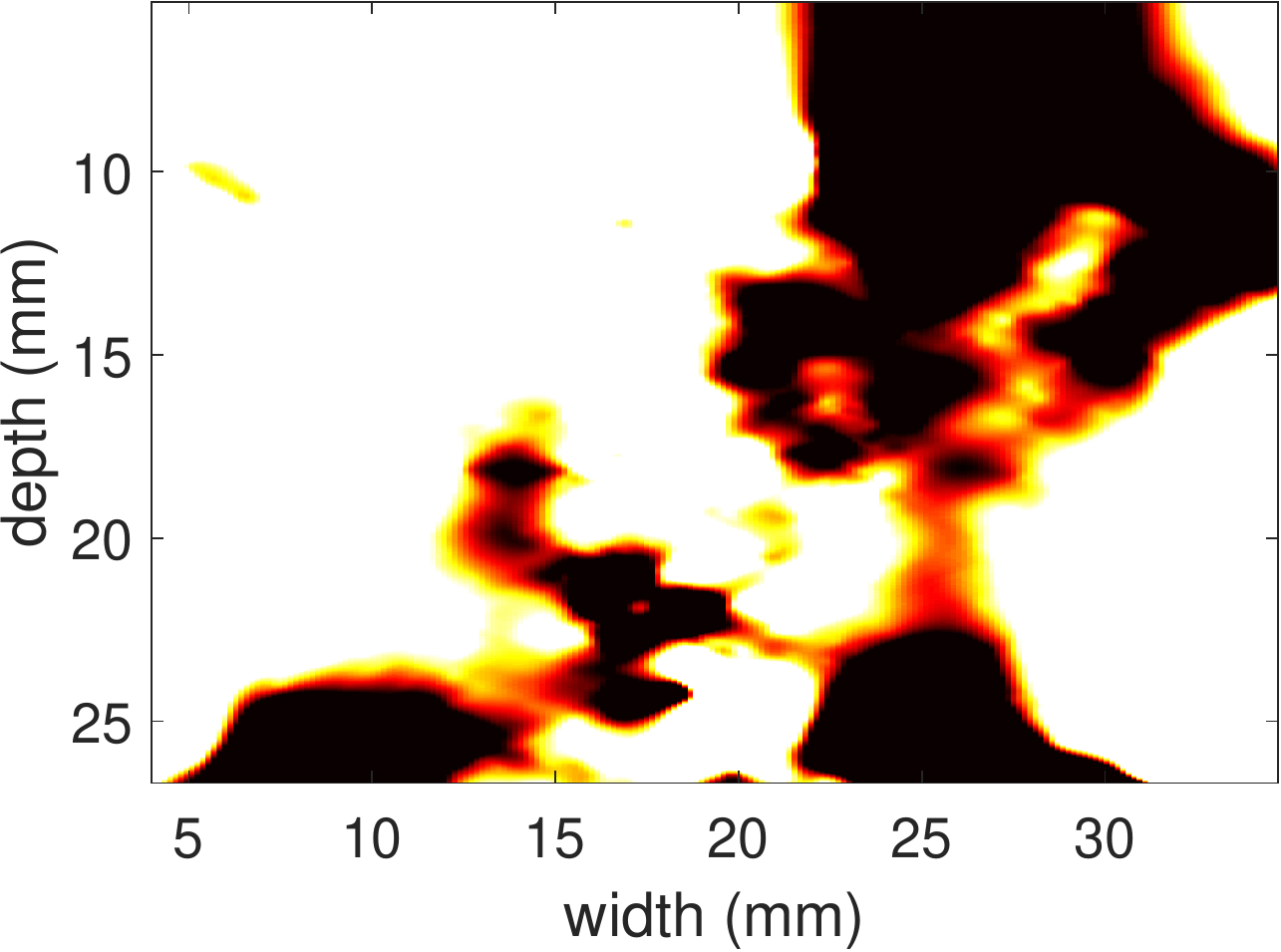}}}%
		\subfigure[MechSOUL]{{\includegraphics[width=0.14\textwidth]{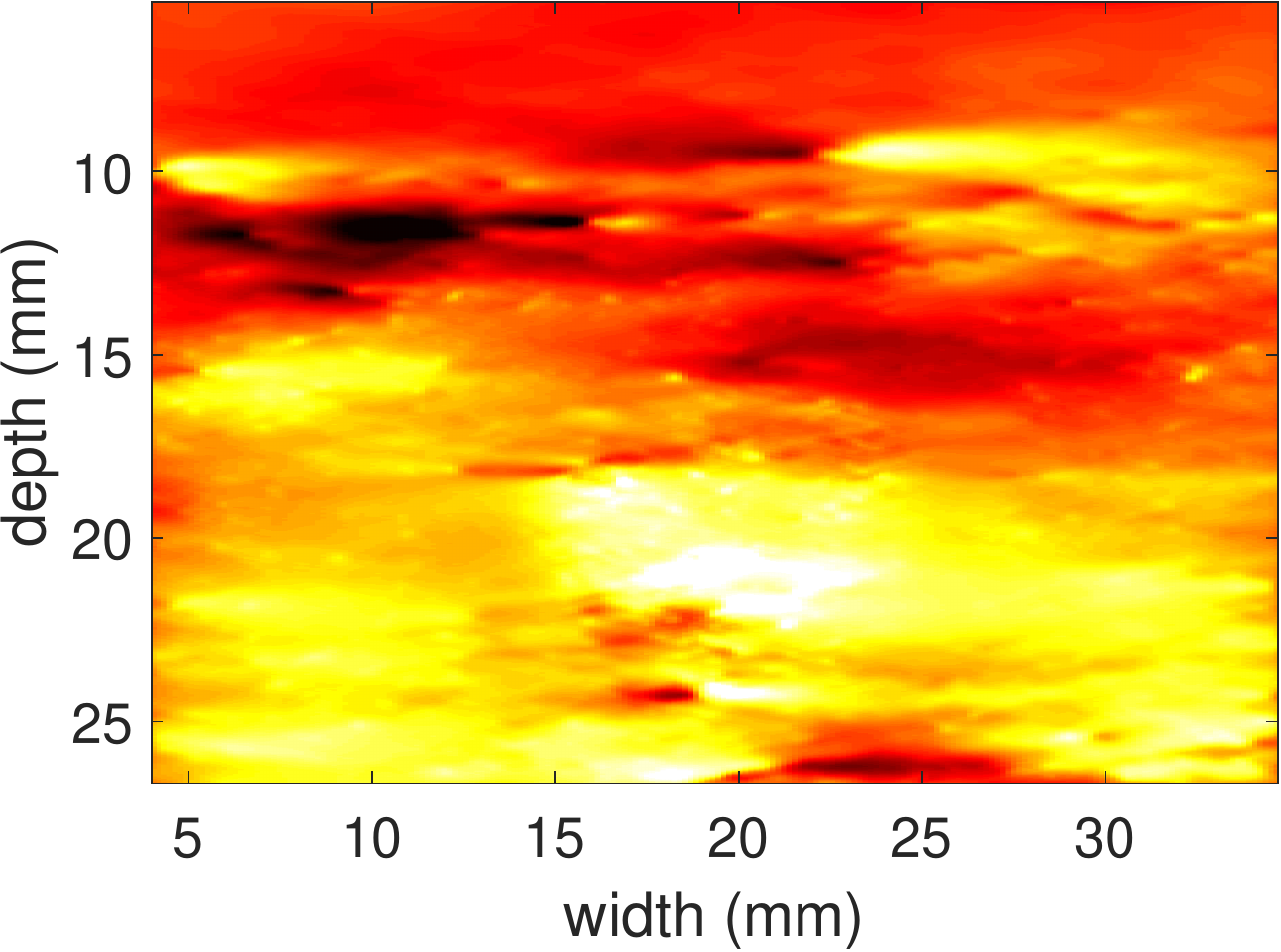}}}%
		\subfigure[$L1$-MechSOUL]{{\includegraphics[width=0.14\textwidth]{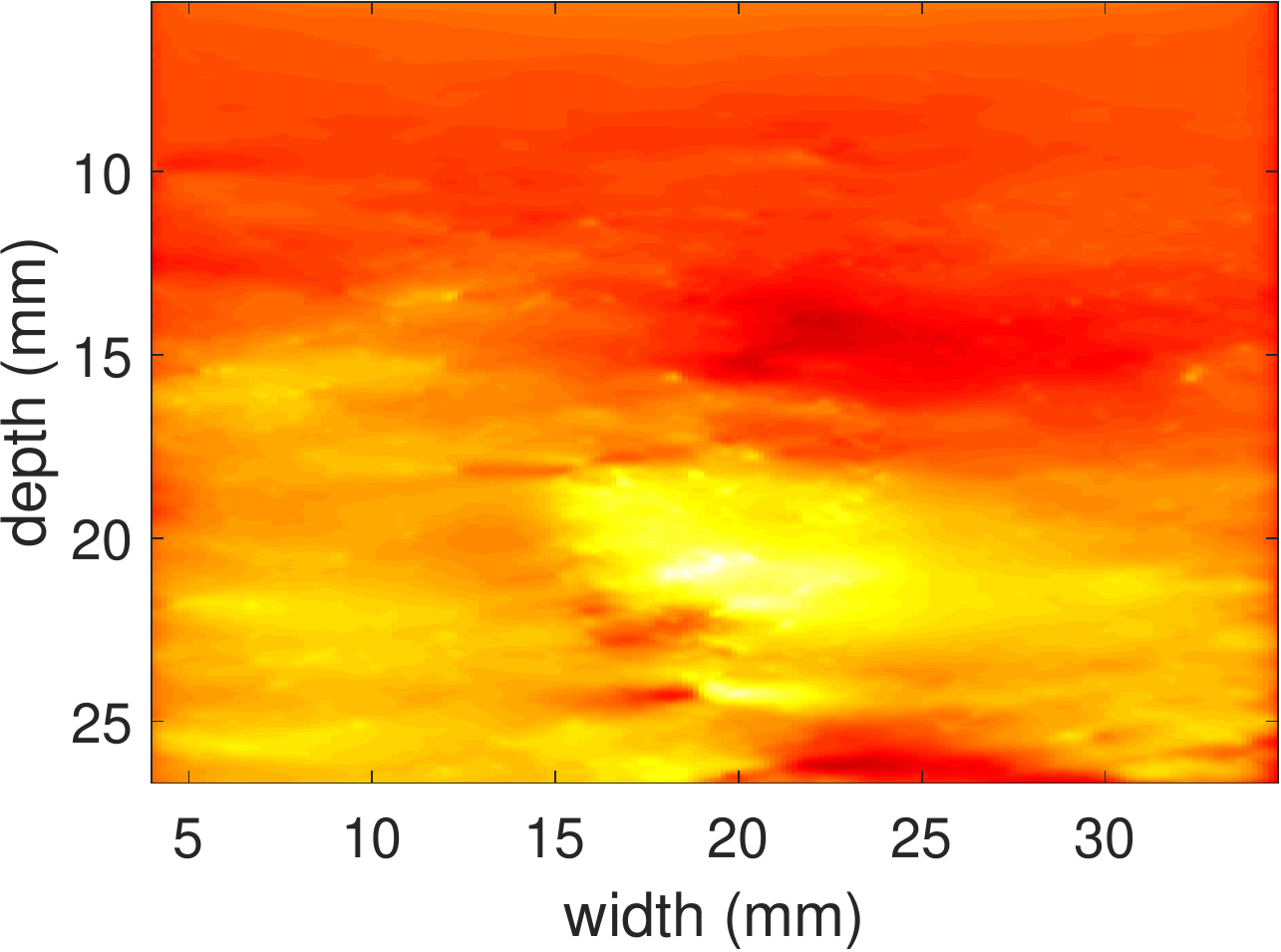}}}
		\subfigure[NCC]{{\includegraphics[width=0.14\textwidth]{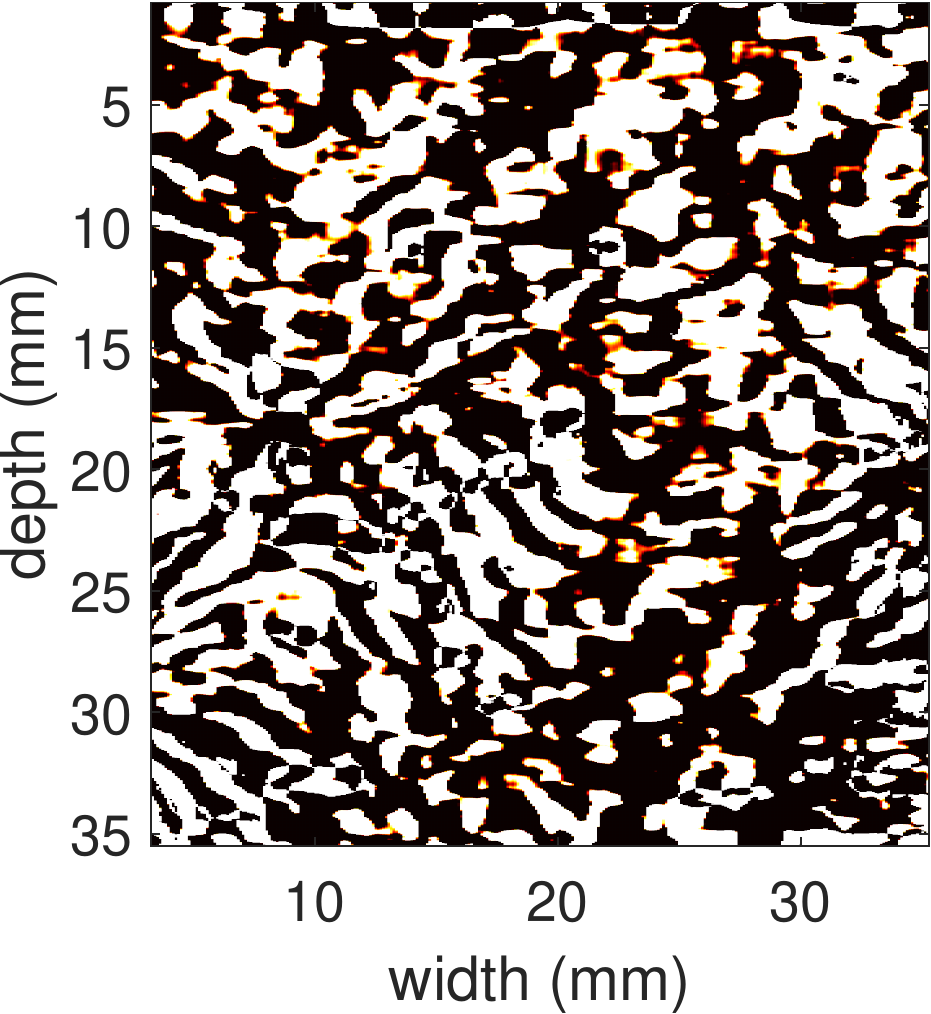}}}%
		\subfigure[NCC + PDE]{{\includegraphics[width=0.14\textwidth]{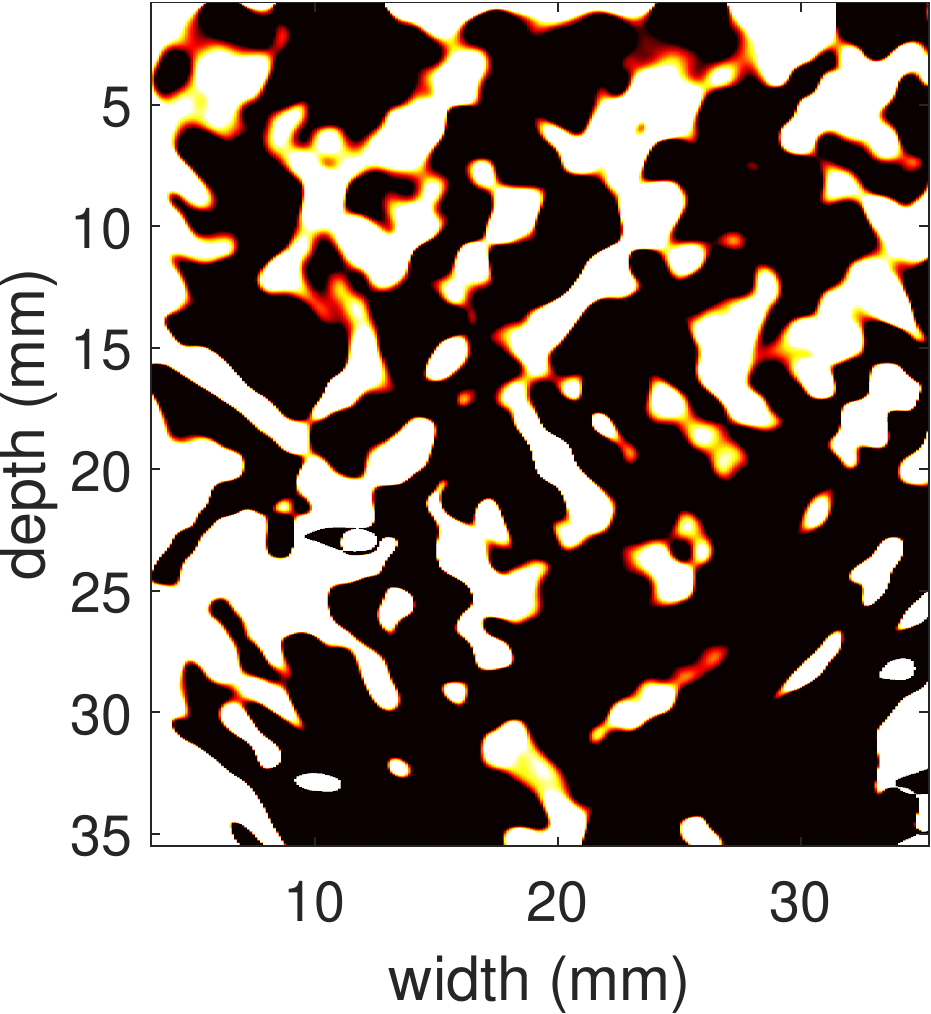}}}%
		\subfigure[SOUL]{{\includegraphics[width=0.14\textwidth]{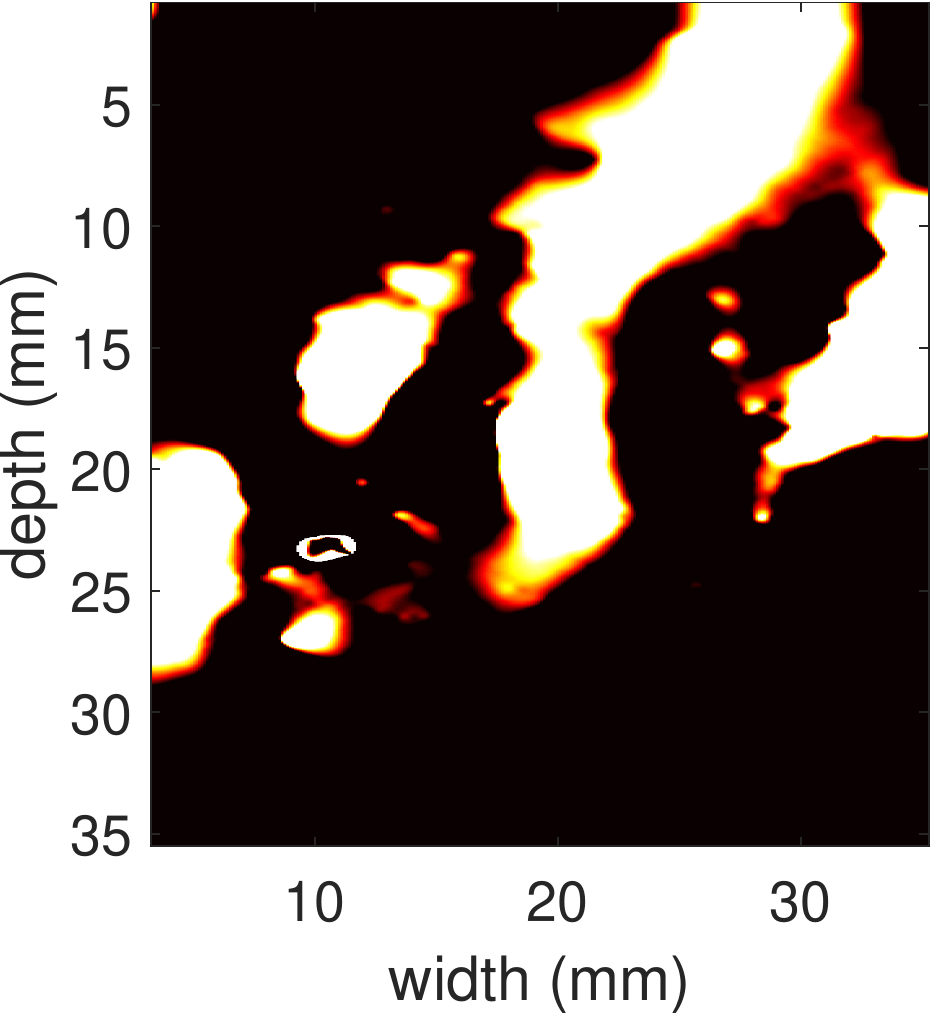}}}%
		\subfigure[$L1$-SOUL]{{\includegraphics[width=0.14\textwidth]{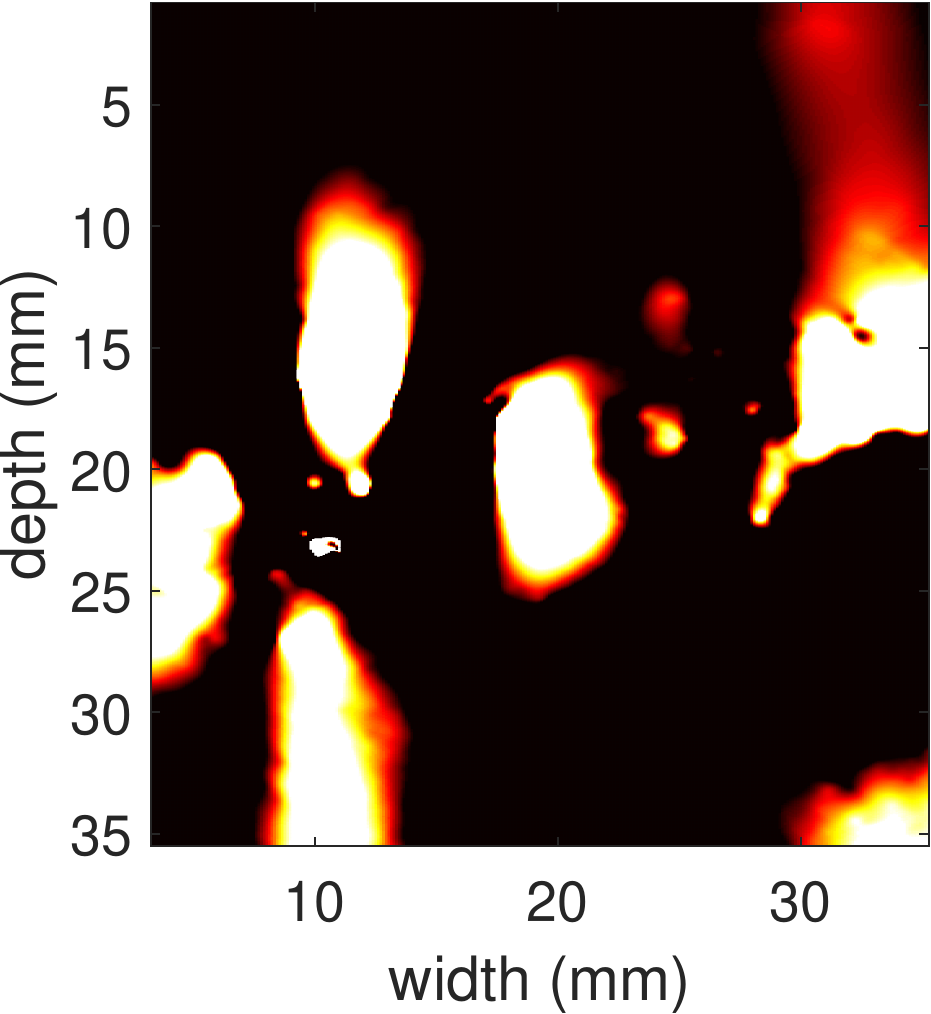}}}%
		\subfigure[MechSOUL]{{\includegraphics[width=0.14\textwidth]{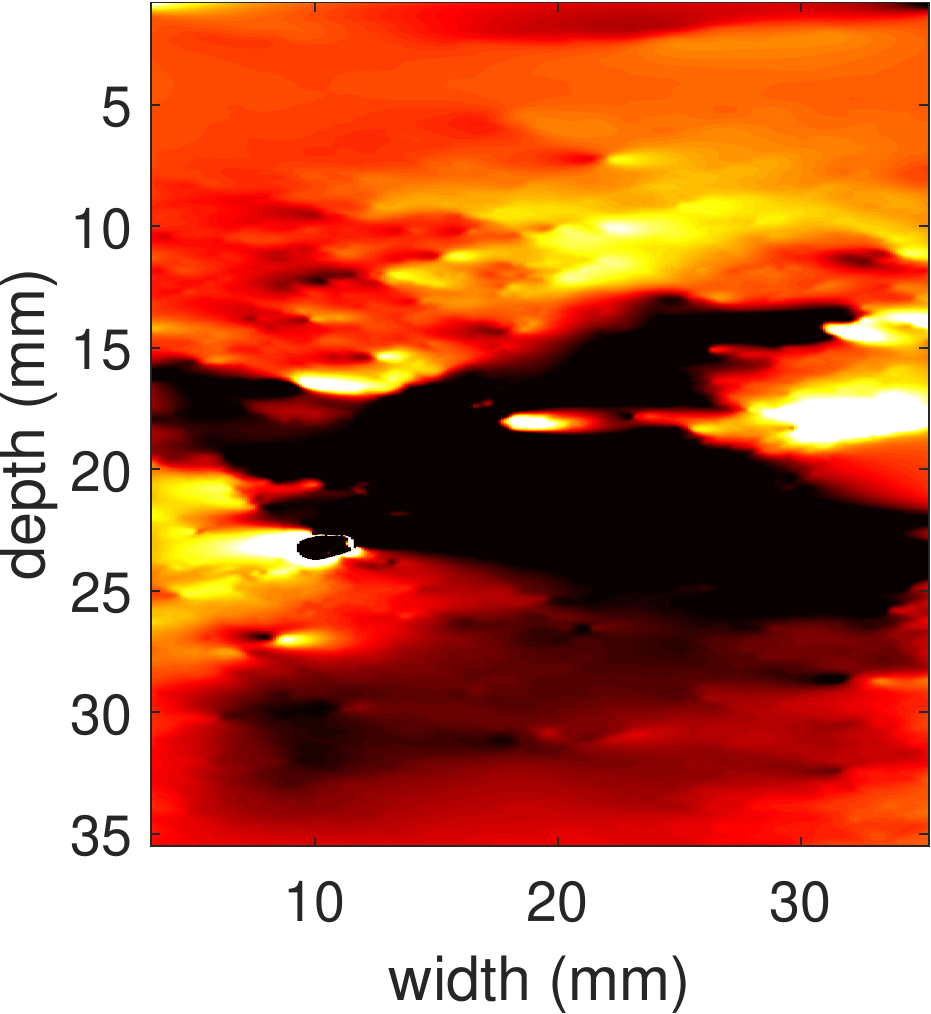}}}%
		\subfigure[$L1$-MechSOUL]{{\includegraphics[width=0.14\textwidth]{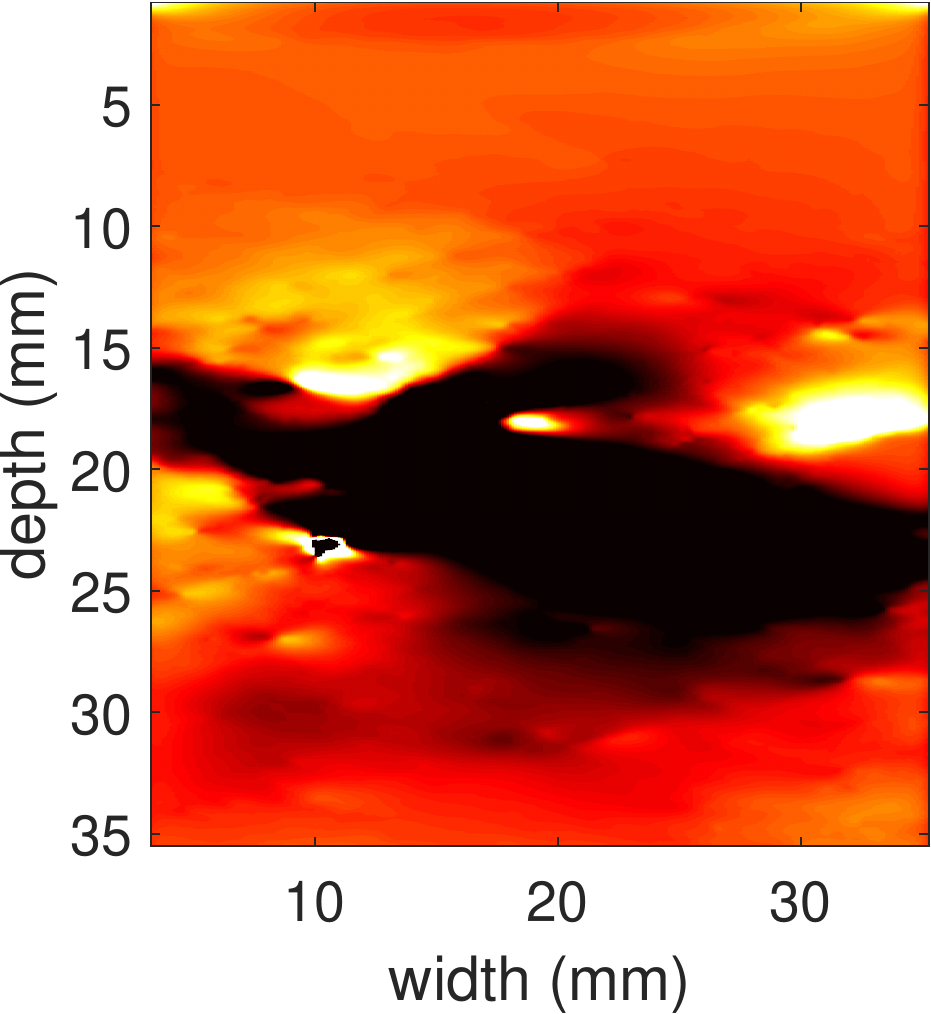}}}
		\subfigure[EPR, patient 1]{{\includegraphics[width=0.25\textwidth]{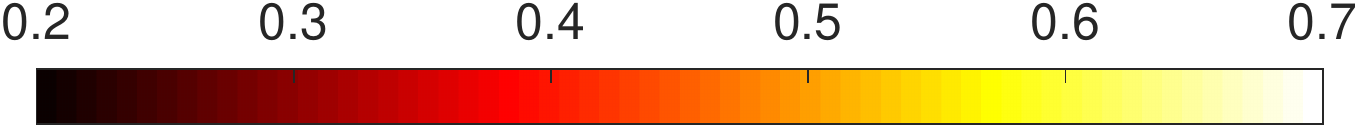}}}%
		\quad
		\subfigure[EPR, patient 2]{{\includegraphics[width=0.25\textwidth]{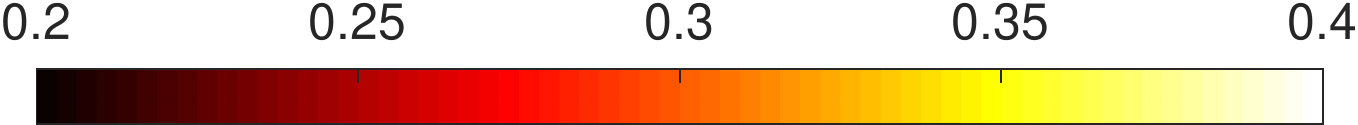}}}%
		\quad
		\subfigure[EPR, patient 3]{{\includegraphics[width=0.25\textwidth]{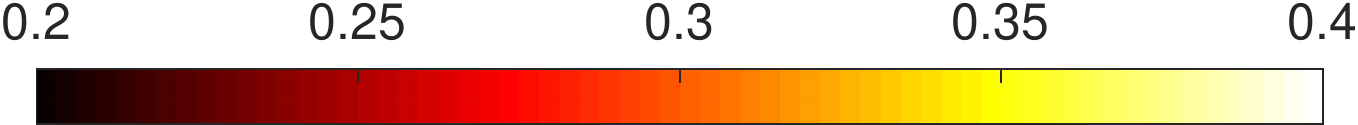}}}%
	\end{center}
	\caption{EPR estimates for the liver datasets. Rows 1, 2, and 3 show the results for patients 1, 2, and 3, respectively. Columns 1 to 6 correspond to NCC, NCC + PDE, SOUL, $L1$-SOUL, MechSOUL, and $L1$-MechSOUL, respectively.}
	\label{liver_epr}
\end{figure*}

\begin{table*}[tb]  
	\centering
	\caption{SNR and CNR values for the first liver cancer dataset.}
	\label{table_liver1}
	\begin{tabular}{c c c c c c c c c c c c c c} 
		\hline
		\multicolumn{1}{c}{} &
		\multicolumn{3}{c}{SNR} &
		\multicolumn{1}{c}{} &
		\multicolumn{3}{c}{CNR} \\
		\cline{2-4} 
		\cline{6-8}
		$ $  $ $&    Axial & Lateral & EPR $ $  $ $&$ $  $ $ &$ $  $ $ Axial & Lateral & EPR\\
		\hline
		NCC &  3.50 $\pm$ 1.57 & 0.46 $\pm$ 0.38 & 0.33 $\pm$ 0.34 && 2.62 $\pm$ 1.45 & 0.32 $\pm$ 0.27 & 0.30 $\pm$ 0.31\\
		NCC + PDE &  11.16 $\pm$ 6.06 & 2.22 $\pm$ 1.58 & 2.48 $\pm$ 1.96 && 8.56 $\pm$ 4.98 & 1.11 $\pm$ 0.74 & 1.84 $\pm$ 1.01\\
		SOUL &  22.90 $\pm$ 9.22 & 5.15 $\pm$ 3.74 & 4.98 $\pm$ 3.80 && 16.53 $\pm$ 6.04 & 2.64 $\pm$ 1.59 & 3.39 $\pm$ 1.94\\
		$L1$-SOUL & \textbf{33.14} $\pm$ 13.13 &  6.35 $\pm$ 2.66 & 6.43 $\pm$ 2.93 && \textbf{22.36} $\pm$ 7.08 & 3.69 $\pm$ 2.15 & 1.83 $\pm$ 1.31\\
		MechSOUL & 21.27 $\pm$ 8.02 & 18.01 $\pm$ 9.11 & 21.16 $\pm$ 11.00 && 16.35 $\pm$ 6.16 & 8.23 $\pm$ 2.95 & 15.20 $\pm$ 4.93\\
		$L1$-MechSOUL & 31.85 $\pm$ 13.11 &  \textbf{36.01} $\pm$ 17.21 & \textbf{30.38} $\pm$ 7.72 && 20.71 $\pm$ 6.50 & \textbf{12.01} $\pm$ 3.60 & \textbf{16.65} $\pm$ 4.78\\
		\hline
	\end{tabular}
\end{table*}

\begin{table*}[tb]  
	\centering
	\caption{SNR and CNR values for the second liver cancer dataset.}
	\label{table_liver2}
	\begin{tabular}{c c c c c c c c c c c c c c c c} 
		\hline
		\multicolumn{1}{c}{} &
		\multicolumn{3}{c}{SNR} &
		\multicolumn{1}{c}{} &
		\multicolumn{3}{c}{CNR} \\
		\cline{2-4} 
		\cline{6-8}
		$ $  $ $&    Axial & Lateral & EPR $ $  $ $&$ $  $ $ &$ $  $ $ Axial & Lateral & EPR\\
		\hline
		NCC &  5.77 $\pm$ 1.58 & 0.75 $\pm$ 0.29 & 0.73 $\pm$ 0.28 && 2.83 $\pm$ 1.23 & 0.67 $\pm$ 0.39 & 0.46 $\pm$ 0.27\\
		NCC + PDE &  14.20 $\pm$ 6.49 & 2.25 $\pm$ 0.72 & 2.41 $\pm$ 0.78 && 5.68 $\pm$ 2.96 & 1.95 $\pm$ 0.89 & 1.34 $\pm$ 0.73\\
		SOUL &  35.18 $\pm$ 21.45 & 3.02 $\pm$ 4.25 & 3.38 $\pm$ 4.92 && \textbf{11.36} $\pm$ 3.60 & 3.30 $\pm$ 2.75 & 2.69 $\pm$ 2.36\\
		$L1$-SOUL & \textbf{57.63} $\pm$ 69.73 &  7.70 $\pm$ 9.39 & 6.92 $\pm$ 10.23 && 10.92 $\pm$ 4.51 & 4.62 $\pm$ 2.89 & 2.99 $\pm$ 1.66\\
		MechSOUL & 31.80 $\pm$ 21.87 & 21.63 $\pm$ 13.24 & 29.36 $\pm$ 15.84 && 10.55 $\pm$ 4.76 & 6.20 $\pm$ 3.44 & 4.01 $\pm$ 1.52\\
		$L1$-MechSOUL & 49.59 $\pm$ 56.93 & \textbf{38.59} $\pm$ 37.35 & \textbf{97.66} $\pm$ 74.50 && 10.29 $\pm$ 4.54 & \textbf{7.28} $\pm$ 3.37 & \textbf{4.86} $\pm$ 1.83\\
		\hline
	\end{tabular}
\end{table*}

\begin{table*}[tb]  
	\centering
	\caption{SNR and CNR values for the third liver cancer dataset. Impractical values are highlighted in red.}
	\label{table_liver3}
	\begin{tabular}{c c c c c c c c c c c c c c c c c c} 
		\hline
		\multicolumn{1}{c}{} &
		\multicolumn{3}{c}{SNR} &
		\multicolumn{1}{c}{} &
		\multicolumn{3}{c}{CNR} \\
		\cline{2-4} 
		\cline{6-8}
		$ $  $ $&    Axial & Lateral & EPR $ $  $ $&$ $  $ $ &$ $  $ $ Axial & Lateral & EPR\\
		\hline
		NCC &  2.59 $\pm$ 1.13 & 0.19 $\pm$ 0.35 & 0.12 $\pm$ 0.32 && 0.99 $\pm$ 0.69 & 0.22 $\pm$ 0.15 & 0.03 $\pm$ 0.03\\
		NCC + PDE &  5.93 $\pm$ 3.09 & 0.48 $\pm$ 0.77 & 0.45 $\pm$ 0.81 && 2.17 $\pm$ 1.69 & 0.79 $\pm$ 0.43 & 0.56 $\pm$ 0.33\\
		SOUL &  40.22 $\pm$ 26.19 & \textcolor{red}{-2.20} $\pm$ 6.99 & \textcolor{red}{-1.70} $\pm$ 6.57 && 2.35 $\pm$ 1.66 & 2.29 $\pm$ 1.97 & 2.27 $\pm$ 1.91\\
		$L1$-SOUL & \textbf{77.68} $\pm$ 58.70 &  5.41 $\pm$ 5.89 & 5.49 $\pm$ 6.13 && \textbf{14.63} $\pm$ 4.33 & 1.23 $\pm$ 0.58 & 1.50 $\pm$ 0.40\\
		MechSOUL & 66.31 $\pm$ 51.25 & 36.96 $\pm$ 23.85 & 57.23 $\pm$ 60.37 && 9.54 $\pm$ 3.87 & 6.23 $\pm$ 1.22 & 4.14 $\pm$ 0.63\\
		$L1$-MechSOUL & 76.94 $\pm$ 56.67 & \textbf{62.54} $\pm$ 61.59 & 123.32 $\pm$ \textbf{136.16} && 13.82 $\pm$ 3.73 & \textbf{8.82} $\pm$ 1.24 & \textbf{5.18} $\pm$ 0.65\\
		\hline
	\end{tabular}
\end{table*}  

\subsection{Ultrasound Simulation and Data Acquisition}
\subsubsection{Hard-inclusion Simulated Phantom}
A homogeneous tissue phantom containing a stiff cylindrical inclusion was simulated, setting the background and inclusion elastic moduli to 20 kPa and 40 kPa, respectively. Both the background and target PRs were set to 0.49. The aforementioned phantom was compressed by $2\%$ using the finite-element (FEM) package ABAQUS (Providence, RI), and the pre- and post-deformed RF frames were simulated with Field II~\cite{field2}. The center and sampling frequencies were set to 5 MHz and 50 MHz, respectively.

\subsubsection{Multi-inclusion Simulated Phantom}
A tissue phantom containing three hard inclusions with different elasticities was simulated. While both the background and inclusion PRs were set to 0.49, Young's moduli corresponding to the background and the three inclusions were fixed at 20 kPa, 40 kPa, 60 kPa, and 80 kPa. Non-uniaxial displacement profiles were created using ABAQUS in two ways: 1) imposing an additional condition that set the lateral displacement of the phantom's left boundary to zero 2) deforming the phantom asymmetrically using a surface traction load containing both axial and lateral components. The pre- and post-compressed RF frames were simulated with Field II~\cite{field2} setting the center frequency and the temporal sampling rate to 7.27 MHz and 40 MHz, respectively.

\subsubsection{Simulated Phantom with Different PRs}
A phantom with the same background and target elasticity moduli (20 kPa) but different Poisson’s ratios (0.45 for background and 0.25 for target) was simulated and compressed using ABAQUS. The RF frames were simulated with Field II using the same imaging setting as the multi-inclusion phantom.  

\subsubsection{Real Breast Phantom}
The experimental phantom data were collected at Concordia University's PERFORM Centre. A hand-held L3-12H linear array probe was used to compress a Zerdine$^{\textregistered}$-made CIRS breast phantom (Model 059). The Young's modulus of the soft tissue-like material was $20 \pm 5$~kPa, whereas the inclusion was at least twice as hard as the background. An Alpinion E-cube R12 research ultrasound system was employed to acquire RF data from the phantom while it was deformed. The transmit frequency and the temporal sampling rate, respectively, were fixed at 10 MHz and 40 MHz.

\subsubsection{\textit{In vivo} Liver Cancer Datasets}
The \textit{in vivo} experiments were conducted at the Johns Hopkins Hospital (Baltimore, MD), where three cancer patients' livers undergoing open-surgical RF thermal ablation were compressed using a hand-held VF 10-5 linear array probe. While the livers were deformed, time-series RF data were collected with an Antares Siemens research ultrasound machine setting the center and sampling frequencies to $6.67$~MHz and $40$~MHz, respectively. The Institutional Review Board approved this study, and written consent was obtained from all patients. Interested readers can find more details of this experiment in~\cite{DPAM}.

\subsection{Parameter Selection}
The two proposed techniques' performances were compared with those of NCC, NCC refined by partial differential equation (PDE)-based technique (NCC + PDE)~\cite{guo2015pde}, SOUL, and $L1$-SOUL. It is worth mentioning that we implemented both NCC and NCC + PDE in this work for comparison purposes. As predecessors of MechSOUL and $L1$-MechSOUL, both SOUL and $L1$-SOUL have been used as comparison benchmarks to assess the impacts of the proposed techniques.

The RF frames were upsampled by a factor of 3 using the MATLAB function \textit{imresize} for the implementation of NCC. The optimal window length and overlap, respectively, were determined as $15 \lambda (= 3 \times 5 \lambda)$ and $86\%$ by manually tuning NCC's performance on a validation set of input frames. The optimal parameter values obtained from the validation frames were used for generating the results for the test frame sets, which are reported in this paper. The estimated axial and lateral displacement fields were resized back to the RF frames' original size with a scaling factor of $\frac{1}{3}$ for the displacement estimates. As suggested in \cite{guo2015pde}, the ratio of the lateral and axial fidelity weights was set to 100 for the PDE-based refinement technique.

The tunable parameters of SOUL, $L1$-SOUL, MechSOUL, and $L1$-MechSOUL were optimized for simulated, phantom, and \textit{in vivo} datasets using validation sets of RF frames according to a cross-validation strategy to avoid any bias and data leakage. The axial and lateral strain images for a large range of possible parameter values were generated. The best parameter set was chosen by visually assessing the strain images’ contrast, background smoothness, and boundary sharpness. This optimal parameter set was used to produce the results for test images, which are reported in this work. The optimal values for $\{\alpha_{1},\alpha_{2},\beta_{1},\beta_{2},w,\gamma\}$ and $\{\alpha_{1s},\alpha_{2s},\beta_{1s},\beta_{2s},w_{f},w_{s},\gamma_{s}\}$ have been shown in Tables I and II of the Supplementary Video. For simulated, phantom, and \textit{in vivo} datasets, respectively, the sharpness controlling parameter $\eta$ was set to 0.001, 0.0006, and 0.008 for the first-order terms and 0.0005, 0.0001, and 0.0013 for the second-order terms. The mechanical constancy weights $\{\alpha_{3},\alpha_{3s}\}$ were set to $\{20,0.045\}$, $\{80,0.072\}$, and $\{5,0.1\}$ for simulated, phantom, and \textit{in vivo} datasets. $\eta_{m}$ was fixed at 0.001, 0.0006, and 0.008, respectively, for the simulated, phantom, and \textit{in vivo} datasets.   

\subsection{Quantitative Metrics}
The ground truth being available, the simulation results have been assessed using root-mean-square error (RMSE) and the peak signal-to-noise ratio (PSNR). RMSE is defined as:

\begin{equation}
\textrm{RMSE}=\sqrt{\frac{\sum\limits_{j=1}^n \sum\limits_{i=1}^m (\hat{q}_{i,j}-q_{i,j})^{2}}{mn}} 
\end{equation}

\noindent
where $\hat{q}_{i,j}$ and $q_{i,j}$ denote the estimated and ground truth values (either strain or EPR) at $(i,j)$. For both simulated and real datasets, elastographic signal-to-noise ratio (SNR) and contrast-to-noise ratio (CNR) have also been reported. SNR and CNR are given by:

\begin{equation}
\textrm{SNR}=\frac{\bar{s_{b}}}{\sigma_{b}} \qquad
\textrm{CNR}=\frac{C}{N}=\sqrt{\frac{2(\bar{{s_{b}}}-\bar{{s_{t}}})^2}{{\sigma_b}^2+{\sigma_t}^2}}
\end{equation}

\noindent
where $\bar{s_{b}}$ and $\bar{s_{t}}$ refer to the mean and $\sigma_b$ and $\sigma_t$ denote the standard deviations of the background and target windows, respectively. Other metrics used in the beamforming community can also be used for quantitative comparisons~\cite{rodriguez2019generalized,bottenus2020histogram,schlunk2022breaking}.

\section{Results}
Calculating the SNR on a single background window and the CNR between a target-background window pair is a common practice in quasi-static ultrasound elastography papers. Nevertheless, elastographic SNR and CNR are highly sensitive to window selection; therefore, single values often fail to quantify the differences among different techniques' performance properly. To tackle this issue, we sweep two 3 mm $\times$ 3 mm spatial windows over the background and the target and calculate 50 SNR (50 background windows) and 150 CNR (3 target and 50 background windows) values. Finally, we summarize the quantitative performance by showing the box plots, mean, and standard deviations of the aforementioned SNR and CNR values.


Substantial improvements in lateral strain and EPR are the main strengths of the proposed algorithms. Therefore, the axial strain images, which are less attractive in this work, are shown in the Supplemental Video for most of the datasets.

\begin{figure*}
	\centering
	\subfigure[Simulated phantom]{{\includegraphics[width=0.2\textwidth]{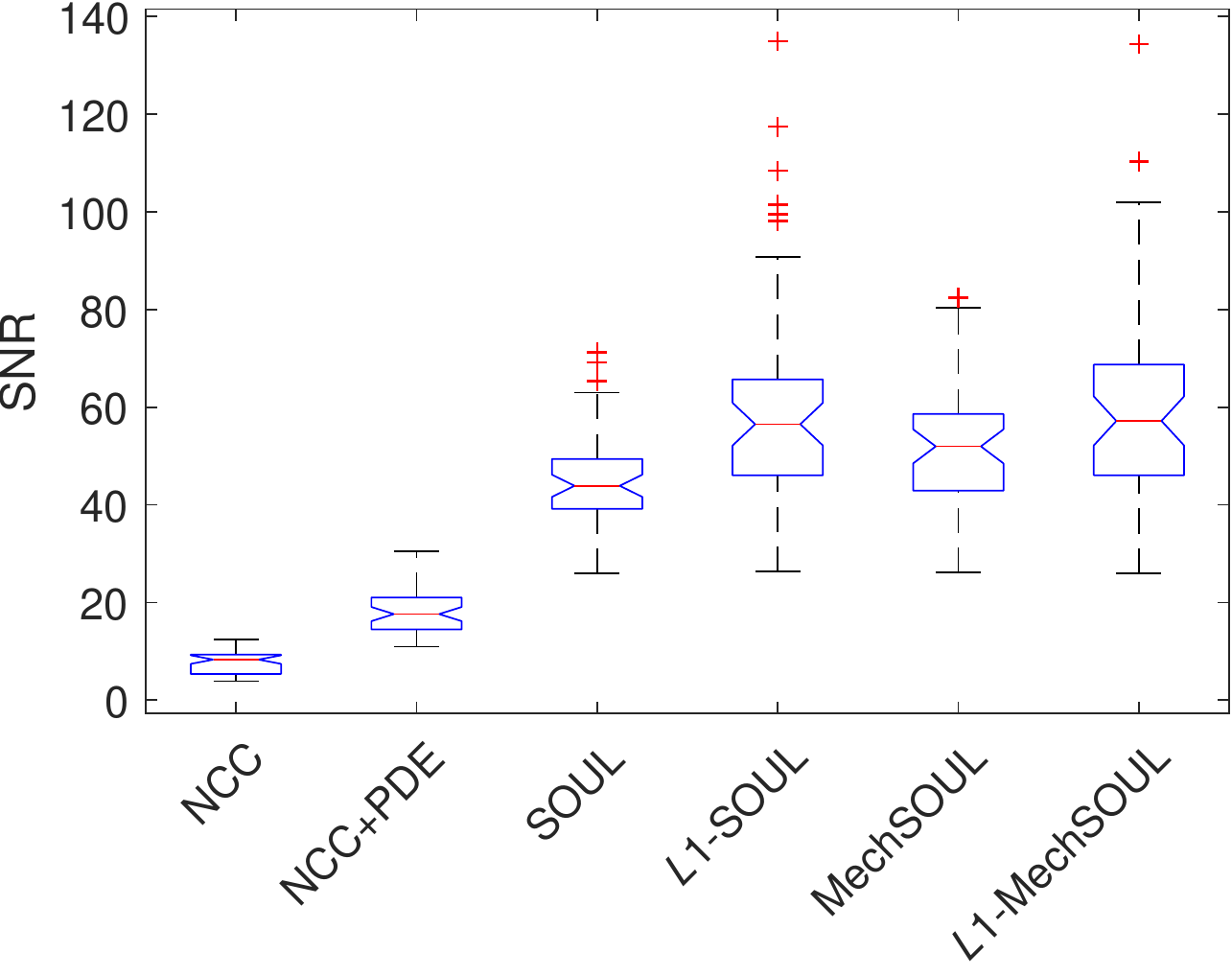}}}%
	\subfigure[Real phantom]{{\includegraphics[width=0.2\textwidth]{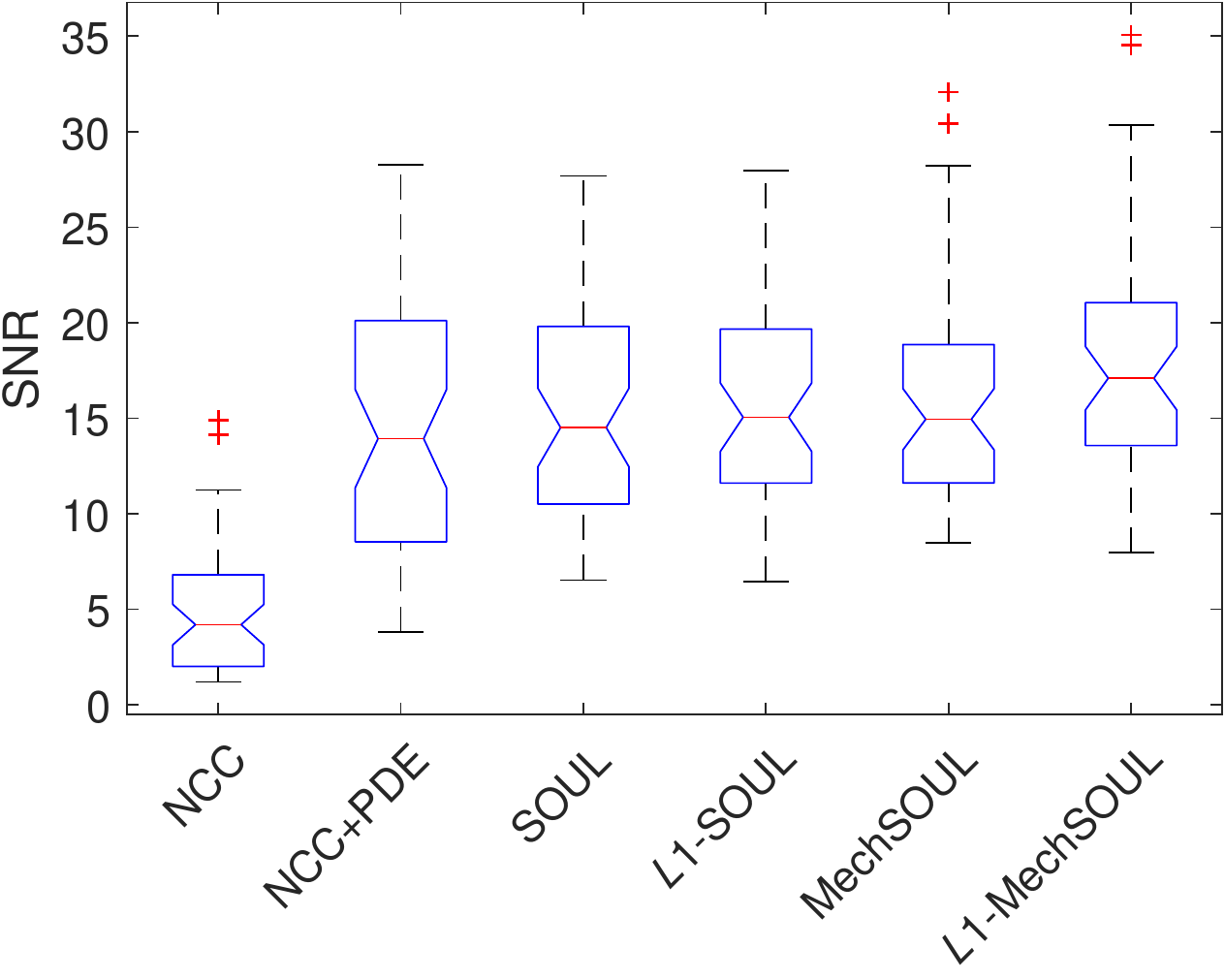}}}%
	\subfigure[Liver patient 1]{{\includegraphics[width=0.2\textwidth]{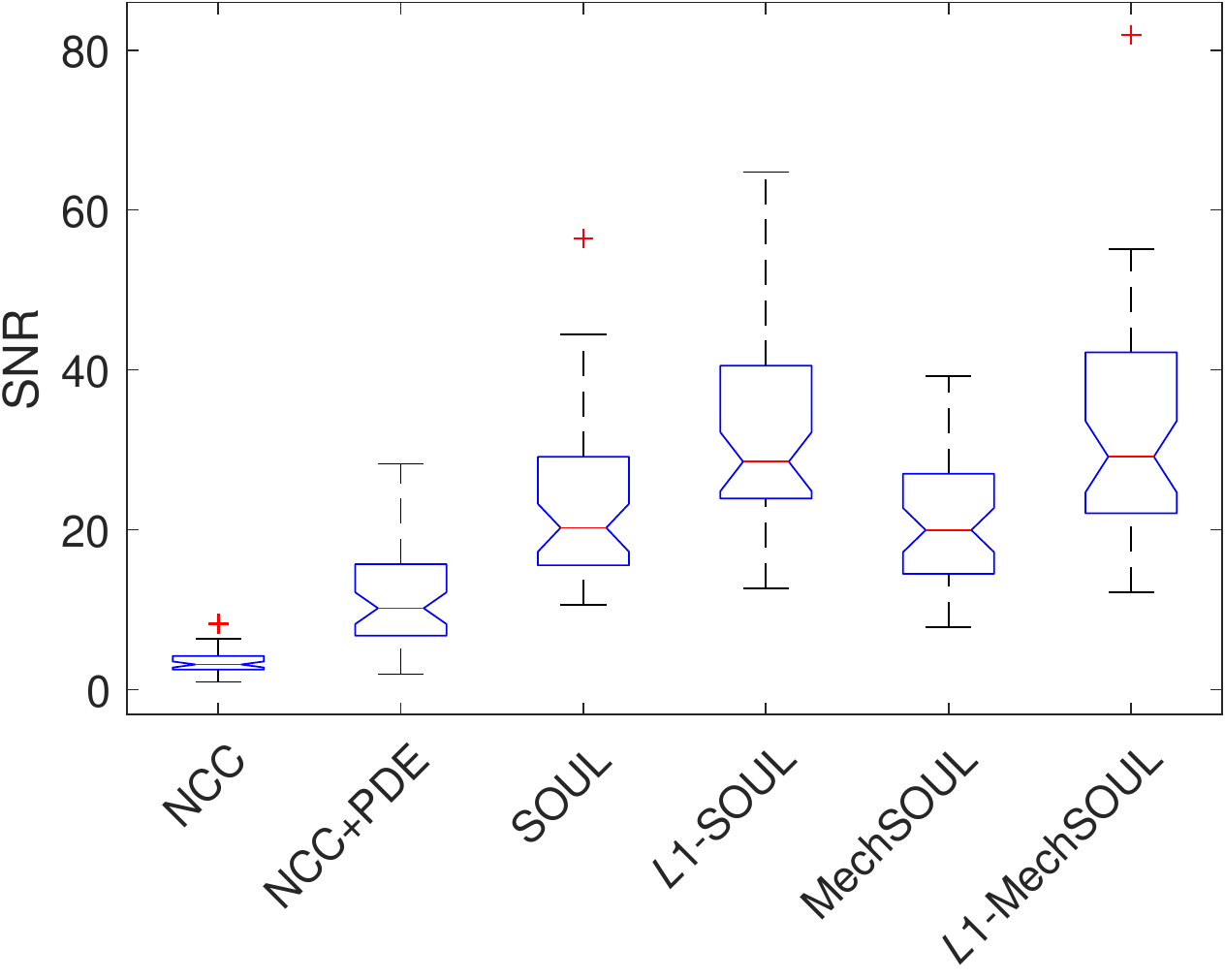}}}%
	\subfigure[Liver patient 2]{{\includegraphics[width=0.2\textwidth]{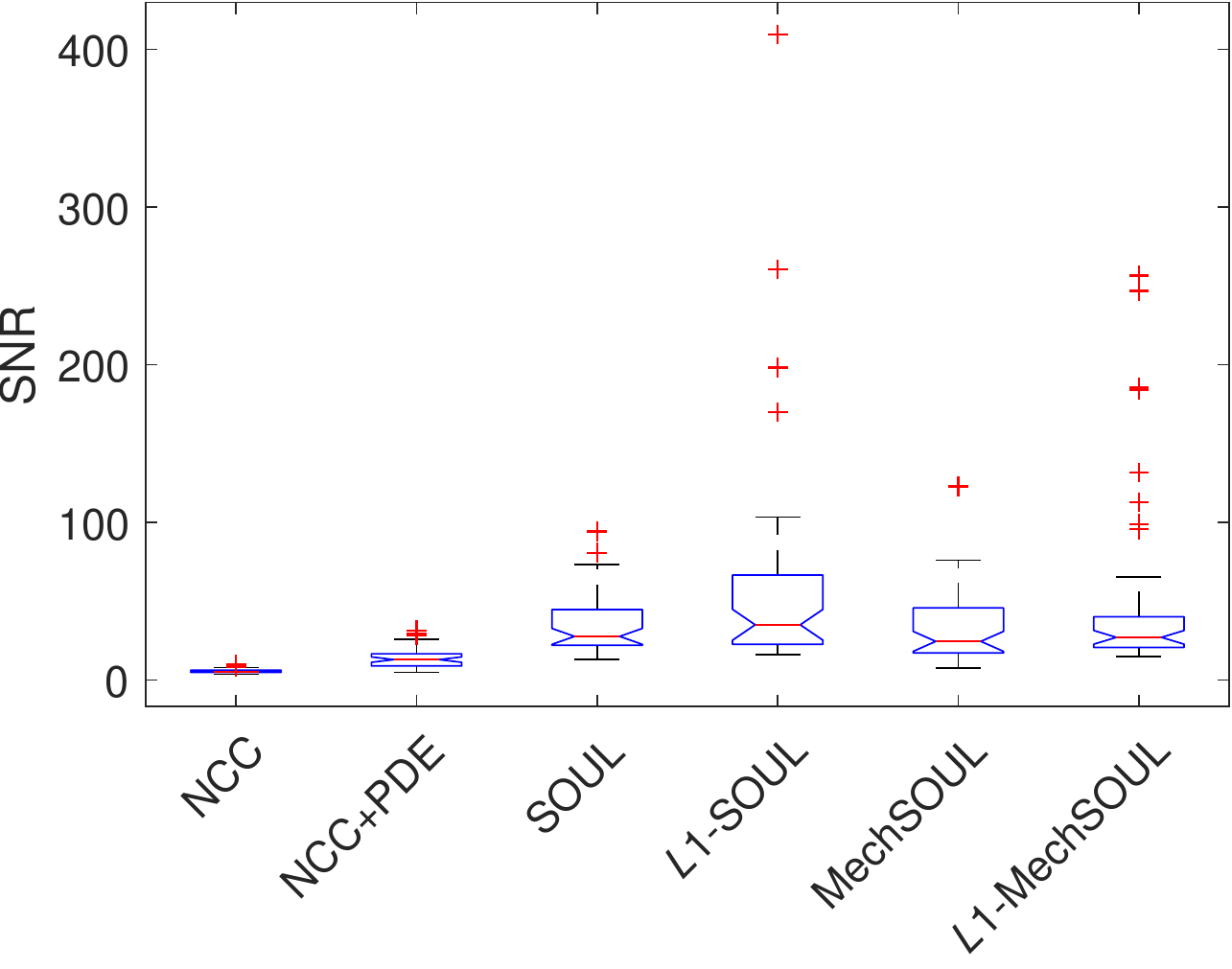}}}%
	\subfigure[Liver patient 3]{{\includegraphics[width=0.2\textwidth]{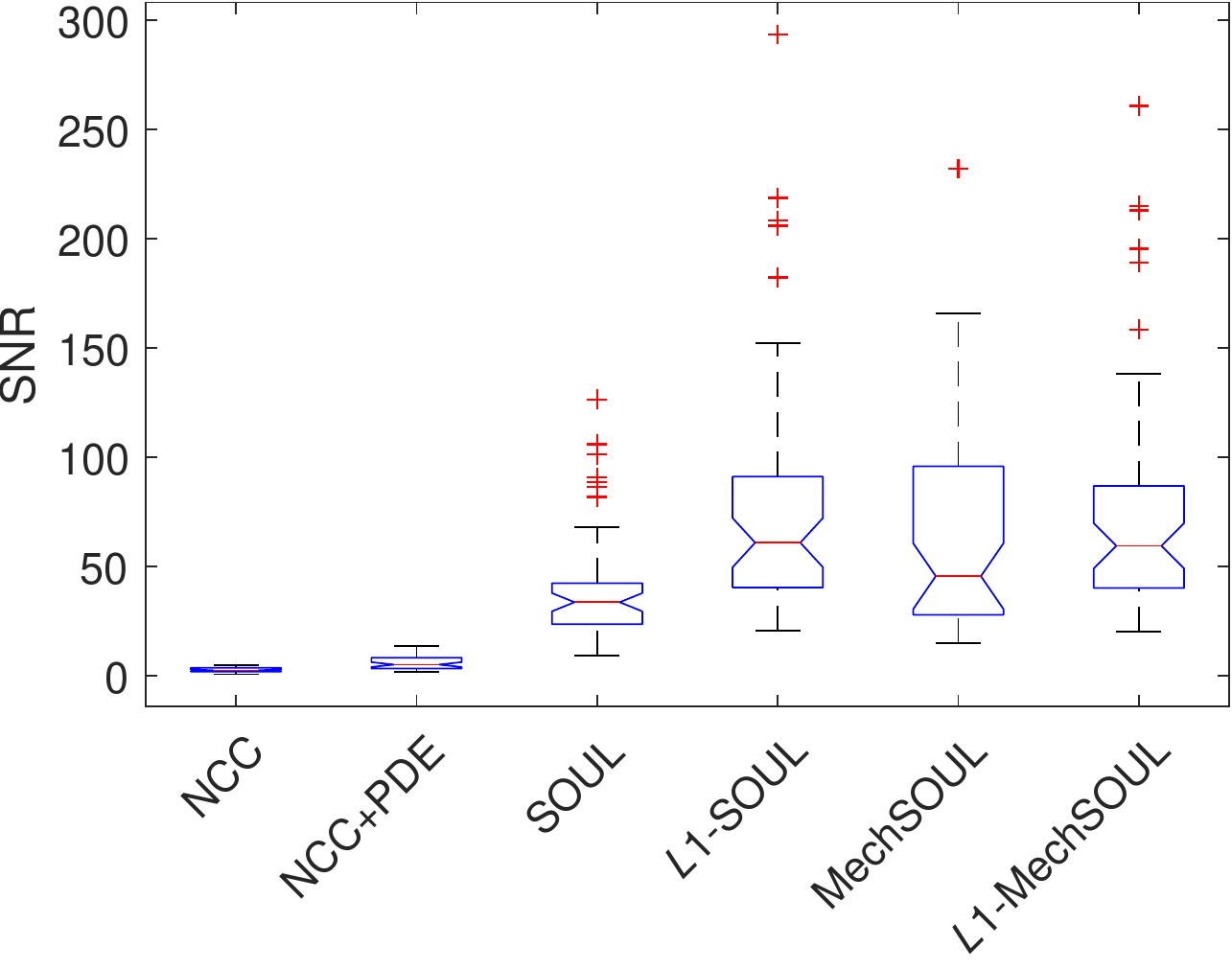}}}
	\subfigure[Simulated phantom]{{\includegraphics[width=0.2\textwidth]{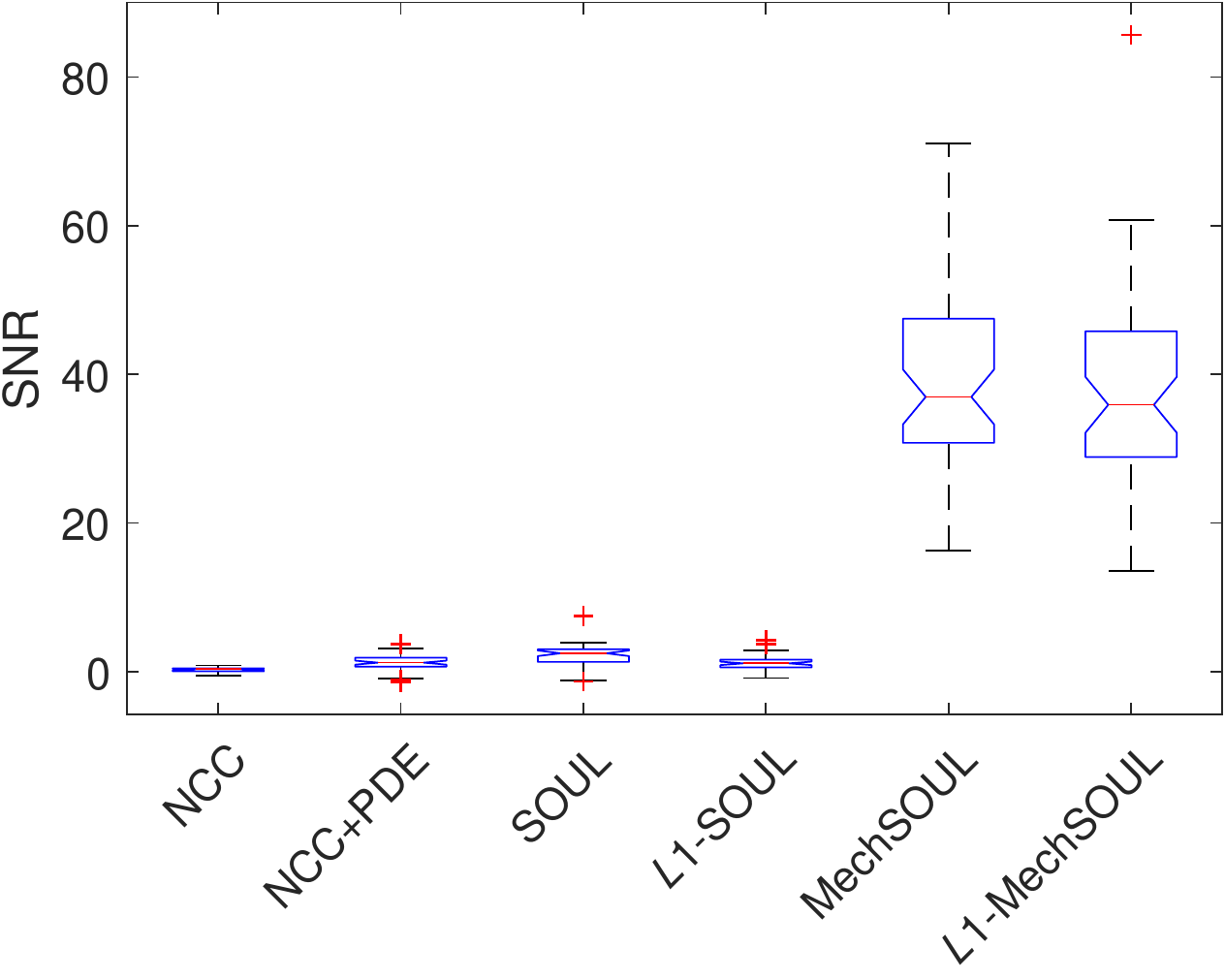}}}%
	\subfigure[Real phantom]{{\includegraphics[width=0.2\textwidth]{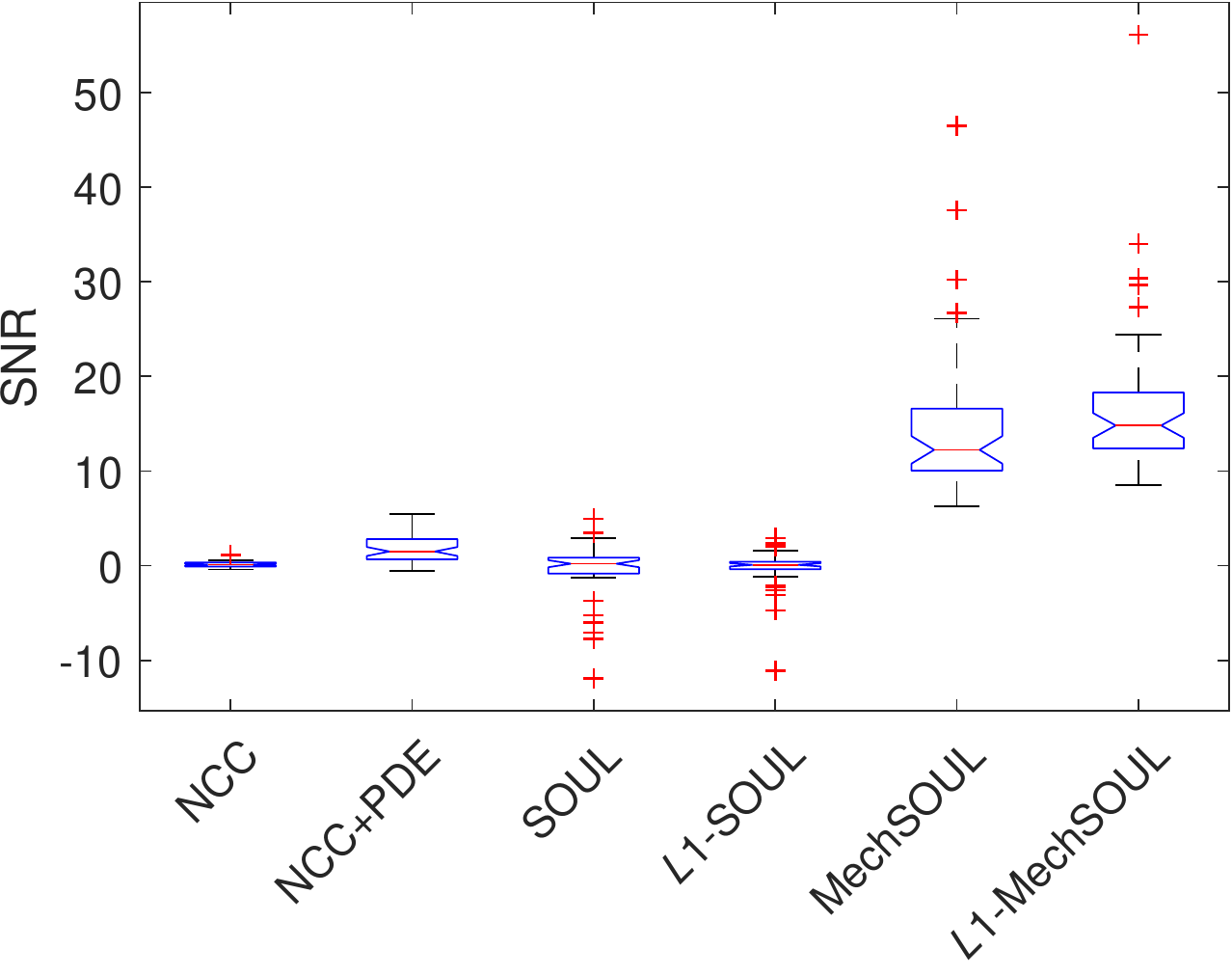}}}%
	\subfigure[Liver patient 1]{{\includegraphics[width=0.2\textwidth]{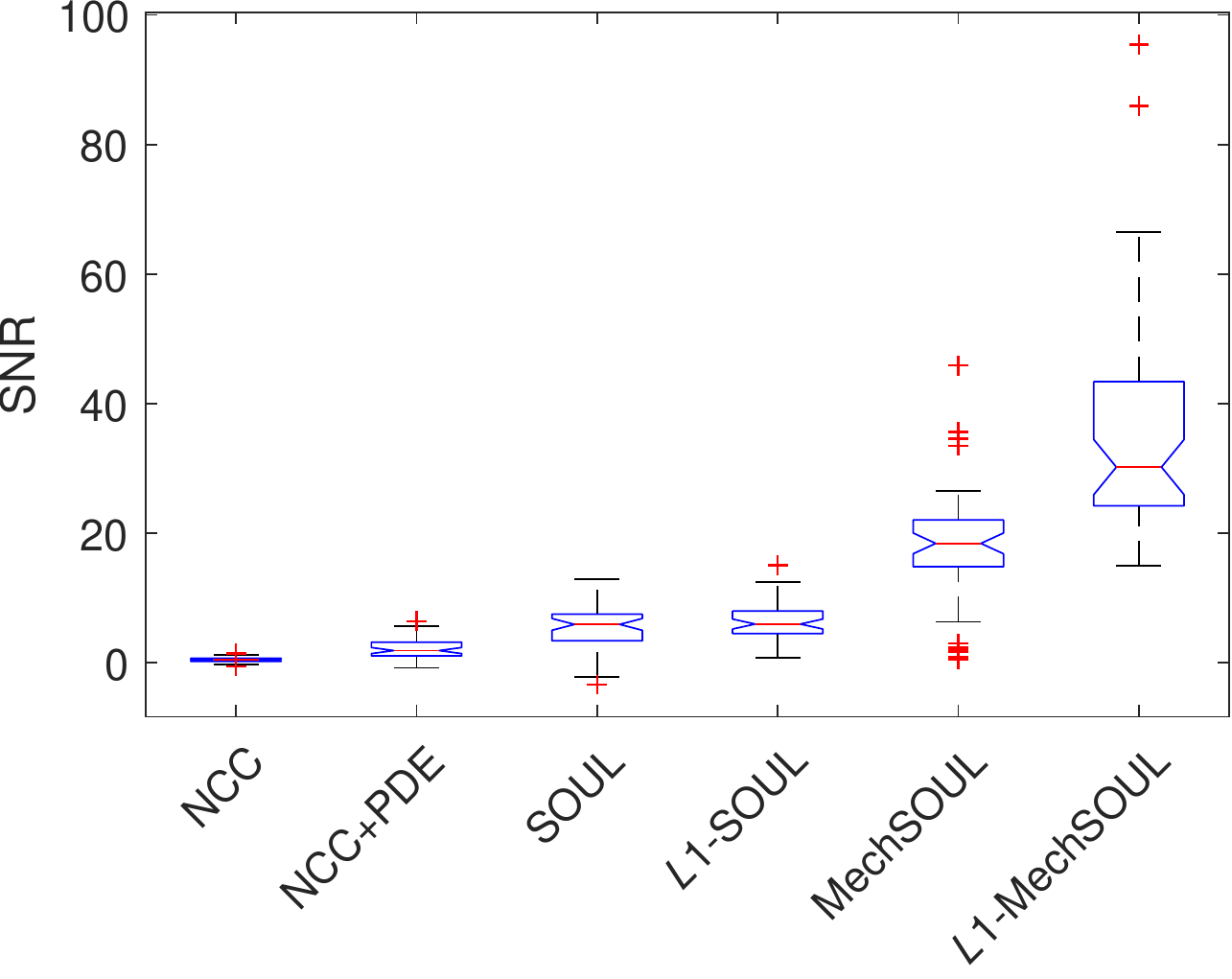}}}%
	\subfigure[Liver patient 2]{{\includegraphics[width=0.2\textwidth]{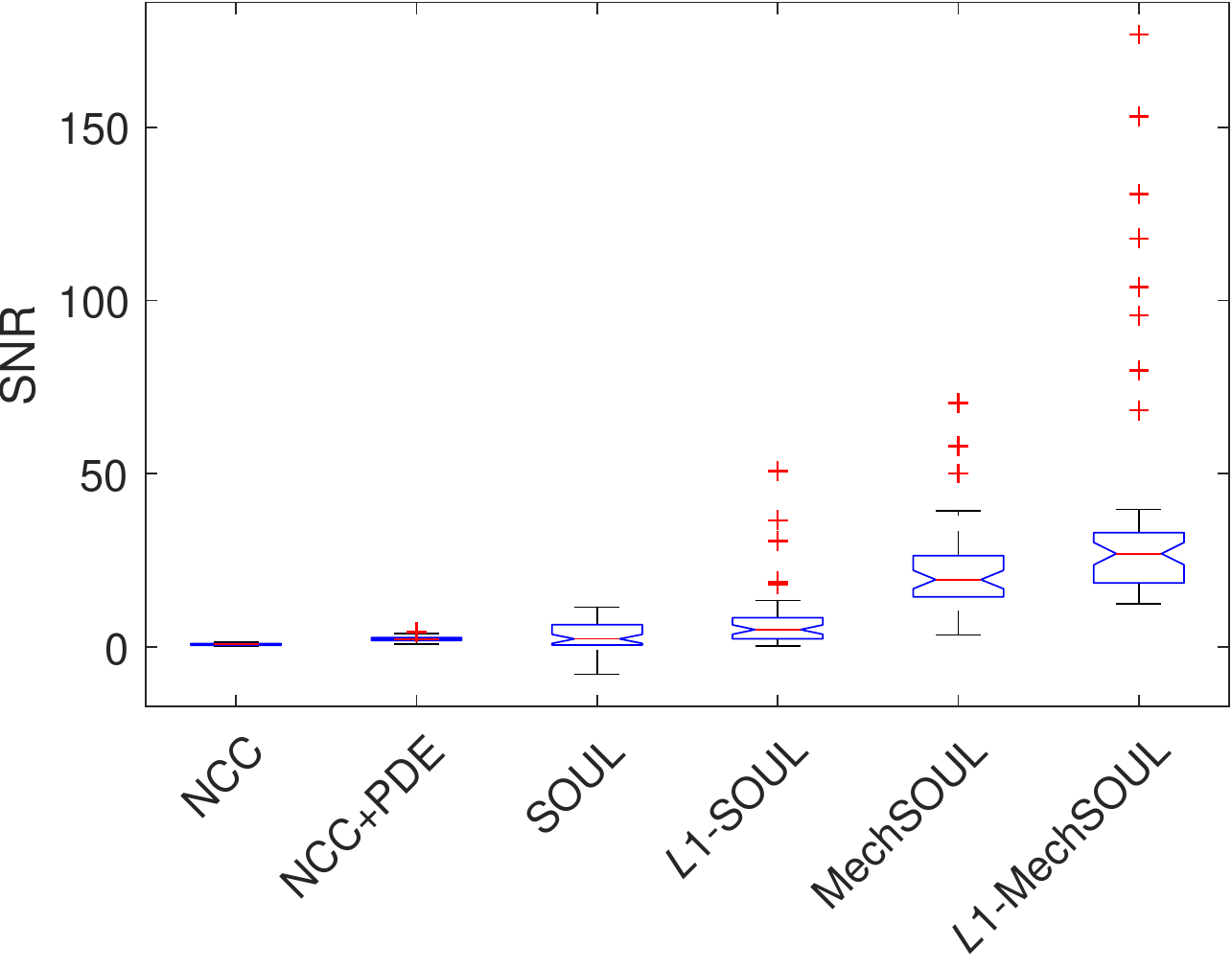}}}%
	\subfigure[Liver patient 3]{{\includegraphics[width=0.2\textwidth]{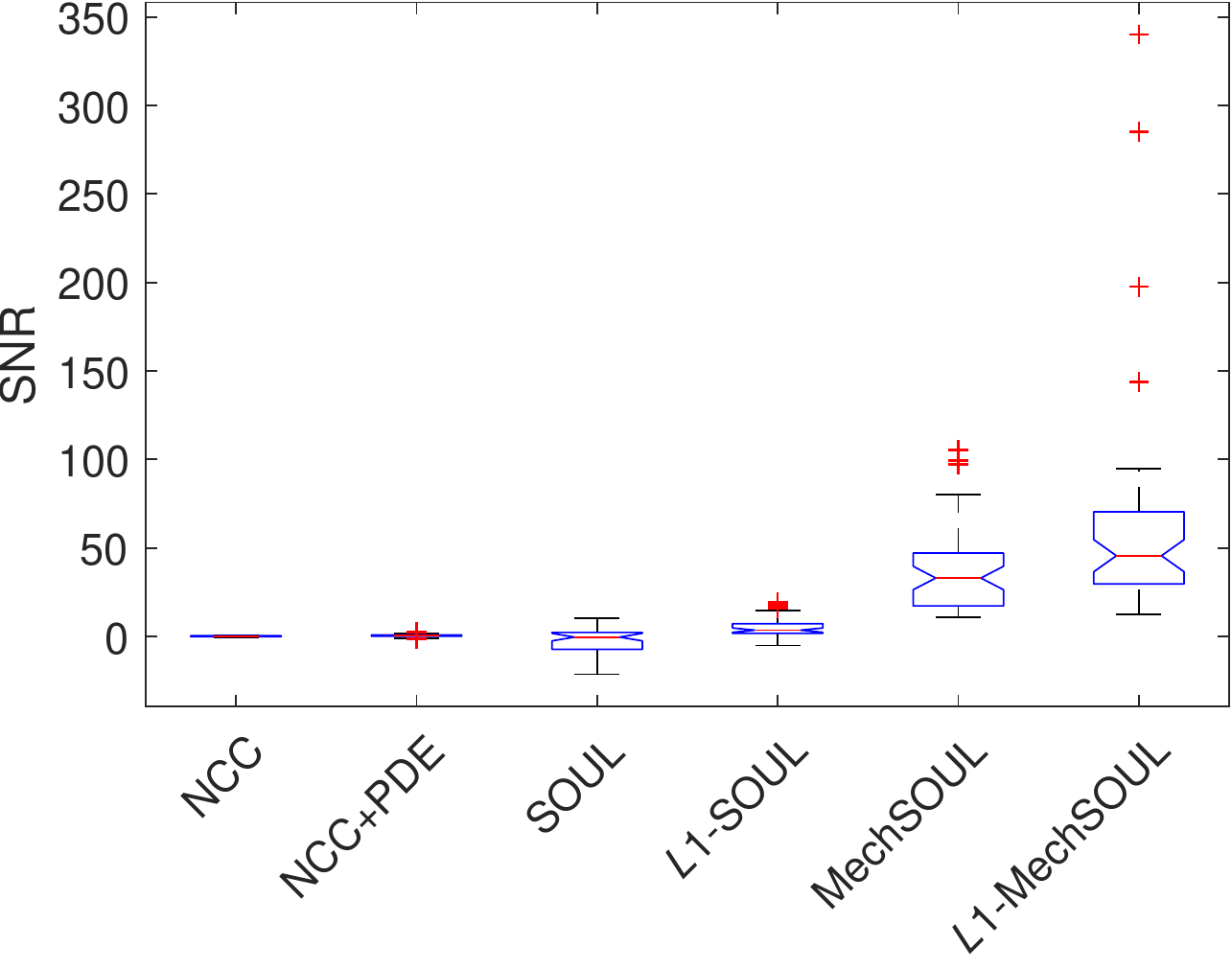}}}
	\caption{Box plots for 50 SNR values. Rows 1 and 2, respectively, correspond to axial and lateral, whereas columns 1 to 5 correspond to hard-inclusion simulated phantom, real phantom, and liver patients 1-3, respectively.}
	\label{snr_plots}
\end{figure*}

\begin{figure*}
	\centering
	\subfigure[Simulated phantom]{{\includegraphics[width=0.2\textwidth]{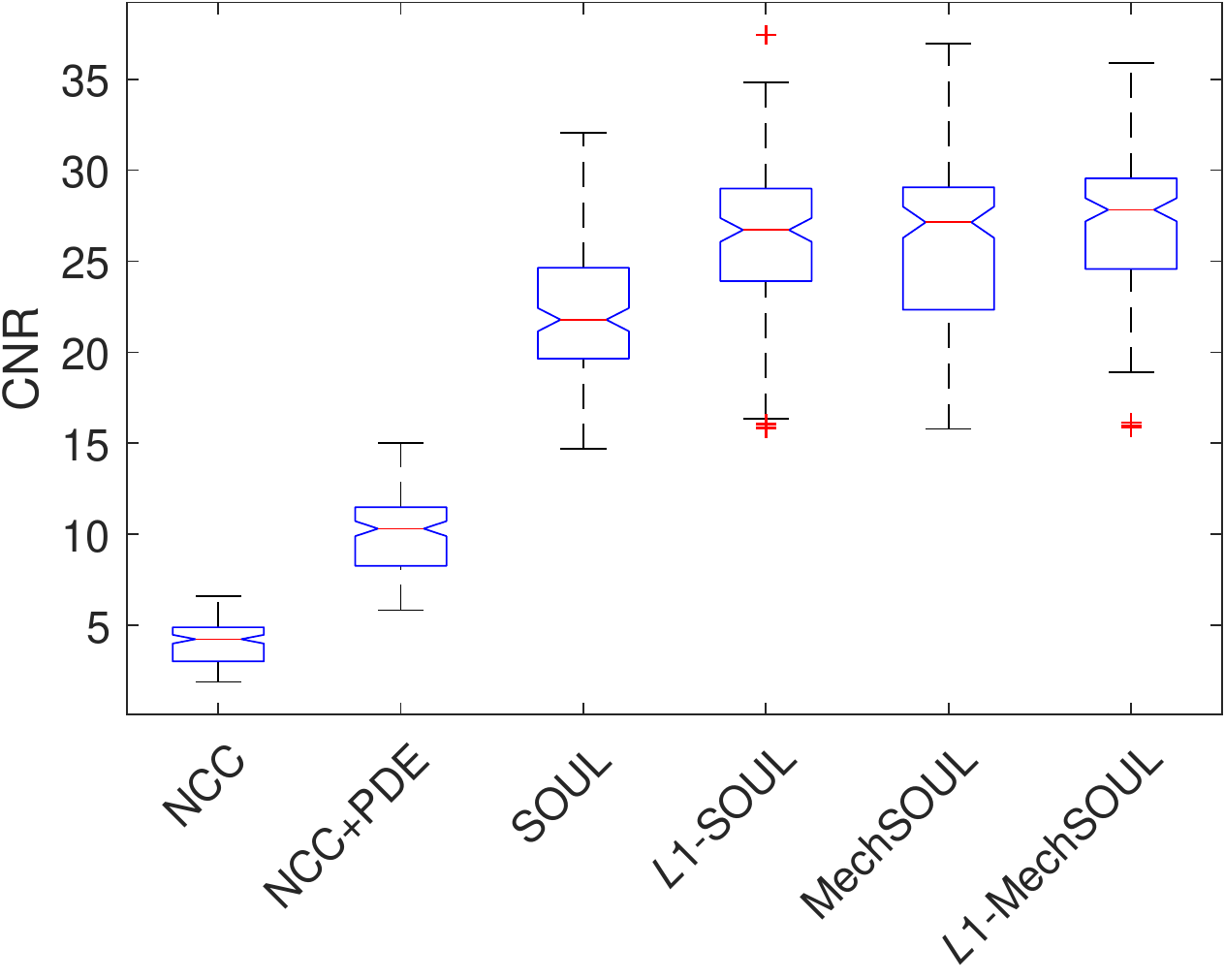}}}%
	\subfigure[Real phantom]{{\includegraphics[width=0.2\textwidth]{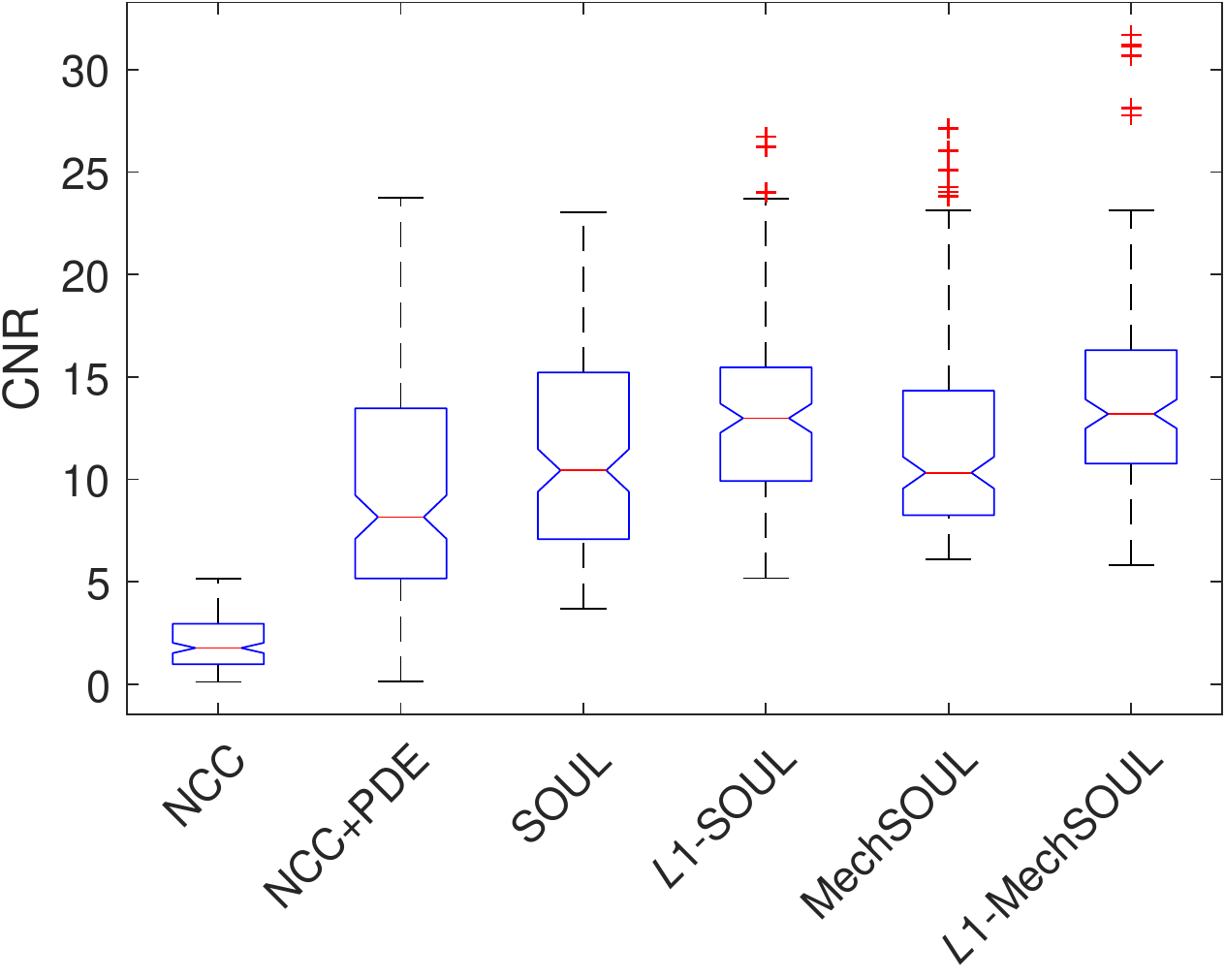}}}%
	\subfigure[Liver patient 1]{{\includegraphics[width=0.2\textwidth]{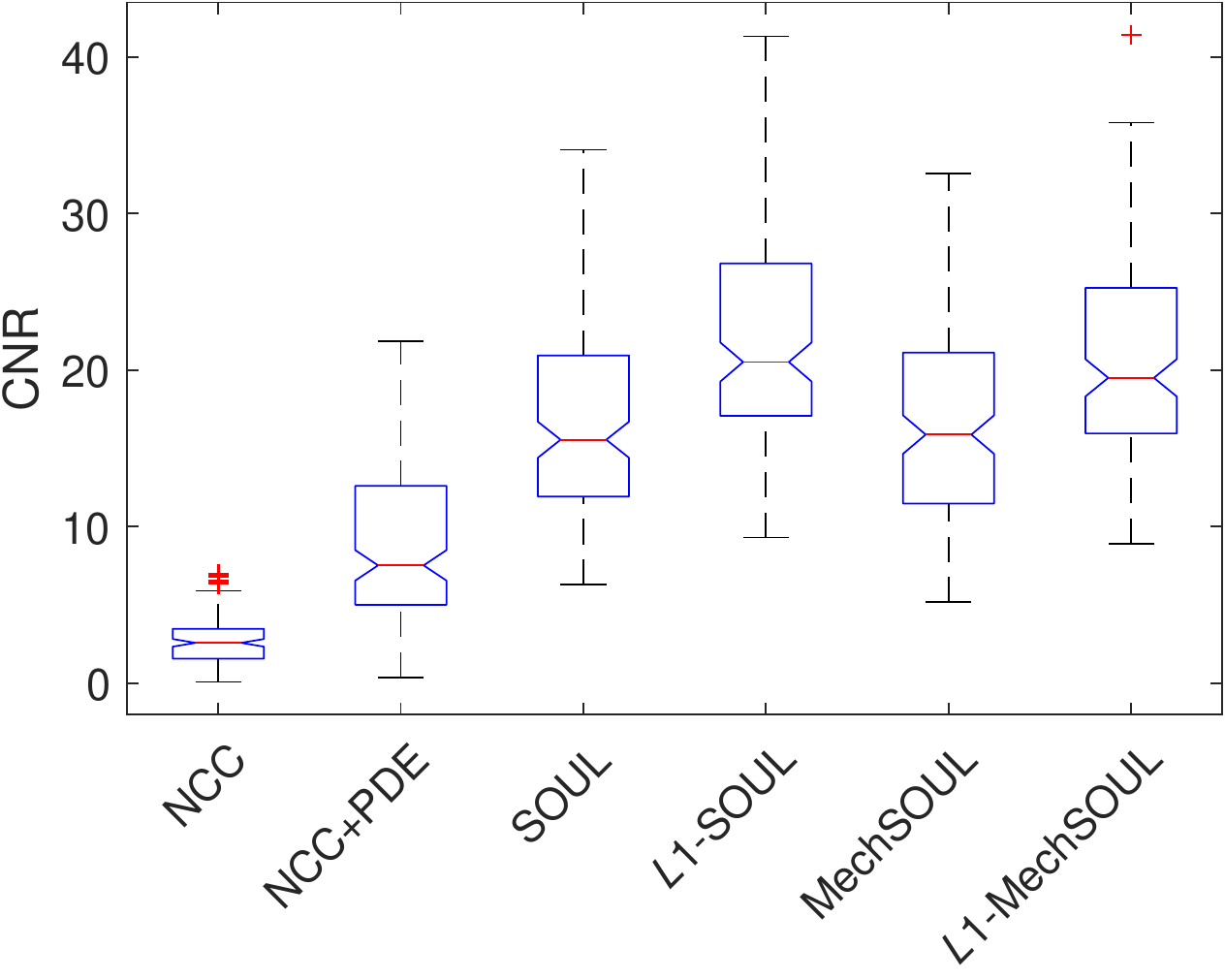}}}%
	\subfigure[Liver patient 2]{{\includegraphics[width=0.2\textwidth]{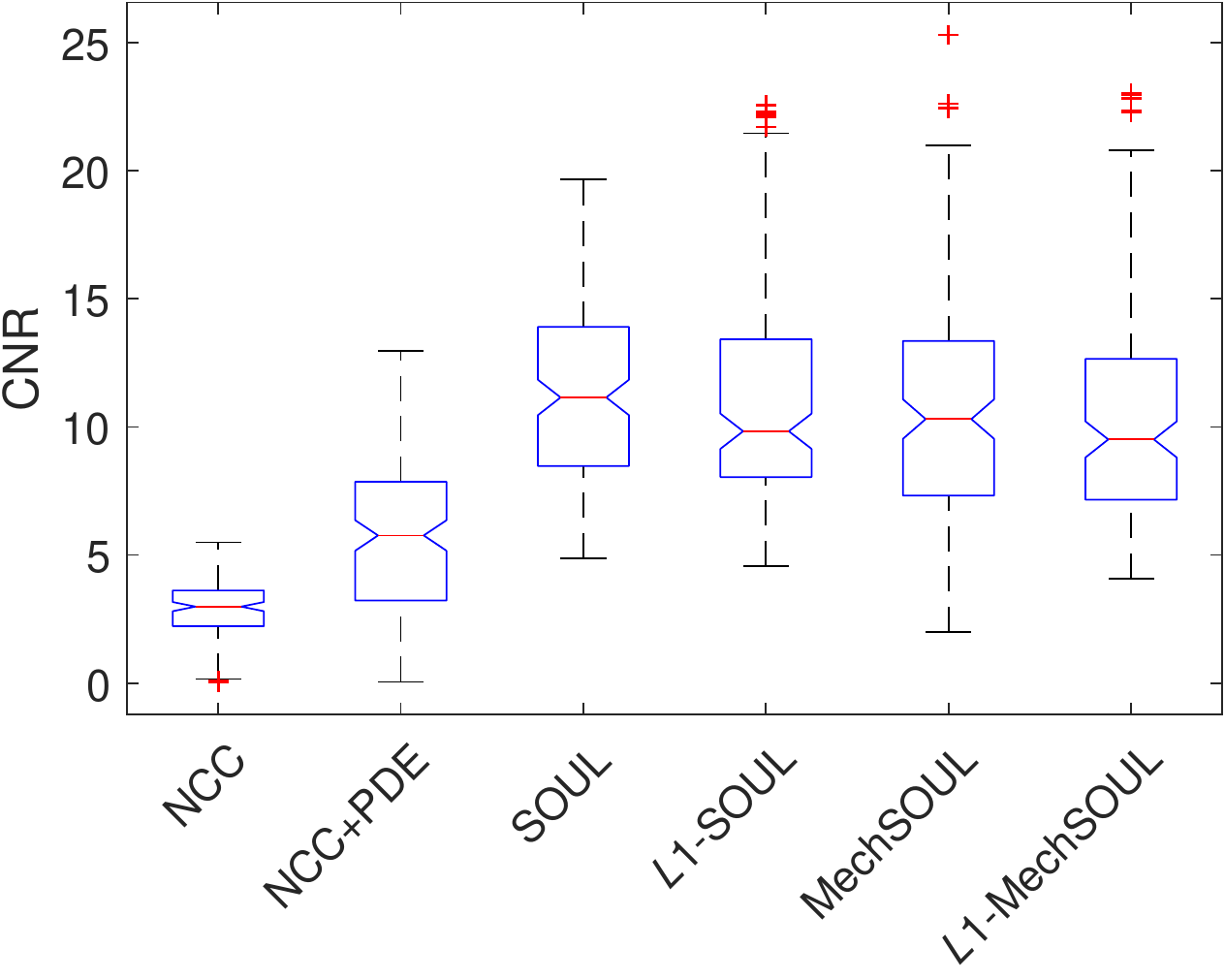}}}%
	\subfigure[Liver patient 3]{{\includegraphics[width=0.2\textwidth]{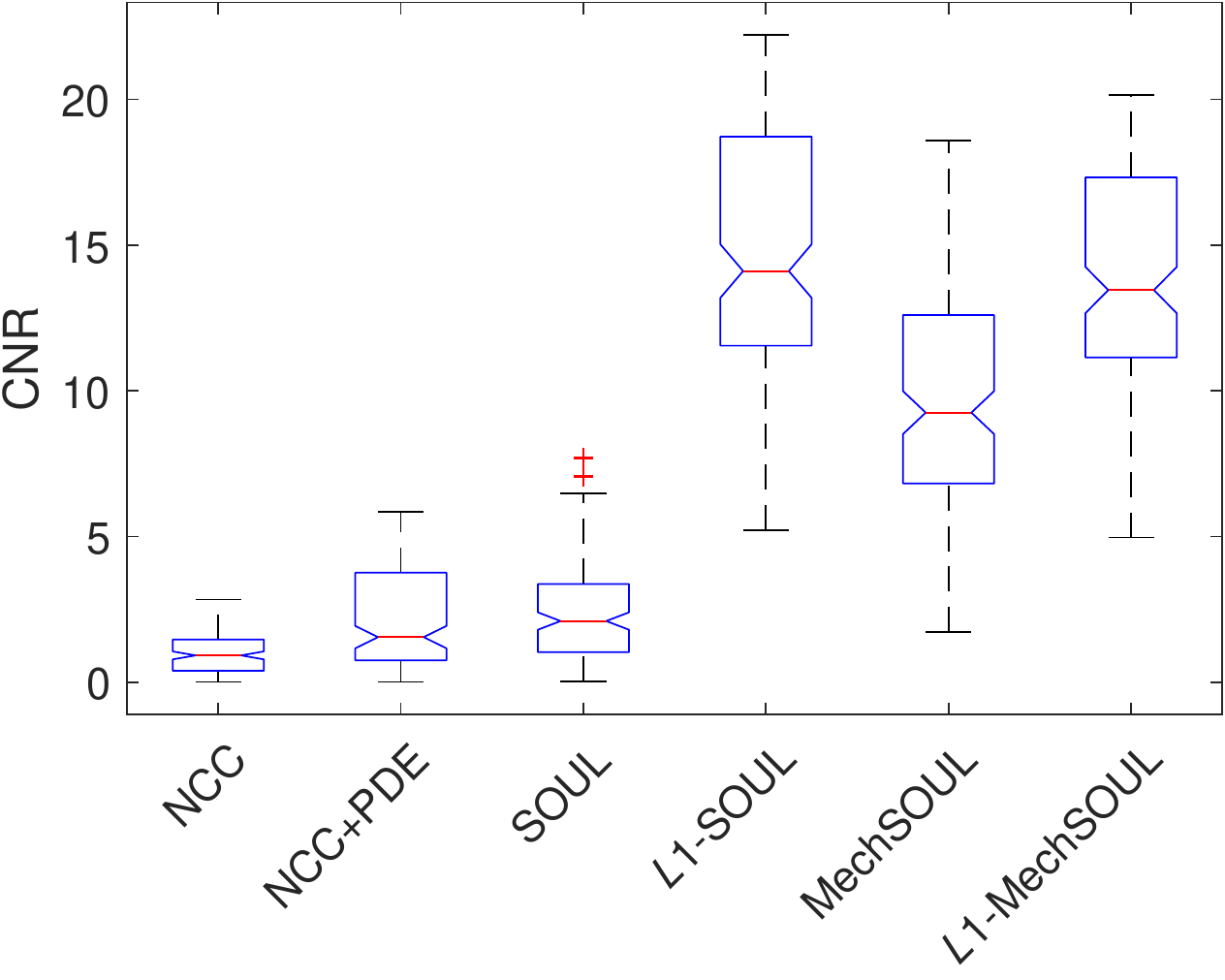}}}
	\subfigure[Simulated phantom]{{\includegraphics[width=0.2\textwidth]{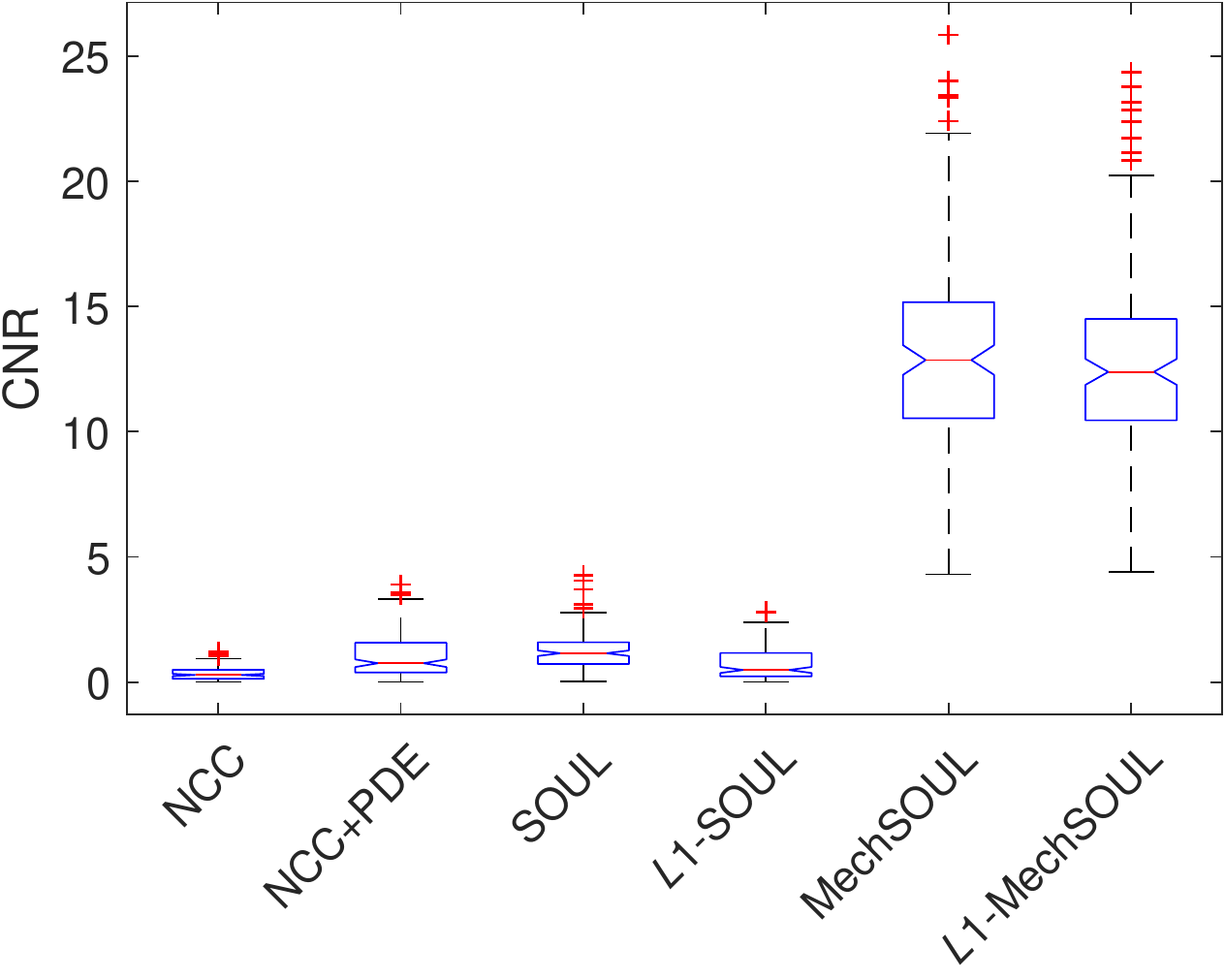}}}%
	\subfigure[Real phantom]{{\includegraphics[width=0.2\textwidth]{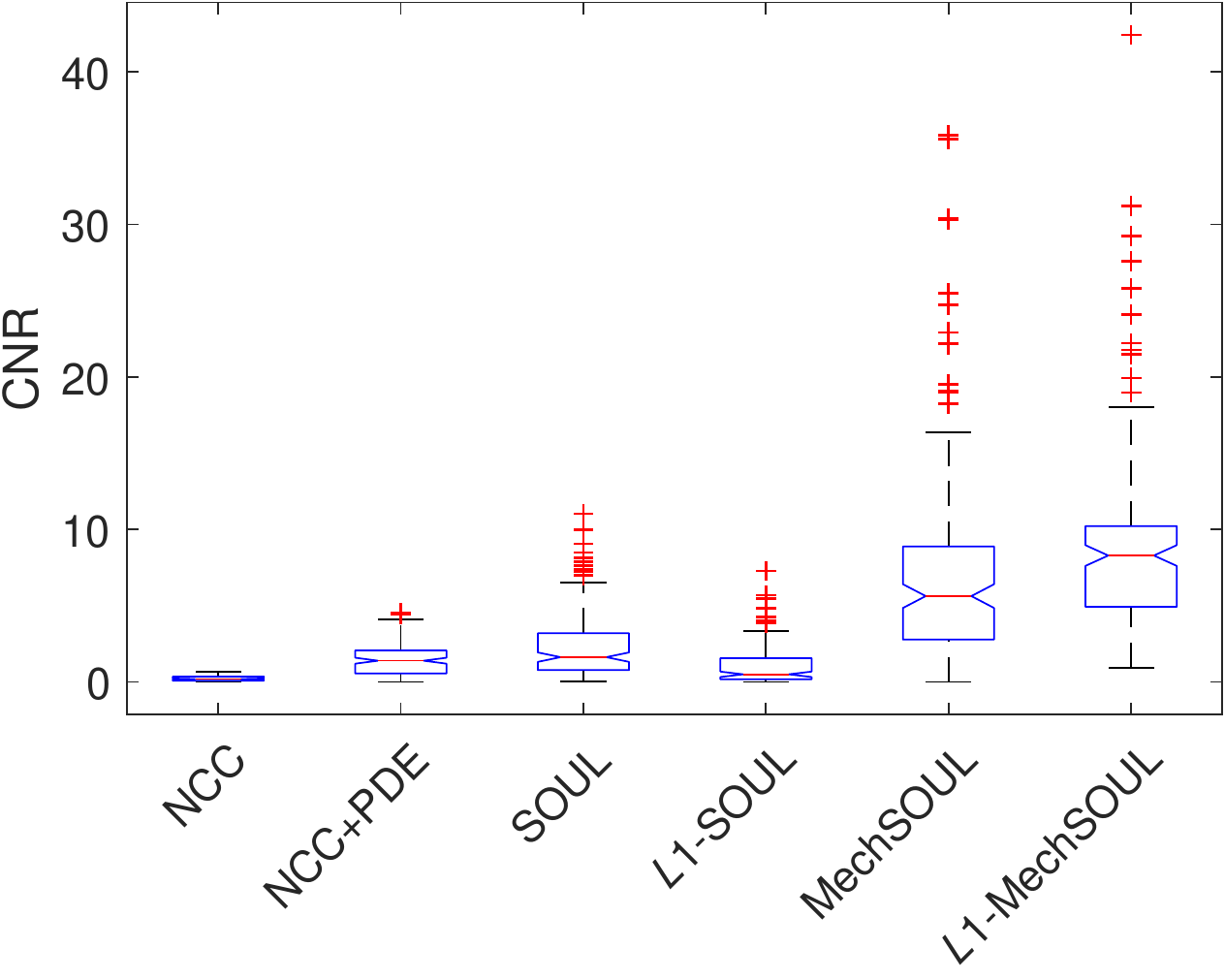}}}%
	\subfigure[Liver patient 1]{{\includegraphics[width=0.2\textwidth]{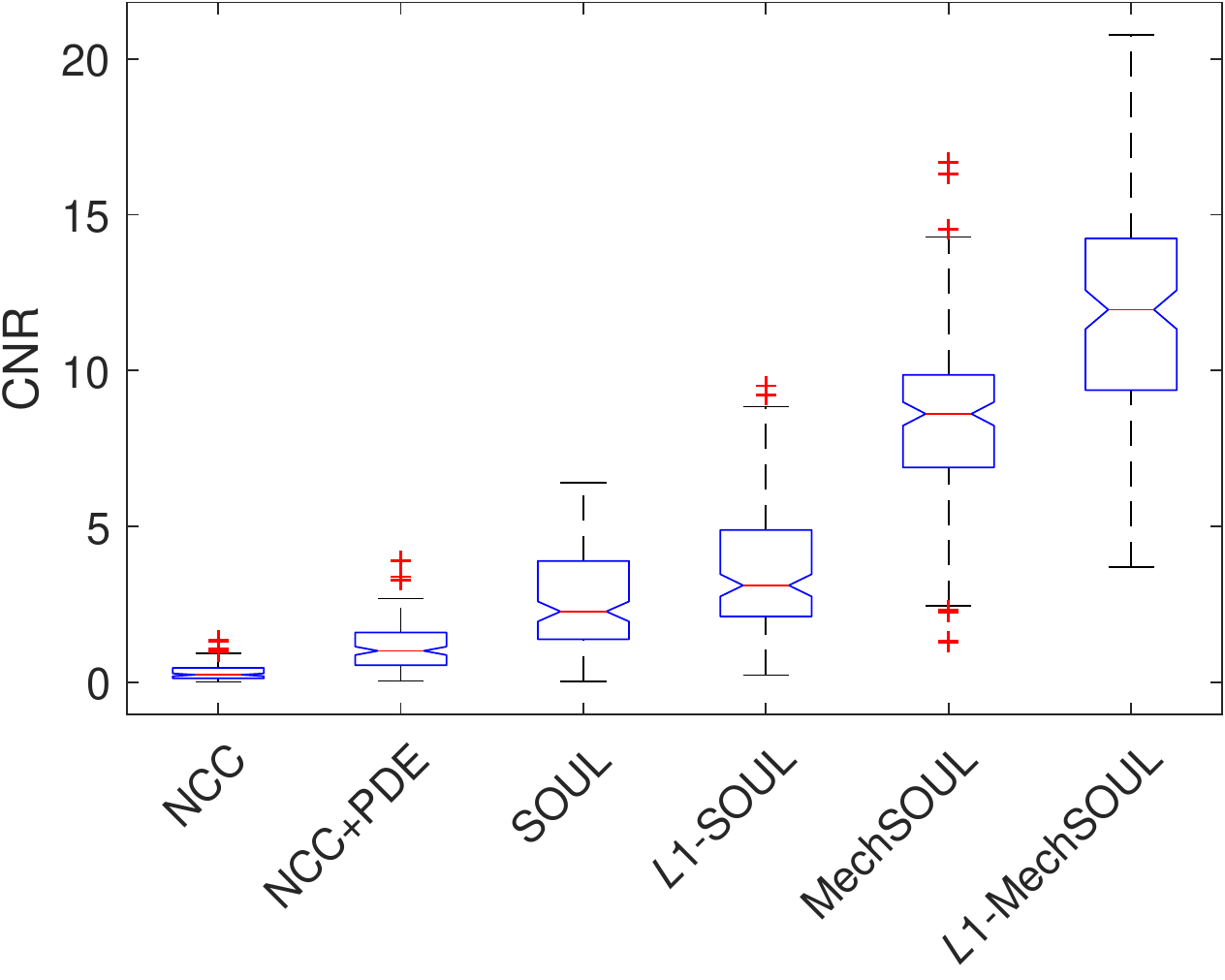}}}%
	\subfigure[Liver patient 2]{{\includegraphics[width=0.2\textwidth]{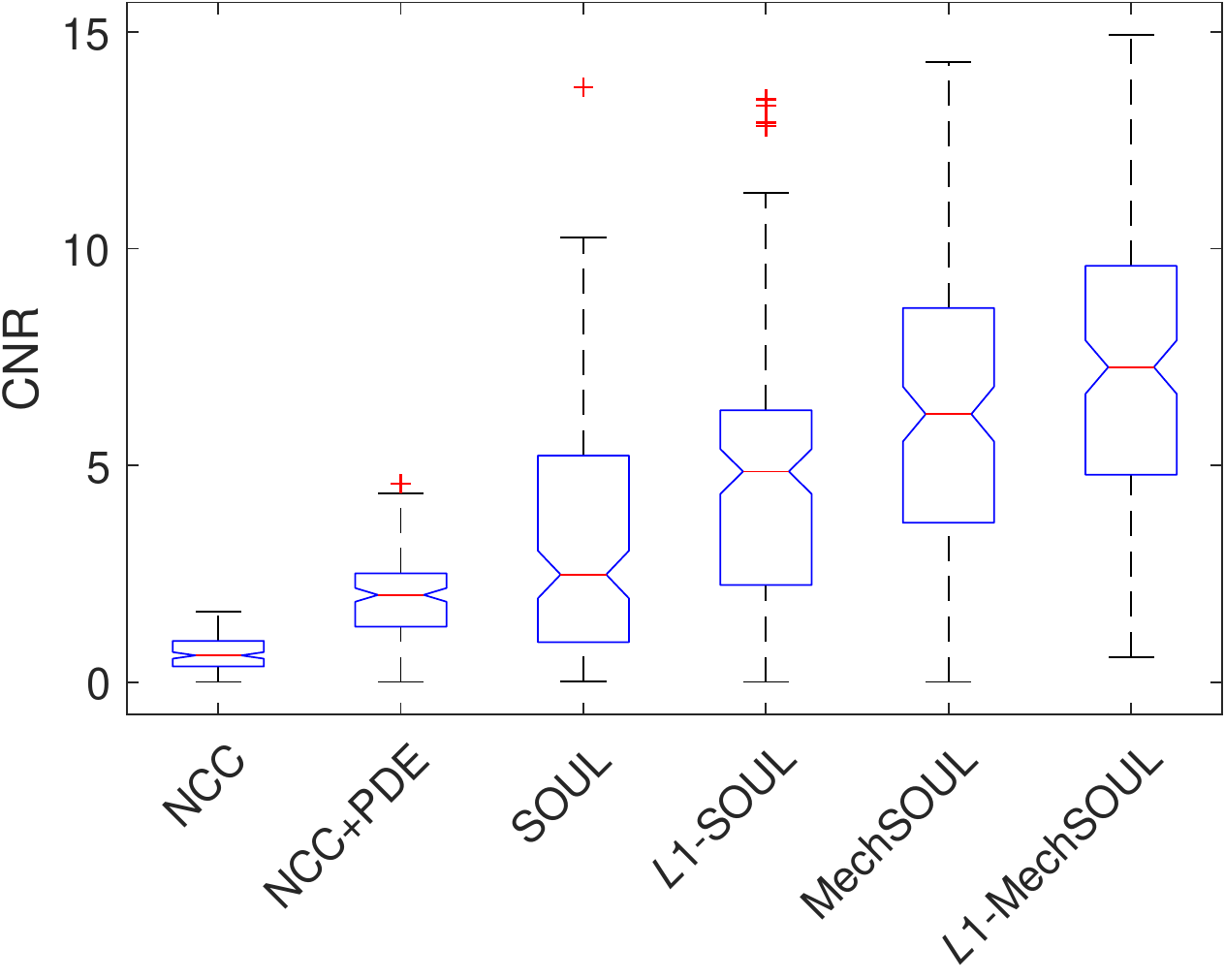}}}%
	\subfigure[Liver patient 3]{{\includegraphics[width=0.2\textwidth]{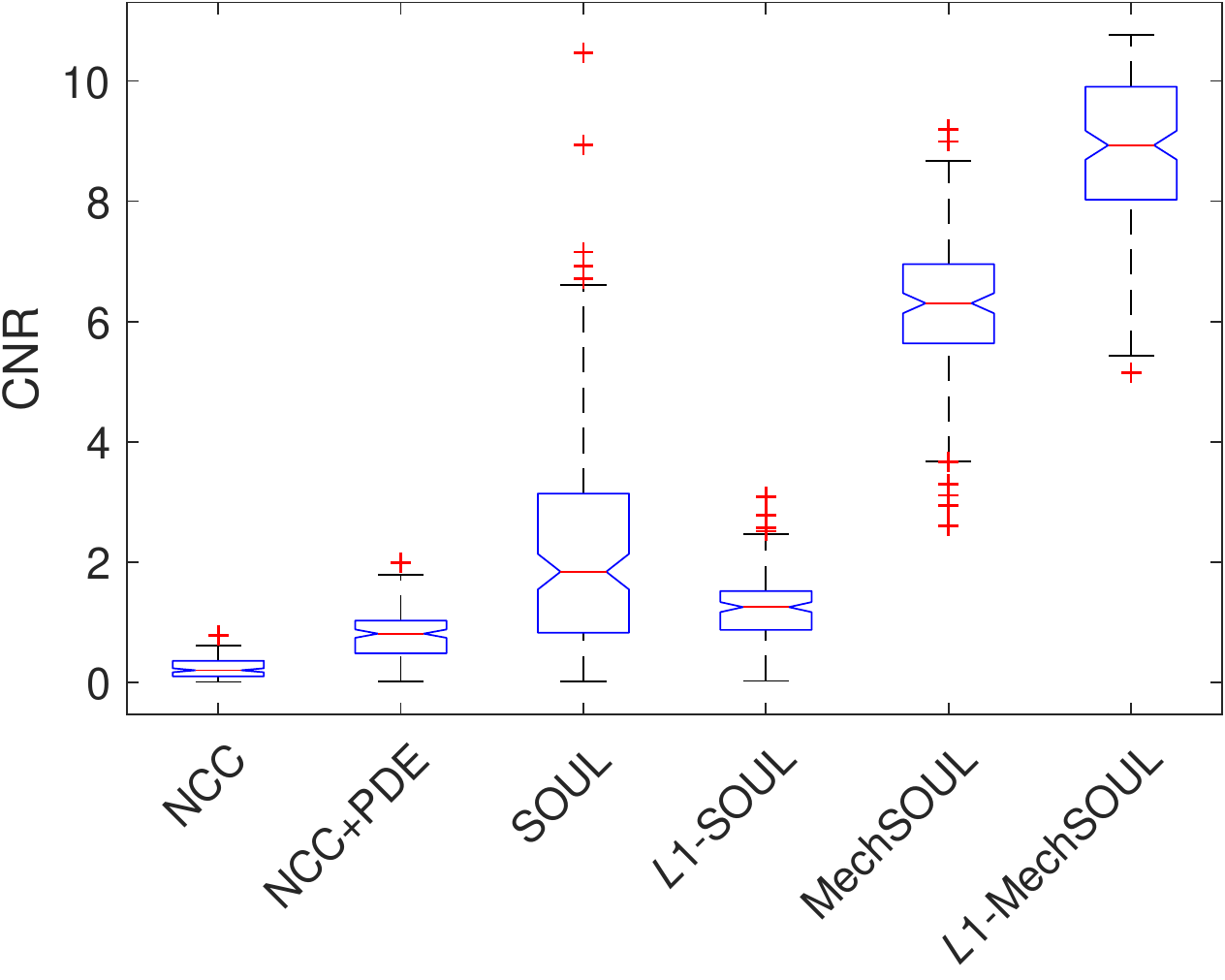}}}
	\caption{Box plots for 150 CNR values. Rows 1 and 2 correspond to axial and lateral, respectively, whereas columns 1 to 5 correspond to hard-inclusion simulated phantom, real phantom, and liver patients 1, 2, and 3, respectively.}
	\label{cnr_plots}
\end{figure*}

\subsection{Hard-inclusion Simulated Phantom Dataset}     
Fig.~\ref{hard_simu} describes that all six techniques successfully distinguish the hard inclusion from the uniform background. NCC produces the noisiest axial strain image. The PDE-based refinement technique substantially improves the output of NCC. $L1$-SOUL and $L1$-MechSOUL obtain sharper axial strain images than the other four techniques. NCC, SOUL, and $L1$-SOUL fail to produce acceptable lateral strain images. However, NCC + PDE turns the lateral estimate of NCC into an acceptable one. The proposed techniques MechSOUL and $L1$-MechSOUL generate high-quality lateral strain maps. Although both MechSOUL and $L1$-MechSOUL show good target-background contrast, $L1$-MechSOUL exploits the power of $L1$-norm regularization to obtain a sharper lateral strain image. The RMSE and PSNR values reported in Tables~\ref{table_rmse} and \ref{table_psnr} indicate substantially higher resemblance of the proposed techniques to the ground truth than NCC, NCC + PDE, SOUL, and $L1$-SOUL. In addition, the SNR and CNR box plots (Figs.~\ref{snr_plots} and \ref{cnr_plots}) and the associated mean and standard deviation values (Table~\ref{table_hard_simu}) demonstrate that MechSOUL and $L1$-MechSOUL substantially outperform the other algorithms in terms of lateral strain estimates.

The EPR maps depicted in Fig.~\ref{hard_simu} reveal that NCC, NCC + PDE, SOUL, and $L1$-SOUL estimate many EPR samples that are beyond the physically possible range (also see Fig. 2 of the Supplemental Video). MechSOUL and $L1$-MechSOUL resolve this issue by estimating EPR maps similar to the ground truth. The EPR RMSE (Table~\ref{table_rmse}), PSNR (Table~\ref{table_psnr}), SNR, and CNR (see Fig. 6 of the Supplementary Video and Table~\ref{table_hard_simu}) substantiate our qualitative assessment.

\subsection{Multi-inclusion Simulated Phantom}
The strain and EPR maps for the multi-inclusion simulated phantom data with an additional lateral boundary condition and surface traction-type loading are reported in Fig.~\ref{boundary_simu} and Fig. 3 of the Supplementary Video, respectively. All six techniques detect the axial strain contrast between the background and the inclusions. PDE-based technique refines NCC's axial estimate to reduce the noise. Due to the TV regularization, $L1$-SOUL and $L1$-MechSOUL obtain sharper axial strain images than SOUL and MechSOUL. NCC, NCC + PDE, SOUL, and $L1$-SOUL fail to render satisfactory lateral strain and EPR maps. MechSOUL and $L1$-MechSOUL produce high-quality lateral strain maps showing proper contrast among the four (background and three inclusions) elastic regions in both loading conditions. The EPR maps generated by MechSOUL and $L1$-MechSOUL also correspond well with the ground truths. Tables~\ref{table_rmse_boundary} and \ref{table_psnr_boundary} and Tables III and IV of the Supplementary Video validate this statement quantitatively. Given the difficulty level of the datasets, this experiment highlights the potential of MechSOUL and $L1$-MechSOUL in simultaneous imaging of axial and lateral strains and the EPR.

\subsection{Simulated Phantom with Different PRs}
Fig.~\ref{different_poisson_simu} demonstrates that all competing techniques generate good-quality uniform axial strain images. However, NCC, SOUL, and $L1$-SOUL fail to visualize the inclusion in the lateral strain images and the EPR maps. NCC + PDE refines NCC estimates to generate good lateral strain and EPR maps. MechSOUL and $L1$-MechSOUL lateral strains do not follow the uniform axial strains blindly and properly delineate the inclusions. Although MechSOUL and $L1$-MechSOUL EPR maps do not replicate the ground truth fully, they are substantially better than the comparison techniques. The RMSE and PSNR values reported in Tables \ref{table_rmse_diffp} and \ref{table_psnr_diffp} substantiate our statements.

\subsection{Real Breast Phantom Dataset}
The axial and lateral strain and the EPR results for the experimental breast phantom are shown in Fig. 4 of the Supplemental Video and Fig.~\ref{phan} of the current document, respectively. All six axial strain images detect the hard inclusion. However, NCC's axial estimate lacks smoothness in the background. NCC + PDE resolves this issue at the cost of visual contrast between the inclusion and the uniform background. The axial strain images obtained by SOUL and $L1$-SOUL are superior to those by NCC-based techniques. MechSOUL and $L1$-MechSOUL axial strain estimates, respectively, marginally outperform the ones generated by SOUL and $L1$-SOUL. The total variation (TV) regularization-based techniques $L1$-SOUL and $L1$-MechSOUL render sharper axial strain images than the other algorithms. NCC, SOUL, and $L1$-SOUL produce noisy lateral strain images with unacceptable target-background contrast. In addition, large spatial regions exhibit lateral strains that are out of physical range when compared to axial strains. PDE refines NCC's lateral result to reduce the noise and visualize the inclusion. MechSOUL and $L1$-MechSOUL successfully estimate high-contrast lateral strain maps with smooth backgrounds and substantially outperform the other four techniques. Note that the lateral strain image provided by $L1$-MechSOUL is visually sharper than the one obtained by MechSOUL. The quantitative metric values reported in Figs. \ref{snr_plots} and \ref{cnr_plots} and Table~\ref{table_phan} corroborate our visual judgement.

Fig.~\ref{phan}, Table~\ref{table_phan}, and Fig. 6 of the Supplementary Video demonstrate that NCC, SOUL, and $L1$-SOUL fail to produce viable EPR distribution. Although NCC + PDE performs better than NCC, it still contains a noticeable amount of EPR samples which are practically impossible. MechSOUL and $L1$-MechSOUL successfully restrict the EPR values to the physically possible range and exhibit higher EPR values around the inclusion than the uniform regions.

\subsection{\textit{In vivo} Liver Cancer Datasets}
Fig. 5 of the Supplemental Video and Fig.~\ref{liver} of the current document, respectively, depict the axial and lateral strain results for the liver cancer datasets collected before the ablation. The B-mode image for patient 1 reveals the tumor by showing a lower echo amplitude than the healthy tissue. However, the target-background echogenic contrasts for the other two patients' B-mode images are negligible. 

The axial strain images clearly distinguish the tumor and healthy tissue for all three patient cases. Similar to the \textit{in silico} and phantom cases, NCC obtains the noisiest axial strain images. The PDE-based refining step resolves this issue of NCC and highlights the important details of the strain images. SOUL and MechSOUL outperform NCC + PDE in terms of background smoothness and the clarity of strain estimation in the shallow tissue region. The TV-regularization feature of $L1$-SOUL and $L1$-MechSOUL enables them to estimate substantially sharper axial strain than SOUL and MechSOUL for patients 1 and 2. In the case of the third liver patient, $L1$-SOUL and $L1$-MechSOUL obtain brighter axial strain images than the other techniques. In general, it is visually evident that the axial strain imaging performance of MechSOUL and $L1$-MechSOUL, respectively, are similar to those of SOUL and $L1$-SOUL. The box plots reported in Figs.~\ref{snr_plots} and \ref{cnr_plots} and the mean and standard deviation values (Tables~\ref{table_liver1}, \ref{table_liver2}, and \ref{table_liver3}) substantiate this observation.

NCC fails to produce acceptable lateral strain images in patients 1 and 3. However, it shows slight target-background contrast for patient 2. NCC + PDE notably improves the lateral estimates of NCC in all three patient cases. The lateral strain images for patients 1 and 2 obtained by SOUL and $L1$-SOUL show minimal contrast between the healthy and pathologic tissues. In addition, the estimated strains are markedly out of the feasible bound. Furthermore, they are highly corrupted by estimation noise. In the case of patient 3, both SOUL and $L1$-SOUL fail to generate appreciable lateral strain images. For all three patients, MechSOUL and $L1$-MechSOUL obtain high-contrast lateral strain maps and substantially outperform the other four algorithms. MechSOUL exhibits a horizontal striking artifact in patient 2's lateral strain images, which is removed by $L1$-MechSOUL. In addition, $L1$-MechSOUL yields sharper lateral strain estimates than MechSOUL in all patient cases. The SNR and CNR box plots (Figs.~\ref{snr_plots} and \ref{cnr_plots}) and their mean and standard deviation values (Tables~\ref{table_liver1}, \ref{table_liver2}, and \ref{table_liver3}) align with our visual perception.

Fig.~\ref{liver_epr} demonstrates that NCC, SOUL, and $L1$-SOUL estimate physically impossible EPR maps. The PDE-based method substantially improves NCC estimates. The proposed techniques estimate smooth EPR maps with the individual EPR values confined to the practical range (also see Fig. 6 of the Supplementary Video and Tables~\ref{table_liver1}-\ref{table_liver3}). MechSOUL and $L1$-MechSOUL yield higher tumor EPR for the first two patients and lower tumor EPR for the third patient. This opposing behavior of tumor EPRs might be related to the complicated deformation physics in patient 3 stemming from multiple blood vessels in the vicinity of the tumor.

\begin{figure}[h]
	\begin{center}
			\subfigure[B-mode]{{\includegraphics[width=0.125\textwidth]{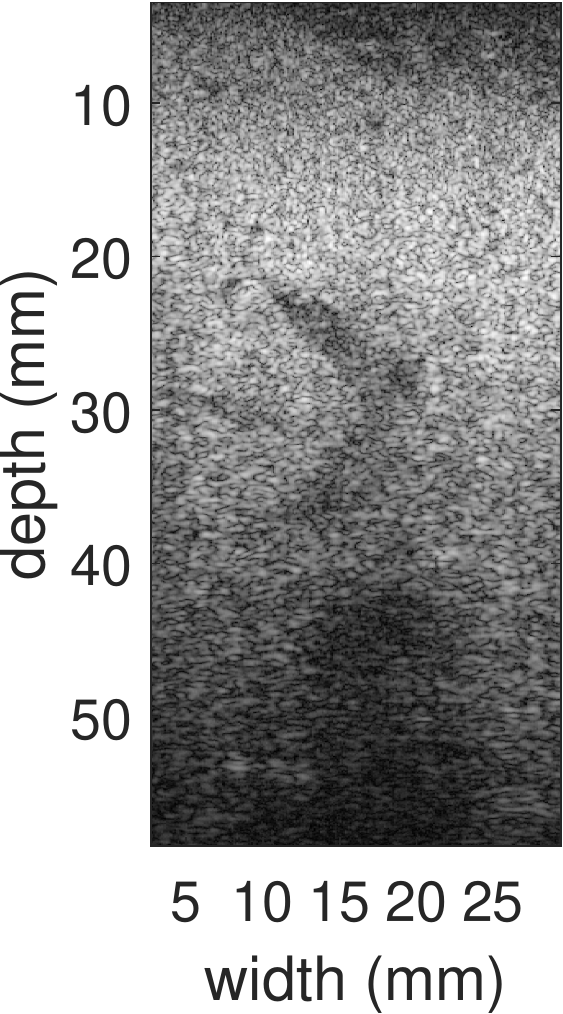}}}%
			\subfigure[MechSOUL]{{\includegraphics[width=0.125\textwidth]{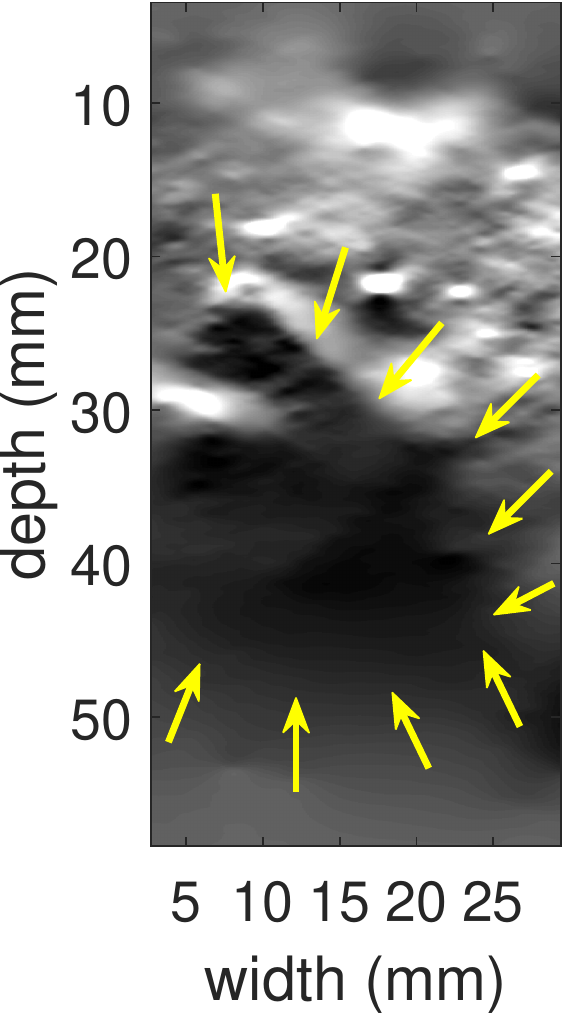}}}%
			\subfigure[MechSOUL]{{\includegraphics[width=0.125\textwidth]{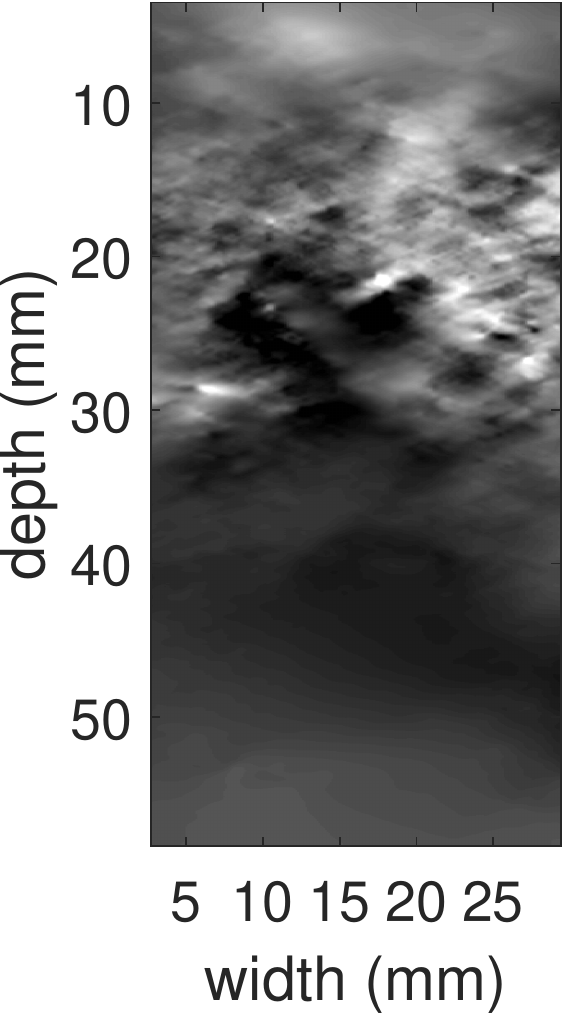}}}%
			\subfigure[MechSOUL]{{\includegraphics[width=0.125\textwidth]{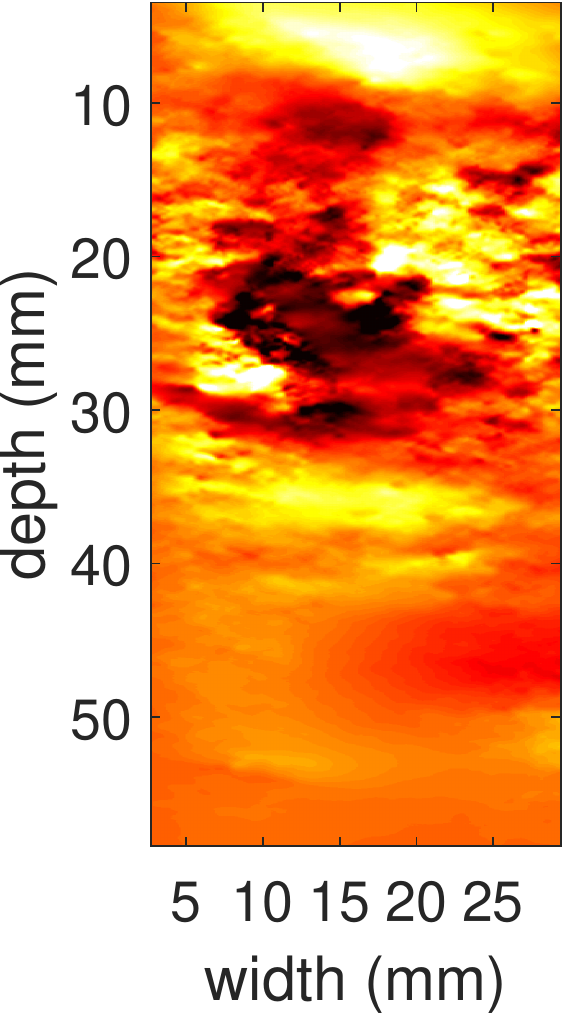}}}
			\begin{flushright}
				\subfigure[$L1$-MechSOUL]{{\includegraphics[width=0.125\textwidth]{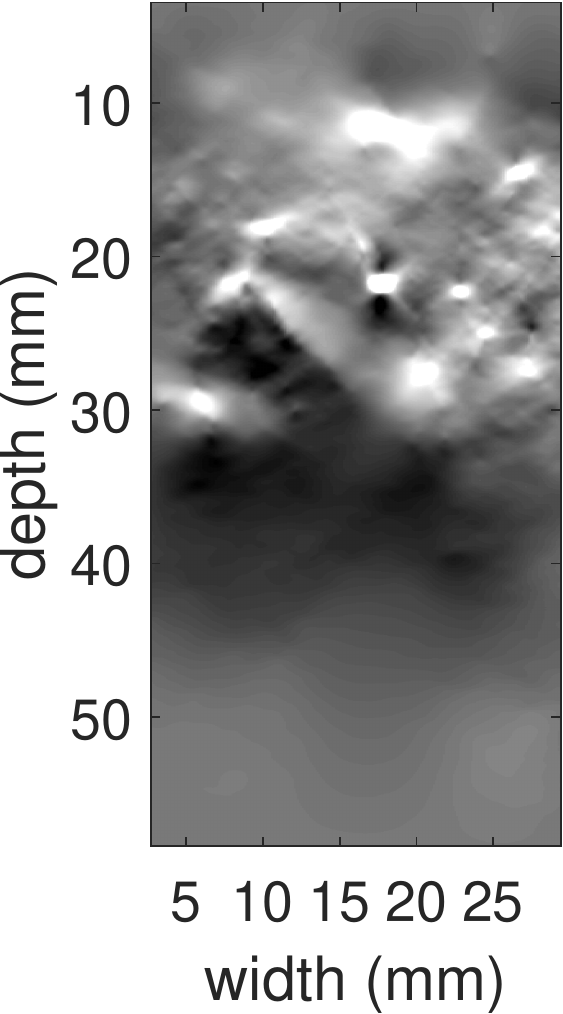}}}%
				\subfigure[$L1$-MechSOUL]{{\includegraphics[width=0.125\textwidth]{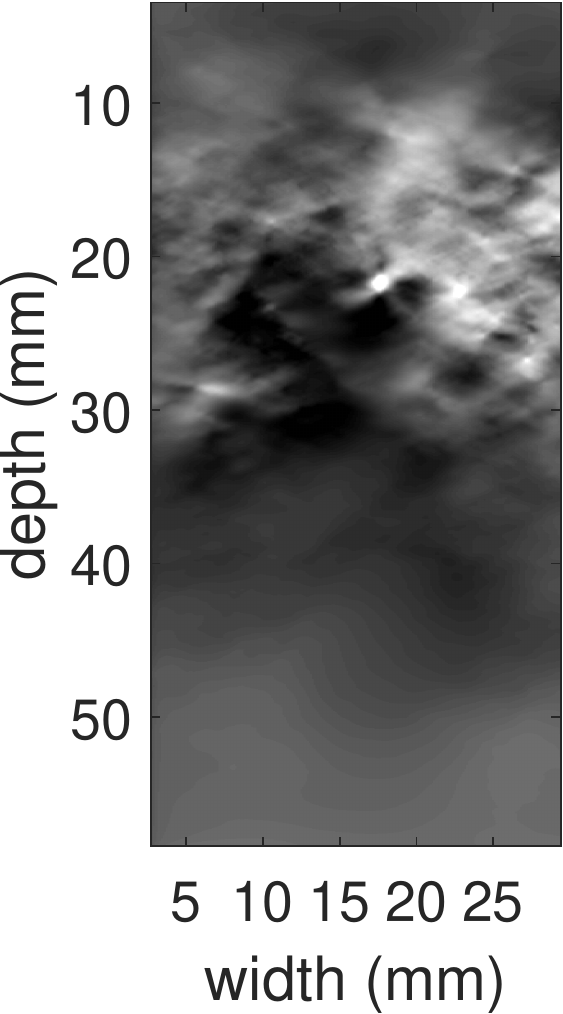}}}%
				\subfigure[$L1$-MechSOUL]{{\includegraphics[width=0.125\textwidth]{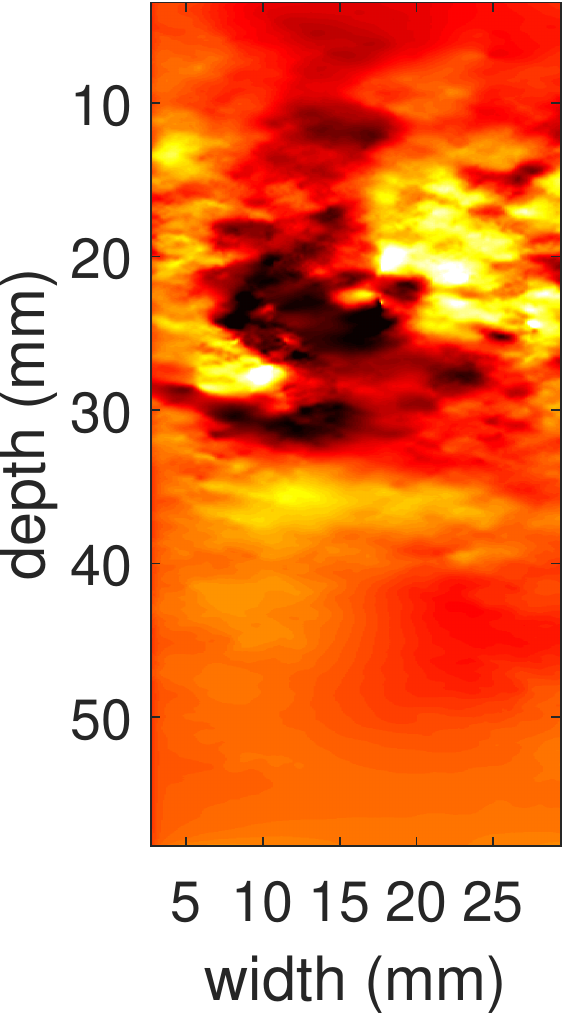}}}	
			\end{flushright}
			\subfigure[Axial strain]{{\includegraphics[width=0.125\textwidth]{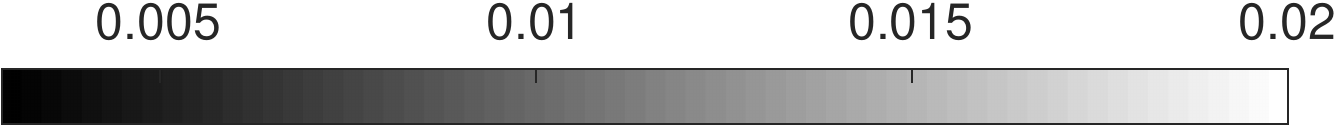}}}%
			\quad
			\subfigure[Lateral strain]{{\includegraphics[width=0.125\textwidth]{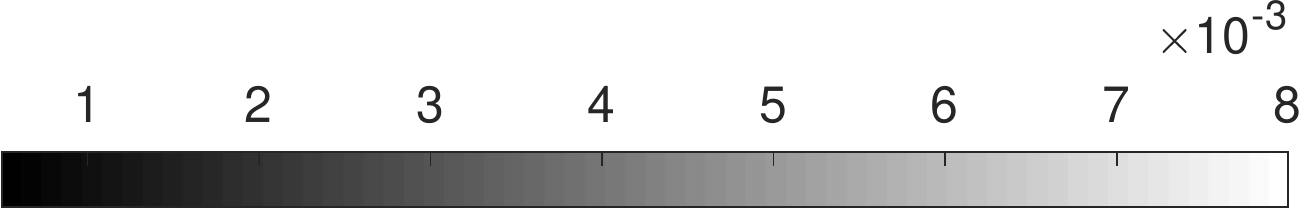}}}%
			\quad
			\subfigure[EPR]{{\includegraphics[width=0.125\textwidth]{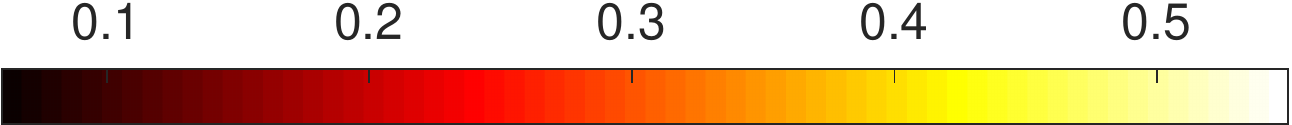}}}
	\end{center}
	\caption{Results for the first liver dataset after ablation. Rows 1 and 2 correspond to MechSOUL and $L1$-MechSOUL, respectively, whereas columns 1 to 4 correspond to B-mode, axial strain, lateral strain, and EPR, respectively.}
	\label{liver_p1_after}
\end{figure}

\begin{figure}[h]
	\centering
		\subfigure[MechSOUL]{{\includegraphics[width=0.22\textwidth]{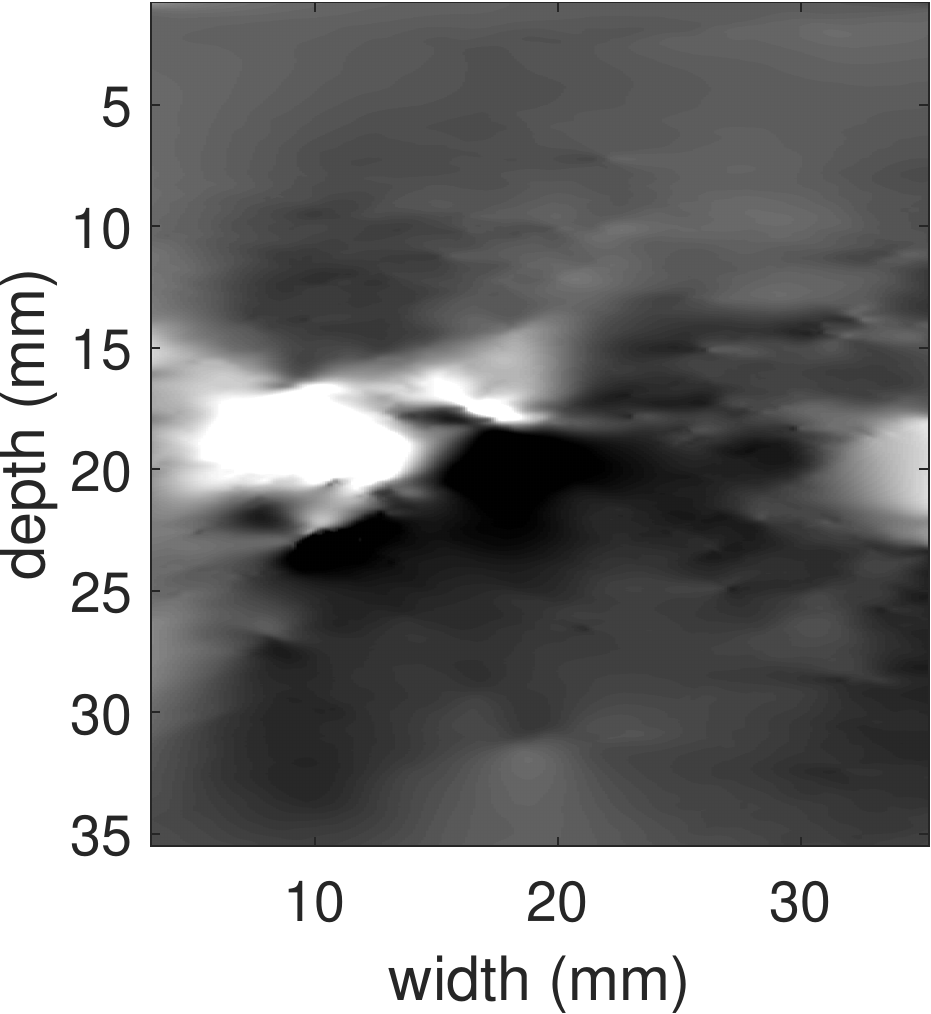}}}%
		\subfigure[$L1$-MechSOUL]{{\includegraphics[width=0.22\textwidth]{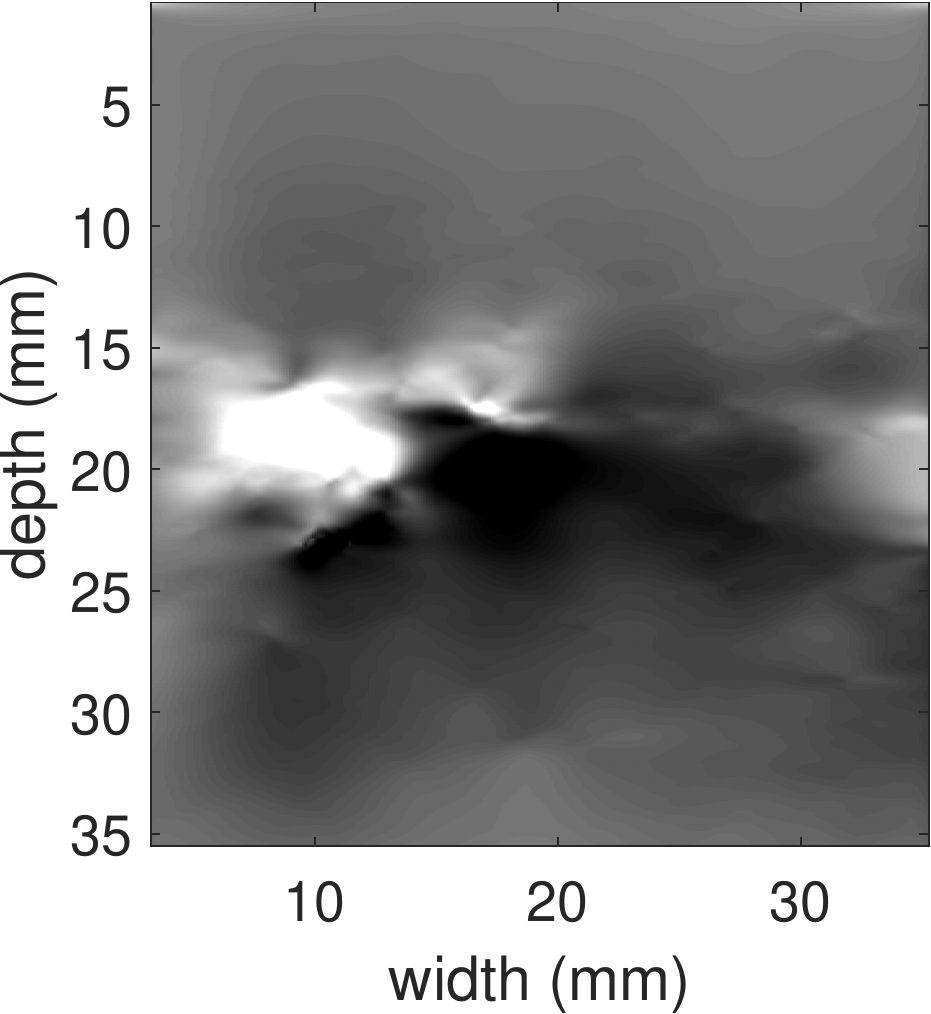}}}	
		\caption{Individually tuned lateral strain results for the liver patient 3. (a) and (b) correspond to MechSOUL and $L1$-MechSOUL, respectively.}
		\label{liver3_indiv}
\end{figure}

\section{Discussion}
The poor lateral displacement or strain estimation capability stemming from low data resolution in this direction is a well-known drawback of the existing ultrasound elastography techniques. Due to the imaging mechanism, ultrasound loses important information associated with the dimension perpendicular to the primary wave propagation. The existing strain imaging frameworks cannot make up for the lost lateral information and, therefore, end up providing lateral estimates substantially inferior to the axial ones. The techniques proposed herein incorporate the tissue deformation mechanics to couple the lateral strain to the axial one and compensate for the information lost by the imaging modality. As demonstrated in the validation examples, this coupled approach dramatically improves lateral strain imaging performance.         

MechSOUL and $L1$-MechSOUL impose an EPR-driven relation between the axial and lateral strains along with data fidelity and spatial smoothness constraints. Employing the aforementioned mechanical constraint is not analogous to calculating the lateral strain as a multiple of the independently estimated axial strain, which is prone to mirroring the accurate axial estimates to the less accurate lateral estimates. Instead, MechSOUL and $L1$-MechSOUL allow the lateral strains to deviate from the axial ones (see Fig.~\ref{different_poisson_simu}) depending on the underlying properties of tissue. Because the proposed techniques solve a unified optimization problem to investigate the mechanical and continuity constraints and the RF data simultaneously. In addition, this work iteratively updates each RF sample's EPR value. 

The proposed techniques introduce new tunable parameters associated with the mechanical constancy terms. This mechanical parameter partly determines how strongly the estimated lateral strain follows the axial strain. On the one hand, a very high mechanical constancy weight might suppress the effect of data fidelity and force the lateral strain to follow the axial one blindly. On the other hand, a tiny parameter value restricts the impact of mechanical constraint, demolishing the sole purpose of this study. Since the mechanical parameters are not correlated with the continuity ones, MechSOUL and $L1$-MechSOUL use the same continuity weights as SOUL and $L1$-SOUL, respectively, and tune only the newly-introduced parameters on validation images. In our experience, the optimality of mechanical constancy weight is unrelated to the RF signal's SNR. While controlled by the material property and the deformation profile, a moderate value of the mechanical constancy parameter leads to a good estimation of the displacement fields. It is worth noting that tuning the mechanical parameters are not cumbersome since the proposed algorithms are not sensitive to reasonable alterations in their values, which is demonstrated in Fig. 7 of the Supplemental Video.

The proposed techniques' parameters were tuned on validation images different from the final test ones. Although parameter values are optimized for simulated, phantom, and \textit{in vivo} datasets, a single parameter set is used for all datasets of the same kind (i.e., same parameter set for all liver datasets). Scatterer size and distribution, attenuation coefficient, imaging settings, noise statistics, and the deformation field's temporal behavior are the main deciding factors for the optimal set of continuity and mechanical weights. Since these properties are different for different types of data used in this study, the optimal parameter values also vary from each other. Note that the simulated datasets employed in this work differ from the real phantom one in terms of both quantitative properties and imaging parameters, which leads to different sets of weights for simulated and real phantoms. Therefore, the parameters can be saved as presets in commercial ultrasound machines for imaging different organs such as thyroid, breast, \textit{etc}. The proposed techniques exhibit good performance for all three simulation experiments (Figs.~\ref{hard_simu}-\ref{different_poisson_simu}) using the same parameter values. A single parameter set leads to high-quality estimations in all \textit{in vivo} cases (Figs.~\ref{liver} and \ref{liver_epr} and Fig. 5 of the Supplementary Video) as well. To further justify our argument, we have conducted sensitivity analyses using two datasets: 1) the first liver patient (before ablation) but with different input frames than Fig.~\ref{liver} 2) the first liver patient after ablation. Note that the second dataset is an entirely new one collected from the first liver patient after a significant clinical procedure that alters the noise statistics. In addition, this dataset was not considered for tuning the parameters. Fig. 8 of the Supplementary Video and Fig.~\ref{liver_p1_after} demonstrate that MechSOUL and $L1$-MechSOUL perform well in both cases for the same parameter sets by properly delineating the tumor (before ablation) or coagulated tissue (after ablation).

\begin{figure*}
	\centering
	\subfigure[B-mode]{{\includegraphics[width=0.2\textwidth]{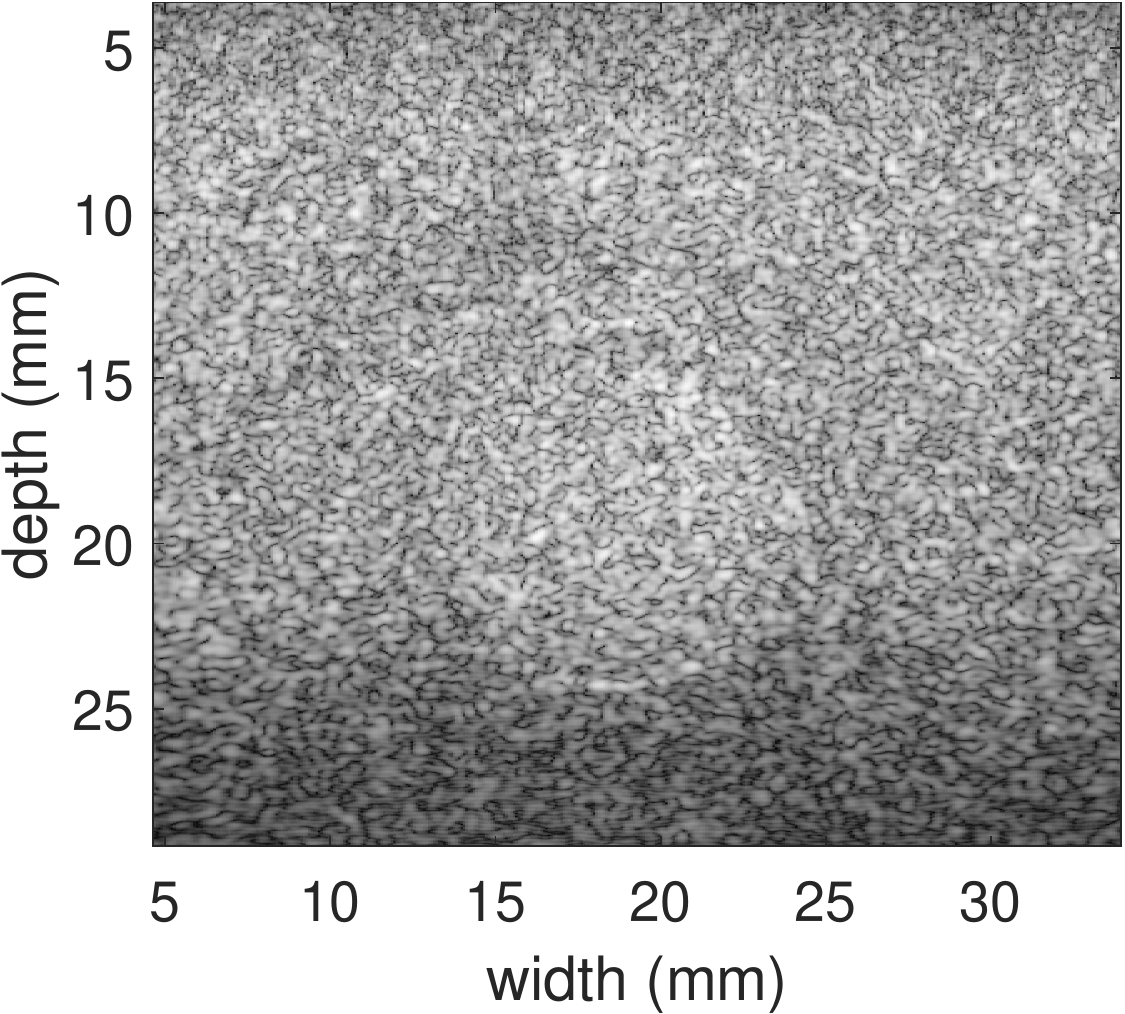}}}%
	\subfigure[SOUL]{{\includegraphics[width=0.2\textwidth]{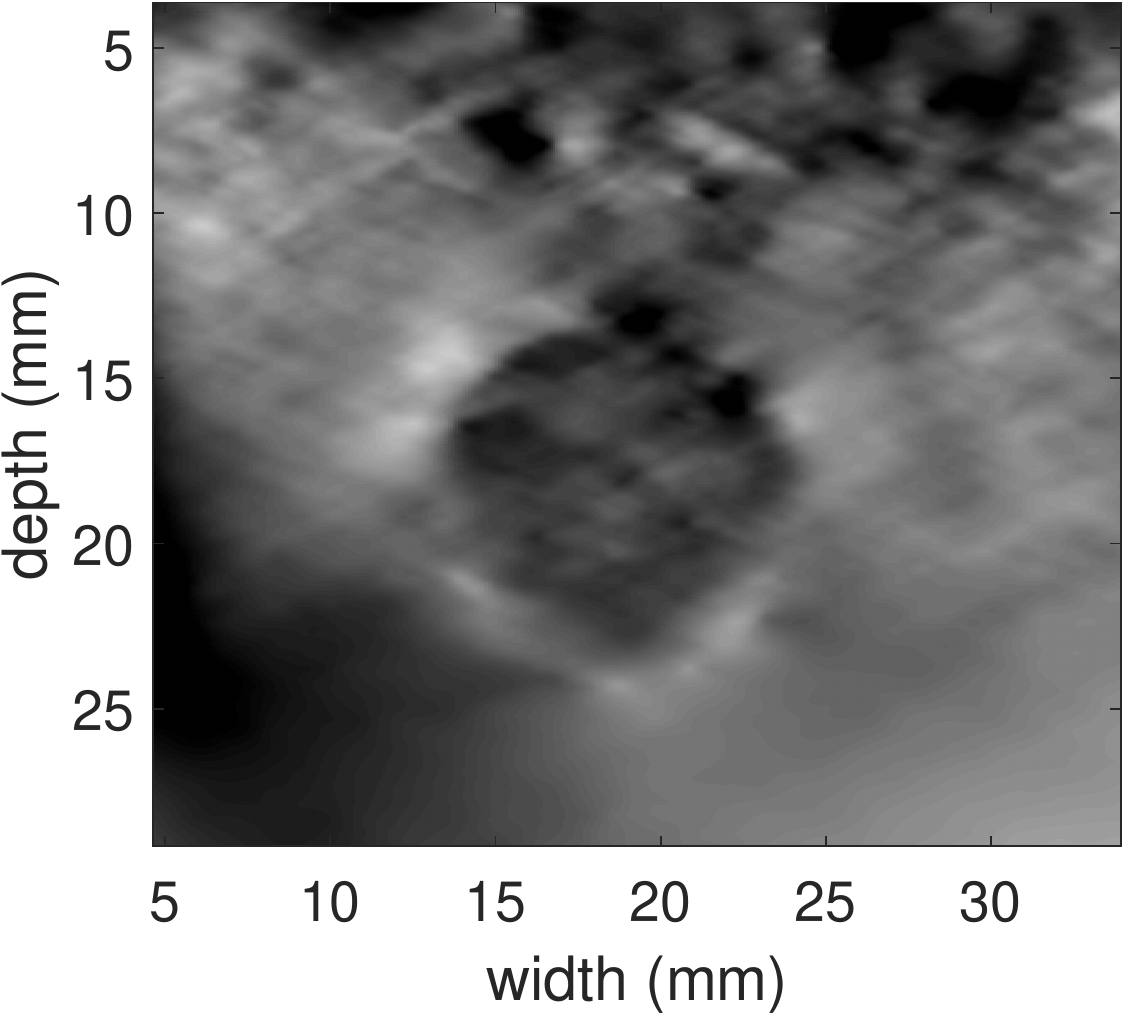}}}%
	\subfigure[$L1$-SOUL]{{\includegraphics[width=0.2\textwidth]{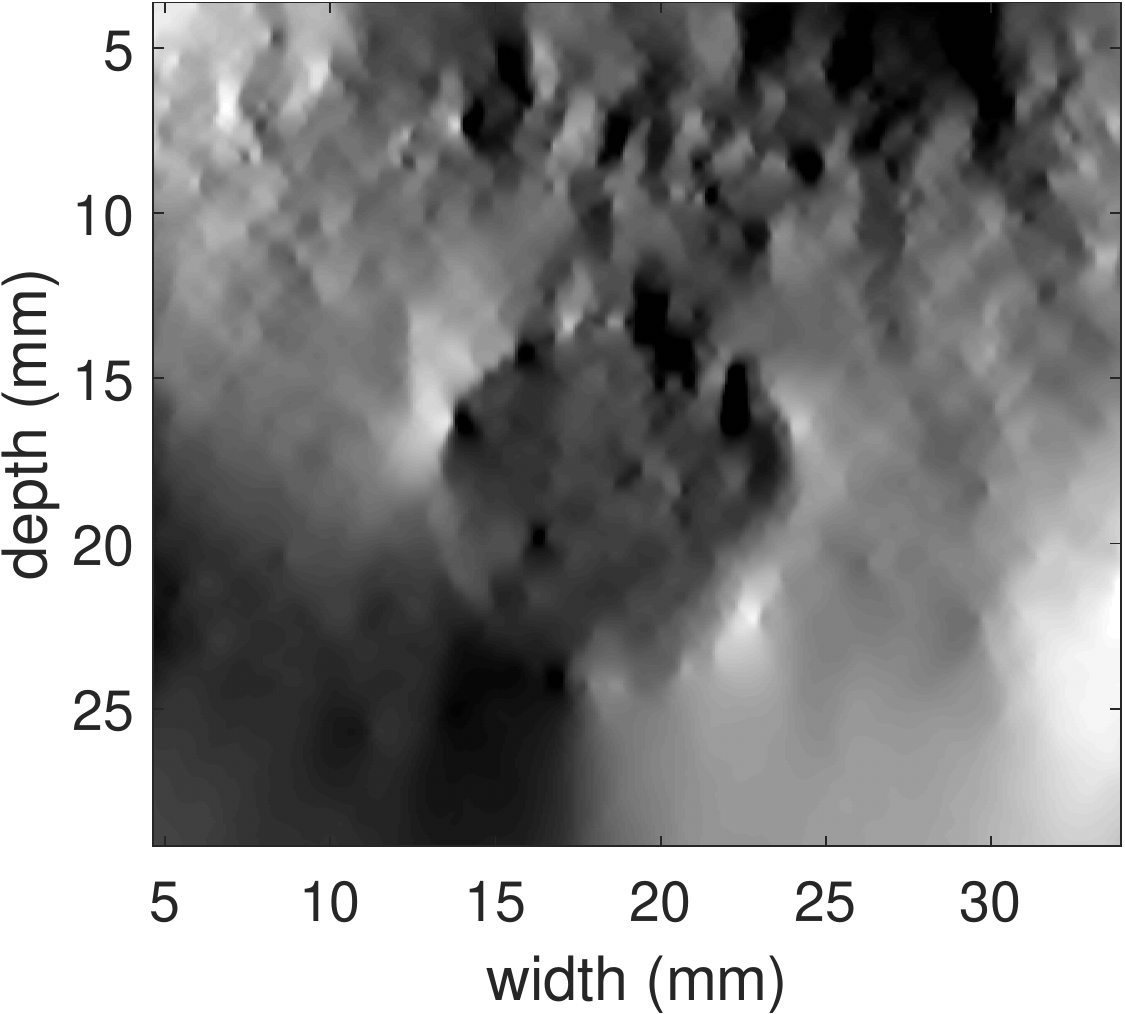}}}%
	\subfigure[MechSOUL]{{\includegraphics[width=0.2\textwidth]{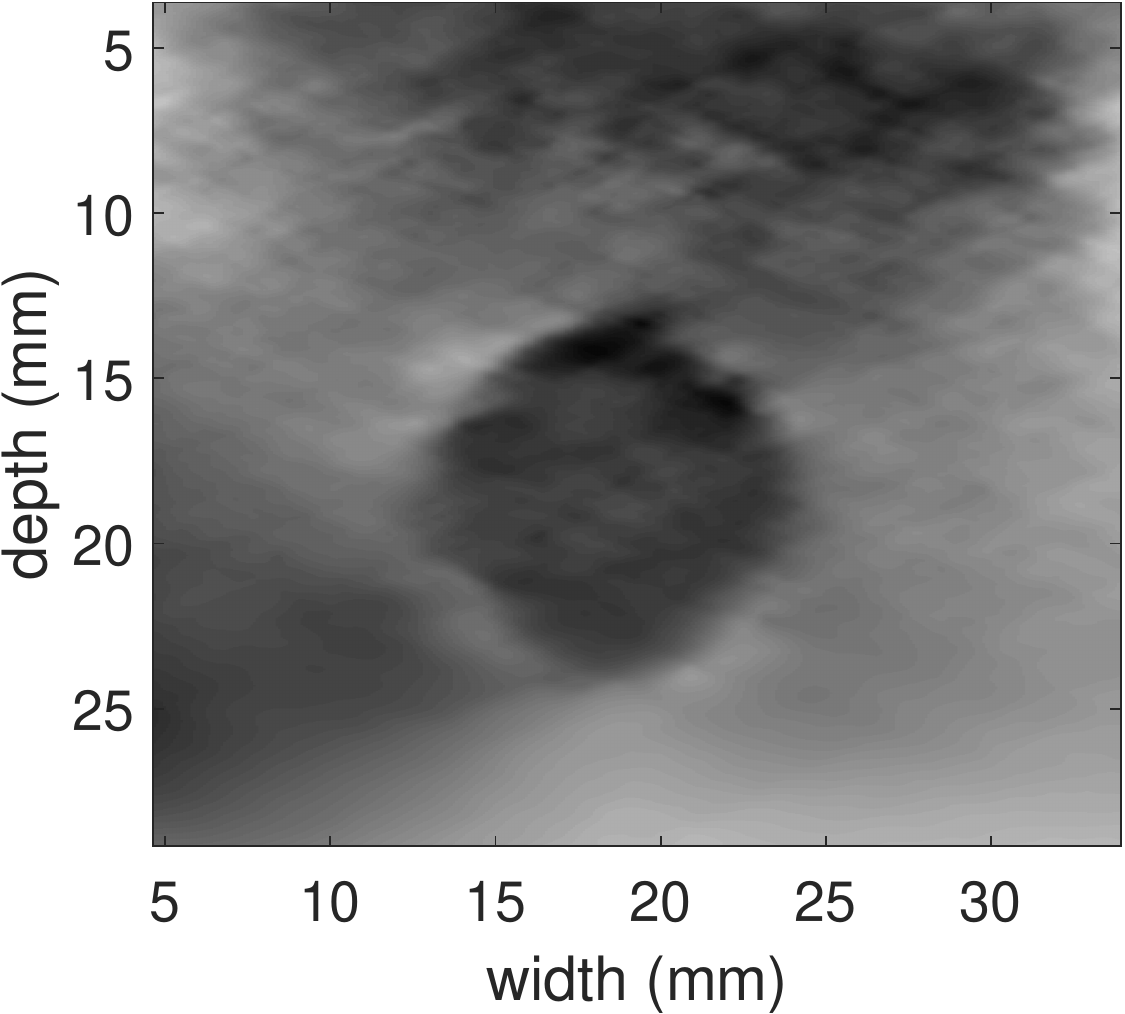}}}%
	\subfigure[$L1$-MechSOUL]{{\includegraphics[width=0.2\textwidth]{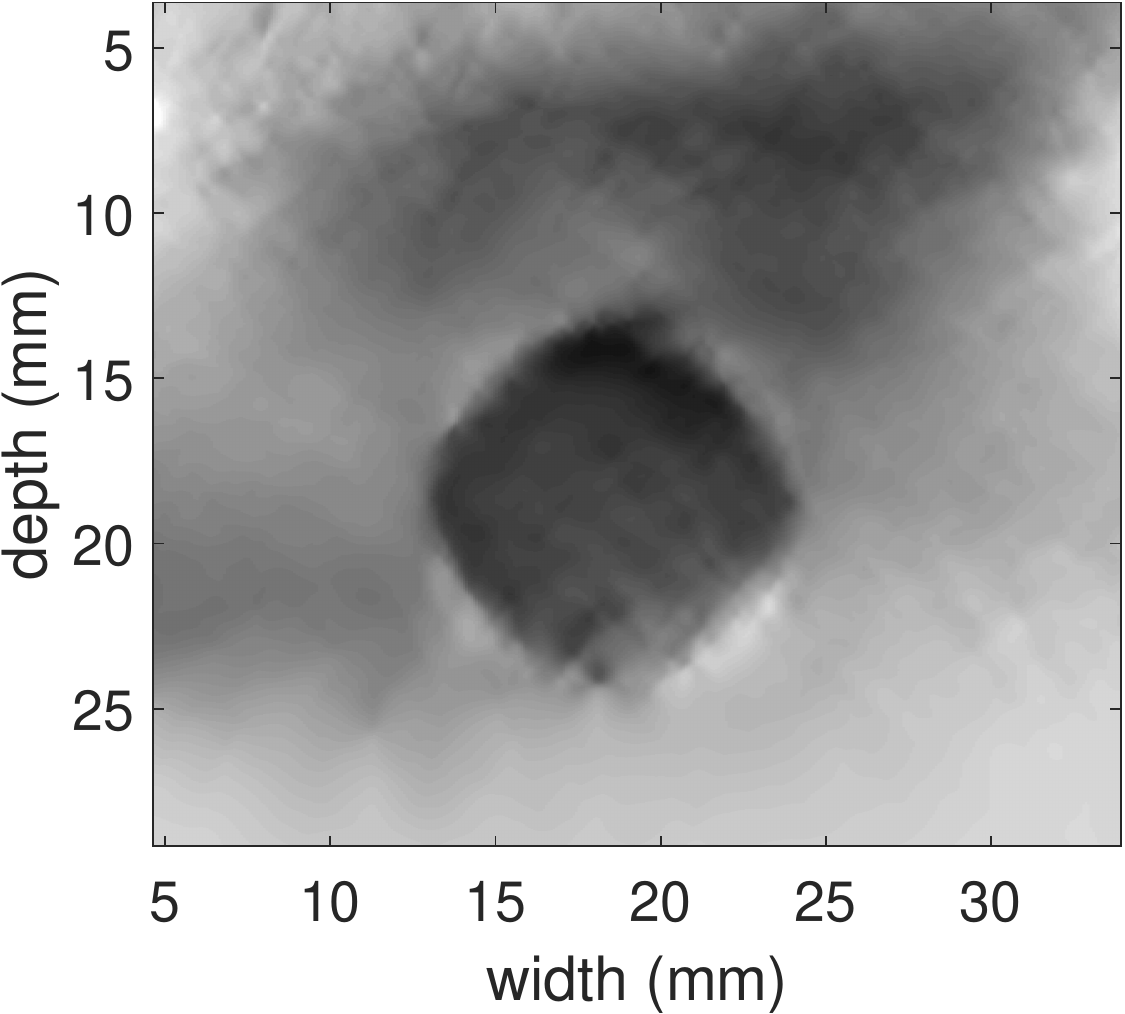}}}
	\subfigure[Lateral strain]{{\includegraphics[width=0.4\textwidth]{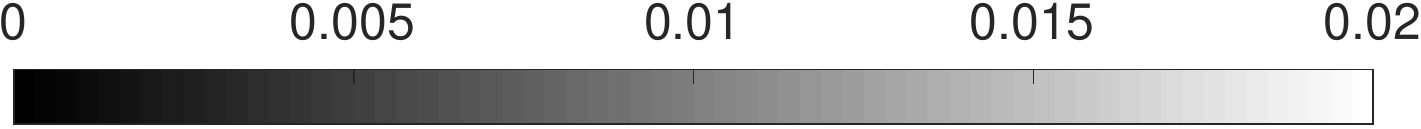}}}%
	\caption{Lateral strain results for an additional phantom dataset. Columns 1 to 5 correspond to B-mode, SOUL, $L1$-SOUL, MechSOUL, and $L1$-MechSOUL, respectively.}
	\label{phan_hopkins}
\end{figure*}

Figs.~\ref{snr_plots} and \ref{cnr_plots} provide the opportunity to conduct a statistical test to determine if the proposed techniques are significantly better than the existing ones. The comparison intervals of the group means obtained from the analysis of variance (ANOVA) followed by a multiple comparison statistical test are reported in Figs. 9 and 10 of the Supplemental Video. The intervals for SOUL, $L1$-SOUL, MechSOUL, and $L1$-MechSOUL are close to each other in most axial cases. However, in the lateral cases, MechSOUL and $L1$-MechSOUL comparison intervals yield significantly higher values than the existing techniques, reassuring the proposed algorithms' superiority in lateral tracking.

The $L1$-norm-based proposed technique $L1$-MechSOUL exhibits sharper strain estimates than MechSOUL in the validation experiments presented in this study. It is worth noting that $L1$-norm regularization does not force a sharp strain map if the underlying strain map is not sharp. It can produce a sharp estimate at the border of two organs where tissue properties display a rapid change or a smooth change where changes in the underlying mechanical properties are gradual. In contrast, $L2$-norm regularization produces a smooth strain map even if mechanical properties have a sharp transition.

Like $L1$-SOUL, $L1$-MechSOUL approximates the $L1$-norm with TVD, establishing a balance between smoothness and sharpness by penalizing the variation and simultaneously allowing sharp transition. An alternating direction method of multipliers (ADMM)-based strategy can eliminate the requirement of TVD approximation by optimizing $L1$-norm's original formulation using the shrinkage function~\cite{boyd2011distributed,eckstein2012augmented} and utilize the full potential of $L1$-norm. ADMM offers this direct optimization feature at the cost of increased complexity and more sensitive parameter tuning. Therefore, ADMM-based optimization of $L1$-MechSOUL's penalty function will be explored in a future extension of this work.

The lateral strain estimation performance of the proposed techniques is substantially better than the existing techniques in all validation experiments conducted in this study. However, the MechSOUL and $L1$-MechSOUL lateral strain images for liver patient 3 are not as good as the other two liver patients. This performance degradation might stem from the complicacy of RF data acquired from patient 3. The field-of-view (FOV) contains blood flow through the annotated vessels, which introduces different types of noise to RF data. In addition, being a combination of several vessels, healthy tissue, and tumor, the FOV poses high-variance distributions of elasticity and EPR. Furthermore, the tumor experiences a complicated deformation physics since it is located underneath the easily-compressible portal vein. It is worth observing that both MechSOUL and $L1$-MechSOUL handle this challenging dataset promisingly and yield perceptible contrast among different tissues. As shown in Fig.~\ref{liver3_indiv}, slightly better performance can be achieved when the strain imaging techniques' parameter sets are dedicatedly optimized for this particular dataset. However, tuning the parameters for each dataset individually is not possible in the clinical context and affects the algorithms' generalizability.

\begin{figure}[h]
		\centering
		\includegraphics[width=0.42\textwidth]{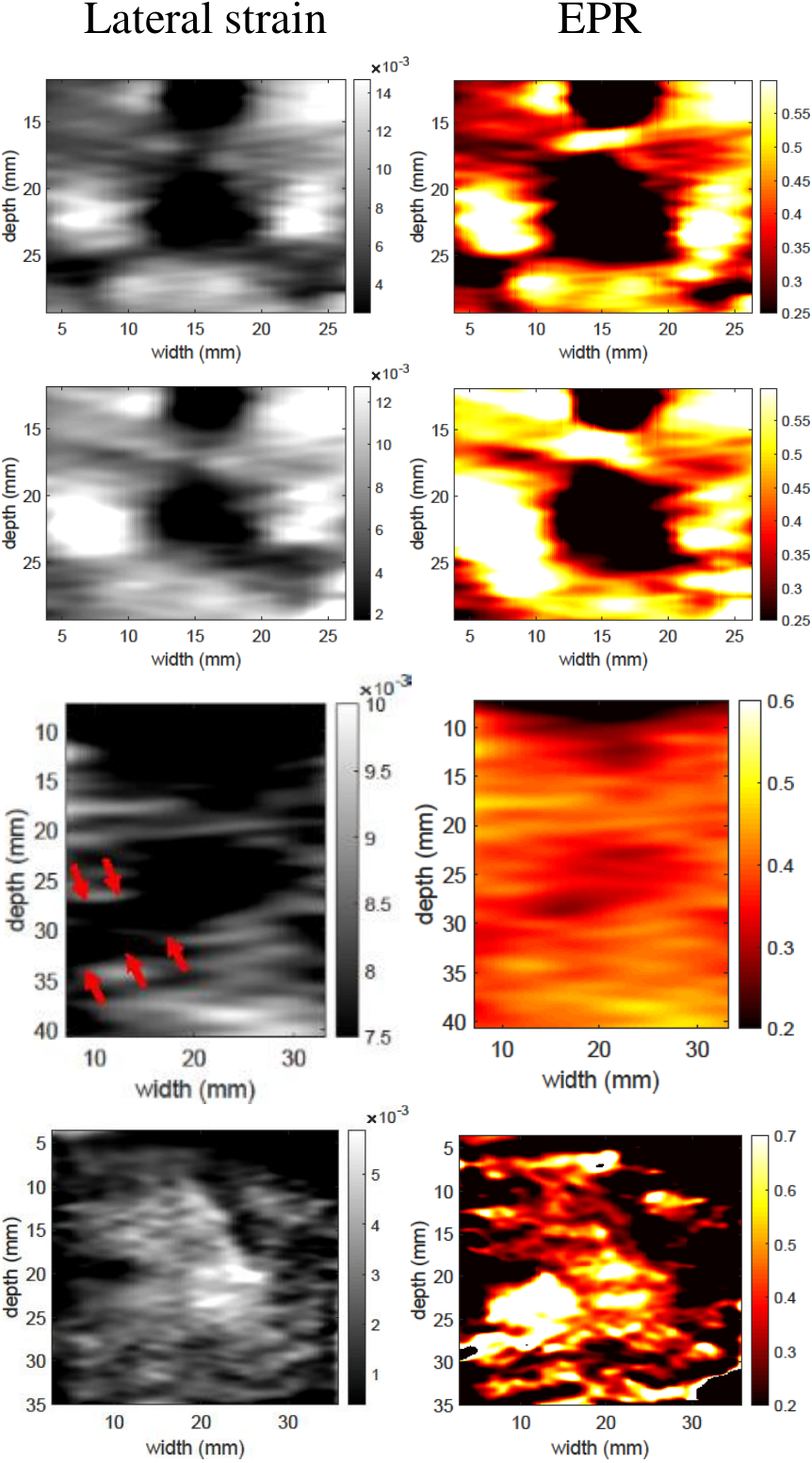}%
		\caption{Lateral strain and EPR results obtained by sPICTURE. Rows 1 to 4 correspond to multi-inclusion simulated phantom (additional boundary condition), multi-inclusion simulated phantom (surface traction), different PR simulated phantom, and the first liver patient, respectively. Columns 1 and 2 represent lateral strain and EPR, respectively. The red arrows indicate estimation artifacts.}
		\label{res_spicture}
\end{figure}

\begin{table}  
	\centering
	\caption{RMSE of sPICTURE for the simulated datasets.}
	\label{table_rmse_spicture}
	\begin{tabular}{c c c c c c c c c} 
		\hline
		$ $  $ $&  Lateral strain & EPR\\
		\hline
		\makecell{Multi-inclusion 1}  & $4 \times 10^{-3}$ & 0.2 \\
		\makecell{Multi-inclusion 2}  & $4 \times 10^{-3}$ & 0.22 \\
		Different PR & $1.3 \times 10^{-3}$ & $6.45 \times 10^{-2}$ \\
		\hline
	\end{tabular}
\end{table}

\begin{table}  
	\centering
	\caption{PSNR (dB) of sPICTURE for the simulated datasets.}
	\label{table_psnr_spicture}
		\begin{tabular}{c c c c c c c c c} 
			\hline
			$ $  $ $&  Lateral strain & EPR\\
			\hline
			\makecell{Multi-inclusion 1}  & 47.87 & 14.50 \\
			\makecell{Multi-inclusion 2}  & 47.96 & 13.23 \\
			Different PR & 57.88 & 23.82 \\
			\hline
		\end{tabular}
\end{table}

SOUL and $L1$-SOUL exhibit poor lateral estimation performance in most of the validation experiments conducted in this work. However, they might produce acceptable lateral strain images when a moderately high strain is applied to the tissue in a highly controlled manner. More specifically, we show results of an experiment where a phantom was uniaxially compressed by approximately $5\%$ using a linear stage mounted on an optical table. Fig.~\ref{phan_hopkins} shows the lateral tracking performance of SOUL, $L1$-SOUL, MechSOUL, and $L1$-MechSOUL for this dataset. Despite being substantially outperformed by MechSOUL and $L1$-MechSOUL, SOUL and $L1$-SOUL generate reasonable lateral strain images in this experiment. We have also conducted a controlled experiment on the hard-inclusion simulated phantom, where the ground truth deformation field for $4\%$ applied strain is obtained from FEM. The pre-deformed frame is generated by warping the Field II-simulated post-deformed RF data based on the ground truth displacements. Fig. 11 of the  Supplementary Video shows that SOUL produces a good lateral strain image in this highly controlled environment where the applied strain is reasonably high. However, MechSOUL substantially outperforms SOUL in this case as well, demonstrating its strength in lateral tracking. These two experiments, in conjunction with the other validation experiments carried out in this work, manifest that SOUL produces reasonable lateral strain maps in a controlled and moderately high-strain scenario, whereas it often fails in realistic cases. Simultaneous exploitation of RF data and tissue deformation physics enables MechSOUL to resolve this issue by performing well in both controlled and realistic settings.

PDE-based refinement~\cite{guo2015pde} is one of the comparison techniques used in this work. Duroy \textit{et al.}~\cite{duroy20212d} also conducted a similar study in a recent work. Both of these techniques refine the initial axial and lateral estimates assuming tissue incompressibility. As demonstrated in this paper, the PDE-driven post-processing strategy improves the lateral estimation quality. However, the incompressibility constraint assumes the Poisson's ratio to be 0.5, which is not true for all biological tissues. In addition, the refinement techniques do not consider RF data and the regularization constraints in a unified manner and, therefore, are prone to failure of the first step. The proposed techniques MechSOUL and $L1$-MechSOUL investigate data, continuity, and mechanical constraints simultaneously and update the EPR value at each sample iteratively to tackle the aforementioned issues.

Our recently accepted deep learning framework self-supervised Physically Inspired ConsTraint for Unsupervised Regularized Elastography (sPICTURE)~\cite{tehrani2022lateral} can be a good competing technique to demonstrate MechSOUL and $L1$-MechSOUL's lateral tracking potential. Fig.~\ref{res_spicture} shows the sPICTURE lateral strain and EPR maps for the multi-inclusion simulated phantoms, different PR simulated phantom, and the first liver patient. For both multi-inclusion simulated phantoms, sPICTURE performs notably better than NCC, NCC+PDE, SOUL, and $L1$-SOUL. However, both MechSOUL and $L1$-MechSOUL substantially outperform sPICTURE in terms of contrast and resemblance to the ground truth. Note that the multi-inclusion phantoms contain non-uniaxial force, and sPICTURE was not trained for such a case during its development. Except for the red-marked outlier region, sPICTURE achieves similar performance as the proposed techniques in the case of the different PR simulated phantom. The RMSE and PSNR values reported in Tables~\ref{table_rmse_spicture} and \ref{table_psnr_spicture} substantiate our visual assessments. sPICTURE shows a contrast between the tumor and the healthy tissue in the case of the first liver patient. Nevertheless, lateral strain and EPR estimation quality are substantially lower than MechSOUL and $L1$-MechSOUL (also see Table~\ref{table_liver1_spicture}). This comparison against sPICTURE, a state-of-the-art deep learning-based lateral estimation technique, is another evidence of MechSOUL and $L1$-MechSOUL's strength in lateral strain imaging.

\begin{table}
	\centering
	\caption{SNR and CNR of sPICTURE for the first liver dataset.}
	\label{table_liver1_spicture}
		\begin{tabular}{c c c c c c c c c} 
			\hline
			$ $  $ $&  Lateral strain & EPR\\
			\hline
			SNR & $3.88 \pm 1.88$ & $2.74 \pm 1.86$ \\
			CNR  & $1.68 \pm 1.43$ & $3.44 \pm 2.08$ \\
			\hline
		\end{tabular}
\end{table} 

The spatial distribution of EPR is directly correlated with tissue compressibility (i.e., ability to change the volume)~\cite{chaudhry2016estimation,tauhidul_2018}. Specific pathologies such as cancer and lymphedema tend to alter the value of this mechanical parameter~\cite{chaudhry2016estimation,tauhidul_2018}. In addition, compressibility often signifies the tissue's sensitivity to treatments or therapies~\cite{chaudhry2016estimation}. Therefore, the EPR contrast between different regions can be used as a marker for tissue's pathologies or susceptibility to treatment. These potential applications of an EPR map make MechSOUL and $L1$-MechSOUL attractive for clinical translation since they substantially improve the EPR image quality.

Both PR and EPR range between 0 and 0.5 for uniform soft materials. An EPR greater than 0.5 in a uniform region for uniaxial compression refers to a negative bulk modulus, which is impossible in thermodynamic equilibrium. Therefore, the validation results showing EPR values greater than 0.5 in uniform and uniaxial cases indicate possible errors in strain estimation.

The success of the proposed algorithms is correlated with the accuracy of the EPR update. Although MechSOUL and $L1$-MechSOUL worked well in all validation experiments conducted in this work, there might be a downfall in their performance in the case of a more challenging dataset where the EPR distribution progresses in a wrong direction. A potential solution to this problem is incorporating an EPR-independent, tensor geometry-driven mechanical constraint such as the compatibility condition. A recent work~\cite{kheirkhah2022novel} has used the compatibility equation to improve lateral strain estimation. However, this work presents a post-processing algorithm that highly depends on the initial axial and lateral estimation accuracy. Since RF data and the mechanical constraint are not investigated simultaneously, this post-processing technique might fail in challenging scenarios like the different PR phantom presented in this study. Simultaneous optimization of data and compatibility constraints in a direct strain imaging framework might resolve this issue.

\section{Conclusion}	
Two novel algorithms, MechSOUL and $L1$-MechSOUL have been proposed for high-accuracy lateral displacement estimation in ultrasonic strain imaging. MechSOUL and $L1$-MechSOUL, respectively, optimize $L2$- and $L1$-norm-based cost functions featuring mechanical as well as data similarity and spatial continuity constraints. The main contribution of the proposed techniques is emphasizing the EPR-inspired sample-wise mechanical congruence between the axial and lateral components of the strain tensor. Integrated optimization of mechanical and data fidelities leads to dramatic improvements of the lateral strain and EPR image quality, as demonstrated in the \textit{in silico}, phantom, and \textit{in vivo} experiments conducted in this study.                            

\section*{Acknowledgment}
This work is funded in part by the Natural Sciences and Engineering Research Council of Canada (NSERC). Md Ashikuzzaman holds PBEEE and B2X Doctoral Research Fellowships granted by the Fonds de Recherche du Québec - Nature et Technologies (FRQNT). The purchase of the Alpinion ultrasound machine was partly funded by Dr. Louis G. Johnson Foundation. The authors thank Drs. E. Boctor, M. Choti, and G. Hager for allowing them to use the liver datasets and the anonymous reviewers for their constructive comments.

\balance
\bibliographystyle{IEEEtran}
\bibliography{ref}

\end{document}